\documentclass[11pt,a4paper,openright]{report}
\usepackage{amsmath,amsthm}
\usepackage{latexsym}
\usepackage{amssymb}
\usepackage{bbold}
\usepackage{citesort}
\usepackage{epsfig}

\def\si{\sigma}
\def\de{\delta}
\def\De{\Delta}
\def\na{\nabla}
\def\la{\langle}
\def\ra{\rangle}
\def\pa{\partial}
\def\fr{\frac}
\def\th{\theta}
\def\al{\alpha}
\def\sgn{{\textrm{sgn}}}

\begin{document}
\hyphenation{wave-func-tion-al}
\setlength\arraycolsep{2pt}

\begin{titlepage}
\renewcommand{\baselinestretch}{1.1}
\large
\begin{center}
\mbox{}\\[-2cm]
\unitlength 1mm

\epsfysize 3cm \epsfclipon\epsffile{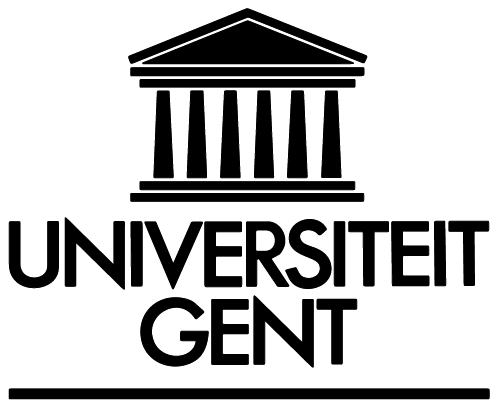}\\
{\large
Universiteit Gent\\
Faculteit Wetenschappen\\
Vakgroep Subatomaire en Stralingsfysica
}\\\vfill
\parbox{14 cm}{
{\huge\bfseries
\begin{center}
The de Broglie-Bohm pilot-wave interpretation of quantum theory
\end{center}
}
}\\\vfill
Ward Struyve\\[4.3cm]

Promotor: Prof.\ dr.\ W.\ De Baere
\\\vfill
Proefschrift ingediend tot het behalen van de graad van\\
Doctor in de Wetenschappen: Natuurkunde\\[1cm]
Oktober 2004
\end{center}
\renewcommand{\baselinestretch}{1}
\end{titlepage}

\normalsize
\thispagestyle{empty}

\newpage

\noindent \\
\thispagestyle{empty}
\newpage
\bibliographystyle{unsrt}
\pagenumbering{roman}
\setcounter{page}{1}
\begin{center}
{\large {\bf Acknowledgements}}
\end{center}
\addcontentsline{toc}{chapter}{Acknowledgements}

\vspace{0.5cm}
\noindent
It is a pleasure to thank the many people who have contributed to this thesis.

In the first place I am grateful to my supervisor Willy de Baere and to the head of department Kris Heyde, who gave me the opportunity to perform research in the fascinating domain of foundations of quantum theory.

I am also grateful to the colleagues in the department of Subatomic and Radiation Physics for their fellowship and aid. In particular I want to thank Stijn De Weirdt for the countless discussions from which I benefited a lot.

Many thanks also to Partha Ghose for the numerous useful discussions and for the hospitality I enjoyed from him and his family during my stay in London. I am grateful to Antony Short for the clarifying discussions on Popper's experiment, to Peter Holland who raised some valuable remarks on the issue of boson trajectories and to Samuel Colin for the stimulating discussions on quantum field theory. 

I want to thank the Perimeter Institute (Canada) for the kind hospitality and the financial support enjoyed during two visits. It was a pleasure to interact with the people there. Especially I want to thank Antony Valentini for the invitations and for the enjoyable and enriching discussions. I have also benefited a lot from the discussions with Hans Westman.

I also want to thank Jean Bricmont, Willy de Baere, Thomas Durt, Kris Heyde, George Horton, Stijn De Weirdt for a careful reading of the thesis and the many valuable suggestions for improvement. 

Finally, I want to express my loving thanks to my wife Fanny for her love, support and patience.

\thispagestyle{empty}
\newpage
\noindent \\

\newpage
\addcontentsline{toc}{chapter}{Table of contents}
\tableofcontents 

\newpage
\noindent \\

\newpage
\pagenumbering{arabic}
\setcounter{page}{1}

\chapter{Introduction}\label{chapter1}
Despite the unsurpassed predictive success of quantum theory, there is, since its inception almost 80 years ago, a persistent problem with its conventional interpretation, namely the {\em measurement problem}. 

The problem arises as follows. Quantum theory was developed in order to explain the behavior of `microscopic' systems. With each microscopic system, quantum theory associates a wavefunction $\psi$. According to the conventional interpretation of quantum theory,{\footnote{With the `conventional interpretation' we mean the Dirac-von Neumann approach \cite{vonneumann54,dirac74} which can be found in most standard textbooks.}} this wavefunction provides the most complete specification of the microscopic system. Further, the dynamics of the wavefunction $\psi$ is governed by two different laws. First, there is the dynamical evolution according to the Schr\"odinger equation, which is deterministic. Given the initial wavefunction one can uniquely determine the wavefunction at a later time. There is also another type of evolution of the wavefunction, which is the {\em collapse} of the wavefunction. The collapse rule is introduced in quantum theory in order to explain the definite outcome that is obtained when a measurement is performed. In this respect, the collapse of the wavefunction is said to occur when a measurement is performed by a `macroscopic observer' (human or not) on the `microscopic system' described by this wavefunction. The result is a replacement of the wavefunction $\psi$ by another wavefunction which from that time on provides the (complete) description of the microscopic system. Contrary to the dynamical law given by the Schr\"odinger equation, the collapse law is not deterministic. 

Considered separately both laws of dynamical evolution are unambiguously defined. On the other hand it is unclear what exactly is meant by a `microscopic' and a `macroscopic' systems, or what exactly is meant by an `observer' and a `system'. Hence it is unclear when the wavefunction evolves according to which of the two dynamical laws. The ambiguity becomes most striking in the following example. When a macroscopic observer performs a measurement on a microscopic system the collapse law should apply. But if the macroscopic system is regarded as a collection of microscopic systems, then the wavefunction of the total system, which consists of the observer and the system under observation, should evolve in time according to the Schr\"odinger equation and the collapse law should not be invoked. It is obvious, that these two ways of describing the measurement process are mutually incompatible if the wavefunction is to be regarded as the most complete specification of the system.

In practical situations the difference between the `macroscopic observer' and the `microscopic system' is of course sufficiently large so that one can often say with certainty whether or not the collapse has occurred. Nevertheless the ambiguous distinction between the `macroscopic observer' and the `microscopic system' presents an obvious logical flaw which is intolerable if one wants to regard quantum theory as a fundamental theory describing Nature. Because the ambiguous distinction is needed for the collapse law, and because the collapse law is invoked to describe the measurement process, the problem is generally referred to as the {\em measurement problem}.  

A possible resolution for the measurement problem resides in the view that the complete specification of a microscopic system is not only provided by the wavefunction, but also by some extra variables.{\footnote{Of course this is not the only way in which the measurement problem can be solved. One could for example also adopt an approach in which the wavefunction is dismissed altogether in the description of a quantum system, or an approach where the wavefunction still gives the complete description of a quantum system, but where the Schr\"odinger equation is modified, as in spontaneous collapse models (for a review see \cite{bassi03}). However, such theories will not be dealt with in the thesis.}} These extra variables should have an objective existence, irrespective of the fact whether or not a measurement is performed. They should also determine the outcome in experiments, so that the collapse law becomes superfluous. There is then no distinction needed between microscopic and macroscopic systems; both are described by these extra variables together with the wavefunction. A theory in which the system is described by such additional variables is called a {\em realistic theory}. If this realistic theory accounts for the same empirical predictions as quantum theory, it is also called an {\em interpretation} of quantum theory. The extra variables are usually termed {\em hidden variables}. However, because the reason for introducing these hidden variables is usually to give a definite account for the outcome in experiments, the term `hidden variables' is a kind of misnomer. For this reason Bell preferred to term these extra variables as {\em beables} \cite{bell87}. This is a term which we shall use frequently further on. 

An example of such a theory was presented by Louis de Broglie in 1927 at the Solvay Congress in Brussels (cf.\ \cite{debroglie60}, and references therein). De Broglie called his theory the {\em pilot-wave theory}. In fact, de Broglie regarded his pilot-wave theory only as a truncated version of his {\em theory of the double solution} which he had been working on since 1923, the time he proposed the idea of associating wave properties to massive particles, the key idea which led to quantum theory. However, because of the unfavorable reception of his pilot-wave approach at the Solvay Congress and because of objections raised by Pauli, de Broglie abandoned his ideas. It was only after David Bohm \cite{bohm1,bohm2} reinvented the ideas of pilot-wave theory in 1952 (although from a different perspective) that de Broglie returned to his original ideas and that he was able to answer Pauli's criticism. 

In de Broglie's pilot-wave theory the description of a quantum system by means of the wavefunction is extended by considering point particles which follow definite trajectories. The velocity field of these particles is fully determined by the wavefunction. Given the initial positions of the particles, their trajectories are fully determined by this velocity field. In this sense the particles are `piloted' by the wavefunction, hence the name pilot-wave theory. With an ensemble of quantum systems (all described by the same wavefunction) there corresponds a distribution of the actual positions of the particles. With a particular assumption on the initial distribution (i.e.\ the particles should initially be distributed according to the quantum distribution), pilot-wave theory reproduces the quantum probabilities for the ensemble. Hence, with this assumption pilot-wave theory can be considered as an interpretation of quantum theory. The hidden variables or the beables in pilot-wave theory are the particle positions.

In order to make the distinction between the notion of a particle in the standard interpretation of quantum theory and the notion of a particle in the theory of de Broglie and Bohm, we will term the latter the {\em particle beable}. Note that Bell himself used the term `beables' to refer to the particle positions instead of to the particles themselves \cite{bell87}, however in the literature the notion of beable is often extended to cover both interpretations.

Not only does the pilot-wave theory provide us with a logically unambiguous theory because it is devoid of the measurement problem, it also provides a clear picture of what the theory is about \cite{durr95}. The standard interpretation is not so clear about what physical entities are associated with the mathematics. The standard interpretation is certainly not about particles, because the most complete description is given by the wavefunction. Pilot-wave theory is unambiguous in this respect. In pilot-wave theory, matter is built of point-particles, the particle beables, moving in three dimensional `physical' space and these particles are causally influenced by the wavefunction which is grounded in configuration space. 

To give credit to both of its inventors we will term the theory of de Broglie and Bohm, the {\em de Broglie-Bohm pilot-wave theory} or for short the {\em pilot-wave theory} \cite{bell87,valentini92,valentini04}. In the literature other names for the theory can be found. Although these different names may also carry nuances in the interpretation of the theory, the basic mathematical structure is the same. For example Bohm and Hiley \cite{bohm1,bohm2,bohm5} and Holland \cite{holland} refer to the theory as the {\em ontological} or {\em causal interpretation}. The group around D\"urr, Goldstein and Zangh\`i prefers to call the theory {\em Bohmian mechanics}. 

The principles of the de Broglie-Bohm pilot-wave formalism are most easily sketched in the case of non-relativistic quantum theory. We do this in the next section. In the following section we then discuss in detail how the pilot-wave interpretation solves the measurement problem. We end the introductory chapter with an outline of the thesis.

\section{The pilot-wave interpretation}\label{principlespwinterpretation}
The standard quantum mechanical description of a system of $N$ spinless particles is given by means of a wavefunction $\psi({\bf x}_1, \dots,{\bf x}_N,t)$ in configuration space ${\mathbb R}^{3N}$, which satisfies the non-relativistic Schr\"odinger equation
\begin{equation}
i \hbar \frac{\partial \psi({\bf x}_1, \dots,{\bf x}_N,t)}{\pa t} =  \left( -\sum^N_{k=1} \frac{\hbar^2 \nabla^2_k}{2m_k} + V \right)  \psi({\bf x}_1, \dots,{\bf x}_N,t),
\label{h0.1}
\end{equation}
with $m_k$ the mass of the $k^{{\textrm{th}}}$ particle and $V({\bf x}_1, \dots,{\bf x}_N)$ a potential. 

In the standard quantum interpretation the wavefunction is used to calculate detection probability distributions for observables. In particular, the probability density to make a joint detection of the $N$ particles at the configuration $({\bf x} _1 ,\dots,{\bf x}_N)$ at a particular time $t$ is given by $|\psi ({\bf x}_1, \dots ,{\bf x}_N,t)|^2$. The continuity equation for this distribution is given by 
\begin{equation}
\frac{\partial |\psi|^2}{\partial t} + \sum^N_{k=1} {\boldsymbol{\nabla}}_k \cdot {\bf{j}}_k  =0,
\label{h0.3}
\end{equation}
with 
\begin{equation}
{\bf{j}}_k = {\bf{v}}_k |\psi|^2
\label{h0.4.01}
\end{equation} 
the 3-vector probability current and 
\begin{equation}
{\bf{v}}_k = \frac{\hbar}{2im_k|\psi|^2} \left( \psi^*  {\boldsymbol{\nabla }}_k \psi -\psi  {\boldsymbol{\nabla }}_k \psi^* \right)  = \frac{\hbar}{m_k} \textrm{Im} \fr{  {\boldsymbol{\nabla}}_k \psi }{\psi}. 
\label{h0.4}
\end{equation} 
The continuity equation expresses the conservation of the detection probability distribution $P$. 

In the pilot-wave interpretation \cite{bohm1,debroglie60}, the $N$-particle wavefunction $\psi$ is not regarded as providing the complete description of a quantum system. One also assumes the existence of $N$ point particles (the particle beables) which have definite positions at all times in physical space ${\mathbb R}^{3}$. If we represent the position of the $k^{{\textrm{th}}}$ particle beable with the 3-vector ${\bf X}_k$, then the trajectories ${\bf X}_k(t)$ are solutions to the differential equations
\begin{eqnarray} 
\frac{{d} {\bf X}_k}{{d} t} &=& {\bf{v}}_k({\bf X}_1, \dots ,{\bf X}_N,t) \nonumber\\
&=& \frac{\hbar}{m_k} \textrm{Im} \fr{ {\boldsymbol{\na}}_k \psi ({\bf x}_1, \dots ,{\bf x}_N,t)}{\psi({\bf x}_1, \dots ,{\bf x}_N,t)} \Big|_{{\bf x}_j={\bf X}_j}.
\label{h0.5}
\end{eqnarray}
In this way the wavefunction acts as a guiding wave, the {\em pilot-wave}, which governs the motion of the particle beables; there is no back-reaction of the particles onto the wavefunction. The equations ({\ref{h0.5}}) are called the {\em guidance equations}.{\footnote{In fact Bohm presented the pilot wave interpretation as a second order formalism \cite{bohm1,bohm2}. This involved a Newtonian-like force law for the particle beable, including an extra potential, the {\em quantum potential}. The reason for Bohm's preference for this second order formalism rests in his observation that the Schr\"odinger equation could be written, by separation of real and imaginary parts, as a Hamilton-Jacobi-like equation together with the continuity equation. In this thesis we adopt the view held by de Broglie \cite{debroglie60}, in which the guidance law (\ref{h0.5}) is regarded as the fundamental dynamical equation for the particle beables. Amongst the main advocates of this view are Bell \cite{bell87}, D\"urr {\em et al.} \cite{durr92} and Valentini \cite{valentini92,valentini04}. Nevertheless, because of the close connection to the classical Hamilton-Jacobi formulation, the second order formulation may be a valuable aid in the study of the emergence of classical mechanics out of quantum theory \cite{valentini92,valentini04,holland,allori01}.}}

According to the pilot-wave interpretation, it is the position of the particle beable that is revealed when a position measurement is performed. In the following section, where we deal with the pilot-wave description of a measurement process, we consider this more carefully.

If we now consider an ensemble of $N$-particle systems, all described by the same wavefunction $\psi$, then this ensemble determines a probability distribution $\rho({\bf X}_1, \dots,{\bf X}_N,t)$ of the actual position vectors of the $N$ particle beables. Because the motion of the particle beables is governed by the guidance equations (\ref{h0.5}), their distribution $\rho$ satisfies the same continuity equation as the quantum mechanical probability density $|\psi|^2$. Therefore, if the densities $\rho$ and $ |\psi|^2$ are equal at a certain time $t_0$, i.e.\
\begin{equation}
\rho({\bf x}_1, \dots,{\bf x}_N,t_0) = |\psi ({\bf x}_1, \dots,{\bf x}_N,t_0)|^2, 
\label{h0.501}
\end{equation}
then the equality will hold for all times $t$, i.e.\
\begin{equation}
\rho({\bf x}_1, \dots,{\bf x}_N,t) =  |\psi ({\bf x}_1, \dots,{\bf x}_N,t)|^2.
\label{h0.502}
\end{equation}
In the pilot-wave interpretation one assumes that initially, before a measurement is performed, the distribution of the particle beables $\rho$ (over the ensemble) is given by the quantum mechanical distribution $|\psi|^2$ (this assumption is also called the {\em quantum equilibrium hypothesis} \cite{durr92} and the distribution $\rho=|\psi|^2$ is called the {\em equilibrium distribution} \cite{valentini912,valentini92,valentini04,durr92}). The densities will then remain equal during the experiment, and pilot-wave theory and standard quantum mechanics will predict the same detection probabilities for the particle positions. 

Because most quantum measurements boil down to position measurements, pilot-wave theory and standard quantum mechanics will in general yield the same detection probabilities. The situation is different, if one considers for example measurements involving time related quantities, such as time of arrival, tunneling times etc. Pilot-wave theory makes unambiguous predictions for such measurements, but in conventional quantum theory there is no consensus about what these quantities should be (see e.g.\ \cite{muga02} and references therein).  

The quantum equilibrium hypothesis is introduced here to match the empirical distributions predicted by quantum theory. However, there exist some possible justifications for the quantum equilibrium hypothesis. One possible justification was presented by D\"urr {et al.}\ \cite{durr92}. But it would take to far to repeat their analysis here. Another possible justification was presented by Valentini \cite{valentini912,valentini92,valentini96,valentini04}. By a sub-quantum $H$-theorem Valentini was able to show that in reasonable circumstances a non-equilibrium distribution $\rho \neq |\psi|^2$ for the particle beables (all guided by the same wavefunction) may approach the equilibrium distribution on a certain coarse grained level.{\footnote{Recently this was illustrated by numerical simulations \cite{valentini042}.}} This suggest that quantum theory can be regarded as an equilibrium theory. Although this yields an interesting research program, we will not consider the possibility of non-equilibrium in this thesis. Our main goal is rather to study in how far a pilot-wave interpretation is possible to cover other domains in quantum theory.

We also want to note that there were recent claims by Ghose \cite{ghose1,ghose2,ghose4,ghose02}, and which were later adopted by Golshani and Akhavan \cite{golshani1,golshani2,golshani3,golshani012,akhavan03,akhavan04}, that standard quantum mechanics and pilot-wave theory would predict incompatible results for some specific experiments. However, we argued elsewhere that these claims are flawed \cite{struyve01,struyve031}. We indicated that a non-equilibrium density for the particle beables is implicitly assumed from the outset. The density of the particle beables then remains in non-equilibrium during the experiment and hence it is obvious that one arrives at incompatible predictions for the two theories. Despite our comment (and that of others \cite{marchildon2,guay03,nikolic03}) the experiment proposed by Ghose was recently performed by Brida {\em et al.} \cite{brida02,brida03,genovese04}. The result of the experiment was that standard quantum theory was confirmed (as expected). A correct analysis of the experiment in terms of pilot-wave theory would have led to the same predictions as standard quantum theory.

Pilot-wave theory has many features which are not present in quantum theory. The most striking property of pilot-wave theory is that it is nonlocal. In fact, as shown by Bell, any realistic theory which leads to the same statistical predictions as standard quantum theory must be nonlocal \cite{bell64}. Yet, quantum theory remains local in the sense that one cannot use quantum theory to send faster than light signals.{\footnote{We can state this more correctly as follows. As shown by Ghirardi {\em et al.} \cite{ghirardi80}, the standard quantum theory of measurement cannot be used for superluminal transmission of signals. Hence, if the velocity of probability flow does not exceed the speed of light, which is for example always the case for the relativistic theory spin-1/2 of Dirac (the velocities along the flowlines of the particle probability density are bounded by the speed of light), then quantum theory does not allow faster than light signals. In non-relativistic quantum theory there is in fact no restriction on the velocity of probability flow. But of course the domain of applicability of non-relativistic quantum theory is limited to `low' speeds.}} In the pilot-wave interpretation, the nonlocality manifests itself by the fact that the position of one-particle beable may depend on the positions of other particle beables (by the guidance law (\ref{h0.5})). This dependence is instantaneous no matter how far the other particle beables may be located. It is also important to note that this nonlocality is not a consequence of dealing with a non-relativistic description of quantum phenomena. The nonlocality of pilot-wave theory is inescapable, even at the level of relativistic quantum theory. A very illustrative example of the nonlocality present in pilot-wave theory was given by Rice \cite{rice96}.

In conclusion, we arrive at pilot-wave theory only by a minor shift in interpretation (although with far reaching consequences). Instead of interpreting 
\begin{equation}
|\psi({\bf x}_1, \dots,{\bf x}_N,t )|^2{d}^3 x_1 \dots {d}^3 x_N
\label{h0.6}
\end{equation}
as the probability of {\em finding} the particles in a volume element ${d}^3 x_1 \dots {d}^3 x_N$ around the configuration $({\bf x}_1, \dots,{\bf x}_N )$, at a certain time $t$, as in the standard interpretation of quantum mechanics, we interpret it in pilot-wave theory as the probability of the particles {\em being} in a volume element ${d}^3 x_1 \dots {d}^3 x_N$ around the configuration $({\bf x}_1, \dots,{\bf x}_N )$ at the time $t$. Inspired by the analogy with the continuity equation in hydrodynamics, we derive the velocity field (\ref{h0.4}) for the particle beables from the quantum continuity equation (\ref{h0.3}) for the density $|\psi|^2$. This is the scheme that we will adopt in the rest of the thesis. When constructing a pilot-wave theory for quantum theory we shall try to identify a continuity equation that can be seen as a conservation equation for a density, be it a density of particles or fields,{\footnote{We will argue that a field ontology seems preferred over a particle ontology in a pilot-wave interpretation for quantum field theory. Instead of introducing {\em particle} beables we will then introduce {\em field} beables.}} and then from this continuity equation we shall try to construct a guidance equation for the particles or fields.

\section{The measurement process}\label{themeasurementprocess}
In this section we describe the standard quantum mechanical measurement process in terms of the pilot-wave interpretation, along the lines of the presentations that can be found in \cite{holland,valentini92,bohm5}. In the pilot-wave interpretation, the measurement process is treated just as any other quantum process; there is no privileged role for the observer or measurement apparatus. 

Suppose a system which is described by the wavefunction $\psi^{(s)}$. The system may consist of $N$ particles, so that the wavefunction lives in $3N$-dimensional configuration space. There also correspond $N$ particle beables with the system, the positions of which we denote by the $3N$-dimensional vector $x^{(s)}$. Suppose similarly an apparatus with wavefunction $\psi^{(a)}$ and a collection of particle beables at the configuration $x^{(a)}$. The apparatus is introduced to measure some property of the system.  In the standard interpretation of quantum theory this property is represented by an operator ${\widehat A}$ and the possible outcomes of the measurement correspond to the eigenvalues of this operator. 

Initially the system under observation and the measurement apparatus have not interacted yet, so that they may be described by the product wavefunction $\psi^{(s)}\psi^{(a)}$. As time evolves the system under observation gets coupled to the apparatus and the total wavefunction $\psi^{(s)}\psi^{(a)}$ evolves to the {\em entangled} wavefunction $\sum_i \psi^{(s)}_i \psi^{(a)}_i$, where the states $\psi^{(s)}_i$ are eigenstates of the operator ${\widehat A}$. This evolution is determined by the Schr\"odinger equation, which should contain a particular interaction Hamiltonian depending on the observable that is being measured (e.g.\ this interaction Hamiltonian could be a von Neumann type of interaction Hamiltonian). The states $\psi^{(a)}_i$ are assumed to be non-overlapping in configuration space, i.e.\ $\psi^{(a)}_i\psi^{(a)}_j \equiv 0$ for $i \neq j$.{\footnote{Note that the condition that the states are non-overlapping is stronger than the condition that they are orthogonal. If states are non-overlapping they are orthogonal, but not vice versa. For example different plane waves are orthogonal but are overlapping.} In fact it is sufficient that the overlap of the states is minimal. This property is generally satisfied in an ordinary measurement. We will present the reason for this below.  

Now if the apparatus would be known to be in one particular state, say $\psi^{(a)}_k$, then the system under observation would be in the state $\psi^{(s)}_k$. In standard quantum theory, the collapse rule is introduced to reduce the state of the total state $\sum_i \psi^{(s)}_i \psi^{(a)}_i$ to the state $ N\psi^{(s)}_k \psi^{(a)}_k$ (with $N$ some normalization factor). The result of the measurement is then the eigenvalue of ${\widehat A}$ corresponding to the eigenstate $\psi^{(s)}_k$.  

In fact, the collapse law does not have to be invoked at this stage yet. A second apparatus may also be introduced, which measures the first apparatus, and so on. This chain of apparatuses getting correlated may then for example be ended by a final observer for which the collapse law may be invoked. As explained in the introduction, the point where the collapse law should apply is not well defined and presents the core of the measurement problem. 

By contrast the pilot-wave description of the measurement process is unambiguous. Because the different terms $\psi^{(s)}_i \psi^{(a)}_i$ are non-overlapping, they can be seen as defining `channels' in configuration space, the channels being the non-overlapping supports of the different terms $\psi^{(s)}_i \psi^{(a)}_i$. The configuration $(x^{(s)},x^{(a)})$, which has the positions of the particle beables as components, enters one of the channels during the interaction.  Suppose the configuration has entered the channel corresponding to $\psi^{(s)}_k \psi^{(a)}_k$. The other terms $\psi^{(s)}_i \psi^{(a)}_i$ with $i \neq k$ are called the {\em empty waves}. If the different terms  $\psi^{(s)}_i \psi^{(a)}_i$ do not overlap again at a later time (this is in principle accomplished by coupling the system with a large number of particles, which leads to decoherence, so that the probability of re-overlap becomes minimal), then as far as the particle beables are concerned, the empty wavepackets have no further influence on the motion of the beables $(x^{(s)},x^{(a)})$ and hence these wavepackets may then be dismissed in the future description of the particle beables. This corresponds to the collapse of the wavefunction in the standard interpretation of quantum mechanics. 

Because we further assume the beables to be distributed according to the quantum distribution, one can easily verify that the probability for the particle beables to enter the channel corresponding to $\psi^{(s)}_k \psi^{(a)}_k$ is given by 
\begin{equation}
\left\langle \psi^{(s)}_k \psi^{(a)}_k  \Bigg| \sum_i \psi^{(s)}_i \psi^{(a)}_i \right\rangle = {\int{ d\Omega {\left|\psi^{(s)}_k \psi^{(a)}_k\right|^2}}} ,
\label{h0.7}
\end{equation}
where the integral on the right hand side ranges over the whole configuration space ($d\Omega$ is the measure on the configuration space). Hence we recover the quantum probabilities in the pilot-wave description of the measurement process. 

So we can describe the measurement process in the context of pilot-wave theory. Essential in our treatment is that at some stage in the measurement chain, the wavefunctions of the apparatus are non-overlapping in configuration space, because this allows us to dismiss the empty wavepackets. This situation is obtained in an ordinary measurement. For example the different states could correspond to macroscopic needles pointing in different directions and one can easily convince oneself that these states are non-overlapping in configuration space. The reason why the total wavefunction of system and apparatus actually evolves to such a superposition is a purely quantum mechanical one.  

\section{Summary and organization of the thesis}
We have seen how we can give a pilot-wave description of a spinless, non-relativistic quantum system. In the main part of the thesis we will study how this pilot-wave interpretation can be extended to cover other domains in quantum theory. We will successively consider the domain of non-relativistic quantum theory, relativistic quantum theory and quantum field theory. More advanced domains, such as quantum gravity and string theory, are not considered. 

The extension of the pilot-wave formulation a spinless, non-relativistic quantum system to include spin does not present any difficulties. We deal with this extension in Chapter \ref{chapter2}. 

The construction of a pilot-wave formulation for relativistic quantum theory is more problematic. In Chapter \ref{chapter3}, we will consider relativistic wave equations and we will consider the question to which extent it is possible to construct a pilot-wave model for these relativistic wave equations. It will turn out that a pilot-wave interpretation with point particle as beables, analogous to the pilot-wave interpretation for non-relativistic quantum mechanics, is in general impossible. The reason is that, already at the standard quantum mechanical level, a particle interpretation in analogy with the one for non-relativistic quantum mechanics is in general not possible. 

It seems that only for the Dirac theory for spin-1/2, and under restricted circumstances, such a quantum mechanical particle interpretation may be provided. For example for sufficiently low energies a one-particle interpretation is possible. This is because there exists a positive density (proportional to the charge density) which is the time component of a future-causal four-vector and which can hence be interpreted as a probability density. For higher energies and if only electromagnetic interaction is considered, one can maintain the particle interpretation, albeit a many-particle one, by the introduction of a Dirac sea. For other types of interaction, such as weak interaction, it is unknown how to extend the notion of a Dirac sea and hence it is unknown how to continue the particle approach. In the domain where the quantum mechanical particle interpretation is applicable for the Dirac theory, a pilot-wave interpretation can be devised. This was already clear to Bohm, who originated the pilot-wave interpretation for the Dirac theory. 

After reviewing Bohm's pilot-wave interpretation for the Dirac theory, we consider the alleged pilot-wave models for the Duffin-Kemmer-Petiau (DKP) theory and the Harish-Chandra (HC) theory, which were initiated by Ghose {\em et al.} The DKP wave equation is a first-order relativistic equation for massive spin-0 and spin-1, but is nevertheless completely equivalent to the second order Klein-Gordon equation in the spin-0 representation and to the Proca equations in the spin-1 representation. The HC equation is the massless counterpart of the DKP equation, which is equivalent to the massless Klein-Gordon equation in the spin-0 representation, and to Maxwell's equations for the electromagnetic field in the spin-1 representation. As is well known there is no {\em quantum mechanical particle interpretation} for these wave equations because of the lack of a conserved, future-causal current (contrary to the Dirac theory, the charge current is not always future-causal for spin-0 and spin-1 bosons). In fact there is not even a {\em quantum mechanical interpretation}, because the lack of a positive definite inner product blocks the setup of a Hilbert space (again this is related to the fact that the charge currents for both spin-0 and spin-1 are not always future-causal). Ghose {\em et al.}\ tried to give a quantum mechanical particle interpretation by constructing a conserved, future-causal current from the energy--momentum tensor. With this quantum mechanical particle interpretation they could also as associate a pilot-wave model. The resulting equations look very similar to the ones for the Dirac theory. However, despite this similarity, we show that the suggested quantum mechanical particle interpretation, and hence also the associated pilot-wave model, suffer from some problems, which make this approach {\em in general} untenable. 

Although the pilot-wave model suggested by Ghose {\em et al.}\ can not be treated as valid model describing physical reality, we think that the model still has value as an illustrative model. For this reason, we further consider the extension of the model to many particles. 

Of course, as is well known, relativistic wave equations are not suitable to describe high energy quantum systems. The theory describing high energy quantum systems is quantum field theory. Quantum field theory is not about {\em particles} in physical 3-space, as was non-relativistic quantum theory; strong localization of particles in physical 3-space leads to problems with causality (i.e.\ superluminal spread of localized states \cite{hegerfeldt74,hegerfeldt80,hegerfeldt85}). Instead quantum field theory can be seen as describing {\em fields} in physical 3-space.{\footnote{Only in momentum space, when using Fock space, one can recover the notion of particles.}} This is clear in the functional Schr\"odinger picture, where the quantum states are described by wavefunctionals, which are defined on a configuration space of fields. Instead of searching for a pilot-wave interpretation for relativistic quantum theory in terms of particle beables, we will therefore consider the possibility of a pilot-wave interpretation in terms of field beables, a view strongly supported by Valentini. 

It will appear that the construction of a pilot-wave theory in terms of field beables presents no difficulty in the bosonic case. This will be illustrated in Chapter \ref{chapter4} with a discussion on the construction of a pilot-wave theory for the massive spin-0 field, the massive spin-1 field, the electromagnetic field and then also for the massive spin-0 field coupled to the electromagnetic field (i.e.\ scalar quantum electrodynamics). In particular we will discuss in detail the two existing models for the electromagnetic field, namely the one by Bohm and Kaloyerou and the one by Valentini. The main difference between the two models is that Bohm and Kaloyerou only introduce beables for gauge independent variables, whereas Valentini also introduces beables for gauge variables. We will show that the guidance equations for the beables corresponding to the gauge variables are rather meaningless, because they only express the fact that these beables are stationary. In addition, inclusion of these beables for gauge dependent variables also makes that the densities of field beables are non-normalizable. In order to avoid this problem, we think it is preferable to adopt the approach by Bohm and Kaloyerou.

A pilot-wave interpretation in terms of field beables for fermionic field theory seems less straightforward. In Chapter \ref{chapter5} we reconsider the idea of Valentini to construct a pilot-wave interpretation in terms of field beables, with the field beables being elements of the Grassmann algebra. This approach looks very promising at first sight, because there exists a functional Schr\"odinger picture for fermionic fields in terms of Grassmann variables. However, closer inspection reveals that it is not possible to associate a pilot-wave interpretation with it.

Hence for fermionic field theory a different approach should be taken. There do exist different approaches. There is for example the pilot-wave approach by Holland and the Bell-type model by D\"urr, Goldstein, Tumulka and Zangh\`\i. The Bell-type model differs from pilot-wave models in the fact that it involves an element of stochasticity. Although these models look very interesting, we do not consider them in detail in the thesis.

It is important to note that when the field approach is taken as fundamental, which we do in this thesis, then this approach is incompatible with the particle approach which was so successful for non-relativistic quantum theory. The field beables that are introduced in field theory do not reside into the particle beables in the non-relativistic limit. Hence, also in the non-relativistic case the actual beables should be regarded to be fields.

Nevertheless, the pilot-wave interpretation in terms of particle beables may serve well for illustrative purposes. An example of this is given in Chapter \ref{chapter6}. There we show that the pilot-wave interpretation in terms of particle beables may serve as a theoretical underpinning for otherwise rather ad hoc trajectories that are used for describing some experiments concerning optical imaging.

\chapter{Particle beables for non-relativistic quantum mechanics}\label{chapter2}
\section{Non-relativistic quantum mechanics}
In the preceding chapter, the pilot-wave interpretation for a system consisting of non-relativistic spinless particles was introduced. In this section we consider the extension of this pilot-wave interpretation to include spin. We only consider the one-particle case. The extension to many particles is straightforward. We consider charged particles which move under the influence of an external electromagnetic field. 

In the standard interpretation of quantum mechanics, a non-relativistic particle with spin $s$ is described by means of a $(2s+1)$-component wavefunction $\psi$, who's index transforms according to the $(2s+1)$-dimensional representation of the rotation group. The wavefunction $\psi$ satisfies the wave equation (cf.\ \cite[p.\ 471]{landau58} and \cite{hurley71}){\footnote{In this chapter, and in subsequent chapters, we adopt the summation convention of Einstein.}}
\begin{equation}
i \hbar \frac{\partial \psi_{\alpha}({\bf x},t)}{\pa t} =  -\frac{\hbar^2 D^2}{2m}  \psi_{\alpha}({\bf x},t) - \frac{eg}{2mc} {\bf S}^{(s)}_{\alpha \beta} \cdot {\bf B}  \psi_{ \beta}({\bf x},t) +(eV_0+ V) \psi_{\alpha}({\bf x},t),
\label{h1.001}
\end{equation}
where $e$ is the charge of the particle. $V_i$ and $V_0$ are the electromagnetic potentials, with corresponding magnetic field ${\bf B} = {\boldsymbol \nabla} \times {\bf V}$ and $V$ denotes an additional scalar potential. $D_i = \partial_i - \frac{ ie}{\hbar c} V_i$ is the covariant derivative. The three $(2s +1)$-dimensional matrices $S^{(s)}_i$ ($i=1,2,3$) are the generators of the rotation group in the $(2s+1)$-dimensional representation. They satisfy the commutation relations
\begin{equation}
\left[S^{(s)}_i,S^{(s)}_j \right] = i \hbar \varepsilon_{ijk} S^{(s)}_k.
\label{h1.00001}
\end{equation}
The constant $g$ is the gyromagnetic factor. Hurley derived the wave equation under the assumptions of Galilean covariance and `minimality' \cite{hurley71}, and he found the gyromagnetic factor 
\begin{equation}
g = \left\{  \begin{array}{ll}
0 & \textrm{for spin }0\\
\frac{1}{s} & \textrm{for spin }s
\end{array} \right. .
\label{h1.002}
\end{equation}
This implies that the correct gyromagnetic factor for an elementary particle can be found even without considering relativistic wave equations (although the correct prediction of the gyromagnetic factor for the electron is generally regarded as a success of the Dirac equation). If the particle has an internal structure, then the gyromagnetic factor may of course differ from $1/s$.

The conservation equation for the particle detection probability density $\psi^{\dagger}  \psi$ reads
\begin{equation}
\frac{\partial \psi^{\dagger}  \psi }{\partial t} + {\boldsymbol \nabla} \cdot {\bf j}=0
\label{h1.003}
\end{equation}
with ${\bf j}$ the 3-vector probability current which can be written as the sum  
\begin{equation}
{\bf j}  = {\bf j}_c + {\bf j}_s,
\label{h1.004}
\end{equation}
of a current which formally resembles the conventional Schr\"odinger current
\begin{equation}
{\bf j}_c = \frac{\hbar}{2mi} \left( \psi^{\dagger} {\boldsymbol D} \psi - \left( {\boldsymbol D} \psi \right)^{\dagger} \psi  \right) =\frac{\hbar}{2mi} \left( \psi^{\dagger} {\boldsymbol \nabla} \psi - \left({\boldsymbol \nabla} \psi^{\dagger} \right)\psi  \right) - \frac{e}{mc}{\bf V}\psi^{\dagger}  \psi 
\label{h1.00401}
\end{equation}
and a `spin current'
\begin{equation}
{\bf j}_s = \frac{g}{2m}  {\boldsymbol \nabla} \times \left( \psi^{\dagger} {\bf  S}^{(s)} \psi \right).
\label{h1.005}
\end{equation}
With the introduction of the {\em spin 3-vector}
\begin{equation}
{\bf s} = \frac{\psi^{\dagger} {\bf S}^{(s)} \psi}{\psi^{\dagger}\psi}
\label{h1.00501} 
\end{equation}
and the {\em magnetic moment 3-vector}
\begin{equation}
{\bf m} = \frac{ge}{2mc} \psi^{\dagger}\psi {\bf s}   =  \frac{ge}{2mc} \psi^{\dagger} {\bf S}^{(s)} \psi,
\label{h1.00502}
\end{equation}
the spin term in the current can also be written as
\begin{equation}
{\bf j}_s = \frac{g}{2m} {\boldsymbol \nabla} \times \left( \psi^{\dagger}  \psi {\bf s} \right) =  \frac{c}{e} {\boldsymbol \nabla} \times {\bf m}.
\label{h1.00503}
\end{equation}
Within a factor $e$, corresponding to a change from the particle probability current to the charge current, the spin current ${\bf j}_s$ formally resembles a magnetization current for a classical polarized medium.

We can construct a pilot-wave interpretation by introducing a structureless particle (the particle beable), who's motion is governed by the wavefunction according to the guidance equation{\footnote{Note that in the previous chapter we used a different notation for the position vector of the particle beable and the argument of the wavefunction. Because this can in fact not lead to possible confusion, we use from now on the same notation for both.}}
\begin{equation}
\frac{{d} {\bf x}}{{d} t} = \frac{ {\bf j}}{ \psi^{\dagger}\psi }.  
\label{h1.008}
\end{equation}
For an ensemble of spin-$s$ particles, all described by the same wavefunction $\psi$, the equilibrium distribution for the particle beables is given by $\psi^{\dagger}\psi$.

\section{Examples}\label{examplespwinonrelativisticlimit}
We now consider some examples:\\

\noindent
{\bf Spin-0:} In the spin-0 case, the generators of the rotation group $S^{(0)}_i$ are zero and the wave equation (\ref{h1.001}) is simply the Schr\"odinger wave equation for a spinless particle
\begin{equation}
i \hbar \frac{\partial \psi({\bf x},t)}{\pa t} =  -\frac{\hbar^2 D^2}{2m}  \psi({\bf x},t)  + (eV_0+ V)  \psi({\bf x},t).
\label{h1.00801}
\end{equation}
The corresponding current is
\begin{equation}
{\bf j} = \frac{\hbar}{2mi} \left( \psi^{*} {\boldsymbol D} \psi - \left( {\boldsymbol D} \psi \right)^{*} \psi  \right) .
\label{h1.0080101}
\end{equation}
The pilot-wave interpretation is the one given originally by de Broglie and Bohm (cf. Section \ref{principlespwinterpretation}). In particular the guidance equation reads
\begin{equation}
\frac{{d} {\bf x}}{{d} t} =  \frac{\hbar}{2mi|\psi|^2} \left( \psi^{*} {\boldsymbol D} \psi - \left( {\boldsymbol D} \psi \right)^{*} \psi  \right) .
\label{h1.00802}
\end{equation}
\\

\noindent
{\bf Spin-1/2:} In the spin-$1/2$ case, the generators of the rotation group are proportional to the $2\times2$ Pauli matrices $\sigma_i$ ($i=1,2,3$), i.e.\ $S^{(1/2)}_i = \hbar \sigma_i/2$. The Schr\"odinger equation (\ref{h1.001}) for the two component wavefunction $\psi$ is then the Pauli equation
\begin{equation}
i \hbar \frac{\partial \psi_{\alpha}({\bf x},t)}{\pa t} =  -\frac{\hbar^2 D^2}{2m}  \psi_{\alpha}({\bf x},t) - \frac{eg\hbar}{4mc} {\boldsymbol \sigma }_{\alpha \beta} \cdot {\bf B}  \psi_{ \beta}({\bf x},t) + (eV_0+ V)  \psi_{\alpha}({\bf x},t).
\label{h1.00803}
\end{equation}
In the corresponding guidance law for the particle beable, there is a spin contribution arising from the nonzero spin term in the current 
\begin{equation}
{\bf j}_s= \frac{g\hbar}{4m}  {\boldsymbol \nabla} \times \left(\psi^{\dagger} {\boldsymbol \sigma} \psi\right),
\label{h1.00804}
\end{equation}
so that the total current reads
\begin{equation}
{\bf j}= \frac{\hbar}{2mi} \left( \psi^{\dagger} {\boldsymbol D} \psi - \left( {\boldsymbol D} \psi \right)^{\dagger} \psi  \right) + \frac{g\hbar}{4m}  {\boldsymbol \nabla} \times \left(\psi^{\dagger} {\boldsymbol \sigma} \psi\right).
\label{h1.00805}
\end{equation}
The corresponding guidance equation reads
\begin{equation}
\frac{{d} {\bf x}}{{d} t} =  \frac{\hbar}{2mi\psi^{\dagger}\psi} \left( \psi^{\dagger} {\boldsymbol D} \psi - \left( {\boldsymbol D} \psi \right)^{\dagger} \psi  \right)  + 
\frac{g\hbar}{4m\psi^{\dagger}\psi}  {\boldsymbol \nabla} \times \left(\psi^{\dagger} {\boldsymbol \sigma} \psi\right).
\label{h1.00806}
\end{equation}
The trajectories for a non-relativistic spin-1/2 particle (with the additional spin term) were recently studied for specific systems. Holland and Philippidis studied the particle paths for spin-1/2 eigenstate for the two slit experiment \cite{holland031}. Colijn and Vrscay studied the spin-1/2 particle paths for hydrogen eigenstates \cite{colijn02,colijn0302} and transitions between them \cite{colijn03}.\\

\noindent
{\bf Spin-1:} In the spin-$1$ case, the generators of the rotation group are the $3\times3$ matrices $( S^{(1)}_j)_{ik} = i \hbar \epsilon_{ijk}$. The Schr\"odinger equation (\ref{h1.001}) for the three component wavefunction $\psi$ is then the non-relativistic spin-1 equation 
\begin{equation}
i \hbar \frac{\partial \psi_{i}({\bf x},t)}{\pa t} =  -\frac{\hbar^2 D^2}{2m}  \psi_{i}({\bf x},t) - \frac{ieg\hbar}{2mc} \varepsilon_{ijk} B_j  \psi_{k}({\bf x},t) + (eV_0+ V)  \psi_{i}({\bf x},t),
\label{h1.00807}
\end{equation}
with $i=1,2,3$. In the corresponding guidance law for the particle beable, there is a spin contribution arising from the nonzero spin term in the current 
\begin{equation}
({\bf j}_s)_i= \frac{g}{2m} \big( {\boldsymbol \nabla} \times (\psi^{\dagger} {\bf    S}^{(1)} \psi )\big)_i = \frac{g\hbar}{2m}\textrm{Im} \big( \psi^{*}_i \partial_j  \psi_j -\psi^{*}_j  \partial_j \psi_i \big),
\label{h1.009}
\end{equation}
so that the total current reads
\begin{equation}
{j}_i= \frac{\hbar}{2mi} \left( \psi^{\dagger} D_i \psi - \left( D_i \psi \right)^{\dagger} \psi  \right) + \frac{g\hbar}{2m}\textrm{Im} \big( \psi^{*}_i \partial_j  \psi_j -\psi^{*}_j  \partial_j \psi_i \big).
\label{h1.0090101}
\end{equation}
The corresponding guidance equation reads
\begin{equation}
\frac{{d} {x}_i}{{d} t} =  \frac{\hbar}{2mi\psi^{\dagger}\psi } \left( \psi^{\dagger} D_i \psi - \left( D_i \psi \right)^{\dagger} \psi  \right)  + \frac{g\hbar}{2m\psi^{\dagger} \psi}\textrm{Im} \big( \psi^{*}_i \partial_j  \psi_j -\psi^{*}_j  \partial_j \psi_i \big).
\label{h1.00901}
\end{equation}

\section{Spin eigenstates}
The Schr\"odinger equation for a spinless particle (\ref{h1.00801}) can be derived from the wave equation (\ref{h1.001}) by considering a spin eigenstate. However the particle current for a spin eigenstate will in general not reduce to the current for a spinless particle (\ref{h1.0080101}); as pointed out before in the case of spin-1/2 \cite{hestenes79,holland99,holland032}, there will be an additional spin contribution to the current. 

Let us consider this in more detail. If we consider a particle with arbitrary spin $s$ for which the wavefunction is a spin eigenstate, then the wavefunction can be written as the product of a space dependent part and a spin dependent part, i.e.\
\begin{equation}
\psi_\alpha({\bf x},t) = \psi'({\bf x},t) \chi_\alpha, \qquad  \chi^{\dagger}_\alpha \chi_\alpha =1,
\label{h1.010}
\end{equation}
with $\psi'({\bf x},t)$ a scalar wavefunction. For vanishing electromagnetic potentials, $V_i=V_0=0$, it follows from the wave equation (\ref{h1.001}) that $\psi'({\bf x},t)$ satisfies the Schr\"odinger equation for a spinless particle. The corresponding particle current for the spin eigenstate reads 
\begin{equation}
{\bf j} =  \frac{\hbar}{2mi} \left( \psi'^{*} {\boldsymbol \nabla} \psi' - \left({\boldsymbol \nabla} \psi'^{*} \right) \psi'  \right) + \frac{g}{2m}{\boldsymbol \nabla} \times (|\psi'|^2{\bf s}) 
\label{h1.01101}
\end{equation}
and the corresponding guidance equation is
\begin{equation}
\frac{{d} {\bf x}}{ {d} t} =  \frac{\hbar}{2mi|\psi'|^2} \left( \psi'^{*} {\boldsymbol \nabla} \psi' - \left({\boldsymbol \nabla} \psi'^{*}  \right)\psi'  \right) + \frac{g}{2m|\psi'|^2} {\boldsymbol \nabla} \times (|\psi'|^2{\bf s}).
\label{h1.012}
\end{equation}
The spin vector ${\bf s}$ is now a constant vector, given by
\begin{equation}
{\bf s} = \chi^{\dagger} {\bf S}^{(s)} \chi .
\label{h1.0011}
\end{equation}
So, even though the spin dependent part $\chi$ can be factored out of the wave equation, leading to the Schr\"odinger equation (\ref{h1.00801}) for $\psi'$, it still appears non-trivially in the current and hence in the guidance equation. As a result, the non-relativistic description of a particle in a spin eigenstate with nonzero spinvector ${\bf s}$, should include the spin term in the current. This spin term potentially plays for example a role in time of arrival measurements (see below).

\section{Note on the uniqueness of the pilot-wave interpretation}
\subsection{On the uniqueness of the particle current and corresponding guidance equation}
Note that the current ${\bf j}$ is not uniquely determined by the continuity equation (\ref{h1.003}). It is determined only up to a divergenceless vector. For example, one can construct a new current ${\bar {\bf j}}$ by adding the divergenceless current ${\bf j}_a$ to the current ${\bf j}$. The newly defined current ${\bar {\bf j}} = {\bf j} + {\bf j}_a$ then also satisfies the continuity equation, with the same probability density $\psi^{\dagger}\psi$. Hence, we are left with an apparent ambiguity in the definition of the particle probability current. This ambiguity is unobservable when the quantum probabilities are derived solely from the density $\psi^{\dagger}\psi$ (such as spin measurements with a Stern-Gerlach setup), because this density is unaltered by the addition of ${\bf j}_a$ to the current. Nevertheless, the additional current may in principle lead to observable effects in measurements involving time (see below). In any case, the guidance equation in the pilot-wave interpretation is derived from the quantum mechanical particle current and the possibility of an additional contribution to the particle current leads to an ambiguity at the level of the pilot-wave interpretation. For example the spin current ${\bf j}_s$ is divergenceless and hence could in principle be relinquished in the definition of the current, and hence also in the definition of the guidance equation. 

Nevertheless, there exist some arguments in favor of (\ref{h1.004}) as the definition for the particle current.{\footnote{Appeal to Noether's theorem in order to derive the correct current is of no use in this case. Because if the charge current is derived as the conserved current corresponding to global phase invariance, then it is only determined up to a total divergence and hence Noether's theorem is not decisive in whether or not we should in include the spin term in the current.}} A first argument rests on the observation that it is the charge current $ej^\mu = e(c\psi^{\dagger}\psi,{\bf j})$ that couples to the electromagnetic field in the Maxwell equations $\partial_\mu F^{\mu \nu}=ej^\mu/c$, with $F^{\mu \nu}$ is the electromagnetic field tensor.{\footnote{The Maxwell equations $\partial_\mu F^{\mu \nu}=j^\mu/c$ can be derived from the fully coupled Lagrangian density 
\begin{equation}
{\mathcal L} = \frac{i\hbar}{2}\left(\psi^{\dagger}  \frac{\partial \psi}{\pa t} - \frac{\partial \psi^{\dagger}}{\pa t}\psi \right) - \frac{\hbar^2 }{2m} ({\bf D} \psi)^{\dagger} \cdot {\bf D} \psi + \frac{eg}{2mc}\psi^{\dagger}   {\bf S}^{(s)} \cdot {\bf B}  \psi - (eV_0+ V) \psi^{\dagger} \psi -\frac{1}{4}F_{\mu \nu}F^{\mu \nu}.
\label{h1.001101}
\end{equation}
}} Hence, if the particle current is chosen to be proportional to the charge current, then it should be given by (\ref{h1.004}). 

In the case of spin-1/2, Holland considered still other arguments to establish the uniqueness of the particle current \cite{holland99,holland032}. Holland first showed that the particle current in the relativistic spin-1/2 Dirac theory is unique (under reasonable assumptions). With the demand that the non-relativistic spin-1/2 particle current should be obtained by taking the non-relativistic limit of the Dirac current, this non-relativistic particle current is also unique. The resulting non-relativistic spin-1/2 current is the one that is presented in Section \ref{examplespwinonrelativisticlimit}. This choice for the non-relativistic current also implies that the pilot-wave interpretation that can be provided for the Dirac equation (see Section \ref{Spin-1/2relquamec}) reduces to the pilot-wave interpretation for non-relativistic quantum theory as presented here in the non-relativistic limit.

In \cite{struyve032} we used similar arguments as Holland's in order to obtain the uniqueness of the particle model for relativistic spin-0 and spin-1 proposed by Ghose {\em et al.}\ \cite{ghose93,ghose94,ghose96,ghose01}. By taking the non-relativistic limit, the particle currents in the particle model of Ghose {\em et al.}\  reduce to the non-relativistic currents given in Section \ref{examplespwinonrelativisticlimit} (with the gyromagnetic factor as given in (\ref{h1.002})). However the model proposed by Ghose {\em et al.}\  contains some features which make it in general untenable to maintain the model as a valid description of physical reality (we will discuss this in detail in Section \ref{Spin-0-1relquamec}, albeit without the discussion on the uniqueness). Therefore, this uniqueness in the case of spin-0 and spin-1 should definitely not be regarded as conclusive. Nevertheless, with Holland's result for spin-1/2, we could require on the basis of uniformity that the spin term should be included for all values of spin.

In fact, Deotto and Ghirardi \cite{deotto98} were among the first to consider different currents compatible with the continuity equation from which different guidance laws for the particle beables could be derived. However, they came to the conclusion that the requirement of Galilean covariance of the current was insufficient to impose its uniqueness.

Although the additional contribution in the guidance equation may not be detectable in quantum probabilities derived from $\psi^{\dagger}\psi$, it plays a role in the case of {\em time of arrival} measurements \cite{finkelstein99,manirul03}. In the pilot-wave interpretation, the distribution of arrival times of the particles at ${\bf x}$ for free particles is given by $|{\bf j}({\bf x},t )|$ and the corresponding mean arrival time at ${\bf x}$ is
\begin{equation}
{\bar t} =\frac{ \int^{+\infty}_0 |{\bf j}({\bf x},t )| t {d} t}{\int^{+\infty}_0 |{\bf j} ({\bf x},t )|  {d} t}.
\label{h1.01202}
\end{equation}
Because these quantities depend on the current ${\bf j}$ and not on the probability density $\psi^{\dagger} \psi$, the spin contribution in the current may {\em in principle} lead to an observable effect. Recently it was argued that for free spin eigenstates, spin contributions with a gyromagnetic factor $g=0$ or $g=1/2$ would in principle be experimentally distinguishable for time of arrival distributions \cite{manirul03}. Hence time of arrival measurements might contribute to the determination of the correct particle current and hence to the correct guidance equation in the pilot-wave interpretation. Of course the question remains whether the difference for mean arrival times, which are calculated with different guidance equations, is {\em experimentally} observable.

The definition of the time of arrival depends in fact not only on the particular choice for the guidance equation. It also depends strongly on the particular pilot-wave model we adhere to. In the next section we give examples of such alternative models. For some of these models the question whether of not we should include a spin term in the guidance equation becomes irrelevant, simply because they involve a different ontology. It may also very well be that for some of these models, the definition of `time of arrival' depends on how exactly we model time measurements. In such a case the definition of time of arrival may not be so unambiguous anymore.    

\subsection{Alternative pilot-wave models}
In the pilot-wave interpretation we presented here, the particle beable is a structureless point particle; it does not carry spin degrees of freedom. Spin is solely a property of the wavefunction. The pilot-wave interpretation still solves the measurement problem, because measurements can in general be reduced to position measurements, the exceptions being measurements involving time.  

There also exist pilot-wave models in which spin degrees of freedom are also attached to the particle beables, with the particles beables being point particles or rigid bodies, see e.g.\ Bohm {\em et al.}\  \cite{bohm551,bohm552} and Holland \cite{holland881,holland882,holland}. Although these models are very interesting and may be very illustrative in some cases, they tend to complicate things. Especially if one only wants the pilot-wave model to reproduce the quantum probabilities, the assumption of an additional structure of the particle beables is unnecessary. 

The view of particle beables as structureless point particles is also the one advocated by Bell \cite{bell66,bell81,bell82}, D\"urr {\em et al.}\ \cite{durr952} and later also by Bohm and Hiley \cite{bohm5}. The only difference with the model presented here is that Bell and D\"urr {\em et al.} do not include the spin part, which arises from the spin current ${\bf j}_s$, in the guidance equation. Bohm and Hiley consider the spin term when they discuss non-relativistic spin-1/2 particles.

There exist still other, completely different, ontologies. For example, in Chapters \ref{chapter4} and \ref{chapter5} we will see that a pilot-wave interpretation with fields as beables seems more natural in quantum field theory. However this field ontology does not reduce to a particle ontology in the non-relativistic limit. This means that if the field ontology is taken as fundamental then even at the level of non-relativistic quantum systems the beables are fields. With a field beable approach, the issue of identifying a unique guidance equation from the continuity equation then of course acquires a new character. We do not know of arguments leading to a preferred choice for the guidance equation in this case. Therefore we will not return to the question of uniqueness when dealing with field theory.  

\chapter{Particle beables for relativistic quantum mechanics}\label{chapter3}
\section{Introduction}
In the preceding chapter we have seen how a pilot-wave interpretation could be constructed for a non-relativistic particle with arbitrary spin. The success of this pilot-wave formulation can be regarded as a consequence of the fact that the non-relativistic wave equations already admitted a quantum mechanical particle interpretation at the level of the standard interpretation. The key feature that allowed for this particle interpretation was the existence of a positive definite density which is conserved. This density could then be successfully identified with a particle density.

The situation is totally different in relativistic quantum theory. When we try to formulate a particle interpretation for relativistic wave equations along the lines of the particle interpretation in non-relativistic quantum theory, we encounter some serious difficulties. The difficulties have nothing to do with the construction of relativistic wave equations itself. Wave equations that are both Lorentz covariant and causal can be found for any value of spin (see e.g.\ the Bhabha wave equations below).{\footnote{As shown by Velo and Zwanziger covariant wave equations may suffer from noncausal propagations \cite{velo691,velo692}, but the Bhabha equations do not suffer from this problem \cite{krajcik761}.}} However, the problem is rooted in the fact that the wave equations in general lack an associated current which has all the mathematical properties required for a particle current. 

A notable exception is the spin-1/2 theory of Dirac. In the Dirac theory there is a conserved current (proportional to the charge current) which has a positive time component in every Lorentz frame. For energies below the threshold of pair creation, this time component can then be interpreted as the particle probability density. If only electromagnetic interaction is considered, then one can extend this particle interpretation to higher energies by using Dirac's original suggestion of the Dirac sea and by passing to the many-particle description. Because one has a quantum mechanical particle interpretation in this case, one can also devise a pilot-wave interpretation, in the same way as the pilot-wave interpretation for non-relativistic quantum mechanics. The first to present this pilot-wave interpretation was Bohm. We recall this pilot-wave interpretation in Section \ref{Spin-1/2relquamec}.

This success for the Dirac equation is not repeated for other types of wave equations. For example if we consider the Bhabha wave equations, which are Dirac-type equations for massive particles with arbitrary spin,{\footnote{For an elaborate review see \cite{krajcik772} and references therein.}} then one can show that the charge density is only positive in the case of spin-1/2, where the Bhabha equation is the Dirac equation. Hence, only in the spin-1/2 representation one can construct a particle interpretation for the Bhabha wave equation starting from the charge current. Because the Bhabha equations are derived from fairly basic principles such as Lorentz covariance and restriction to first-order space and time derivatives,{\footnote{In fact the Bhabha equations can be seen as the relativistic counterparts of the non-relativistic wave equations ({\ref{h1.001}}), which can be derived from general principles such as Galilean covariance and `minimality' \cite{hurley71}.}} we do not think that the issue can be easily resolved by resorting to other relativistic wave equations.

If we consider the energy density for the class of Bhabha wave equations, then one can show that the energy density is only positive in the case of spin-0 and spin-1.{\footnote{The proof that the charge density is not positive definite for integer spin representations and that the energy density is not positive definite for half-integer spin representations can be found in \cite{bhabha49,akhiezer65}. That the charge density is not positive definite for spin higher than 1/2, is discussed in \cite{krajcik761}. Akhiezer and Berestetskii mention without proof that for spins higher than one, neither the charge density nor energy density is positive definite \cite[p.\ 240]{akhiezer65}.}} In the spin-0 case and the spin-1 case the Bhabha wave equation reduces to the first-order Duffin-Kemmer-Petiau (DKP) wave equation \cite{kemmer39}, which is completely equivalent with the familiar second order Klein-Gordon equation and Proca equations. 

Hence, it could be tempting to try to construct a particle interpretation from the energy--momentum tensor for spin-0 and spin-1. This is indeed what has been done by Ghose {\em et al.}\ \cite{ghose93,ghose94,ghose96,ghose01}. Together with a quantum mechanical particle interpretation, Ghose {\em et al.}\ then also devised a pilot-wave model for the DKP equation. Both the quantum mechanical particle interpretation and the pilot-wave model display formal similarities with the equations for the Dirac theory.

Ghose {\em et al.}\ also extended the particle interpretation to massless spin-0 and spin-1 particles. In this case the particles are described by the first-order Harish-Chandra theory \cite{harish46}, which can be obtained from the DKP theory only by minimal modifications, and which is equivalent with the massless Klein-Gordon theory and Maxwell's theory. 

We will discuss the particle interpretation of Ghose {\em et al.}\ in detail in Section \ref{Spin-0-1relquamec}. In particular, we will indicate some features which make it hard to maintain this particle interpretation as a valid description for relativistic spin-0 and spin-1 particles. We argue that the quantum mechanical particle interpretation for the DKP equation can at best be regarded valid for sufficiently low energies, because in the non-relativistic limit it reduces to the particle interpretation for non-relativistic spin-0 and spin-1 particles. This reasoning does of course not apply to the massless case, for which the wavefunctions propagate at the speed of light. 

As a result, also the corresponding pilot-wave interpretation for the DKP equation can only be considered as a valid approximation in the non-relativistic limit. In general the pilot-wave interpretation, or we better call it a {\em trajectory model} from now on, should rather be regarded as describing the tracks of energy flow and all the results reported for this trajectory model should be reinterpreted as such. In this way the tracks of energy flow merely provide a visualization of processes, they do not entail some new interpretation for quantum theory. But because such visualizations are interesting on their own, we extend the trajectory model to many particles.

Of course, as is well known, the problem of assigning particle probability densities to relativistic wave equations and the associated difficulties with the interpretation of the negative energy states (localization of a particle within its Compton wavelength implies the appearance of negative energy states), led to the conception of field theory. In field theory the notion of fields rather than the notion of particles is fundamental. Hence, as in the standard interpretation, a pilot-wave interpretation in terms of field beables might be better suited for dealing with high energy phenomena. We will see in the following chapter that such an interpretation in terms of fields is perfectly possible. Although the pilot-wave interpretation for fermionic fields still has to be developed further, the simplicity of the pilot-wave interpretation for bosons in terms of field beables is in striking contrast with the difficulties that are encountered when trying to develop a pilot-wave interpretation in terms of particle beables. Because of the striking simplicity of the field description and because a particle interpretation has even been abandoned in favor of a field interpretation already at the level of standard quantum theory,{\footnote{Of course one still has the notion of particles in Fock space. But in Fock space states are composed of states with definite momenta and if one tries to construct states which are localized in physical 3-space one encounters violations of causality \cite{hegerfeldt74,hegerfeldt80,hegerfeldt85}.}} we consider the pilot-wave approach in terms of fields as fundamental. Nevertheless, a particle interpretation may still serve well as an illustrative model for the description of low energy phenomena.

\section{Spin-1/2 relativistic quantum mechanics} \label{Spin-1/2relquamec}
\subsection{Massive spin-1/2: The Dirac formalism}
The Dirac equation reads{\footnote{In this chapter we work in units in which $\hbar=c=1$. The Lorentzian indices, which are denoted by $\mu,\nu,\dots$, are raised and lowered by the metric $g_{\mu \nu} = \textrm{diag}(1,-1,-1,-1)$. The index $0$ denotes the time index and the indices $i,j,\dots$ denote the spatial index.}}
\begin{equation}
(i\gamma^{\mu} \partial_{\mu} - m )\psi = 0, 
\label{h2.001.1}
\end{equation}
with $m$ the mass of the particle and $\gamma^\mu$ the Dirac matrices which satisfy the commutation relations $\gamma^{\mu}\gamma^{\nu} + \gamma^{\nu} \gamma^{\mu} = 2 g^{\mu \nu}$. Equivalently one can write the Dirac equation in the Schr\"odinger form
\begin{equation}
i \partial_{0} \psi =  (-i \alpha^{i} \partial_{i} + m\beta   ) \psi,
\label{h2.0013.2} 
\end{equation}
with $\alpha^{i}= \gamma^0 \gamma^i$ and $\beta = \gamma^0$. 

The current $j^{\mu}= {\bar \psi} \gamma^{\mu}\psi$ (which differs from the charge current by a factor $e$) is a conserved and future-causal Lorentz 4-vector \cite{holland}. Hence, this current has a positive time component in every Lorentz frame. This allowed Dirac to interpret the density $j^0({\bf x},t) =\psi^{\dagger} ({\bf x},t)\psi({\bf x},t)$ as the probability density for a spin-1/2 particle to be detected at the position ${\bf x}$ at the time $t$. The integral curves of the 4-vector $j^{\mu}$ can then be seen as the flowlines of the `particle detection probability'. Because the current $j^{\mu}$ is future-causal it is guaranteed that these probability flowlines are time-like. Hence, the charge current can be a given a particle interpretation, which meets all basic relativistic requirements. 

The pilot-wave interpretation now proceeds in the same way as in the non-relativistic case \cite{bohm53,holland92,holland,bohm5,bohm871,durr99}. A structureless point particle (the particle beable) is introduced for which the 4-vector velocity field is given by $u^\mu =j^{\mu}/\sqrt{j^\nu j_\nu}$. The possible trajectories $x^\mu(\tau)$ are the integral curves of the velocity field, i.e.\ they are solutions to the guidance equation
\begin{equation}
\frac{{d} x^\mu }{ {d} \tau } = u^\mu.
\label{h2.0013.3}
\end{equation}
The trajectory of a particle beable is uniquely determined by the specification of an initial configuration $x^\mu(\tau_0)$. Equivalently one can write the trajectories as curves ${\bf x}(t)$ which are found by solving the guidance equation
\begin{equation}
\frac{{d} x^i}{ {d} t} = \frac{u^{i}}{u^0} = \frac{j^i}{j^0}.
\label{h2.00302}
\end{equation}
The trajectory of a particle beable is then uniquely determined by the specification of an initial position ${\bf x}(t_0)$. Because the vector $u^\mu$ is future-causal, the motions are future-causal. In an ensemble the probability density of the particle beables is given by $j^0= \psi^{\dagger} \psi$.

In the non-relativistic limit the pilot-wave model for the Dirac equation reduces to the pilot-wave interpretation for non-relativistic quantum mechanics as presented in Chapter \ref{chapter1} \cite{bohm5}.

If an interaction with an electromagnetic field $V_{\mu}$ is introduced, through the minimal coupling prescription $\partial_{\mu} \to D_{\mu} = \partial_{\mu} + ieV_{\mu} $, then the pilot-wave interpretation as described above is still applicable \cite{bohm5,holland}. This is because the charge current is still conserved (an electromagnetic field does not posses charge and hence cannot exchange charge with a charged particle) and because the charge current is still future-causal (the charge current contains no derivatives and hence this current retains its form after minimal coupling).

It now seems that the trajectory interpretation presents no problem at all. As noted by Holland \cite{holland92,holland}, the particle trajectories are always well defined, regardless of whether or not the state contains negative energy contributions. However, this success is only a deceptive appearance. Also in the pilot-wave interpretation negative energy states require a meaningful interpretation. This is because the problem of a possible radiation catastrophe manifests itself already at the level of the wavefunction. Hence, in order to prevent a positive energy wavefunction to lose energy by radiative transitions to lower and lower energies, one has to give a meaningful interpretation to the negative energy states. One could for example adopt the original idea by Dirac and assume a Dirac sea where every negative energy state is occupied, so that due to the Pauli principle, a transition to a negative energy state is impossible. This then requires a many-particle approach and in the corresponding pilot-wave model we should then consider a many-particle wavefunction describing an infinite, although fixed, number of particles, i.e.\ all the particles in the Dirac sea (the negative energy particles) plus the number of positive energy particles that one wants to describe. This is also the way Bohm and Hiley viewed the pilot-wave interpretation for the Dirac theory \cite{bohm871}. When dealing with specific problems, the total wavefunction will factorize and it will be sufficient to consider a finite number of particles (the many-particle case is reviewed in the following section).{\footnote{Recently the pilot-wave interpretation which explicitly incorporates every particle in the Dirac sea was re-derived by Colin as the continuum limit of the stochastic Bell model \cite{colin031,colin032,colin033}.}} However, although the idea of a Dirac sea may serve well to describe spin-1/2 particles with electromagnetic interaction, it remains the question of how this model could be adopted to give an account for other types of interaction such as weak interaction (recall the standard example of beta decay \cite[p.\ 144]{sakurai67}). Of course, particle creation and annihilation finds a natural home in quantum field theory (there is no need for a Dirac sea), and hence a pilot-wave interpretation for quantum field theory is desired. 

We turn to quantum field theory in the next two chapters. For the moment we continue to discuss the non-quantized Dirac theory a little more. In the rest of the section we recall the many-particle Dirac formalism. In the context of this formalism we can make some statements about the possibility of formulating a Lorentz invariant pilot-wave model. Similar statements will also apply in the context of quantum field theory.

\subsection{The many-particle Dirac formalism}\label{manyparticledirac}
The one-particle Dirac formalism is extended to many-particles by the introduction of an $N$-particle wavefunction $\psi_{r_1 \dots r_N }({\bf x}_1, \dots,{\bf x}_N, t)$ with $N$ spin indices. The wavefunction is assumed to be anti-symmetric in order to satisfy the Pauli principle. We also introduce operators $\gamma^{\mu}_{(r)},\alpha_{(r)}^{i}$ and $\beta_{(r)}$ which operate only on the $r^{\textrm{th}}$ spin index, belonging to the $r^{\textrm{th}}$ particle, e.g.\ $\gamma^{\mu}_{(r)} = {\mathbb 1} \otimes \dots \otimes\gamma^{\mu} \otimes \dots \otimes {\mathbb 1} $ with $\gamma^{\mu}$ at the $r^{\textrm{th}}$ place in the product \cite{bohm5,holland99,durr99}. The Schr\"odinger form of the Dirac equation for the $N$-particle system then reads
\begin{equation}
i \partial_{0} \psi = \sum^N_{r = 1} \Big(-i { \alpha}_{(r)}^{i} \partial^{(r)}_{i} + \beta_{(r)} m_{r}   \Big) \psi 
\label{h2.0072} 
\end{equation}
where $\partial^{(r)}_{i} = \partial / \partial({\bf x}_{r})^i$ and $m_{r}$ is the mass of the $r^{\textrm{th}}$ particle. The tensor current is defined as 
\begin{equation}
j^{\mu_1 \dots \mu_N} = \psi^{\dagger} \gamma^{0}_{(1)}\gamma^{\mu_1}_{(1)} \dots \gamma^{0}_{(N)}\gamma^{\mu_N}_{(N)} \psi
\label{h2.0073}
\end{equation}
and satisfies the conservation equation
\begin{equation}
\partial_0 j^{0_1 \dots 0_N} + \sum^N_{r = 1}  \partial^{(r)}_{i_r} j^{0_1 \dots i_{r} \dots 0_N} = 0.
\label{h2.0074}
\end{equation}
with $j^{0_1 \dots 0_N}= \psi^{\dagger} \psi$ a positive quantity and $ j^{0_1 \dots i_{r} \dots 0_N}=\psi^{\dagger} {\alpha}_{(r)}^i \psi$.

In the pilot-wave interpretation the velocity field for the $r^{\textrm{th}}$ particle beable is given by \cite{bohm5,holland99,durr99}
\begin{equation}
u^{\mu_{r}}_{r} = \frac{j^{0_1 \dots \mu_{r} \dots 0_N} }{\sqrt{j^{0_1 \dots \nu_{r} \dots 0_N}j_{0_1 \dots \nu_{r} \dots 0_N} }},
\label{h2.007401}
\end{equation}
so that the corresponding guidance equation reads
\begin{equation}
\frac{{d} x_{r}^i}{ {d} t} = \frac{\psi^{\dagger} {\alpha}_{(r)}^i \psi}{\psi^{\dagger} \psi}.
\label{h2.0075}
\end{equation}
Because 
\begin{equation}
j^{0_1 \dots \mu_{r} \dots 0_N} j_{0_1 \dots \mu_{r} \dots 0_N} \ge 0
\label{h2.0076}
\end{equation}
the motion of the particles is future-causal. The distribution of the particle beables in an ensemble is assumed to be the equilibrium distribution, i.e.\ $j^{0_1 \dots 0_N}=\psi^{\dagger}\psi$.

\subsection{Note on Lorentz covariance}\label{noteonlorentzinvariance}
In the one-particle case, the pilot-wave interpretation is covariant. First, the trajectories have a covariant meaning because the defining velocity field $u^\mu$ transforms as a Lorentz 4-vector under Lorentz transformations. Second, also the probability interpretation is covariant. To see this, consider an arbitrary spacelike hypersurface $\sigma (x)$, with $n_{\mu}(x)$ the future-oriented unit normal. The positive Lorentz scalar $j^{\mu}(x) n_{\mu}(x)$ is then interpreted as the probability for a particle to cross the spacelike hypersurface $\sigma (x)$ at $x$ in any frame \cite{durr99}. In the case of an equal-time hyperplane the crossing probability is given by $j^0$. 

In the multi-particle case, the pilot-wave interpretation is not Lorentz covariant. The velocity fields ({\ref{h2.007401}}) do not transform as 4-vectors. In addition, as shown by Berndl {\em et al.} \cite{berndl96}, equilibrium can in general not hold simultaneously in all Lorentz frames. By this is meant the following. If we assume the density of crossings through an equal-time hyperplane in a particular frame to be given by $\psi^{\dagger} \psi$, then the density of crossings of crossings will be given by $\psi^{\dagger} \psi$ for any other equal-time hyperplane in {\em this} frame. But the density of crossings of crossings through an equal-time hyperplane in {\em another} Lorentz frame will in general not equal $\psi'^{\dagger}\psi'$, with $\psi'$ the wavefunction in the other frame. Berndl {\em et al.} showed this feature must hold, not only for the pilot-wave model presented above, but for any pilot-wave model for the many-particle Dirac theory. An exception occurs when the wavefunction is a product wavefunction of one-particle wavefunctions.

Bohm and Hiley accepted the idea of a preferred  Lorentz frame to which the pilot-wave interpretation should be formulated \cite{bohm5}. The same opinion is hold by Valentini. Valentini argues that the natural symmetry of pilot-wave theory is Aristotelian invariance (because of the first-order character of the guidance law) and that hence the search for a Lorentz covariant pilot-wave model is misguided \cite{valentini92,valentini96,valentini97}. According to this view, the Lorentz covariance of the one-particle Dirac theory and the Galilean covariance of the pilot-wave interpretation of non-relativistic quantum mechanics are,{\footnote{The Galilean invariance of pilot-wave theory for non-relativistic quantum theory is discussed in \cite[pp.\ 122-124]{holland}.}} despite appearances, not the fundamental symmetries. With Aristotelian invariance as the fundamental symmetry, there should be a preferred class of Aristotelian inertial frames{\footnote{This is a class of reference frames which are connected by Aristotelian transformations. Aristotelian transformations being time-independent transformations of 3-space, such as translations or rotations.}} relative to which the pilot-wave interpretation should be formulated. In this class of Aristotelian inertial frames, the interactions between the different particles may be instantaneous (due to the nonlocality of pilot-wave theory at the subquantum level), but no causality paradoxes arise, because the notions of cause and effect only make sense with respect to this frame. 

On the other hand, Berndl {\em et al.}\ express the desire for a Lorentz invariant pilot-wave model \cite{berndl96,durr99}. In \cite{durr99} they provide an example of how some notion of covariance could be introduced. In their model they introduce a particular space-time foliation as an additional space-time structure. This particular space-time foliation would then be determined by some covariant law, possibly depending on the Dirac wavefunction. Then, instead of writing the pilot-wave interpretation with respect to a particular frame as in the view of Bohm, Hiley and Valentini, it should be written with respect to this particular space-time foliation. If the equilibrium density is assumed on one of the leaves of the foliation, it is guaranteed that the model reproduces the predictions of standard quantum theory.

There exist still other pilot-wave models which are Lorentz covariant \cite{squires93,dewdney02,goldstein03}. However these models lack a probability interpretation, which makes it difficult to relate these models to standard quantum theory. The model by Squires \cite{squires93} could be ruled out from the start simply because it is a local model and hence in contradiction with the empirically verified violations of the Bell inequalities.

We will further not consider the possibility of constructing a Lorentz covariant pilot-wave model. The merit of the model by Berndl {\em et al.}\ is that it provides a counter example to claims that a covariant pilot-wave interpretation is impossible. But as Berndl {\em et al.}\ indicate themselves \cite{berndl96}, any theory can be made Lorentz covariant by the incorporation of additional structure. If the pilot-wave interpretation is then devised to reproduce the quantum statistics, it remains a mere guessing what the additional structure should look like. In the pilot-wave models for field theory that we consider similar remarks apply. The pilot-wave models are not Lorentz covariant at the subquantum level, but at the quantum level they will yield the same statistics as the conventional interpretation and hence at the quantum level the models are Lorentz covariant.

\section{Spin-0 and spin-1 relativistic quantum mechanics} \label{Spin-0-1relquamec}
In this section we consider the particle interpretation for relativistic spin-0 and spin-1 bosons proposed by Ghose {\em et al.} Because the Duffin-Kemmer-Petiau (DKP) formalism and the Harish-Chandra (HC) formalism are perhaps less known, we review the essential elements in the following two sections. A more elaborate discussion of the DKP formalism and the HC formalism can respectively be found in \cite{kemmer39} and \cite{harish46}, or in the review by Ghose \cite{ghose96}.

\subsection{Massive spin-0 and spin-1: The Duffin-Kemmer-Petiau formalism}\label{massivespin0-spin1dkpformalism}
The DKP equation for a particle with mass $m$ reads
\begin{equation}
(i\beta^{\mu} \partial_{\mu} - m )\psi = 0, 
\label{h2.1.1}
\end{equation}
with adjoint equation
\begin{equation}
i\partial^{\mu}  {\bar \psi} \beta_{\mu} + m {\bar \psi} = 0,
\label{h2.1.2}
\end{equation}
where ${\bar \psi} =\psi^{\dagger} \eta_0 $ with $\eta_0 = 2\beta^2_0 - 1 $, and the DKP matrices $\beta^\mu$ satisfy the commutation relations
\begin{equation}
\beta^{\mu}\beta^{\nu}\beta^{\lambda} + \beta^{\lambda}\beta^{\nu}\beta^{\mu} = \beta^{\mu}g^{\nu \lambda} + \beta^{\lambda} g^{\nu \mu}.
\label{h2.2}
\end{equation}
There are three inequivalent irreducible representations of the $\beta^{\mu}$, one is $10 \times 10$ and describes spin-1 bosons, another one is $5 \times 5$ which describes spin-0 bosons and the third one is the trivial $1 \times 1$ representation. 

The DKP equations can be derived from the Lagrangian density 
\begin{equation}
\mathcal{L}_{DKP} = \frac{i}{2} ({\bar \psi} \beta_{\mu}\partial^{\mu} \psi  - \partial^{\mu}  {\bar \psi} \beta_{\mu} \psi ) -m{\bar \psi}\psi.
\label{h2.3}
\end{equation}
The corresponding conserved symmetrized energy--momentum tensor reads
\begin{equation}
\Theta^{\mu \nu}_{DKP} =  m{\bar \psi} (\beta^{\mu}\beta^{\nu} +\beta^{\nu}\beta^{\mu} - g^{\mu \nu})\psi.
\label{h2.6}
\end{equation}
The conserved charge current is
\begin{equation}
s^{\mu}_{DKP} = e{\bar \psi} \beta^{\mu} \psi
\label{h2.6.101}
\end{equation}
with $e$ the charge of the particle.

The DKP equation ({\ref{h2.1.1}}) can be written in the following equivalent Hamiltonian form 
\begin{eqnarray}
i \partial_{0} \psi &=& (-i{\tilde \beta}^{i} \partial_{i} + m\beta_{0}   ) \psi, \label{h2.13.1}\\
i \beta^{i} \beta^2_{0}\partial_{i}\psi &=& m (1 -  \beta^2_{0}) \psi
\label{h2.13.2} 
\end{eqnarray}
with ${\tilde \beta}^{i} = \beta^0 \beta^i - \beta^i \beta^0$. The first equation is a Schr\"odinger-like equation and the second equation has to be regarded as an additional constraint on the wavefunction $\psi$. Only when the two equations ({\ref{h2.13.1}}) and ({\ref{h2.13.2}}) are taken together, they are equivalent with the covariant form ({\ref{h2.1.1}). 

That the constraint equation is compatible with the equations of motion can be seen by writing ({\ref{h2.13.1}}) and ({\ref{h2.13.2}}) in operator form. From the Schr\"odinger-like equation ({\ref{h2.13.1}}) we find the free Hamiltonian operator
\begin{equation}
{\hat H} = {\tilde {\boldsymbol \beta}} \cdot {\hat {\bf p}} + m \beta_0,
\label{h2.p2}
\end{equation}
with ${\hat {\bf p}}=-i {\boldsymbol \nabla}$ the momentum operator. Physical states have to satisfy the additional constraint ({\ref{h2.13.2}), which can be written as
\begin{equation}
{\hat C} \psi = 0,
\label{h2.p2.01}
\end{equation}
with ${\hat C}$ an idempotent operator, i.e.\ ${\hat C}^2={\hat C}$, given by  
\begin{equation}
{\hat C} = \frac{1}{m}  {\boldsymbol \beta} \cdot {\hat {\bf p}} \beta^2_{0} + 1 - \beta^2_{0}    = 1 - \frac{1}{m}  {\hat H} \beta_0.
\label{h2.p2.0101}
\end{equation}
Because $ {\hat H} \beta_0 {\hat H} = m {\hat H}$, we have that ${\hat C}{\hat H}=0$. If a wavefunction initially, say at time $t=0$, satisfies the constraint ({\ref{h2.13.2}), i.e.\ ${\hat C} \psi({\bf x},0)$, then the wavefunction will also satisfy the constraint at a later time, because in the Heisenberg picture we have ${\hat C}\psi({\bf x},t) = {\hat C} \exp\left(-i{\hat H}t \right) \psi({\bf x},0)=0 $. Hence the constraint ({\ref{h2.13.2}) is compatible with the equation of motion ({\ref{h2.13.1}}).

With the help of ({\ref{h2.p2.01}}) one can further show that 
\begin{equation}
{\hat H}^2 \psi = ({\hat {\bf p}}^2 + m^2)\psi.
\label{h2.p2.1}
\end{equation}
Hence every component of the DKP wavefunction satisfies the massive Klein-Gordon equation. Contrary to the Dirac case, the equality is not valid on the operator level, i.e.\ ${\hat H}^2 \neq ({\hat {\bf p}}^2 + m^2)$. Rather one has the property
\begin{equation}
{\hat H}^3 = {\hat H} ({\hat {\bf p}}^2 + m^2).
\label{h2.p2.12}
\end{equation}

The equivalence of the DKP wave equation with the Klein-Gordon equation in the spin-0 representation and the Proca equations in the spin-1 representation can be shown in a representation independent way \cite{harish46,lunardi00,umezawa56}. However, it is instructive to see how this equivalence arises in the explicit matrix representations which are given in Appendix \ref{appa}. 

In the spin-0 representation, the constraint equation ({\ref{h2.13.2}}) implies a DKP wavefunction $\psi$ with the following five components: $\omega$, $\partial_{1} \phi/\sqrt{m}$, $\partial_{2} \phi/\sqrt{m}$, $\partial_{3} \phi/\sqrt{m}$, $\sqrt{m}\phi$ with $\omega$ and $\phi$ two scalar wavefunctions. Equation ({\ref{h2.13.1}}) further implies $\omega= \partial_{0} \phi/\sqrt{m}$. In this way the number of independent components of the wavefunction $\psi$ is reduced to one and we can write 
\begin{equation}
\psi = \frac{1}{\sqrt{m}}\left( \begin{array}{c}
 \partial_\mu \phi   \\
m \phi \\
\end{array} \right).
\label{h2.13.21}
\end{equation}
The Schr\"odinger equation ({\ref{h2.13.1}}) then reduces to the massive Klein-Gordon (KG) equation for $\phi$
\begin{equation}
\square \phi + m^2 \phi=0.
\label{h2.13.3}
\end{equation}
By substitution of the wavefunction ({\ref{h2.13.21}}) into the DKP Lagrangian ({\ref{h2.3}}), the DKP energy--momentum tensor ({\ref{h2.6}}) and the DKP charge current ({\ref{h2.6.101}}), we obtain respectively the KG Lagrangian for the KG wavefunction $\phi$ 
\begin{equation}
\mathcal{L}_{KG}  = \partial^{\alpha} \phi \partial_{\alpha} \phi^* - m^2 \phi^* \phi,
\label{h2.16}
\end{equation}
the KG energy--momentum tensor 
\begin{equation}
\Theta^{\mu \nu}_{KG} = \partial^{\mu} \phi \partial^{\nu} \phi^* + \partial^{\mu} \phi^* \partial^{\nu} \phi - g^{\mu \nu} \mathcal{L}_{KG} 
\label{h2.15}
\end{equation}
and the KG charge current
\begin{equation}
s^{\mu}_{KG} = ie\big(\phi^* \partial^{\mu}\phi -\phi \partial^{\mu}\phi^*  \big). 
\label{h2.1601}
\end{equation}

In the spin-1 representation we can take the ten components of the DKP wavefunction as: 
\begin{equation}
\psi = (-{\bf E}, {\bf B},m{\bf A}, -mA_0)^T/\sqrt{m}.
\label{h2.18.1}
\end{equation}
Equation ({\ref{h2.13.2}}) then implies the following relations
\begin{equation}
 {\boldsymbol \nabla} \cdot {\bf E} = - m^2 A_0, \qquad {\bf B} =  {\boldsymbol \nabla} \times {\bf A}.
\label{h2.19}
\end{equation}
The equation ({\ref{h2.13.1}}) leads to the relations
\begin{equation}
\partial_0 {\bf E}  =  {\boldsymbol \nabla} \times {\bf B}+ m^2 {\bf A}, \quad \partial_0 {\bf B} = -  {\boldsymbol \nabla} \times {\bf E} ,\quad
{\bf E}= - {\boldsymbol \nabla} A_0 - \partial_0 {\bf A}. 
\label{h2.20}
\end{equation}
The equations in ({\ref{h2.19}}) and ({\ref{h2.20}}) are recognized as the Proca equations
\begin{equation}
\partial_\mu G^{\mu \nu} = - m^2 A^{\nu}
\label{h2.22}
\end{equation}
for $A^\mu=(A_0,{\bf A})$, with $G^{\mu \nu} = \partial^{\mu} A^{\nu} - \partial^{\nu} A^{\mu}$.

If the Kemmer wavefunction ({\ref{h2.18.1}}) is substituted into the DKP Lagrangian ({\ref{h2.3}}), the DKP energy--momentum tensor ({\ref{h2.6}}) and the DKP charge current ({\ref{h2.6.101}}), we obtain with the help of the relations ({\ref{h2.19}}) and ({\ref{h2.20}}) respectively the Proca Lagrangian
\begin{equation}
\mathcal{L}_P  = - \frac{1}{2} G^*_{\mu \nu}G^{\mu \nu} + m^2 A^*_{\mu}A^{\mu},
\label{h2.23}
\end{equation}
the symmetrized Proca energy--momentum tensor 
\begin{equation}
\Theta^{\mu \nu}_P = -G^{* \mu \alpha} G^{\nu}_{\ \alpha} - G^{ \mu \alpha} G^{*\nu}_{\ \ \alpha}  + m^2 (A^\mu A^{* \nu} + A^{* \mu} A^\nu )- g^{\mu \nu} \mathcal{L}_P
\label{h2.24}
\end{equation}
and the Proca charge current
\begin{equation}
s^{\mu}_{P} = ie\big(G^{* \mu \nu}A_{\nu}  - G^{\mu \nu} A^*_\nu \big).
\label{h2.24.02}
\end{equation}

\subsection{Massless spin-0 and spin-1: The Harish-Chandra formalism}\label{masslessspin-0andspin-1harishchandra}
We cannot just take the limit $m \to 0$ in the DKP theory to describe massless bosons. Nevertheless, one can describe massless spin-0 and spin-1 bosons in a first-order formalism in close analogy with the DKP formalism. This formalism was developed by Harish-Chandra (HC). In this section we will only review the essential elements of this theory. This theory can be cast in a representation independent form, but as in the case of the DKP theory, we will often fall back to the explicit representation given in Appendix \ref{appa}. 

In order to describe massless spin-0 and spin-1 bosons, Harish-Chandra proposed a modification of the DKP equations by replacing the mass $m$ by $m\gamma$,
\begin{eqnarray}
(i\beta^{\mu} \partial_{\mu} - m\gamma )\psi &=& 0, \label{h2.m.1}\\
i\partial^{\mu}  {\bar \psi} \beta_{\mu} + m {\bar \psi}\gamma &=& 0,
\label{h2.m.2}
\end{eqnarray}
with $\gamma$ a matrix that satisfies 
\begin{eqnarray}
\gamma^2  &=&  \gamma, \label{h2.m.3}\\
\gamma \beta^{\mu} + \beta^{\mu} \gamma &=& \beta^{\mu}.
\label{h2.m.4}
\end{eqnarray}

From ({\ref{h2.m.1}), ({\ref{h2.m.3}) and ({\ref{h2.m.4}) one can derive the second order wave equation 
\begin{equation}
\square (\gamma \psi) = 0.
\label{h2.m.5}
\end{equation}
Hence the wavefunction $(\gamma \psi)$ describes a massless boson.

In both the 10-dimensional and 5-dimensional representation we have two inequivalent choices for $\gamma$. For each representation we only presented one particular choice in Appendix \ref{appa}. With our choice for $\gamma$ in the 10-dimensional representation, massless spin-1 particles are described. If the wavefunction $\psi$ is real then $\gamma \psi$ describes an uncharged, massless spin-1 particle.{\footnote{In a general representation we cannot describe uncharged particles simply by assuming $\psi$ is real. A more general condition is needed then, which can be found in \cite{harish46}. Harish-Chandra called this condition the {\em reality condition}.}} In this case the HC theory is equivalent with the Maxwell's theory for the electromagnetic field. With our choice for $\gamma$ in the 5-dimensional representation, massless spin-0 particles are described. As in the spin-1 case, an uncharged particle is described by a real wavefunction. The other choice for $\gamma$ in the 10-dimensional representation also describes massless spin-0 particles. The other choice for $\gamma$ in the 5-dimensional representation is unphysical.

The wave equations are invariant under transformations $\psi \to \psi + (1-\gamma) {\tilde \psi}$ with ${\tilde \psi}$ satisfying
\begin{equation}
i\beta^{\mu} \partial_{\mu} {\tilde \psi} = 0.
\label{h2.m.51}
\end{equation}
This invariance corresponds to the gauge invariance in the spin-1 case. The wave equations can be derived from the following gauge invariant Lagrangian density 
\begin{equation}
\mathcal{L}_{HC} = \frac{i}{2} ({\bar \psi} \beta_{\mu}\partial^{\mu} \psi  - \partial^{\mu}  {\bar \psi} \beta_{\mu} \psi ) -m{\bar \psi}\gamma \psi.
\label{h2.m.6}
\end{equation}
The symmetrized energy--momentum tensor is given by 
\begin{eqnarray}
\Theta^{\mu \nu}_{HC} &=&  m{\bar \psi} (\beta^{\mu}\beta^{\nu} +\beta^{\nu}\beta^{\mu} - g^{\mu \nu})\gamma \psi \nonumber\\ 
&=&  m\psi^{\dagger} \gamma \eta_0 (\beta^{\mu}\beta^{\nu} +\beta^{\nu}\beta^{\mu} - g^{\mu \nu})\gamma \psi.
\label{h2.m.701}
\end{eqnarray}
Note that the factor $m$ does not represent the mass of the particle (the mass is zero), but a constant that can be removed by a suitable normalization of $\gamma \psi$. The conserved charge current is also formally the same as in the massive case
\begin{equation}
s^{\mu}_{HC} = e{\bar \psi} \beta^{\mu} \psi.
\label{h2.m.702}
\end{equation}
For uncharged bosons $\psi$ the charge current is zero. Not only the charge $e$ is then zero, but because $\psi$ satisfies the reality condition also the current ${\bar \psi} \beta^{\mu} \psi$ is identically zero.

The HC equation ({\ref{h2.m.1}}) can be written in the following equivalent Hamiltonian form 
\begin{eqnarray}
i \partial_{0} (\gamma \psi) &=& -i{\tilde \beta}^{i} \partial_{i}  (\gamma \psi), \label{h2.m.8}\\
i \beta^{i} \beta^2_{0}\partial_{i}\psi &=& m (1 -  \beta^2_{0})\gamma  \psi.
\label{h2.m.9} 
\end{eqnarray}
Similarly as in the massive case, the first equation is a Schr\"odinger-like equation and the second equation has to be regarded as an additional constraint on the wavefunction $ \psi$. Although it follows from ({\ref{h2.m.9}}) that 
\begin{equation}
i \beta^{i} \beta^2_{0}\partial_{i}(\gamma \psi) = 0,
\label{h2.m.10}
\end{equation}
only the set ({\ref{h2.m.8}}), ({\ref{h2.m.9}}) and not the set ({\ref{h2.m.8}}), ({\ref{h2.m.10}}) is equivalent with the wave equation ({\ref{h2.m.1}). Similarly as in the massive case, one can show that the constraint ({\ref{h2.m.9}}) is compatible with the Schr\"odinger equation ({\ref{h2.m.8}}) by using ({\ref{h2.m.10}}). The Hamiltonian operator ${\hat H}$ and the constraint operator ${\hat C}$ can be obtained from the corresponding operators in the massive case, cf.\ ({\ref{h2.p2}}) and ({\ref{h2.p2.01}}), simply by putting $m=0$.

Just as in the massive case the equivalence of the HC theory with the massless Klein-Gordon theory in the spin-0 representation and with the Maxwell theory in the uncharged spin-1 representation is easily shown with the explicit representation for the matrices $\gamma$ in Appendix \ref{appa}. The action of the matrices $\gamma$ on the wavefunctions ({\ref{h2.13.21}}) and ({\ref{h2.18.1}}), results in a projection on the mass independent components of the wavefunctions, i.e.\ the massless states $\gamma\psi $ are obtained from the massive states $\psi$ (given by ({\ref{h2.13.21}}) and ({\ref{h2.18.1}}) just by putting the mass $m$ equal to zero. This simplicity is the reason why we only presented the particular representations for $\gamma$ given in Appendix \ref{appa}.

\section{Trajectory models for spin-0 and spin-1 bosons}\label{pwiinterpretation}
In the context of the massive Klein-Gordon theory, de Broglie initially entertained the idea of constructing a pilot-wave model for spin-0 particles by giving a particle interpretation to the charge current \cite{debroglie60}. Later, Vigier gave a particle interpretation to the charge current in the DKP formalism, thereby extending the pilot-wave model of de Broglie to account also for spin-1 particles \cite{vigier56}. 

However, as is well known, a quantum mechanical particle interpretation for the charge current for spin-0 or spin-1 bosons is in general untenable because the current is not always future-causal. Even a restriction of the positive energy part of the Hilbert space presents no solution. There exist examples of superpositions of positive energy eigenstates for which the charge current is spacelike, or for which the charge density may become negative in certain space-time regions \cite{kyprianidis85,holland,horton02}. In the case of an uncharged boson the situation is even worse because then the charge current is zero.

In order to circumvent the problems in associating a particle interpretation to the charge current, Ghose {\em et al.}\ \cite{ghose93,ghose94,ghose96,ghose01} proposed to start instead from the energy--momentum tensor, from which future-causal 4-vectors can be constructed. They did this as follows. Let the tensor $\Theta^{\mu \nu}$ represent the DKP energy--momentum tensor $\Theta^{\mu \nu}_{DKP}$ in the massive case and the CH energy--momentum tensor $\Theta^{\mu \nu}_{HC}$ in the massless case and let $n^{\mu}$ be a constant future-causal 4-vector (below possible examples for the vector $n^{\mu}$ are presented). By contraction of the energy--momentum tensor $\Theta^{\mu \nu}$ and the 4-vector $n^{\mu}$ we obtain a 4-vector
\begin{equation}
j^{\mu} = \Theta^{\mu \nu} n_{\nu}
\label{h2.7}
\end{equation}
which is future-causal and conserved.{\footnote{The fact that $j^{\mu}$ is future-causal can be derived as follows. We use the notation ${\tilde \psi}$ to represent either the wavefunction $\psi$ for massive bosons or $\gamma \psi$ for massless bosons. Because $\Theta^{00} =  m{\tilde \psi}^{\dagger} {\tilde \psi} \geqslant 0$ and $\Theta^{0\mu } \Theta_{0\mu } \geqslant 0$ (which can be verified in the explicit representations used in Appendix \ref{appa}), the vector $\Theta^{0 \mu } = \delta^0_\nu \Theta^{\nu \mu } $ is future-causal. Because the product of two future-causal vectors is positive, $j^0 = \Theta^{0\mu } n_{\mu}\geqslant 0$ for $n^{\mu}$ future-causal. The fact that $j^{\mu} j_{\mu} \geqslant 0$ can be seen if we perform a Lorentz transformation such that $\Lambda^\mu_{\ \nu} n^{\nu} =\delta^\mu_0$ because then $j^{\mu} j_{\mu} =\Theta'^{0\mu } \Theta'_{ 0 \mu }$, with $\Theta'^{\mu_1 \mu_2 } = \Lambda^{\mu_1}_{\ \nu_1} \Lambda^{\mu_2}_{\ \nu_2} \Theta^{\nu_1  \nu_2 }$ and $\Theta'^{0\mu } \Theta'_{ 0 \mu }$ is positive as it has the same form as the positive quantity $\Theta^{0\mu } \Theta_{ 0 \mu}$ (just replace ${\tilde \psi}({\bf x},t)$ in the energy--momentum tensor by ${\tilde \psi}'({\bf x}',t')$, where the accents refer to quantities in the new Lorentz frame).}}

Hence, the current $j^{\mu}$ satisfies all the properties required for a particle current. Ghose {\em et al.}\ interpret the current $j^{\mu}$ as a particle current, and associate a pilot-wave interpretation to it along the lines of the pilot-wave interpretation of the Dirac theory. The velocity field of the particle beable is given by $u^\mu =j^{\mu}/\sqrt{j^\nu j_\nu}$, so that the possible trajectories ${\bf x} (t)$ are solutions to the guidance equation
\begin{equation}
\frac{{d} x^i}{ {d} t} = \frac{u^{i}}{u_0} = \frac{j^i}{j^0}.
\label{h2.8}
\end{equation}
For an ensemble of particles all described by the same DKP or HC wavefunction, the probability density of the particle beables is then given by $j^0$. 

In the case that $n^{\mu} = \delta^\mu_0$, the probability density is given by the energy density $j^0=\Theta^{00}= m\psi^{\dagger} \psi$ and the guidance equation reads
\begin{displaymath}
\frac{{d} x^i}{{d} t} = \frac{\Theta^{i0}}{\Theta^{00}} = \left\{  \begin{array}{ll} \frac{\psi^{\dagger} {\tilde \beta}^i \psi}{\psi^{\dagger} \psi}  & \qquad \textrm{in the massive case}  \\ 
 \frac{\psi^{\dagger}\gamma {\tilde \beta}^i\gamma \psi}{\psi^{\dagger} \gamma \psi}  & \qquad \textrm{in the massless case} \end{array} \right. 
\label{h2.9}
\end{displaymath}
where ${\tilde \beta}^i = \beta^0 \beta^i - \beta^i \beta^0$. Hence, in the case that $n^{\mu} = \delta^\mu_0$ the pilot-wave equations look similar to the pilot-wave equations in the Dirac theory.

Let us now turn to the definition of $n^{\mu}$. Ghose {\em et al.}\ proposed the vector $n^{\mu}$ to be the vector normal to a particular spacelike foliation \cite{ghose93}. The vector $n^{\mu}$ is then generally space-time dependent, but the program of associating trajectories to the current $j^{\mu}$ can still be carried out, provided some minor modifications are taken into account, such as the introduction of a covariant derivative on the foliation. However, in the following we will restrict our attention to foliations in terms of hyperplanes so that the vector $n^{\mu}$ is a constant vector. In this case one can alternatively regard the constant vector $n^{\mu}$ as the 4-velocity $a^{\mu}$ of some observer.

In fact it was Holland who initiated the construction of a conserved current from the energy--momentum tensor by contracting it with the four-velocity of an observer, but in the context of Maxwell's theory and the KG theory \cite{holland,holland93}. However, Holland was reluctant to regard the conserved current as a particle probability current and regarded the guidance equation in ({\ref{h2.8}) as the defining formula for the tracks of energy flow. Holland gave a series of arguments supporting this view. Most of the arguments were against the notion of a photon as a localized object and do not apply to massive bosons for which the particle aspect is well accepted. An other argument by Holland is that, if for example $a^{\mu} = \delta^\mu_0$, then the boson density $j^0$ (for photons or massive spin-0 particles) would be given by the energy density, which it is manifestly not in the analogous formula for the quantized boson field.

We agree with Holland that the current $j^\mu$ should not be interpreted as a particle current. But we think that the main reasons for not doing so are the following. 

Even if we adopt the view that the energy density can be seen as the particle density (we can take the energy density as an observable quantity, whether or not it is interpreted as a particle density), it is only in the frame at rest relative to the observer with velocity $a^{\mu}$ that we have that $a^\mu= \delta^\mu_0$ and that the probability density is the energy density. In general the quantity $j^{0}=\Theta^{0 \nu} a_{\nu}$ does not correspond to a known observable quantity. Even for another choice of the vector $n^\mu$ it is unclear to which known observable quantity $j^{0}=\Theta^{0 \nu} n_{\nu}$ could correspond.

Second, the current depends on the arbitrary choice of the observer (or for that matter on the arbitrary choice of the foliation), hence although the current $j^{\mu} =\Theta^{\mu \nu} a_\nu$ is written in a covariant form it is not covariant in content. For example if we consider two observers, $O$ and $O'$, which describe the same system relative to the frame at which they are at rest, then observer $O$ associate the following velocity field to the system
\begin{equation}
u^{\mu}(x) = \frac{\Theta^{\mu 0}(x)}{\sqrt{\Theta^{\nu 0}(x) \Theta_{\nu 0}(x)}} =  \frac{\Theta'^{\mu \rho}(x') h_{\rho}}{\sqrt{\Theta'^{\nu \alpha}(x') h_{\alpha} \Theta'_{\nu {\beta} }(x')h^{\beta} }},
\label{h2.9.01}
\end{equation}
where in the last equality we have written the vector $u^{\mu}(x)$ with respect to the observer $O'$; i.e.\ if $\Lambda$ denotes the passive Lorentz transformation from the frame of observer $O$ to that of observer $O'$, we have $\Theta'^{\mu_1 \mu_2}(x') = \Lambda^{\mu_1}_{\ \nu_1}  \Lambda^{\mu_2}_{\ \nu_2} \Theta^{\nu_1 \nu_2}(x)$ and $h^\mu = \Lambda^{\mu}_{\ \nu} \delta^{\nu}_0$. It is clear that this last expression does not equal the velocity field of the system relative to observer $O'$, which is 
\begin{equation}
u'^{\mu}(x') = \frac{\Theta'^{\mu 0}(x')}{\sqrt{\Theta'^{\nu 0}(x') \Theta'_{\nu 0}(x')}}. 
\label{h2.9.02}
\end{equation}
Similarly the two observers would also not agree on the particle probability densities if they described the system each relative to their rest frame. This implies that even at the standard quantum level the notion of the preferred observer would be present, which is in fact a sufficient reason to abandon the approach. A similar objection was raised by Bohm {\em et al.}\ in the context of the electromagnetic field \cite{bohm871,bohm5}.

This last objection could be removed by trying to find a `covariant' determination for the preferred observer or the preferred frame, in a way similar as was done in the multi-particle Dirac case by D\"urr {\em et al.}\ (cf.\ Section \ref{manyparticledirac}). However with the distinction that in the model of D\"urr {\em et al.}\ the notion of a preferred foliation was only present at the subquantum level of the pilot-wave interpretation (at the level of the particle beables) and not at the statistical level of the theory. In the model presented above, the current $j^{\mu}$ is fully determined by the choice of the 4-vector $n^{\mu}$, and hence although there may be some `covariant law' which determines this vector, it plays an important role in the statistical predictions of the model through the distribution $j^0= \Theta^{0 \mu}n_{\mu}$.

An example of such a covariant vector, is the total energy--momentum 4-vector
\begin{equation}
P^\mu = \int_\sigma {d} \sigma_\nu \Theta^{\mu \nu} = \int {d}^3 x \Theta^{\mu 0},
\label{h2.9.0201}
\end{equation}
with $\sigma$ an arbitrary spacelike hypersurface (the vector is independent of the choice of hypersurface due to the fact that the energy--momentum tensor is conserved). This vector transforms as a Lorentz 4-vector under Lorentz transformations and is conserved and future-causal (as the continuous sum over the future-causal vectors $\Theta^{\mu 0}$). We could then use the normalized constant 4-vector $n^{\mu}=P^\mu/\sqrt{P_\nu P^\nu}$ to contract the energy--momentum tensor with. Although the vector $P^\mu$ is nonlocally determined, we can introduce some notion of covariance in this way.{\footnote{For the massive Klein-Gordon field, Dewdney, Horton and Nesteruk developed still another covariant model starting from the energy--momentum tensor \cite{dewdney96,horton00,horton01,dewdney02}. In their model the flowlines were defined as the integral curves of the 4-vector $W^\mu$, with $W^\mu$ a future-causal eigenvector of the energy--momentum tensor and the probability density is given by the {\em intrinsic energy density}.}}

Finally, there is the problem that if we consider a massive boson interacting with an electromagnetic field, then although the total energy--momentum tensor of the massive boson and the electromagnetic field is conserved, they are not conserved separately. We discuss this in detail in Section \ref{minimallycoupleddkp}. Hence, if we would then construct the current $j^{\mu}= \Theta^{\mu \nu}_{DKP} n_\nu$ just as in the free case ($\Theta^{\mu \nu}_{DKP}$ is the energy--momentum tensor for the massive boson, which is unaltered in form after minimal coupling), then this current is no longer conserved. Although a particle interpretation of the current $j^{\mu}$ is still formally possible (the density is positive and the flowlines are future-causal), it would require the notion of particle creation and annihilation along the flowlines and hence this would contravene with the current experimental knowledge.{\footnote{This has of course nothing to do with pair creation, because even for energies below the threshold of pair creation, particle creation and annihilation as implied by a particle interpretation of the current $j^{\mu}$ would occur.}} 

We can conclude that, although it is already hard to maintain a particle interpretation of $j^{\mu}= \Theta^{\mu \nu} n_\nu$ even in the free case (due to the problems with Lorentz covariance and due to the fact that in general there does not correspond an empirically known quantity associated with the density $j^{0}$), it becomes in general untenable in the interacting case. A particle interpretation may perhaps only be maintained in the non-relativistic limit, with an electromagnetic field that is sufficiently weak. This is because in the non-relativistic limit, the current $j^{\mu}$ reduces to the same non-relativistic current as in the free current (see Section \ref{non-rellimtrajmodphot}), and hence in the non-relativistic limit the current $j^{\mu}$ is conserved again. Of course this reasoning does not apply for the electromagnetic field (or the equivalent Harish-Chandra field). However, we hold the view that in general the trajectory model should rather be regarded as describing the tracks of energy flow, as proposed by Holland, and that all the results reported for this trajectory model should be reinterpreted as such. For example, the trajectories for the electromagnetic field drawn in \cite{ghose01} for the double slit experiment{\footnote{Similar trajectories were presented in \cite{prosser76}, but outside the context of pilot-wave theory.}} and for reflection through a glass slab should be seen as the lines of energy flow.

The rest of the chapter is organized as follows. In the following section we give some remarks on the definition of an inner product. In Section \ref{non-rellimtrajmodphot} we consider the non-relativistic limit of the trajectory model. In Section \ref{many-particledkpformalism} we extend the trajectory model to many-particles. Finally, in Section \ref{minimallycoupleddkp} we consider the minimal coupling in the DKP theory.

\section{Note on the definition of the inner product}
For the moment we have mainly focussed on the possibility of constructing a conserved, future-causal current which could then be interpreted as a particle current. Related to the problem of finding such a current is the problem to define an inner product. Usually the inner product for the Kemmer theory is defined as \cite{kemmer39,krajcik74}
\begin{equation}
\langle \psi_1 | \psi_2 \rangle = \int_\sigma {d} \sigma_\mu {\bar \psi}_1 \beta^{\mu} \psi_2 = \int d^3 x {\bar \psi}_1 \beta^{0} \psi_2,
\label{h2.9.0201001}
\end{equation}
where $\sigma$ is an arbitrary spacelike hypersurface. The quantity $\langle \psi_1 | \psi_2 \rangle$ is inspired by the form of the charge current ({\ref{h2.6.101}}). This quantity is Lorentz invariant (it is hypersurface independent) and further satisfies all the requirements for an inner product, except for one, namely positivity. For an uncharged system the situation is somehow worse because then $\langle \psi | \psi \rangle=0$ for all wavefunctions $\psi$. The problem in defining a positive inner product is another problem which dissolves when passing to quantum field theory (e.g.\ in the functional Schr\"odinger picture in quantum field theory, the inner product of two wavefunctionals is well defined). 

In the previous section we have seen that the construction of a conserved, future-causal current from the energy--momentum tensor, as opposed to using the charge current, also has its problems. One can now consider the question whether one encounters the same problems when one tries to construct an inner product, starting from the energy--momentum tensor, i.e.\ inspired by the the current $j^\mu= \Theta^{\mu \nu}n_{\nu}$. The answer to this question is yes. Suppose we would define 
\begin{equation}
\langle \psi_1 | \psi_2 \rangle = \int_\sigma {d} \sigma_\mu {\bar \psi}_1  (\beta^{\mu}\beta^{\nu} +\beta^{\nu}\beta^{\mu} - g^{\mu \nu}) \psi_2 n_{\nu}.
\label{h2.9.0201002}
\end{equation}
In the case $n^{\mu}$ represents the normal on some preferred foliation, then one can check that the quantity $\langle \psi_1 | \psi_2 \rangle$  satisfies all the requirements for an inner product, however it is obviously not Lorentz-invariant because $n^{\mu}$ does not transform as a 4-vector under Lorentz transformations. Now if we would take a Lorentz 4-vector for $n^{\mu}$, e.g.
\begin{equation}
n^{\mu} \sim \int_\sigma {d} \sigma_\nu {\bar \psi}_1 (\beta^{\mu}\beta^{\nu} +\beta^{\nu}\beta^{\mu} - g^{\mu \nu}) \psi_2
\label{h2.9.0201003}
\end{equation}
then the quantity $\langle \psi_1 | \psi_2 \rangle$ would be Lorentz invariant and it would satisfy all requirements for an inner product, except for linearity.

\section{Non-relativistic limit of the trajectory model for massive spin-0 and spin-1 bosons}\label{non-rellimtrajmodphot}
In this section we consider the non-relativistic limit of the trajectory model for massive bosons. Because the DKP energy--momentum tensor reduces to the Klein-Gordon (KG) energy--momentum tensor in the spin-0 representation and to the Proca energy--momentum tensor in the spin-1 representation (cf.\ Section \ref{massivespin0-spin1dkpformalism}), we can simply consider the non-relativistic limits of the KG and Proca energy--momentum tensor. 

\subsection{The spin-0 case}
In order to take the non-relativistic limit of the Klein-Gordon equation ({\ref{h2.13.3}}), we substitute $\phi = e^{-imt} \psi'/\sqrt{2}m$, where the energy of the wavefunction $\psi'$ is much smaller than the rest energy $m$, in the KG equation. As is well known this leads to the Schr\"odinger equation for $\psi'$ (e.g.\ \cite{greiner90}). The KG energy--momentum tensor reduces to
\begin{eqnarray}
\Theta^{00}_{KG} &=& |\psi'|^2,\quad \Theta^{i0}_{KG} =-\frac{1}{m} \textrm{Im} \big( \psi'^* \partial_i \psi' \big),  \nonumber\\
\Theta^{ij}_{KG} &=& \frac{1}{m^2} \textrm{Re}(\partial_i \psi' \partial_j \psi'^*) + \delta_{ij} \bigg(\frac{\textrm{Im}(\psi'\partial_0\psi'^* ) }{m} - \frac{\partial_k \psi' \partial_k \psi'^*}{2m^2} \bigg)
\label{h2.17}
\end{eqnarray}
in the non-relativistic limit. Because $\Theta^{00}_{KG}$, $\Theta^{i0}_{KG}$, $\Theta^{ij}_{KG}$ are respectively of zero order, of first order and of third order in $p/m$ and because the components $n^i$ are at least of the same order as $n^0$ ($n^\mu$ is future-causal), the current $j^\mu= \Theta^{\mu \nu}_{KG} n_\nu$ reduces to the conventional Schr\"odinger current in the non-relativistic limit (up to the constant factor $n^0$, which can be removed by renormalizing $\psi'$)
\begin{eqnarray}
j^0  \! &=& \Theta^{00}_{KG} = |\psi'|^2,  \nonumber\\
j^i &=& \Theta^{i0}_{KG} = \frac{1}{m} \textrm{Im} ( \psi'^*  \partial_i \psi' ).
\label{h2.17.01}
\end{eqnarray}
In this way the trajectory model associated with the energy--momentum tensor reduces to the pilot-wave interpretation for the non-relativistic Schr\"o\-ding\-er equation originally presented by de Broglie and Bohm (cf. Section \ref{examplespwinonrelativisticlimit}).

\subsection{The spin-1 case}
In order to take the non-relativistic limit in the massive spin-1 case, we write the Proca equations as a KG equation for each component of the field $A^\mu$
\begin{equation}
\square A^\mu + m^2 A^\mu=0
\label{h2.24.0}
\end{equation}
with subsidiary condition
\begin{equation}
\partial_\mu A^\mu =0.
\label{h2.24.01}
\end{equation}
We again separate the rest energy by putting $A^\mu = e^{-imt}\phi^\mu /\sqrt{2} m$.
The condition $\partial_\mu A^\mu =0$ then reduces to $\phi^0 = \partial_i \phi^i/im$, which implies that we can take $\phi^0$ as the {\em small component} of the wavefunction $\phi^\mu$. The non-relativistic limit of equation ({\ref{h2.24.0}}) results in the Schr\"odinger equation for each component $\phi^\mu$. If we define the wavefunction $\Phi = (\phi^1,\phi^2,\phi^3)^T$, then the non-relativistic limit of the Proca equations can be written as
\begin{equation}
i\frac{\partial \Phi}{\partial t}  = -\frac{\nabla^2 \Phi}{2m} .
\label{h2.24.02}
\end{equation}  
With similar assumptions on the vector $n^{\mu}$ as in the spin-0 case, the spin-1 current $j^{\mu} = \Theta^{\mu \nu}_P n_{\nu}$ reduces to the non-relativistic current for a spin-1 particle as given in ({\ref{h1.0090101}}) in the non-relativistic limit:
\begin{eqnarray}
j^0  \! &=& \Theta^{00}_{P} = \Phi^{\dagger} \Phi \nonumber\\
j^i &=& \Theta^{i0}_{P} = \frac{1}{m} \textrm{Im} \big(\Phi^{\dagger}  \partial_i  \Phi  + \phi^{i*} \partial_j  \phi^j -\phi^{j*}  \partial_j \phi^i \big).
\label{h2.24.031001012}
\end{eqnarray}

As a side remark, we note that in the non-relativistic limit, the DKP charge current also reduces to the non-relativistic currents which were presented in Section \ref{examplespwinonrelativisticlimit}, in both the spin-0 case and the spin-1 case.

\section{The many-particle Duffin-Kemmer-Petiau formalism}\label{many-particledkpformalism}
In this section we generalize the one-particle trajectory model to many particles. The trajectory model is constructed in close analogy with the pilot-wave model for the many-particle Dirac equation. The difference is that the trajectory model for bosons is constructed from the energy--momentum tensor and not from the multi-particle charge tensor. As indicated in the one-particle case, these boson trajectories should not be regarded as genuine particle trajectories but as representing flowlines of energy. The multi-particle generalization of the HC theory can be constructed analogously, but we will not consider this here.

The single-particle DKP theory can be extended to a $N$-particle system as follows. The wavefunction for the $N$-particle system is defined to be $\psi_{r_1 \dots r_N }({\bf x}_1, \dots,{\bf x}_N, t)$, with the $r_i$ denote $N$ spin indices. The wavefunction is assumed to be symmetric under permutations of the particle labels. We also introduce the operators $\beta_{(r)}^{\mu}$ which operate only on the $r^{\textrm{th}}$ spin index. The wave equation for the $N$-particle wavefunction is defined to be  
\begin{eqnarray}
i \partial_{0} \psi &=& \sum^N_{r = 1} \Big(-i {\tilde \beta}_{(r)}^{i} \partial^{(r)}_{i} + \beta_{0}^{(r)} m_{r}   \Big) \psi, \label{h2.26.1}\\
i \beta_{(r)}^{i} \big(\beta^{(r)}_{0} \big)^2 \partial^{(r)}_{i}\psi &=&  m_{r} \left[ 1 -  \big(\beta^{(r)}_{0}\big)^2 \right] \psi,
\label{h2.26.2} 
\end{eqnarray}
where $\partial^{(r)}_{i} = \frac{\partial}{\partial({\bf x}_{r})^i}$ and $m_{r}$ is the mass of the $r^{\textrm{th}}$ particle. 

The equations ({\ref{h2.26.1}}) and ({\ref{h2.26.2}}) are a straightforward generalization of the Schr\"odinger form of the one-particle DKP equation. It can easily be verified that a wavefunction $\psi$, constructed as an arbitrary superposition of direct products of one-particle DKP wavefunctions at equal time, obeys the many-particle DKP equations ({\ref{h2.26.1}}) and ({\ref{h2.26.2}}). Conversely, $\psi$ can only be written in such a form. 

In order to construct the multi-particle generalization of the one-particle energy--momentum tensor ({\ref{h2.6}}) we first define the following operator:
 \begin{equation}
\Gamma_{(r)}^{\mu \nu} =m_r \eta^{(r)}_0 (\beta_{(r)}^{\mu}\beta_{(r)}^{\nu} +\beta_{(r)}^{\nu}\beta_{(r)}^{\mu} - g^{\mu \nu}).
\label{h2.26.3}
\end{equation}
The multi-particle energy--momentum tensor of rank $2N$ then reads (see also \cite{ghose01})
\begin{equation}
\Theta^{\mu_1 \dots \mu_{2N}}_{DKP} = \psi^{\dagger}\Gamma_{(1)}^{\mu_1 \mu_2} \dots \Gamma_{(N)}^{\mu_{2N-1} \mu_{2N}}\psi
\label{h2.27}
\end{equation}
and satisfies the conservation equation
\begin{equation}
\partial_0 \Theta^{0_1 \nu_1 \dots 0_N \nu_{N}}_{DKP} + \sum^N_{r = 1}  \partial^{(r)}_{i_r} \Theta^{0_1 \nu_1 \dots i_{r} \nu_{r} \dots 0_N \nu_N}_{DKP} = 0.
\label{h2.27.1}
\end{equation}

By contracting the energy--momentum tensor $\Theta^{\mu_1 \dots \mu_{2N}}$ with a constant tensor $n^{\mu_1 \dots \mu_N}$ of rank $N$ we can construct a rank $N$ tensor current
\begin{equation}
j^{\mu_1 \dots \mu_N} = \Theta^{\mu_1 \nu_1 \dots \mu_{N} \nu_{N}}_{DKP} n_{\nu_1 \dots \nu_N}
\label{h2.28}
\end{equation}
which satisfies the conservation equation
\begin{equation}
\partial_0 j^{0_1 \dots 0_N} + \sum^N_{r = 1}  \partial^{(r)}_{i_r} j^{0_1 \dots i_{r} \dots 0_N} = 0.
\label{h2.29}
\end{equation}
This is the multi-particle generalization of the one-particle current $j^{\mu}$ defined in ({\ref{h2.7}}). In order to be able to define causal trajectories we require that the tensor $n^{\mu_1 \dots \mu_N}$ is such that the vectors $j^{0_1 \dots \mu_{r} \dots 0_N}$ are future-causal for every $r = 1, \dots ,N$. We will give examples below. The trajectories $x^{\mu}_k(\tau)$  $(k=1,\dots,N)$ can then be defined as solutions to the `guidance equations'
\begin{equation}
\frac{{d} x^\mu_{r}}{ {d} \tau} = \frac{j^{0_1 \dots \mu_{r} \dots 0_N}}{j^{0_1 \dots 0_N}}.
\label{h2.30}
\end{equation}
Because the vectors $j^{0_1 \dots \mu_{r} \dots 0_N}$ are assumed to be future-causal, the trajectories will be time-like or null. The density of crossings through constant time hyperplanes is defined to be the positive quantity $j^{0_1 \dots 0_N}$.

Examples for the tensor $n^{\mu_1 \dots \mu_N}$ are generated by considering generalizations of the vector $n^{\mu}$ in the one-particle case. If $a^{\mu}$ is the constant future-causal 4-velocity of a particular observer, then we can take $n^{\mu_1 \dots \mu_N} = a^{\mu_1} \dots a^{\mu_N}$. The proof that $j^{0_1 \dots \mu_{r} \dots 0_N}$ is future-causal for every $r = 1, \dots ,N$ runs as follows. In Section \ref{pwiinterpretation} it was shown that the operator $\Gamma^{(r)}_{0 \mu} a^{\mu}$ is positive in spin space $\mathbb{C}^M$, where $M$ is the dimension of the representation of the $\beta$ matrices. As a result the operator
\begin{equation}
\Gamma= \Gamma^{(1)}_{0 \nu_1} a^{\nu_1} \dots \widehat{\Gamma^{(r)}_{0 \nu_{r}} a^{\nu_{r}}} \dots \Gamma^{(N)}_{0 \nu_{N}}a^{\nu_{N}},
\label{h2.30.00101}
\end{equation}
where the hat indicates that the term should be omitted from the product, is a positive operator in $N-1$ particle spin space $(\mathbb{C}^M )^{\otimes (N-1)}$. Consequently there exists an operator $\Omega$ in $N-1$ particle spin space such that $\Gamma = \Omega^{\dagger} \Omega$. As a result one can write
\begin{equation}
j^{0_1 \dots \mu_{r} \dots 0_N} = (\Omega \psi)^{\dagger} \Gamma_{(r)}^{\mu_{r}\nu_{r} } a_{\nu_{r}}  (\Omega \psi).
\label{h2.30.00102}
\end{equation}
This shows that $j^{0_1 \dots \mu_{r} \dots 0_N}$ is the sum of $M(N-1)$ (sum over all but the $r^{\textrm{th}}$ spin index of $\Omega \psi$) vectors of the form $\Psi^{\dagger}\Gamma_{(r)}^{\mu_{r}\nu_{r} } a_{\nu_{r}} \Psi$. Because each such term is future-causal the sum $j^{0_1 \dots \mu_{r}\dots 0_N}$ is also future-causal. 

The one-particle vector $P^{\mu}$ defined in ({\ref{h2.9.0201}}) is generalized to the many-particle case by
\begin{equation}
P^{\mu_1 \dots \mu_N} = \int {d} x^3_1 \dots \int {d} x^3_N  \Theta^{\mu_1 0_1 \dots \mu_{N} 0_N}_{DKP}. 
\label{h2.27.2}
\end{equation}
The proof that $P^{\mu_1 \dots \mu_N}$ leads to future-causal vectors $j^{0_1 \dots \mu_{r} \dots 0_N}$ proceeds essentially in the same way as in the first example, however we will not present it here.

One can easily show that for both definitions of $n^{\mu_1 \dots \mu_N}$ the trajectory laws for a product state 
\begin{equation}
\psi_{r_1 \dots r_N }({\bf  x}_1, \dots,{\bf x}_N, t) =\psi_{1,r_1} ({\bf  x}_1,t) \dots \psi_{N,r_N} ({\bf  x}_N,t)
\label{h2.27.201}
\end{equation}
reduce to the one-particle trajectory laws. I.e.\ for a product state, the density of crossings through constant time hyperplanes is given by the product of one-particle densities, i.e.\ $j^{0_1 \dots 0_N} =j^{0_1}_1 \dots j^{0_1}_N$ with $j^{0_\alpha}_\alpha =\psi^{\dagger}_{\alpha} \Gamma^{0\mu} n_\mu \psi_{\alpha}$, with $n^{\mu}$ the vector defined in the one-particle case. Second, the velocity field for each particle reduces to the one-particle velocity field defined in ({\ref{h2.8}}).

In the special case that $n^{\mu_1 \dots \mu_N} = \delta^{\mu_1}_0 \dots \delta^{\mu_N}_0$ the density of crossings becomes $j^{0_1 \dots 0_N} = (\Pi_r m_r)\psi^{\dagger} \psi$ and the guidance equations become
\begin{equation}
\frac{{d} x^i_{r}}{ {d} t} = \frac{\psi^{\dagger} {\tilde \beta}_{(r)}^i \psi}{\psi^{\dagger} \psi}.
\label{h2.30.002}
\end{equation}
The resulting conservation equation ({\ref{h2.29}}), which can also be directly derived from ({\ref{h2.26.1}}), turns into 
\begin{equation}
\partial_0(\psi^{\dagger} \psi)  + \sum^N_{r = 1}  \partial^{(r)}_{i} (\psi^{\dagger} {\tilde \beta}_{(r)}^i \psi)  = 0.
\label{h2.30.1}
\end{equation}
In this case the trajectory model displays a formal resemblance to the pilot-wave equations for the multi-particle Dirac equation as presented in Section \ref{manyparticledirac}.

We also consider the non-relativistic limit of the trajectory model. The non-relativistic limit of the multi-particle DKP equations reduces to the non-relativistic Schr\"odinger equation for $N$ spinless particles in the spin-0 representation and to the non-relativistic Schr\"odinger equation for $N$ spin-1 particles in the spin-1 representation. As in the one-particle case, the trajectory model constructed through the energy--momentum tensor reduces to the many-particle pilot-wave interpretations for spin-0 and spin-1 in the respective representations. In particular, in the spin-0 representation, the `guidance equations' ({\ref{h2.30}}) reduce to the guidance equations originally presented by de Broglie and Bohm (cf. Section \ref{examplespwinonrelativisticlimit})
\begin{equation}
\frac{{d} {\bf x}_{r}}{ {d} t} = \frac{\textrm{Im} \big( \psi'^* {\boldsymbol \nabla}_{r} \psi' \big)}{m |\psi'|^2}, 
\label{h2.35}
\end{equation}
with $\psi'( {\bf x}_1 ,\dots , {\bf x}_N)$ the non-relativistic $N$-particle wavefunction. In the spin-1 representation the guidance equations become
\begin{equation}
\frac{{d} {\bf x}_{r}}{ {d} t} =  \frac{\textrm{Im} ( \Phi^{\dagger}  {\boldsymbol \nabla}_{r}  \Phi)}{m \Phi^{\dagger} \Phi} + \frac{{\boldsymbol \nabla}_{r} \times (\Phi^{\dagger} {\bf  S}^{(1)}_{r} \Phi )}{2m\Phi^{\dagger} \Phi},
\label{h2.36}
\end{equation}
with $ {\bf  S}^{(1)}_{r}$ operating only on the $r^{\textrm{th}}$ spin index of the non-relativistic multi-particle wavefunction $\Phi_{r_1 ,\dots ,r_N} ( {\bf x}_1 ,\dots , {\bf x}_N,t)$, where each spin index $r_i$, $i=1,\dots,N$, runs from 1 to 3. Note the spin contribution, which is similar to the spin contribution in the one-particle spin-1 guidance equation ({\ref{h1.00901}}).

The multi-particle DKP charge tensor current is defined as 
\begin{equation}
s^{\mu_1 \dots \mu_N}_{DKP} = \psi^{\dagger}\eta_{(1)}^0 \beta_{(1)}^{\mu_1} \dots \eta_{(N)}^0 \beta_{(N)}^{\mu_N}  \psi
\label{h2.37}
\end{equation}
and satisfies the conservation equation
\begin{equation}
\partial_0 s^{0_1 \dots 0_N}_{DKP} + \sum^N_{r = 1}  \partial^{(r)}_{i} s^{0_1 \dots i_{r} \dots 0_N}_{DKP} = 0.
\label{h2.38}
\end{equation}
Just as in the one-particle case, the vectors $s^{0_1 \dots \mu_r \dots 0_N}_{DKP}$ have the same non-relativistic limit as the vectors $j^{0_1 \dots \mu_r \dots 0_N}$. 

\section{The minimally coupled Duffin-Kemmer-Petiau theory}\label{minimallycoupleddkp}
The minimally coupled DKP Lagrangian is obtained from the free DKP Lagrangian by applying the minimal coupling prescription $\partial_{\mu} \to D_{\mu} = \partial_{\mu} + ie V_{\mu}$, with $V^\mu=(V_0,{\bf V})$, and yields \cite{kemmer39}
\begin{equation}
\mathcal{L}_{DKP} = \frac{i}{2} ({\bar \psi} \beta_{\mu}D^{\mu} \psi  - D^*_{\mu}  {\bar \psi} \beta^{\mu} \psi ) -m{\bar \psi} \psi.
\label{h2.25.071001}
\end{equation}
The corresponding coupled wave equation reads
\begin{equation}
(i\beta^{\mu} D_{\mu} - m )\psi = 0.
\label{h2.25.07100101}
\end{equation}
The symmetrized energy--momentum tensor is the same as in the free case
\begin{equation}
\Theta^{\mu \nu}_{DKP} =  m\psi^{\dagger}  \eta_0 (\beta^{\mu}\beta^{\nu} +\beta^{\nu}\beta^{\mu} - g^{\mu \nu}) \psi.
\label{h2.25.071002}
\end{equation}
However, the energy--momentum tensor is no longer conserved, but instead satisfies the equation
\begin{equation}
\partial_\mu \Theta^{\mu \nu }_{DKP} = F^{ \nu}_{\ \mu} s_{DKP}^\mu,
\label{h2.25.071003}
\end{equation}
with $s^{\mu}_{DKP}$ the charge current, which is formally the same as in the free case, and $F^{\mu  \nu }= \partial^{\mu} V^{\nu} - \partial^{\nu} V^{\mu}$ the electromagnetic tensor. The right hand side of the equation ({\ref{h2.25.071003}}) is recognized as the Lorentz force. Hence, if we would now construct the current $j^{\mu}= \Theta^{\mu \nu}_{DKP} n_\nu$, as in the free case, then this current is no longer conserved and a particle interpretation becomes untenable.

It is interesting to note that Ghose {\em et al.}\ \cite{ghose93,ghose94,ghose96} considered the energy--momentum tensor $\Theta^{\mu \nu}_{DKP}$ to be conserved when an interaction with an electromagnetic field $V^{\mu}$ is introduced via minimal coupling. Therefore they thought that even in the case of an interaction a particle interpretation was possible. The reason for the discrepancy is that they introduced minimal coupling in the DKP theory in an other way than presented above. 

Let us consider this in more detail. As explained above one can introduce minimal coupling at the level of the covariant form of the DKP equation ({\ref{h2.25.07100101}}). The covariant equation can also be written in the following Schr\"odinger form
\begin{eqnarray}
i D_{0} \psi &=& (-i{\tilde \beta}^{i} D_{i} + m\beta_{0} ) \psi  - \frac{ie}{2m} F^{\mu \nu}(\beta_{\nu}\beta_{0}\beta_{\mu} + \beta_{\nu} g_{\mu 0})\psi, \label{h2.25.071004}\\
i \beta^{i} \beta^2_{0} D_{i}\psi &=& m (1 -  \beta^2_{0}) \psi.
\label{h2.25.071005} 
\end{eqnarray}
On the other hand, as proposed by Ghose {\em et al.}, one can introduce minimal coupling at the level of the Schr\"odinger form of the DKP theory (cf.\ ({\ref{h2.13.1}}) and ({\ref{h2.13.2}})) which results in 
 \begin{eqnarray}
i D_{0} \psi &=& (-i{\tilde \beta}^{i} D_{i} + m\beta_{0}   ) \psi, \label{h2.25.071006}\\
i \beta^{i} \beta^2_{0} D_{i}\psi &=& m (1 -  \beta^2_{0}) \psi.
\label{h2.25.071007} 
\end{eqnarray}
The difference is clear, the term involving $F^{\mu \nu}$ in the Schr\"odinger form ({\ref{h2.25.071004}}) is not present in ({\ref{h2.25.071006}}). This additional term has no equivalent in the spin-1/2 Dirac theory and is hard to interpret \cite{kemmer39,nowakowski98,lunardi00}. It has recently been argued that this additional term is irrelevant \cite{nowakowski98,lunardi00}. The argument is that when the DKP theory is reduced to its physical components, then the DKP theory reduces to the minimally coupled Klein-Gordon theory in the spin-0 case and to the minimally coupled Proca theory in the spin-1 case, where the anomalous term containing $F^{\mu \nu}$ disappears. However, this does not settle the question whether or not we can safely introduce minimal coupling at the level of the Schr\"odinger form in the DKP theory. We argue that in general the only correct way to introduce minimal coupling is at the level of the covariant form of the DKP equations. 

 The minimally coupled Schr\"odinger form implies the `covariant form' ({\ref{h2.25.07100101}}), which can be seen by multiplying ({\ref{h2.25.071006}}) with $\beta_0$ and by adding ({\ref{h2.25.071007}}) to it \cite{ghose96}. Hence, if the minimally coupled Schr\"odinger form is regarded as fundamental, then both the equations ({\ref{h2.25.071004}}) and ({\ref{h2.25.071006}}) should be valid. This implies that the additional term containing the tensor $F^{\mu \nu}$ should be zero. Without considering an explicit representation we can already show that this implies that the introduction of minimal coupling at the level of the Schr\"odinger form is in general untenable. 

If minimal coupling is introduced at the level of the Schr\"odinger form of the DKP equation, we may derive from ({\ref{h2.25.071006}}) (by multiplying ({\ref{h2.25.071006}}) by $\psi^{\dagger}$ from the right, multiplying the conjugate of ({\ref{h2.25.071006}}) by $\psi$ from the left, and subtracting the two from each other \cite{ghose96}) that
\begin{equation}
\partial_0 (\psi^{\dagger}\psi ) +\partial_i (\psi^{\dagger}  {\tilde \beta}^i  \psi ) = 0
\label{h2.25.9}
\end{equation}
or differently written
\begin{equation}
\partial_\mu \Theta^{\mu 0 }_{DKP} = 0.
\label{h2.34.02}
\end{equation}
If we assume that the the Schr\"odinger form of the DKP equation is covariant,{\footnote{Although a pilot-wave model may be non-covariant at the subquantum level, it is desired that we have covariance at the quantum level, therefore the wave equation should be covariant.}} i.e.\ if we assume that the wave equation has the form ({\ref{h2.25.071006}}) in every Lorentz frame, then we have $\partial_\mu \Theta^{\mu 0 }_{DKP} = 0$ in every Lorentz frame. Because $\partial_\mu \Theta^{\mu \nu }_{DKP}$ transforms as a 4-vector under Lorentz transformations, this implies that the time component of $\partial_\mu \Theta^{\mu \nu }_{DKP}$ is zero in every Lorentz frame. Because the only 4-vector which satisfies this property is the zero vector, we have
\begin{equation}
\partial_\mu \Theta^{\mu \nu }_{DKP} = 0.
\label{h2.34.021}
\end{equation}
It was this equation which led Ghose {\em et al.}\ \cite{ghose93,ghose94,ghose96} to conclude that a particle interpretation associated with the energy--momentum tensor was still possible in the presence of an external field. However, because the mi\-ni\-mally coupled Schr\"odinger form implies the minimally coupled covariant form, both ({\ref{h2.25.071003}}) and ({\ref{h2.34.021}}) should be valid when the minimally coupled Schr\"odinger form is regarded as fundamental. This implies however that the Lorentz force is zero, i.e.\
\begin{equation}
F^{ \nu}_{\ \mu} s_{DKP}^\mu = 0,
\label{h2.34.023}
\end{equation}
which is in general not the case. From this we may conclude that minimal coupling should be introduced at the level of the covariant form of the DKP equation.

The discrepancy between the introduction of minimal coupling at the level of the covariant form of the DKP equation and the introduction of minimal coupling at the level of the Schr\"odinger form of the DKP equation does not disappear when the theory is reduced to its physical components. Consider for example the case of spin-0. In the same way as was explained in Section \ref{massivespin0-spin1dkpformalism}, using the explicit representation in Appendix \ref{appa}, we can reduce the theory to its physical components. By using the explicit representation, it follows from both equations ({\ref{h2.25.071004}}) and ({\ref{h2.25.071006}}) that the DKP wavefunction $\psi$ can be written in terms of a remaining physical component $\phi$:
\begin{equation}
\psi = \frac{1}{\sqrt{m}}\left( \begin{array}{c}
 D_\mu \phi   \\
m \phi \\
\end{array} \right)
\label{h2.34.03}
\end{equation}
where $\phi$ satisfies the minimally coupled Klein-Gordon equation
\begin{equation}
(D_{\mu}D^{\mu} + m^2)\phi = 0.
\label{h2.34.031}
\end{equation}
Hence, both ways of introducing minimal coupling then lead to the minimally coupled Klein-Gordon theory \cite{nowakowski98}. When the equation ({\ref{h2.25.071003}}) is written in terms of the physical component $\phi$ by making the substitution ({\ref{h2.34.03}}), then we obtain
\begin{equation}
\partial_\mu \Theta^{\mu \nu }_{KG} = F^{\nu}_{\ \mu} s_{KG}^\mu,
\label{h2.34.033}
\end{equation}
with  
\begin{equation}
\Theta^{\mu \nu}_{KG} =D^{\mu} \phi (D^{\nu} \phi)^* + (D^{\mu} \phi)^* D^{\nu} \phi - g^{\mu \nu} \big(D_{\alpha} \phi (D^{\alpha} \phi)^* - m^2 \phi^* \phi \big)
\label{h2.34.04}
\end{equation}
the Klein-Gordon energy--momentum tensor and 
\begin{equation}
s^{\mu}_{KG} =  ie\big(\phi^* D^{\mu}\phi - (D^{\mu}\phi)^* \phi \big) 
\label{h2.34.041}
\end{equation}
the charge current in the minimally coupled Klein-Gordon theory. On the other hand, when ({\ref{h2.34.021}}) is written in terms of the physical component $\phi$ we obtain 
\begin{equation}
\partial_\mu \Theta^{\mu \nu }_{KG} =0.
\label{h2.34.042}
\end{equation}
Because only ({\ref{h2.34.033}}), and not ({\ref{h2.34.042}}), can be derived from the minimally coupled Klein-Gordon equation ({\ref{h2.34.031}}), we should introduce minimal coupling at the level of the covariant form of the DKP equation and not at the level of the Schr\"odinger form of the DKP equation. 

Note that it is a general property for charged matter, that the matter energy--momentum tensor (even for fermions) is not conserved when an external electromagnetic field is introduced. This is because a charged particle exchanges energy and momentum when interacting with an electromagnetic field. On the other hand, the charge current is always conserved for bosons and for fermions, because the electromagnetic field does not carry charge. This is the reason why it presented no problem to maintain the particle interpretation in the Dirac theory, which was associated with the charge current and not with the energy--momentum tensor, after an electromagnetic field was introduced.

From our analysis it also follows that the physical interpretation of the term containing $F^{\mu \nu}$ in the Schr\"odinger form ({\ref{h2.25.071004}}) may perhaps be sought in the fact that it contributes to the Lorentz force.

\newpage

\noindent \\
\chapter{Field beables for bosonic quantum field theory}\label{chapter4}
\section{Introduction}
The construction of a pilot-wave for bosonic quantum field theories has been discussed before in a number of papers. Already in his seminal paper \cite{bohm2} in 1952, Bohm presented a pilot-wave interpretation for the free electromagnetic field. Later, in the 80's and 90's this pilot-wave interpretation was further elaborated on by Kaloyerou \cite{kaloyerou85,kaloyerou94,kaloyerou96}. In 1992, Valentini \cite{valentini92,valentini96,valentini04} presented a different approach for the electromagnetic field. In the meanwhile the free massless scalar field was treated by Bohm and Hiley \cite{bohm84} (later reviews of their treatment can be found in \cite{bohm872,bohm5,holland,holland93}). The extension to a free massive scalar field was given independently, and along the same lines, by Kaloyerou \cite{kaloyerou85} and Valentini \cite{valentini92,valentini96}. In \cite{valentini92,valentini96,valentini04}, Valentini further considered the pilot-wave interpretation for the massive scalar field coupled to the electromagnetic field.

All these authors introduced fields as beables in the pilot-wave interpretation. As explained in the previous chapter, we also favour the field beable approach for quantum field theory. So it is along these lines that we try to develop the pilot-wave interpretation further. 

We start with considering the pilot-wave interpretation for the quantized massive spin-0 and spin-1 field coupled to a non-quantized external electromagnetic field. Instead of simply presenting the pilot-wave interpretation for the quantized Klein-Gordon theory or the quantized Proca theory, we will take a different approach. We will start with the Duffin-Kemmer-Petiau (DKP) theory. First we will run through the quantization procedure, i.e.\ we will first apply the rules of canonical quantization, as set out by Dirac, in order to quantize the DKP theory. Then only afterwards we will present the corresponding pilot-wave theory. The reason to do so is twofold. 

First, it was reported in the literature that the equivalence of the quantized DKP theory and the quantized Klein-Gordon theory in the spin-0 representation or the quantized Proca theory in the spin-1 representation is not that obvious, not even when only electromagnetic interaction is considered \cite{krajcik772,fainberg001} (although the equivalence as wave equations is well established). However, we show that, by using Dirac's recipe of canonical quantization, it is straightforward to show the equivalence (in the case of spin-0 the equivalence was also shown by Fainberg and Pimentel by using similar arguments \cite{fainberg001}). Once the DKP theory is quantized it is no problem to provide a pilot-wave interpretation in terms of field beables. But because of the equivalence, we could equally well have started from the quantized Klein-Gordon theory or quantized Proca theory from the start. In the case of a free spin-0 field the pilot-wave interpretation then of course reduces to the one originally presented by Kaloyerou and Valentini.

The second reason to go through the canonical quantization procedure for the DKP theory, is that it is a good example of how the quantization program of Dirac works. When we discuss the pilot-wave interpretation for the electromagnetic field, we will need to appeal to the canonical quantization program again. Only, in this case, slight complications arise due to the fact that we are dealing with a gauge theory. The canonical quantization of the electromagnetic field is of course well know, but it is instructive to recall it. This mean reason to do this is that the two existing approaches to a pilot-wave interpretation for the electromagnetic field, namely the one by Bohm and Kaloyerou, and the one by Valentini, find a natural home in two different ways of quantizing theories with gauge symmetries. Against the background of canonical quantization it is then easy to compare the two approaches. 

A careful look at Valentini's model reveals that it suffers from a problem. Namely the densities of field beables are non-normalizable. This problem is not present in the model by Bohm and Kaloyerou. The reason is that Valentini introduces beables corresponding to gauge degrees of freedom, whereas Bohm and Kaloyerou only introduce beables for gauge independent degrees of freedom. 

After presenting the pilot-wave interpretation for the free electromagnetic field we then consider the pilot-wave interpretation for the quantized Klein-Gordon field coupled to quantized electromagnetic field (scalar quantum electrodynamics). Scalar quantum electrodynamics was first treated by Valentini \cite{valentini92}, but in the model that we present here, beables are introduced only for gauge invariant degrees of freedom and hence our model does not suffer from the problem of non-normalizable field beable densities.

This is the organization of the chapter. We start with a review of Dirac's procedure of canonical quantization in Section \ref{quantizingconstrainedsystems}. In Sections \ref{Massive spin-0} and \ref{Massive spin-1}, we consider the quantization of the quantized DKP field coupled to a non-quantized electromagnetic field in respectively the spin-0 and the spin-1 representation, together with the corresponding pilot-wave interpretation. In Section \ref{maxwell}, we treat the quantization of the electromagnetic field and we discuss in detail the model of Bohm and Kaloyerou and the model of Valentini. Then, in Section \ref{scalarqed}, we consider the pilot-wave interpretation for scalar quantum electrodynamics. In Section \ref{anoteonthequantizationnon-Abeliangaugetheories} we discuss the possibility of constructing a pilot-wave interpretation for non-Abelian gauge theories. We end the chapter with a discussion on how the pilot-wave interpretation in terms of field beables solves the measurement problem.

The main conclusion of this chapter is that it presents no problem to construct a pilot-wave interpretation in terms of field beables for bosonic quantum field theory. Even for gauge theories we can develop a pilot-wave interpretation. This stands in sharp contrast with the problems we encountered when we tried to develop a pilot-wave interpretation for relativistic wave equations for bosons. However, in the following chapter, where we consider fermionic field theory, it will appear much more difficult to construct a field beable model for fermionic fields.

\section{Canonical quantization of constrained systems}\label{quantizingconstrainedsystems}
In this section we review Dirac's procedure of canonical quantization of a constrained system. This review is mainly based on Dirac's original presentation \cite{dirac64} and the book by Henneaux and Teitelboim \cite{henneaux91} (which closely follows Dirac's original presentation). We also consulted the book by Sundermeyer \cite{sundermeyer82} and the one by Gitman and Tyutin \cite{gitman90}. A short introduction on canonical quantization can be found in Weinberg's book \cite[pp.\ 325-330]{weinberg95}. Although this section is self-contained, we only review the essential ingredients of canonical quantization. We refer the interested reader to the aforementioned books for a more detailed treatment. The reader which is familiar with Dirac's procedure of canonical quantization can skip this review.

\subsection{Hamiltonian formulation of a constrained system}\label{introductionquantizingconstrainedsystems}
For simplicity we present Dirac's analysis for a system with a finite number of degrees of freedom. In Section \ref{quantizationofafieldtheory} we indicate how the transition to a continuous number of degrees of freedom can be made.

We assume that the dynamics can be derived from the action
\begin{equation}
S= \int d t L(q,\dot{q}). 
\label{h30.0000001}
\end{equation}
Here $L(q,\dot{q})$ is a Lagrangian which is function of the coordinates $q_n$, where $n=1,\dots,N$, and the corresponding velocities $\dot{q} = d q/ d t$. By requiring that the action $S$ be stationary with respect to variations in the coordinates $q_n$ we obtain the Euler-Lagrange equations of motion
\begin{equation}
\frac{d }{dt} \frac{\partial  L}{\partial  {\dot q}_n } - \frac{\partial  L}{\partial  q_n} = 0.
\label{h30.000001}
\end{equation}

In order to quantize the system, we have to move from the Lagrangian formulation, where the dynamical variables are the velocity phase-space variables $q_n$ and $\dot{q}_n$, to the Hamiltonian formulation, where the dynamical variables are the momentum phase-space variables $q_n$ and $p_n$. Canonical quantization then proceeds by associating operators with the momentum phase-space variables and by imposing certain commutation relations for these operators. Dirac's procedure of canonical quantization provides a scheme for imposing these commutation relations in a way consistent with the various constraints that may arise in the Hamiltonian formulation.

In order to arrive at the Hamiltonian formulation we first need the momenta $p_n$ canonically conjugate to the coordinates $q_n$. These are defined as
\begin{equation}
p_n = \frac{\partial  L}{\partial {\dot q}_n }(q,\dot{q}).
\label{h30.00001}
\end{equation}
If the rank of the Hessian matrix 
\begin{equation}
\frac{\delta^2  L}{\partial {\dot q}_{n'} \partial {\dot q}_n }(q,\dot{q})
\label{h30.00003}
\end{equation}
is maximal at each point in velocity phase-space, then we can write {\em all} the velocities $\dot{q}_n$ as functions of the momenta and the coordinates. In this case the mapping from velocity phase-space variables to momentum phase-space variables is invertible. If the rank is not maximal, then this mapping is not invertible. In this case the momenta ({\ref{h30.00003}}) are not all independent, but there are, rather, some relations 
\begin{equation}
\chi_m(q,p) = 0, \quad m=1,\dots,M  \leqslant N.
\label{h30.00004}
\end{equation}
These relations are called the {\em primary constraints}. 

For simplicity we have hereby assumed that the rank of the Hessian matrix is constant throughout velocity phase-space so that the constraints can be written in the particular form ({\ref{h30.00004}}). For future convenience we also assume that the rank of $\partial \chi_m/\partial(q_n, p_{n'})$ is $M$ throughout velocity phase-space. This condition ensures that all primary constraints are lineary independent. It also excludes the use of equivalent sets of constraints, such as e.g.\ $\chi^2_m(q,p) = 0$, $m=1,\dots,M$. 

We proceed by defining the {\em canonical Hamiltonian} 
\begin{equation}
H_C = p_n \dot{q}_n - L,
\label{h30.00004001}
\end{equation}
which is a function of the $q_n$ and the $\dot{q}_n$. By using the definition of the momenta and by using the Euler-Lagrange equations of motion, we find that the canonical Hamiltonian varies as
\begin{equation}
\delta H_C = \sum^N_{n=1} (\delta p_n  \dot{q}_n - \delta q_n \dot{p}_n)
\label{h30.00005}
\end{equation}
under infinitesimal variations of the coordinates and the velocities. Because $\delta H_C$ does not contain variations in the velocities $\dot{q}_n$, we can express $H_C$ solely in terms of the variables $q_n$ and $p_n$, i.e.\ $H_C = H_C(q,p)$. 

The momentum phase-space variables cannot be varied independently because of the primary constraints. As a result the equations of motion which are derived from ({\ref{h30.00005}}) contain arbitrary functions $u_m$ of the momentum phase-space variables:
\begin{eqnarray}
\dot{q}_n &=& \frac{\partial H_C}{\partial p_n } + \sum^M_{m=1} u_m \frac{\partial \chi_m}{\partial p_n }, \\
\dot{p}_n &=& -\frac{\partial H_C}{\partial q_n } - \sum^M_{m=1} u_m \frac{\partial \chi_m}{\partial q_n }.
\label{h30.00006}
\end{eqnarray}
At this point it is useful to introduce the Poisson bracket $[F,G]_P$ for two momentum phase-space functions $F(q,p)$ and $G(q,p)$:
\begin{equation}
 [F,G]_P = \sum^N_{n=1} \frac{\partial F}{\partial q_n} \frac{\partial G}{\partial p_n} - \frac{\partial F}{\partial p_n} \frac{\partial G}{\partial q_n}.
\label{h30.00008}
\end{equation}
Using the Poisson bracket, we can write the equation of motion for any function $F(q,p)$ as
\begin{equation}
 \dot{F}= [F,H_C]_P + \sum^M_{m=1} u_m [F,\chi_m]_P.
\label{h30.00007}
\end{equation}

The equations of motion can be written in an even more concise form. In order to do so let us first introduce Dirac's equality sign `$\approx$' which is defined by
\begin{equation}
F(q,p) \approx G(q,p) \Leftrightarrow F(q,p)\big|_{\chi_m=0;\ m=1,\dots, M} = G(q,p)\big|_{\chi_m=0;\ m=1,\dots, M}.
\label{h30.00009}
\end{equation}
If $F \approx G$, one says that $F$ {\em weakly equals} $G$. By further introducing the {\em total Hamiltonian} 
\begin{equation}
H_T = H_C + \sum^M_{m=1} u_m \chi_m,
\label{h30.00010}
\end{equation}
the equations of motion ({\ref{h30.00007}}) can be written as
\begin{equation}
\dot{F} \approx [F,H_T]_P.
\label{h30.00011}
\end{equation}

Now we can further examine the consequences of these equations of motion. In the first place there will be some consistency conditions. The constraints $\chi_m \approx 0$ have to be weakly conserved in time. Hence we have the requirement 
\begin{equation}
\dot{\chi}_{m'} \approx [\chi_{m'},H_T]_P \approx [\chi_{m'},H_C]_P + \sum^M_{m=1} u_m [\chi_{m'},\chi_m]_P \approx 0
\label{h30.00012}
\end{equation}
for $m'=1,\dots, M$. Some of these equations may be equations which do not contain any $u_m$ and hence they may lead to new constraints that have to be satisfied. These constraints are called {\em secondary constraints}. In turn, the secondary constraints should also be conserved in time and hence may lead to further constraints. We can repeat this procedure until no further constraints are found. In each step of evaluating the consistency condition $\dot{\chi}_{m'} \approx 0$, the equality sign $\approx$ refers to the full set of constraints obtained at that stage. The newly obtained constraints are denoted by 
\begin{equation}
\chi_k \approx 0, \quad k=M+1, \dots, M+K. 
\label{h30.00013}
\end{equation}
So that the full set of constraints can then be written as
\begin{equation}
\chi_j \approx 0, \quad j=1, \dots, J = M+K. 
\label{h30.00014}
\end{equation}

Not only may the consistency requirements that the constraints are conserved in time lead to new constraints, they may also determine some of the coefficients $u_m$. To see this consider again the consistency conditions 
\begin{equation}
[\chi_{j},H_T]_P \approx [\chi_{j},H_C]_P + \sum^M_{m=1} u_m [\chi_{j},\chi_m]_P \approx 0
\label{h30.00015}
\end{equation} 
where $j$ ranges from $j=1, \dots, J$, and let us now regard them as equations for the $u_m$. Let $u_m=U_m$ be a particular solution for ({\ref{h30.00015}}). The most general solution is then
\begin{equation}
u_m = U_m + \sum^A_{a=1} v_a V_{am},
\label{h30.00016}
\end{equation} 
where the $V_{am}$, with $a=1,\dots,A$, are $A$ independent solutions of the homogeneous equation 
\begin{equation}
\sum^M_{m=1} V_{am} [\chi_{j},\chi_m]_P \approx 0
\label{h30.00017}
\end{equation}
 and the $v_a$ are arbitrary coefficients. 

We may substitute these expression for the $u_m$ in the total Hamiltonian to obtain
\begin{equation}
H_T = H' + \sum^A_{a=1} v_a \chi_a
\label{h30.00018}
\end{equation}
where
\begin{eqnarray}
H' &=& H_C + \sum^M_{m=1}  U_m \chi_m, \\
\chi_a &=&\sum^M_{m=1} V_{am} \chi_m.
\label{h30.00019}
\end{eqnarray}
We see that the equation of motion for an arbitrary function $F(q,p)$ now reads
\begin{equation}
 \dot{F} \approx [F,H_T]_P \approx [F,H']_P + \sum^A_{a=1}v_a [F,\chi_a]_P
\label{h30.00020}
\end{equation}

Notice that the equation of motion for $F$ may depend on the arbitrary coefficients $v_a$. This remaining arbitrariness denotes the presence of some gauge invariance. In this context it is said that the $\chi_a$, $a=1,\dots,A$, generate infinitesimal gauge transformations. This can be seen as follows. Let $F_v$ and $F_{v'}$ be phase-space functions evolving form the initial value $F_0$ with two different sets of coefficients, $v_a$ and $v'_a$. To first order we have 
\begin{equation}
 F_v(\delta t) = F_0 + [F_0,H_T]_P\delta t \approx F_0 + [F_0,H']_P\delta t + \sum^A_{a=1} v_a [F_0,\chi_a]_P\delta t
\label{h30.00021}
\end{equation}
 and hence
\begin{equation}
 \delta F(\delta t) = F_v(\delta t) - F_v'(\delta t) \approx  \sum^A_{a=1} (v_a - v'_a)[F_0,\chi_a]_P\delta t.
\label{h30.00022}
\end{equation}
So the functions $F_v(\delta t)$ and $F_{v'}(\delta t)$ are related by an infinitesimal canonical transformation generated by $\sum^A_{a=1}(v_a - v'_a)\chi_a\delta t$.{\footnote{A transformation $Q(q,p,t)$,  $P(q,p,t)$ is canonical if \label{canonicaltransformation}
\begin{equation}
[Q_{n},Q_{n'}]_P = [P_{n},P_{n'}]_P=0, \quad [Q_{n},P_{n'}]_P=\delta_{n n'},
\label{h30.0009101}
\end{equation} 
where the Poisson bracket is calculated with respect to the variables $q$ and $p$. This means that in the case of a canonical transformation, the Poisson brackets of phase-space functionals are the same, whether calculated with old or new phase-space variables. The new Hamiltonian $H'(Q,P)$ corresponding to the transformed system is 
\begin{equation}
H' =  \sum^N_{n=1} {P}_{n}{\dot Q}_{n} - L + \frac{d F}{d t}, 
\label{h30.000910102}
\end{equation}
where the Lagrangian $L$ is expressed in terms of the primed variables. The function $F(Q,P)$ is the {\em generating function}, which is determined by
\begin{eqnarray}
 \frac{\delta F}{\delta Q_{n}} &=& \sum^N_{n'=1}p_{n'} \frac{\delta q_{n'}}{\delta Q_{n}} - P_{n},  \nonumber\\
\frac{\delta F}{\delta P_{n}} &=& \sum^N_{n'=1}  p_{n'} \frac{\delta q_{n'}}{\delta P_{n}}.
\label{h30.000910101}
\end{eqnarray}
\label{canonicaltransformation}}}
The constraints $\chi_a$, with $a=1,\dots,A$, are not the only generators of infinitesimal gauge transformations. One can for example show that $[\chi_a,\chi_{a'}]_P$ and $[H',\chi_{a'}]_P$, with $a,{a'}=1,\dots,A$, also generate gauge transformations. In practice this means that some of the non-primary constraints will also generate infinitesimal gauge transformations (one can show that $[\chi_a,\chi_{a'}]_P$ and $[H',\chi_a]_P$ weakly vanish and hence they are linear combinations of the constraints $\chi_j$, $j=1,\dots,J$). 

Suppose we have some additional lineary independent constraints $\chi_{a}$, with $a=A+1,\dots,A'$, which generate infinitesimal gauge transformations. One can then make the gauge invariance of the dynamics explicit by using the {\em extended Hamiltonian}
\begin{equation}
H_E = H_T +  \sum^{A'}_{a=A+1} v_{a}\chi_{a}. 
\label{h30.00023}
\end{equation}
The corresponding equation of motion for a function $F$ reads
\begin{equation}
 \dot{F} \approx [F,H_E]_P \approx [F,H']_P + \sum^{A'}_{a=1}v_a [F,\chi_a]_P.
\label{h30.0002301}
\end{equation}
It is clear that a function $F$ for which the equation of motion depends on the arbitrary functions $v_a$, $a=1,\dots,A'$ can not be an observable quantity. Hence an observable function should have Poisson brackets zero with all the generators of gauge transformations $\chi_a$, $a=1,\dots,A'$; in other words an observable function is gauge invariant.

The constraints $\chi_{b}$ which will be added to the Hamiltonian are the {\em first class constraints}. A function $F$ of the momentum phase-space variables is called {\em first class} if it has zero Poisson brackets with all the constraints, i.e.\
 \begin{equation}
[F,\chi_j]_P \approx 0 , \quad j=1,\dots,J. 
\label{h30.00024}
\end{equation}
Using their definition ({\ref{h30.00019}}), one can check that the constraints $\chi_a$, $a=1,\dots,A$, which were defined in ({\ref{h30.00019}}), are first class, as well as $H'$. One can also show that the Poisson bracket of two first class functions is also first class. Hence $[\chi_a,\chi_{a'}]_P$ and $[H',\chi_a]_P$ are also first class. So it seems that the set of first class constraints corresponds to the set of generators of infinitesimal gauge transformations. This was indeed conjectured by Dirac. However, later, counter examples of this conjecture were presented (although these had no physical relevance). Although the set of first class constraints does hence not correspond to the set of infinitesimal gauge transformations, we will have to treat the first class as such when we try to quantize the system. I.e.\ the first class constraints should be added to the total Hamiltonian to yield the extended Hamiltonian.

The constraints which are not first class are called {\em second class constraints}. The distinction between first class constraints and second class constraints is important if we want to quantize the system. We now discuss two distinct cases separately. In the first case we assume a system which only has second class constraints and in the second case we assume a system which has only first class constraints. 

\subsection{Canonical quantization of systems with second class constraints}\label{scconstraint}
Assume that all the constraints are second class constraints. This implies that the matrix
\begin{equation}
C_{jj'} = [\chi_j, \chi_{j'}]_P,
\label{h30.0007}
\end{equation}
with $j,j'=1,\dots,J$, is non-singular (for every point in momentum phase-space). Otherwise there would exist a linear combination of the constraints $\chi_j$ which has Poisson brackets zero with every constraint and hence this linear combination would constitute a first class constraint. Because $C$ is anti-symmetric the number of second class constraints must necessarily be even, otherwise the determinant of $C$ would be zero. The inverse of $C$ is denoted by $C^{-1}$.

If we would now quantize the system, by associating the operators ${\widehat F}$ and ${\widehat G}$ with some momentum phase-space functions $F(q,p)$ and $G(q,p)$, and by imposing the commutation relations{\footnote{One cannot associate quantum operators with all momentum phase-space functions, because some phase-space functions may correspond to more than one quantum operator. Therefore applying the prescription ({\ref{h30.0008}}) to any momentum phase-space function would lead to contradictions. This is the operator ordering ambiguity. Different choices for operator orderings may correspond to different quantum theories.}}
\begin{equation}
[{\widehat F},{\widehat G}] = i{\widehat{[F,G]_P}},
\label{h30.0008}
\end{equation}
then these commutation relations would not always be consistent with the constraints.{\footnote{For fermionic theories anti-commutation relations can be imposed.}} I.e.\ the operations of imposing the constraints and taking the commutator of the operators would not always commute. For this reason Dirac introduced the {\em Dirac bracket} 
\begin{equation}
[F,G]_D = [F,G]_P - [F,\chi_j]_P C^{-1}_{jj'} [\chi_{j'}, G]_P,
\label{h30.0009}
\end{equation}
which is defined for any two phase-space functions $F(q,p)$ and $G(q,p)$. The basic properties of the Dirac bracket are the same as for the Poisson bracket: linearity, antisymmetry, Leibnitz rule and Jacobi identity. In addition, if $F$ or $G$ is a linear combination of constraints then $[F,G]_D=0$. Hence the operations of imposing the constraints and taking the Dirac bracket commute. The system can now be quantized by imposing the following equal-time commutation relations for the operators ${\widehat F}$ and ${\widehat G}$  
\begin{equation}
[{\widehat F},{\widehat G}]=i{\widehat{[F,G]_D}}.
\label{h30.00091}
\end{equation}

One can easily verify that equations of motion for any phase-space function $F$ can be written as 
\begin{equation}
{\dot F} \approx [F,H_C]_D \approx [F,H]_D, 
\label{h30.000901}
\end{equation}
with 
\begin{equation}
H=H_C\big|_{\chi_j=0;j=1,\dots,J}.
\label{h30.000902}
\end{equation}
Hence, once the Dirac bracket is found, the total Hamiltonian is of no further use. By using the Dirac bracket we can calculate the equations of motion using the Hamiltonian $H$. So the total Hamiltonian is only needed to find the secondary and further $n$-ary constraints. 

\subsubsection{The true degrees of freedom}
By using the Dirac bracket we can quantize a system with second class constraints. However, because of the constraints the phase-space variables are not all independent; we are in fact dealing with {\em too many} phase-space variables. Nevertheless it is in principle possible to reduce the number of phase-space variables by isolating the {\em true degrees of freedom}. This is due to a theorem by Maskawa and Nakajima \cite[pp.\ 329-330]{maskawa76,weinberg95}.{\footnote{The same result was also presented in \cite[pp.\ 82-85]{sundermeyer82} and in \cite[p.\ 30]{gitman90}.}} It will appear crucial for the construction of a pilot-wave interpretation, at least for the field theories we consider, to separate out the true degrees of freedom.

The Maskawa-Nakajima theorem states that if there are $J=2R$ second class constraints (and no first class constraints), then we can, {\em at least locally}, perform a canonical transformation such that the new canonical variables can be written in terms of two sets $Q_{l}$ and ${\bar Q}_{k}$ and their respective conjugate momenta $P_{l}$ and ${\bar P}_{k}$, with $l=1,\dots, N-R$ and $k=N-R+1, \dots, N$, such that the constraints in terms of the new variables read ${\bar Q}_{k}={\bar P}_{k}=0$ for $k=N-R+1, \dots, N$. The theorem further states that 
\begin{equation}
[F,G]_D \big|_{\chi_j =0}  =  \sum^{n-r}_{l=1}\frac{\partial F^*}{\partial Q_{l}}\frac{\partial G^*}{\partial P_l} - \frac{\partial G^*}{\partial Q_{l}}\frac{\partial F^*}{\partial P_l},
\label{h30.00092}
\end{equation}
with $F^*(Q,P)=F\left(q(Q,P,{\bar Q},{\bar P}\right),q\left(Q,P,{\bar Q},{\bar P})\right)|_{{\bar Q}_{k}={\bar P}_{k}=0}$ and a similar definition for $G^*$. This means that the Dirac bracket equals the Poisson bracket `restricted to the unconstrained variables', when the constraints are imposed. The Hamiltonian $H$ defined ({\ref{h30.000902}}) reduces to the following Hamiltonian for the variables $Q_{l}$ and $P_{l}$, with $l=1,\dots, n-r$:
\begin{equation}
H(P,Q) = \sum^{n-r}_{l=1}  {P}_{l}{\dot Q}_l - L\big|_{{\bar Q}_{k}={\bar P}_{k}=0} + \frac{d F}{d t}\Big|_{{\bar Q}_{k}={\bar P}_{k}=0}.
\label{h30.0009301}
\end{equation}
$F$ is the generation function of the canonical transformation (cf.\ the footnote on p.\ \pageref{canonicaltransformation}). Gitman and Tyutin call the Hamiltonian $H(P,Q)$ the {\em physical Hamiltonian} \cite[p.\ 31]{gitman90}. 

In this way, the theory can at least locally be recast in terms of unconstrained variables (the true degrees of freedom) $Q_{l}$ and $P_{l}$ with $l=1,\dots, n-r$, for which the Dirac bracket equals the Poisson bracket, and for which the dynamics is governed by the Hamiltonian ({\ref{h30.0009301}}). The canonical variables ${\bar Q}_{k}$ and ${\bar P}_{k}$ ($k=n-r+1, \dots, n$) are the constraints and can hence be omitted in the description of the system. In this way, the dimension of the phase-space is reduced from $2n$ to $2n-2r$, this is the number of phase-space variables we started with, minus the number of constraints. 

One can also show that the true degrees of freedom are, at least locally, unique up to a canonical transformation \cite[p.\ 31]{gitman90}. I.e.\ given another set of true degrees of freedom $({\tilde Q}_{l},{\tilde P}_{l})$, then there exists a canonical transformation from $(Q_{l},P_{l})$ to $({\tilde Q}_{l},{\tilde P}_{l})$, which maps the physical Hamiltonian $H(P,Q)$ to the physical Hamiltonian ${\tilde H}({\tilde Q},{\tilde P})$ for the true degrees of freedom $({\tilde Q}_{l},{\tilde P}_{l})$. 

If we now quantize the system, by associating operators with the remaining $2n-2r$ canonical variables, the commutation relations ({\ref{h30.00091}}) for these operators become
\begin{equation}
[{\widehat Q}_{l_1},{\widehat Q}_{l_2}] = [{\widehat P}_{l_1},{\widehat P}_{l_2}] = 0, \quad [{\widehat Q}_{l_1},{\widehat P}_{l_2}]  = i\delta_{l_1 l_2} , 
\label{h30.00093}
\end{equation}
with $l_1,l_2 = 1 ,\dots ,  n-r$. 

The quantum description of the system then runs as follows. A quantum system is described by a vector $| \psi \rangle$ (the state vector) in a Hilbert space, with inner product $\langle \psi_2 | \psi_1 \rangle$. The operators ${\widehat Q}_{l}$ and ${\widehat P}_{l}$ $(l = 1 ,\dots ,  n-r)$ now act on these state vectors. In the Heisenberg picture, the operators ${\widehat Q}_{l}$ and ${\widehat P}_{l}$ are the dynamical objects and the states are time independent. However, in order to construct a pilot-wave interpretation we will not need the Heisenberg picture, but the Schr\"o\-ding\-er picture. In the Schr\"o\-ding\-er picture, the states and not the operators, are the dynamical objects. Because we have the standard canonical commutation relations for the operators ${\widehat Q}_{l}$ and ${\widehat P}_{l}$, we can use the standard representation
\begin{equation}
{\widehat Q}_{l} = Q_{l}, \qquad   {\widehat P}_{l} = -i\frac{\partial}{\partial {Q_{l}}},
\label{h30.00094}
\end{equation} 
for the operators in the Schr\"o\-ding\-er picture. In this representation, the operators act on the wavefunction $\Psi(Q_1,\dots,Q_{n-r},t)=\langle Q_1,\dots,Q_{n-r} |\Psi (t)\rangle$, with $| Q_1,\dots,Q_{n-r}\rangle$ the simultaneous eigenstates of the operators ${\widehat Q}_{l}$. In the representation ({\ref{h30.00094}}) the Hamiltonian operator is written as $ {\widehat H}\!\left( Q,\!-i\partial / \partial {Q} \right)$ and the dynamical evolution of $\Psi(Q_1,\dots,Q_{n-r},t)$ is given by the Schr\"o\-ding\-er equation
\begin{equation}
i\frac{\partial }{\partial t} \Psi(Q_1,\dots,Q_{n-r},t) = {\widehat H}\left( Q,-i\partial / \partial {Q} \right) \Psi(Q_1,\dots,Q_{n-r},t).
\label{h30.0009402}
\end{equation}
A pilot-wave interpretation can then be devised by looking at the conservation equation for the probability density $|\Psi(Q_1,\dots,Q_{n-r},t)|^2$.

For a system described by a continuum number of degrees of freedom, the scheme to construct a pilot-wave interpretation proceeds along the same lines. The transition to a continuum number of degrees of freedom is discussed in Section \ref{quantizationofafieldtheory}. In the case of fermionic degrees of freedom we have to use a different representation from the one above. We discuss this in the next chapter.

\subsection{Canonical quantization of a system with first class constraints}\label{firstclassconstraints}
Suppose now that all constraints $\chi_j$ are first class constraints. If there were also second class constraints, these could be dealt with separately, in the way described in the previous section. In the case a system has first class constraints, the matrix $C_{jj'}$ is singular and Dirac's method of quantization of systems with only second class constraints cannot be applied. There are two ways to proceed\footnote{We mention here only two methods of dealing with first class constraints, there are still other methods \cite[p.\ 110]{sundermeyer82}. However, these are less frequently used and, moreover, it is unclear whether these approaches may lead to a pilot-wave interpretation.}:
\begin{itemize}
\item
{\bf Constraints as operator identities:} We have seen that the presence of first class constraints indicates the presence of some gauge invariance. This gauge invariance means that the evolution of the coordinates $q$ and $p$ is not uniquely fixed by their initial values; at each time one can perform a gauge transformation which yields a physical equivalent state. Only phase-space functions $F(q,p)$ which are gauge invariant are physically observable. 

One can eliminate the gauge variables by adding further restrictions on the canonical variables. This is done by imposing further constraints, called {\em gauge constraints}. It is permissible to bring in these further constraints because they merely remove the arbitrary elements in the theory and do not affect the gauge invariant quantities.

A good set of gauge constraints
\begin{equation}
C_{j'}(q,p) \approx 0, 
\label{h30.0009403}
\end{equation}
called a {\em canonical gauge} by Henneaux and Teitelboim \cite[p.\ 27]{henneaux91} and an {\em admissible} gauge by Sundermeyer \cite[p.\ 102]{sundermeyer82}, satisfies the following two properties:
\begin{itemize}
\item
The gauge must be attainable. I.e.\ given a set of canonical variables $q$ and $p$ there exists a gauge transformation which brings the given set into one which satisfies ({\ref{h30.0009403}}). The transformation must be obtained by iteration of infinitesimal transformations of the form $\epsilon_{j} [F,\chi_j]$, where $F$ represents the canonical variables $q$ and $p$.
\item
Second, the conditions ({\ref{h30.0009403}}) must fix the gauge completely. As long as there would be a residual gauge freedom, the initial conditions on the canonical variables would be insufficient to uniquely determine their future evolution. 

This condition implies that no gauge transformations but the identity preserve ({\ref{h30.0009403}}) or in other words that the equations $\sum_j \epsilon_{j} [C_{j'},\chi_{j}] \approx 0$ must imply that $\epsilon_{j}=0$. 
\end{itemize}

One can show that the two conditions taken together imply that the number of gauge constraints must equal the number of first class constraints \cite[p.\ 27]{henneaux91}. The second condition further implies that the set of constraints $\{C_{j'},\chi_j\}$ is second class. This means that by adding the gauge constraints, the first class constraints turned second class so that we can quantize the system by using the Dirac bracket. 

Once we have a suitable set of gauge constraints, we can, at least locally, perform a Mas\-ka\-wa-Nakajima canonical transformation to new canonical coordinates. In terms of these new coordinates the constraints form a set of canonical pairs. The other canonical pairs, which are unconstrained, then form the true degrees of freedom and the system can be expressed solely in terms of these true degrees of freedom. By definition, the true degrees of freedom have zero Poisson brackets with the constraints. In particular, the true degrees of freedom will have zero Poisson brackets with the first class constraints (the first class constraints in the new coordinates are found by performing the canonical transformation to the constraints $\chi_j$). This means that the true degrees of freedom are gauge independent variables. Therefore we will often refer to the true degrees of freedom as the gauge independent variables. The true degrees of freedom are unique up to a canonical transformation; in particular they are independent of the particular choice of admissible gauge (an extensive discussion of this can be found in \cite[pp.\ 36-60]{gitman90}).

\item
{\bf Constraints as conditions on states:} We can also quantize the canonical variables as if there were no constraints. The commutation relations for the operators associated with the canonical variables, are determined by the Poisson bracket as in equation ({\ref{h30.0008}}). The constraints are reintroduced by demanding that physical states $|\Psi \rangle$ satisfy 
\begin{equation}
{\widehat \chi}_j |\Psi \rangle =0 .
\label{h30.0201}
\end{equation}

The advantage of this method of dealing with constraints is that, because the commutation relations are simply the standard commutation relations, we can use the standard representation for the operators. If we quantize by treating the constraints as operators, the commutation relations are derived from the Dirac bracket. This can make it more complicated to find a suitable representation, because it requires the identification of the true degrees of freedom.  

The disadvantage of dealing with constraints as conditions on states is that in general we will have to introduce a non-trivial measure on the configuration space in order to construct an inner product which yields finite numbers. A suitable measure can be found by applying the Faddeev-Popov formalism. 

If we want to construct a pilot-wave model in the context of this method of dealing with constraints, we encounter a similar problem. The density of field beables will be non-normalizable. We will illustrate this explicitly in Section \ref{maxwell}, where we will try to construct a pilot-wave theory for the quantized electromagnetic field by starting from this scheme. It will turn out that when we try to solve this problem, we are naturally led to the pilot-wave interpretation which may be obtained by quantizing the electromagnetic field by treating constraints as operator identities. We will discuss these issues in more detail in Section \ref{constrconsta}.
\end{itemize}

Finally, we want to note that these two schemes of dealing with first class constraints are very well suited for the quantization of the electromagnetic field because it presents no problem to fix the gauge globally. On the other hand, if we consider the quantization of non-Abelian gauge theories (Yang-Mills theories), it is more difficult to find a suitable gauge. This will lead to difficulties in both schemes of dealing with first class constraints. In the first scheme this is because a gauge is imposed from the start (as additional constraints). In the second scheme, this is because a gauge is needed to perform the Faddeev-Popov trick.

\subsection{Quantization of a field theory}\label{quantizationofafieldtheory}
So far we only have considered system which can be described by a finite number of degrees of freedom. The transition to a system which is described by a continuum number of degrees of freedom is straightforward. One can think of the transition as a replacement of the discrete label $n$ of the coordinates $q_n$ by a continuum label ${\bf x}$, i.e.\ $q_n(t)=q(t,n) \to \psi_{{\bf x}} (t) = \psi(t,{\bf x})$. Usually the fields also carry an additional discrete label. Throughout this chapter, we will assume that the fields $\psi_i({\bf x})$ and their derivatives vanish sufficiently fast at spatial infinity. In this way possible boundary terms that arise when performing partial integration may be omitted. 

Sums which appeared for systems with a finite number of degrees of freedom change to integrals. Derivatives with respect to the canonical coordinates $\partial / \partial q_n$ are replaced by functional derivatives $\delta / \delta \psi_i({\bf x})$. 

As an example we can consider the definition of the Poisson bracket for fields. Let $\psi_i(t,{\bf x})$ be the fields with canonically conjugate momenta $\Pi_{\psi_i}(t,{\bf x})$. The Poisson bracket for two functions $F(\psi_i,\Pi_{\psi_i})$ and $G(\psi_i,\Pi_{\psi_i})$ then reads
\begin{equation}
[F,G]_P = \sum_i \int d^3 x \left( \frac{\delta F}{\delta \psi_i(t,{\bf x})} \frac{\delta G}{\delta \Pi_{\psi_i} (t,{\bf x})} - \frac{\delta G}{\delta \psi_i(t,{\bf x})} \frac{\delta F}{\delta \Pi_{\psi_i} (t,{\bf x})} \right).
\label{h30.02010001}
\end{equation}
In the expression all the fields are considered at the time $t$.

In order to construct a pilot-wave theory we will work in the functional Schr\"odinger picture. This is realized by using the representation
\begin{equation}
{\widehat \psi}_i({\bf x})= \psi_i({\bf x}),\quad {\widehat \Pi}_{\psi_i}({\bf x}) =-i\frac{\delta }{\delta \psi_i({\bf x})}.
\label{h30.02010002}
\end{equation}
The Schr\"odinger equation is then a functional differential equation for the wavefunctional $\Psi(\psi_i({\bf x}),t)$.{\footnote{The wave\-func\-tion\-al is also called the {\em super-wavefunction} and the corresponding Schr\"o\-ding\-er equation is called the {\em super-Schr\"o\-ding\-er equation} \cite{bohm2,kaloyerou85,bohm872}.}}

If the fields $\psi_i$ represent the true degrees of freedom then we can define the inner product of two wavefunctionals as
\begin{equation}
\langle \Psi_1 |  \Psi_2 \rangle = \int \left( \Pi_j {\mathcal D}\psi_j \right)   \Psi_1^*(\psi_i,t)   \Psi_2(\psi_i,t),
\label{h30.00901}
\end{equation}
with ${\mathcal D}\psi_i = \Pi_{{\bf x}} d\psi_i({\bf x})$. 

The wavefunctional can be written as $\Psi(\psi_i({\bf x}),t)= \langle \psi_i({\bf x}) | \Psi \rangle $ where the $| \psi_i({\bf x}) \rangle$ form a basis of the Hilbert space. They are the joint eigenstates of the operators ${\widehat \psi}_i({\bf x})$, i.e.\  ${\widehat \psi}_j({\bf x}) | \psi_i({\bf x}) \rangle = \psi_j({\bf x}) | \psi_i({\bf x}) \rangle$. In the standard quantum mechanical interpretation, the quantity 
\begin{equation}
|\Psi(\psi_i({\bf x}),t)|^2=|\langle \psi_i ({\bf x}) |\Psi(t)\rangle|^2
\label{h30.00902}
\end{equation}
 can be interpreted as the probability density to find the system with wavefunctional $\Psi$ in the field configuration $(\psi_i({\bf x}))$. In the pilot-wave interpretation, we will introduce field beables $\psi_i(t,{\bf x})$ which are distributed according to this density $|\Psi(\psi_i({\bf x}),t)|^2$ and for which the dynamics is governed by the guidance equations. As in the preceding chapters, the guidance equations will be derived from the continuity equation for $|\Psi(\psi_i({\bf x}),t)|^2$, by considering the analogy with the continuity equation in hydrodynamics.

We want to stress that in using the functional Schr\"odinger picture and in deriving the pilot-wave interpretation we will not adopt the greatest possible mathematical rigour. There are several problems associated to dealing with an infinite number of degrees of freedom, which will not be addressed in this thesis. First, there is the problem of infinities which plagues quantum field theory. Here, we will not make an effort to incorporate some renormalization scheme in devising the pilot-wave interpretations. Second there is the problem how to make mathematical sense out of the measure ${\mathcal D}\psi_i$; because it is a measure on an infinite dimensional configuration space it cannot be a Lebesgue measure. Third, the fields $\psi_i({\bf x})$ in the representation ({\ref{h30.02010002}}) can in fact not be treated as smooth functions, but should be distributions and accordingly they should be smeared \cite[p.\ 56]{haag96}. 

Although we do not elaborate on these problems, they certainly need attention in the future. The reader which is uncomfortable with our neglection of these problems can implicitly assume that we describe fields confined to a box of finite volume and with periodic boundary conditions, for which a cutoff is introduced for large momenta. Under these assumptions the fields are described by a finite number of degrees of freedom in momentum space, so that the above problems dissolve.  

We can also make some general notes on locality and covariance of the pilot-wave field models. In Section \ref{principlespwinterpretation} we noted the pilot-wave interpretation for non-relativistic quantum theory is nonlocal at the subquantum level. I.e.\ the motions of the particle beables are nonlocally correlated. But the nonlocality can not be used for superluminal signaling (at least not if the particles are distributed according to the quantum equilibrium hypothesis, which we assume in this thesis). Then in Section \ref{noteonlorentzinvariance} we noted that the pilot-wave interpretation is not Lorentz invariant either. But, as with the locality, Lorentz invariance is satisfied on the empirical level. These properties also apply to the field beable approach. I.e.\ on the subquantum level the pilot-wave interpretations for quantum field theories, will be nonlocal and not Lorentz invariant. But at the empirical level the pilot-wave interpretation makes the same (statistical) predictions as the standard interpretation, so that at this level we regain locality and Lorentz invariance. These issues are thoroughly discussed in \cite{bohm872,valentini92,kaloyerou94} and we will not re-address them here.

\section{The Duffin-Kemmer-Petiau theory in the \mbox{spin-0} representation}\label{Massive spin-0}
In this section, we consider the construction of a pilot-wave interpretation for the Duffin-Kemmer-Petiau (DKP) theory in the spin-0 representation. First we show the equivalence of the quantized DKP theory and the quantized Klein-Gordon theory. Then afterwards we present the corresponding pilot-wave interpretation.

\subsection{Equivalence with the canonically quantized Klein-Gordon theory}
We start with quantizing the DKP theory by applying Dirac's canonical quantization procedure. We consider the DKP field coupled to a non-quantized electromagnetic field $V^{\mu}=(V_0,{\bf V})$, which is introduced via the minimal coupling prescription $\partial_{\mu} \to D_{\mu} = \partial_{\mu} + ieV_{\mu}$.{\footnote{Lorentzian indices will be denoted by Greek letters $\mu,\nu,\dots$ and Euclidean indices will be denoted by Latin letters $i,j,\dots$ The Lorentzian indices are raised and lowered by the metric $g_{\mu \nu} = \textrm{diag}(1,-1,-1,-1)$ and the Euclidean indices are raised and lowered by the metric $\delta_{ij} = \textrm{diag}(1,1,1)$. In this chapter we will often start from a Lagrangian density written in terms of Lorentzian vectors, but when we pass to the Hamiltonian formulation we will use Euclidean vectors. Note that in the previous chapter, the indices $i,j,\dots$ were used to denote the space index of tensorial objects.}} The equivalence of the quantized DKP theory in the spin-0 representation, coupled to a quantized electromagnetic field, and the quantized Klein-Gordon theory, coupled to a quantized electromagnetic field, can be shown in the same way. We will not do this explicitly here. In Section \ref{scalarqed}, where we discuss the pilot-wave interpretation for a massive bosonic field interacting with a quantized electromagnetic field, we will start from the coupled Klein-Gordon theory instead of from the coupled DKP theory.

If we write the five component DKP field as $\psi= m^{-1/2} (\phi_{\mu}, m\phi)^T$ with $\phi^{\mu}=(\phi_0,{\boldsymbol \phi})$, then in the spin-0 representation given in Appendix {\ref{appa}}, the minimally coupled DKP Lagrangian (which is obtained from the Lagrangian density ({\ref{h2.25.071001}})) reads
\begin{eqnarray}
L_K = \int d^3 x \mathcal{L}_K &=& \int d^3 x \bigg( \frac{1}{2}\big(\phi^{*}_{\mu} D^{\mu} \phi - \phi^{*} D_{\mu} \phi^{\mu} + (D_{\mu} \phi)^{*}\phi^{\mu} - (D_{\mu} \phi^{\mu})^{*} \phi \big) \nonumber\\ 
&&\qquad\qquad - m^2\phi^* \phi - \phi^{*}_{\mu}\phi^{\mu}\bigg).
\label{h30.001}
\end{eqnarray}
The equations of motion are
\begin{eqnarray}
D_\mu \phi^\mu + m^2 \phi &= 0, \quad &\left( D_\mu \phi^\mu \right)^* + m^2 \phi^* = 0,  \nonumber\\
D_\mu \phi -  \phi_\mu &= 0, \quad &\left( D_\mu \phi \right)^* -  \phi^*_\mu = 0.
\label{h30.00102}
\end{eqnarray}
The canonically conjugate momenta are{\footnote{Often the space dependence of the fields will not be written explicitly. E.g.\ instead of writing $\Pi_{\psi}({\bf x}) = \frac{\delta L }{\delta {\dot \psi({\bf x})}  }$ we write $\Pi_{\psi} = \frac{\delta L }{\delta {\dot \psi}  }$. Similarly instead of writing the constraints as $\chi_{\psi}({\bf x}) =0$ we write them as $\chi_{\psi} =0$. In this way the constraint $\chi_{\psi} =0$ represents in fact an infinite number of constraints; one corresponding to each point in space.}}
\begin{eqnarray}
\Pi_{\phi_0} =& \frac{\delta L }{\delta {\dot \phi}_0  } = -\frac{\phi^{*}}{2} , \qquad 
\Pi_{\phi^{*}_0} &=  \frac{\delta L }{\delta {\dot \phi}^{*}_0  } = -\frac{\phi}{2} , 
\nonumber\\
\Pi_{\phi_i} =& \frac{\delta L }{\delta {\dot \phi}_i  }=0, \qquad \quad \;\; 
\Pi_{\phi^*_i} &= \frac{\delta L }{\delta {\dot \phi}^*_i  }=0,  
\nonumber\\
\Pi_{\phi} =& \frac{\delta L }{\delta {\dot \phi}  } = \frac{ \phi^{*}_0}{2}, \qquad \quad \;
\Pi_{\phi^{*}} &=  \frac{\delta L }{\delta {\dot \phi}^{*}  } = \frac{\phi_0}{2} .
\label{h30.002}
\end{eqnarray}
We can immediately identify the primary constraints
\begin{eqnarray}
\chi_{\phi_0} =&\Pi_{\phi_0} +  \frac{\phi^{*}}{2} , \qquad 
\chi_{\phi^{*}_0} &= \Pi_{\phi^{*}_0} +  \frac{\phi}{2} , 
\nonumber\\
\chi_{\phi_i} =& \Pi_{\phi_i}, \qquad \qquad \;\;
\chi_{\phi^*_i} &= \Pi_{\phi^*_i}, \qquad 
\nonumber\\
\chi_{\phi} =& \Pi_{\phi} -  \frac{\phi^{*}_0}{2}  , \qquad \;\;
\chi_{\phi^{*}} &= \Pi_{\phi^{*}} -  \frac{\phi_0}{2} . 
\label{h30.004}
\end{eqnarray}
The corresponding canonical Hamiltonian reads 
\begin{equation}
H_C  =\!\!  \int \!d^3 x \Big(\phi^{*} D_{i} \phi_{i} + \phi D^{*}_{i} \phi^{*}_{i} +  m^2 \phi^*\phi + \phi^{*}_{0}\phi_{0}- \phi^{*}_{i}\phi_{i} +ie V_0 \left(\phi^{*}\phi_0  - \phi\phi^{*}_0 \right)\!  \Big) .
\label{h30.003}
\end{equation}
The total Hamiltonian reads
\begin{equation}
H_T  = H_C + \sum_{\gamma} \int d^3 x u_{\gamma}({\bf x}) \chi_{\gamma}({\bf x}),
\label{h30.0031}
\end{equation}
where the label $\gamma$ takes the values ${\phi_0},{\phi^*_0},\phi,\phi^*,\phi_i,\phi^*_i$ and the $u_{\gamma}$ are arbitrary fields. In order to find out whether there are secondary constraints we impose the consistency conditions that the primary constraints are weakly conserved in time, i.e.\ $[\chi_{\gamma},H_T]_P\approx 0$. The conditions that the constraints $\chi_{\phi_0},\chi_{\phi^*_0},\chi_{\phi},\chi_{\phi^*}$ are conserved yield respectively
\begin{eqnarray}
u_{\phi^*} &=& \phi^*_0  + ieV_0 \phi^*, \nonumber\\
u_{\phi} &=& \phi_0  - ieV_0 \phi, \nonumber\\
u_{\phi^*_0} &=& -D^*_i \phi^*_i - m^2\phi^* + ieV_0 \phi^*_0,\nonumber\\
u_{\phi_0} &=& -D_i \phi_i - m^2\phi - ieV_0 \phi_0.
\label{h30.00311}
\end{eqnarray}
Hence these consistency conditions determine some of the arbitrary fields $u_{\gamma}$ and do not lead to further constraints. The conditions that $\phi_i$ and $\phi^*_i$ are weakly conserved lead to the secondary constraints
\begin{equation}
\chi_{s\phi_i } = D_{i} \phi + \phi_i, \qquad \chi_{s\phi^*_i } = D^*_{i} \phi^* + \phi^*_i .
\label{h30.005}
\end{equation}
In turn, the requirement that the secondary constraints $\chi_{s\phi_i }$ and $\chi_{s\phi^*_i }$ are conserved, determine the fields $u_{\phi^*_i}$ and $u_{\phi_i}$
\begin{equation}
u_{\phi^*_i} = -   D^*_i u_{\phi^*}, \qquad u_{\phi_i} = -   D_i u_{\phi}.
\label{h30.0050101}
\end{equation}
Hence we see that all the fields $u_\gamma$ are determined. This means that all the constraints are second class constraints and the system can be quantized by using the Dirac bracket. 

In order to construct the Dirac bracket we need the inverse of the matrix (cf.\ Section {\ref{scconstraint}})
\begin{equation}
C_{\gamma \gamma'}({\bf x},{\bf y}) = [\chi_{\gamma}({\bf x}), \chi_{\gamma'}({\bf y})]_P
\label{h30.00601}
\end{equation}
where the labels $\gamma$ and $\gamma'$ take the values $\phi_0,\phi^{*}_0,\phi_i,\phi^*_i,s\phi_i,s\phi^*_i $. This matrix has nonzero components
\begin{eqnarray}
C_{{\phi^{*}_0}, \phi }({\bf x},{\bf y}) &= & -C_{ \phi, {\phi^{*}_0} } ({\bf y},{\bf x})= \delta({\bf x}-{\bf y}), \nonumber\\ 
C_{{\phi_0}, \phi^{*} } ({\bf x},{\bf y})& =& -C_{ \phi^{*}, {\phi_0}} ({\bf y},{\bf x})=  \delta({\bf x}-{\bf y}), \nonumber\\ 
C_{\phi ,s\phi_i }({\bf x},{\bf y})&=&  -C_{ s\phi_i ,\phi }({\bf y},{\bf x})= D^*_{x_i} \delta({\bf x}-{\bf y}), \nonumber\\ 
C_{ \phi^*, s\phi^*_i}({\bf x},{\bf y})&=&  -C_{s\phi^*_i  ,\phi^* }({\bf y},{\bf x}) = D_{x_i} \delta({\bf x}-{\bf y}), \nonumber\\ 
C_{ s\phi_i ,\phi_j  }({\bf x},{\bf y})&=& -C_{ \phi_i ,s\phi_j  }({\bf y},{\bf x})=  \delta_{ij}\delta({\bf x}-{\bf y}), \nonumber\\ 
C_{ s\phi^*_i ,\phi^*_j  }({\bf x},{\bf y})&=& -C_{\phi^*_i  , s\phi^*_j }({\bf y},{\bf x})=  \delta_{ij}\delta({\bf x}-{\bf y}).
\label{h30.006}
\end{eqnarray}
The inverse $C^{-1}_{\gamma \gamma'}$ has the following nonzero components
\begin{eqnarray}
C^{-1}_{\phi,\phi^*_0}({\bf x},{\bf y}) &=&-C^{-1}_{\phi^*_0 ,\phi}({\bf y},{\bf x}) =  \delta({\bf x}-{\bf y}),\nonumber\\
C^{-1}_{\phi^*,\phi_0 }({\bf x},{\bf y}) &=&-C^{-1}_{\phi_0 ,\phi^*}({\bf y},{\bf x}) = \delta({\bf x}-{\bf y}),\nonumber\\
C^{-1}_{\phi_i^*,\phi_0 }({\bf x},{\bf y}) &=&-C^{-1}_{\phi_0 ,\phi_i^*}({\bf y},{\bf x}) = D_{x_i} \delta({\bf x}-{\bf y}),\nonumber\\ 
C^{-1}_{\phi_0^*,\phi_i }({\bf x},{\bf y}) &=&-C^{-1}_{\phi_i ,\phi^*_0}({\bf y},{\bf x})= D_{x_i} \delta({\bf x}-{\bf y}),\nonumber\\
C^{-1}_{\phi_i,s\phi_j}({\bf x},{\bf y}) &=&-C^{-1}_{s\phi_i ,\phi_j}({\bf y},{\bf x}) =  \delta_{ij} \delta({\bf x}-{\bf y}),\nonumber\\
C^{-1}_{\phi^*_i,s\phi^*_j}({\bf x},{\bf y}) &=&-C^{-1}_{s\phi^*_i ,\phi^*_j}({\bf y},{\bf x}) =  \delta_{ij} \delta({\bf x}-{\bf y}).
\label{h30.0062}
\end{eqnarray}
The Dirac bracket for the fields $\phi,\phi^{*},\phi_0,\phi^{*}_0$ now reads
\begin{eqnarray}
[\phi ({\bf x}),\phi^{*}_0 ({\bf y})]_D &=&\delta({\bf x}-{\bf y}), \nonumber \\  
 {[\phi^{*}({\bf x}), \phi_0 ({\bf y})]_D} &=&\delta({\bf x}-{\bf y})
\label{h30.0063}
\end{eqnarray}
and all the other Dirac brackets involving the fields $\phi,\phi^{*},\phi_0,\phi^{*}_0$ are zero. Because it is a property of the Dirac bracket that one can impose the constraints before evaluating the Dirac bracket, the commutation relations involving other fields can be derived from the commutation relations of $\phi,\phi^{*},\phi_0,\phi^{*}_0$, by using the constraints (\ref{h30.004}) and (\ref{h30.005}). Therefore there is no need to give them explicitly.

In order to find the true degrees of freedom, we perform the Maskawa-Nakajima transformation to new canonical variables, which we denote with an additional twidle:
\begin{eqnarray}
{\widetilde \phi} =&  \frac{\phi}{2} -\Pi_{\phi^{*}_0} ,  \qquad  \qquad \quad \qquad {\widetilde \phi^{*}} &=  \frac{\phi^{*}}{2} - \Pi_{\phi_0}, \nonumber\\
\Pi_{{\widetilde \phi}} =& \frac{\phi^{*}_0}{2}  + \Pi_{\phi} + D^{*}_i \Pi_{\phi_i},  \qquad  \;\; \Pi_{{\widetilde \phi}^{*}} &= \frac{\phi_0}{2}  + \Pi_{\phi^{*}} + D_i \Pi_{\phi^{*}_i}, \nonumber\\
{{\widetilde \phi}_0} =&\frac{\phi_0}{2}  - \Pi_{\phi^{*}} - D_i \Pi_{\phi^{*}_i} ,  \qquad \;\; {{\widetilde \phi}^{*}_0} &= \frac{\phi^{*}_0}{2}  - \Pi_{\phi} - D^{*}_i \Pi_{\phi_i},\nonumber\\
\Pi_{{\widetilde \phi}_0} =& \Pi_{\phi_0} +  \frac{\phi^{*}}{2} ,  \qquad \;\;  \quad \quad \quad \Pi_{{\widetilde \phi}^{*}_0} &= \Pi_{\phi^{*}_0} +  \frac{\phi}{2} , \nonumber\\
{{\widetilde \phi}_i} =& \phi_i + D_i \phi , \qquad \qquad \quad \; \quad {{\widetilde \phi}^{*}_i} &= \phi^{*}_i + D^{*}_i \phi^{*}, \nonumber\\
\Pi_{{\widetilde \phi}_i} =& \Pi_{\phi_i} , \qquad \qquad \qquad \qquad \Pi_{{\widetilde \phi}^{*}_i} &=  \Pi_{\phi^{*}_i}.
\label{h30.007001}
\end{eqnarray}
For the new variables the constraints read 
\begin{equation}
{\widetilde \phi}_0 = \Pi_{{\widetilde \phi}_0} = {{\widetilde \phi}^{*}_0} = \Pi_{{\widetilde \phi}^{*}_0} = {\widetilde \phi}_i = \Pi_{{\widetilde \phi}_i} = {\widetilde \phi}^{*}_i = \Pi_{{\widetilde \phi}^{*}_i} = 0,
\label{h30.007002}
\end{equation}
so that the true degrees of freedom are ${\widetilde \phi},{\widetilde \phi^{*}},\Pi_{{\widetilde \phi}},\Pi_{{\widetilde \phi}^{*}}$. The physical Hamiltonian $H$ is found by performing the canonical transformation (\ref{h30.007001}) on the canonical Hamiltonian $H_C$ and by imposing the constraints:
\begin{equation}
H =  \int d^3 x \left( \Pi_{{\widetilde \phi}^{*}}  \Pi_{{\widetilde \phi}} + \left(D^*_{i} {\widetilde \phi}^{*}\right)  D_{i} {\widetilde \phi} + m^2{\widetilde \phi}^{*}{\widetilde \phi}   +ie V_0 \left({\widetilde \phi}^{*}  \Pi_{{\widetilde \phi}^{*}}  -{\widetilde \phi}  \Pi_{{\widetilde \phi}}  \right)\right).
\label{h30.00501}
\end{equation}
This Hamiltonian is recognized as the Klein-Gordon Hamiltonian. Because the Dirac bracket for the true degrees of freedom is simply the Poisson bracket, the theory is quantized by imposing the commutation relations
\begin{eqnarray}
[{\widehat {\widetilde \phi}} ({\bf x}), {\widehat \Pi}_{{\widetilde \phi}}  ({\bf y})] &=& i[{\widetilde \phi}  ({\bf x}), \Pi_{{\widetilde \phi}} ({\bf y})]_P =i\delta({\bf x}-{\bf y}), \nonumber \\  
{[{\widehat {{\widetilde \phi}^{*}}} ({\bf x}), {\widehat \Pi}_{{\widetilde \phi}^{*}}  ({\bf y})]} &=& i [{\widetilde \phi}^{*}({\bf x}),  \Pi_{{\widetilde \phi}^{*}} ({\bf y})]_P =i\delta({\bf x}-{\bf y}).
\label{h30.007}
\end{eqnarray}
The other fundamental commutation relations involving the operators ${\widehat {\widetilde \phi}},{\widehat \Pi}_{{\widetilde \phi}},{\widehat {{\widetilde \phi}^{*}}}$ and ${\widehat \Pi}_{{\widetilde \phi}^{*}}$ are zero. The commutation relations are realized by the representation 
\begin{eqnarray}
{\widehat {\widetilde \phi}  }  ({\bf x}) &=&  \phi ({\bf x}), \quad  {\widehat \Pi}_{{\widetilde \phi}} ({\bf x}) = -i\frac{\delta }{\delta \phi({\bf x})}, \nonumber\\
{\widehat {{\widetilde \phi}^{*}}} ({\bf x}) &=& \phi^{*} ({\bf x}), \quad   {\widehat \Pi}_{{\widetilde \phi}^{*}}  ({\bf x}) = -i\frac{\delta }{\delta \phi^*({\bf x})}.
\label{h30.008}
\end{eqnarray}
In this representation the Hamiltonian (\ref{h30.00501}) reads{\footnote{In fact there appears an operator ordering ambiguity at this point. The term proportional to $V_0$ in the Hamiltonian arises from associating operators to ${\widetilde \phi}^{*}  \Pi_{{\widetilde \phi}^{*}}  -{\widetilde \phi}  \Pi_{{\widetilde \phi}} $ (this quantity is proportional to the charge density). We have chosen ${\widehat {{\widetilde \phi}^{*}}} {\widehat \Pi}_{{\widetilde \phi}^{*}} - {\widehat  {\widetilde \phi}} {\widehat \Pi}_{{\widetilde \phi}}$. This operator ordering is also the Weyl ordering \cite[p.\ 347]{greiner96}. Another operator ordering choice is ${\widehat {{\widetilde \phi}^{*}}} {\widehat \Pi}_{{\widetilde \phi}^{*}} -{\widehat \Pi}_{{\widetilde \phi}}  {\widehat  {\widetilde \phi}} $. Although this choice is not Weyl ordered, it is Hermitian. The same operator ordering ambiguity arises when we quantize the coupled Klein-Gordon theory. So one can always choose operator orderings so that the quantized DKP theory is equivalent with the quantized Klein-Gordon theory. A similar remark will apply in the spin-1 case.}}
\begin{equation}
{\widehat H} =  \int d^3 x \left(- \frac{\delta}{\delta \phi^* } \frac{\delta}{\delta \phi} + |D_i \phi|^2 + m^2|\phi|^2 + eV_0\left(\phi^* \frac{\delta}{\delta \phi^*} - \phi \frac{\delta}{\delta \phi}  \right)\right). 
\label{h30.009}
\end{equation}
The operators act on wave\-func\-tion\-als $\Psi(\phi,\phi^*,t)=\langle \phi,\phi^* |  \Psi  (t)\rangle$, so that we obtain the functional Schr\"o\-ding\-er equation 
\begin{equation}
i\frac{\partial \Psi(\phi,\phi^*,t)}{\partial t} = {\widehat H} \Psi(\phi,\phi^*,t). 
\label{h30.010}
\end{equation}
This quantum field theory can also be derived from quantizing the Klein-Gordon theory.{\footnote{Fainberg and Pimentel \cite{fainberg001} also established the result that the canonical quantization of the DKP theory in the spin-0 representation leads to the canonically quantized Klein-Gordon theory. They also give a strict proof of equivalence of the theories for the method of path-integral quantization.}} 

In fact there was also a shorter route to get to the quantized theory. From the constraints we can already read of that the fields $\phi,\phi^{*},\phi_0,\phi^{*}_0$ can be taken as independent degrees of freedom, because all the other canonical variables can be expressed in terms of these fields using the constraints. We can then quantize the theory by imposing the commutation relations 
\begin{eqnarray}
[{\widehat \phi}({\bf x}),{\widehat\phi}^{*}_0 ({\bf y})] &=& i[\phi ({\bf x}),\phi^{*}_0 ({\bf y})]_D =i\delta({\bf x}-{\bf y}), \nonumber \\  
{[{\widehat \phi}^{*}({\bf x}),{\widehat \phi}_0 ({\bf y})]} &=& i [\phi^{*}({\bf x}), \phi_0 ({\bf y})]_D =i\delta({\bf x}-{\bf y}).
\label{h30.01001}
\end{eqnarray}
Operators corresponding to the other degrees of freedom can be expressed in terms of the operators ${\widehat \phi},{\widehat \phi}^{*},{\widehat\phi}_0$ and ${\widehat\phi}^{*}_0$. If we then realize the commutation relations by the representation
\begin{eqnarray}
{\widehat  \phi}  ({\bf x}) &=&  \phi ({\bf x}), \quad {\widehat  \phi}^{*}_0 ({\bf x}) = -i\frac{\delta }{\delta \phi({\bf x})}, \nonumber\\
{\widehat  \phi}^{*} ({\bf x}) &=& \phi^{*} ({\bf x}), \quad  {\widehat  \phi}_0 ({\bf x}) = -i\frac{\delta }{\delta \phi^*({\bf x})},
\label{h30.01002}
\end{eqnarray}
we obtain the same quantum theory as the one presented above. If we look at the Maskawa-Nakajima canonical transformation (\ref{h30.007001}), then we see that on the constraint space, the true degrees of freedom read
\begin{eqnarray}
{\widetilde \phi} &=& \phi,  \qquad  {\widetilde \phi^{*}} =  \phi^{*}, \nonumber\\
{\Pi_{{\widetilde \phi}}} &=& \phi^{*}_0 ,  \qquad  \Pi_{{\widetilde \phi}^{*}} = \phi_0.
\label{h30.01003}
\end{eqnarray}
Hence the fields $\phi,\phi^{*},\phi_0,\phi^{*}_0$ could be regarded as the true degrees of freedom from the start. In the rest of the chapter we will often take this shortcut to obtain the quantized theory. The independent degrees of freedom will be identified as the true degrees of freedom, and the commutation relations of the corresponding operators will be derived from the Dirac bracket.

\subsection{Pilot-wave interpretation}
Now that we have the quantum field theory, we can construct a pilot-wave interpretation. The conservation equation corresponding to the functional Schr\"o\-ding\-er equation (\ref{h30.010}) reads
\begin{equation}
\frac{\partial |\Psi|^2}{\partial t}+\int{ d^3x \bigg( \frac{\delta J_{\phi}}{\delta \phi}+ \frac{\delta J_{\phi^*}}{\delta \phi^*}\bigg)}  =0,
\label{h30.01001}
\end{equation}
with
\begin{eqnarray}
J_{\phi}&=&\frac{1}{2i} \left(\Psi^* \frac{\delta}{\delta \phi^*}\Psi - \Psi \frac{\delta}{\delta \phi^*} \Psi^* \right) - ieV_0 |\Psi|^2  \phi, \nonumber\\
J_{\phi^*}&=&\frac{1}{2i} \left(\Psi^* \frac{\delta}{\delta \phi}\Psi - \Psi \frac{\delta}{\delta \phi} \Psi^* \right) + ieV_0 |\Psi|^2  \phi^*. 
\label{h30.011}
\end{eqnarray}
In the pilot-wave interpretation we introduce the field beables $\phi$ and $\phi^*$ whose motion is governed by the wavefunctional via the guidance equations 
\begin{equation}
{\dot \phi} =  J_{\phi}/ |\Psi|^2, \qquad  {\dot \phi}^* =  J_{\phi^*}/ |\Psi|^2.
\label{h30.012}
\end{equation}
The density of field beables is given by the equilibrium density $|\Psi|^2$. In the free case this pilot-wave interpretation reduces to the one originally presented by Kaloyerou \cite{kaloyerou85} and by Valentini \cite{valentini92,valentini96}.

\section{The Duffin-Kemmer-Petiau theory in the spin-1 representation}\label{Massive spin-1}
In this section, we consider the construction of a pilot-wave interpretation for the Duffin-Kemmer-Petiau (DKP) theory in the spin-1 representation. First we show the equivalence of the quantized DKP theory and the quantized Proca theory. Then afterwards we present the corresponding pilot-wave interpretation.

\subsection{Equivalence with the canonically quantized Proca theory}
As in the spin-0 case we consider the DKP field coupled to a non-quantized electromagnetic field $V^{\mu}$. The treatment of the DKP field coupled to a quantized electromagnetic field is completely analogous and will not be treated explicitly. 

If we write the ten component DKP field $\psi$ as 
\begin{equation}
\psi= {m}^{-1/2} (-{\bf E}, {\bf B},m{\bf A}, -mA_0)^T, 
\label{h30.01201}
\end{equation}
then in the spin-1 representation given in Appendix {\ref{appa}}, the minimally coupled DKP Lagrangian reads
\begin{eqnarray}
L_K = \int d^3 x \mathcal{L}_K &=& \int d^3 x \bigg(\frac{1}{2} (A^*_i D_0 E_i + A_i D^*_0 E^*_i -  E^*_i D_0 A_i -  E_i D^*_0 A^*_i)\nonumber\\
&&\qquad +A^*_0D_i E_i -  E^*_i  D_i A_0 -  \varepsilon_{ijk} (B^*_i D_j A_k + A^*_i D_j B_k) \nonumber\\
&&\qquad  - E^*_i E_i + B^*_i B_i + m^2 A^*_0 A_0 - m^2 A^*_i A_i \bigg)
\label{h30.0121}
\end{eqnarray}
The equations of motion are
\begin{eqnarray}
D_0  E_i =& \varepsilon_{ijk} D_j B_k + m^2 A_i, \qquad D^*_0  E^*_i &= \varepsilon_{ijk} D^*_j B^*_k + m^2 A^*_i ,  \nonumber\\
D_0  A_i =& -  E_i -  D_i A_0 , \qquad  \qquad D^*_0  A^*_i &= -  E^*_i -  D^*_i A^*_0,  \nonumber\\
D_i  E_i =& - m^2 A_0, \qquad  \qquad \qquad D^*_i  E^*_i &= - m^2 A^*_0,  \nonumber\\
B_i =& \varepsilon_{ijk} D_j A_k, \qquad \qquad \qquad \;\; B^*_i &= \varepsilon_{ijk} D^*_j A^*_k.
\label{h219}
\end{eqnarray}
The canonically conjugate momenta are
\begin{eqnarray}
\Pi_{A_i} =& \frac{\delta L }{\delta {\dot A_i}  } = -\frac{1}{2} E^*_i, \qquad 
\Pi_{A^*_i} =& \frac{\delta L }{\delta {\dot A^*_i}} = -\frac{1}{2} E_i,\nonumber\\
\Pi_{A_0} =& \frac{\delta L }{\delta {\dot A_0}  } = 0,\qquad \qquad
\Pi_{A^*_0} =& \frac{\delta L }{\delta {\dot A^*_0} } = 0,\nonumber\\
\Pi_{E_i} =& \frac{\delta L }{\delta {\dot E_i}  } = \frac{1}{2} A^*_i, \qquad \quad
\Pi_{E^*_i} =& \frac{\delta L }{\delta {\dot E^*_i}  } = \frac{1}{2} A_i,\nonumber\\
\Pi_{B_i} =& \frac{\delta L }{\delta {\dot B_i}  } = 0, \qquad \qquad
\Pi_{B^*_i} =& \frac{\delta L }{\delta {\dot B^*_i}  } = 0.
\label{h30.0122}
\end{eqnarray}
Hence the primary constraints are
\begin{eqnarray}
\chi_{A_i}=& \Pi_{A_i}  + \frac{1}{2} E^*_i,\qquad  \chi_{A^*_i} &= \Pi_{A^*_i}  + \frac{1}{2} E_i,\nonumber\\
\chi_{A_0}=& \Pi_{A_0},\qquad \qquad \quad \chi_{A^*_0} &= \Pi_{A^*_0}  , \nonumber\\
\chi_{E_i}=& \Pi_{E_i}  - \frac{1}{2} A^*_i,\qquad \: \chi_{E^*_i} &= \Pi_{E^*_i}  - \frac{1}{2} A_i,\nonumber\\
\chi_{B_i} =& \Pi_{B_i}  ,\qquad \qquad \quad \; \chi_{B^*_i} &= \Pi_{B^*_i} .
\label{h30.0124}
\end{eqnarray}
The canonical Hamiltonian reads
\begin{eqnarray}
H_C\!\! &=& \!\!\int d^3 x \Big( E^*_i D_i A_0 + \varepsilon_{ijk} B^*_i D_j A_k + E_i D^*_i A^*_0 + \varepsilon_{ijk} B_i D^*_j A^*_k \nonumber\\ 
\!\!&&\!\!\! + E^*_i E_i - B^*_i B_i + m^2 (A^*_i A_i  - A^*_0 A_0) +  ieV_0(A_i E^*_i - A^*_i E_i)\! \Big).
\label{h30.0123}
\end{eqnarray}
The total Hamiltonian reads 
\begin{equation}
H_T  = H_C + \sum_{\gamma} \int d^3 x u_{\gamma}({\bf x}) \chi_{\gamma}({\bf x}),
\label{h30.0031}
\end{equation}
where the label $\gamma$ takes the values $A_i ,A^*_i, A_0,A^*_0 , E_i,E^*_i, B_i ,B^*_i$ and the $u_{\gamma}$ are arbitrary fields. The requirement that the constraints $\chi_{A_i},\chi_{A^*_i},\chi_{E_i}$ and $\chi_{E^*_i}$ are weakly conserved in time respectively yield
\begin{eqnarray}
u_{E^*_i} &=& \varepsilon_{ijk} D^*_k B^*_j +m^2 A^*_i + ieV_0 E^*_i,          \nonumber\\ 
u_{E_i} &=& \varepsilon_{ijk} D_k B_j +m^2 A_i - ieV_0 E_i,          \nonumber\\
u_{A^*_i}  &=& - D^*_i A^*_0 + ie V_0 A^*_i - E^*_i,          \nonumber\\ 
u_{A_i}  &=& - D_i A_0 - ie V_0 A_i - E_i.
\label{h30.0032}
\end{eqnarray}
The consistency requirements that the constraints $\chi_{A_0},\chi_{A^*_0},\chi_{B_i}$ and $\chi_{B^*_i}$ are weakly conserved in time yield the secondary constraints
\begin{eqnarray}
\chi_{sA^*_0}&=& - D^*_i E^*_i - m^2 A^*_0,\qquad \chi_{sA_0}=  - D_i E_i - m^2 A_0 , \nonumber\\
\chi_{sB^*_i}&=&  \varepsilon_{ijk}D^*_j A^*_k - B^*_i,\qquad \chi_{sB_i}=  \varepsilon_{ijk}D_j A_k - B_i.
\label{h30.0125}
\end{eqnarray}
The requirement that the constraints $\chi_{sA_0},\chi_{sA^*_0},\chi_{sB_i}$ and $\chi_{sB^*_i}$ are weakly conserved in time fix the the remaining fields $u_{A_0},u_{A^*_0},u_{B_i}$ and $u_{B^*_i}$
\begin{eqnarray}
u_{A_0}&=& - D^*_i u_{E^*_i}    /m^2, \qquad u_{A^*_0}= - D_i u_{E_i}    /m^2,\nonumber\\
u_{B_i}&=& \varepsilon_{ijk} D_j u_{A_k}, \qquad u_{B^*_i}= \varepsilon_{ijk} D^*_j u_{A^*_k}.
\label{h30.012501}
\end{eqnarray}
Because all the fields $u_{\gamma}$ are determined by the consistency conditions, all constraints are second class constraints. 

At this stage we can already identify the true degrees of freedom. From the constraints we see that we can take the fields $A_i,A^*_i,E_i,E^*_i$ as the true degrees of freedom. When the constraints are imposed, the canonical Hamiltonian can be written in terms of these fields{\footnote{The distribution $h_{ij}$ is introduced for notational convenience. By using this distribution the Hamiltonian seems to depend nonlocally on the fields. However, this is only apparently because one can explicitly perform the integration over $h_{ij}({\bf y},{\bf z})$ in order to obtain a local Hamiltonian.}}
\begin{eqnarray}
H  &=&   \int d^3 y d^3 z   h_{ij}({\bf y},{\bf z}) E^*_i({\bf y}) E_j({\bf z}) +  \int d^3 x \bigg(\frac{1}{2} G^*_{ij}({\bf x}) G_{ij}({\bf x})\nonumber\\
&&  +m^2 A^*_i({\bf x})A_i({\bf x}) + i eV_0 \big(A_i({\bf x}) E^*_i({\bf x})  - A^*_i({\bf x}) E_i({\bf x})  \big)\bigg) ,
\label{h30.013}
\end{eqnarray}
with
\begin{eqnarray}
G_{ij}  &=& D_{i} A_j - D_{j} A_i, \nonumber\\
h_{ij} ({\bf y},{\bf z}) &=& \left(-\frac{1}{m^2} D_{y_i} D_{y_j} + \delta_{ij}\right) \delta({\bf y}-{\bf z}).
\label{h30.0130001}
\end{eqnarray}

In order to quantize the system we need the Dirac bracket. The matrix $C_{\gamma \gamma'}({\bf x},{\bf y})$, where the labels $\gamma$ and $\gamma'$ take the values $A_0,A^*_0 ,A_i ,A^*_i, E_i,E^*_i, B_i ,B^*_i,$ $sA_0 , sA^*_0,sB_i,sB^*_i$, has the following nonzero components
\begin{eqnarray}
C_{ A_i ,E^*_j  }({\bf x},{\bf y}) &=& -C_{ E^*_j ,  A_i }({\bf y},{\bf x}) =  \delta_{ij} \delta({\bf x} -{\bf y} )   , \nonumber\\     
C_{ A^*_i , E_j }({\bf x},{\bf y}) &=& -C_{ E_j , A^*_i }({\bf y},{\bf x}) = \delta_{ij} \delta({\bf x} -{\bf y} )    , \nonumber\\     
C_{ A_0 , sA_0 }({\bf x},{\bf y}) &=& -C_{sA_0  ,A_0  }({\bf y},{\bf x}) =  m^2 \delta({\bf x} -{\bf y} )   , \nonumber\\     
C_{ A^*_0 ,sA^*_0  }({\bf x},{\bf y}) &=& -C_{ sA^*_0 , A^*_0 }({\bf y},{\bf x}) =   m^2 \delta({\bf x} -{\bf y} )   , \nonumber\\     
C_{ B_j , sB_i}({\bf x},{\bf y}) &=& -C_{sB_i  ,B_j  }({\bf y},{\bf x}) =   \delta_{ij} \delta({\bf x} -{\bf y} )    , \nonumber\\     
C_{ B^*_j ,sB^*_i  }({\bf x},{\bf y}) &=& -C_{sB^*_i  , B^*_j }({\bf y},{\bf x}) =    \delta_{ij}\delta({\bf x} -{\bf y} )     , \nonumber\\     
C_{ E_i , sA_0 }({\bf x},{\bf y}) &=& -C_{sA_0  , E_i }({\bf y},{\bf x}) =   D^*_{x_i} \delta({\bf x} -{\bf y} )  , \nonumber\\     
C_{ E^*_i , sA^*_0 }({\bf x},{\bf y}) &=& -C_{ sA^*_0 , E^*_i }({\bf y},{\bf x}) =   D_{x_i} \delta({\bf x} -{\bf y} )  , \nonumber\\     
C_{ sB_j, A_i  }({\bf x},{\bf y}) &=& -C_{ A_i,sB_j }({\bf y},{\bf x}) =  \varepsilon_{ijk} D^*_{x_k} \delta({\bf x} -{\bf y} )   , \nonumber\\     
C_{ A^*_i ,sB^*_j  }({\bf x},{\bf y}) &=& -C_{ sB^*_j ,A^*_i  }({\bf y},{\bf x}) =  \varepsilon_{ijk} D_{x_k} \delta({\bf x} -{\bf y} )   .   
\label{h30.0126}
\end{eqnarray}
Its inverse $C^{-1}_{\gamma \gamma'}$ has the following nonzero components
\begin{eqnarray}
C^{-1}_{E^*_j,A_i}({\bf x},{\bf y})&=& - C^{-1}_{A_i,E^*_j} ({\bf y},{\bf x})= \delta_{ij} \delta({\bf x}-{\bf y}), \nonumber\\ 
C^{-1}_{E_i,A^*_j}({\bf x},{\bf y})&=& - C^{-1}_{A^*_j,E_i}({\bf y},{\bf x}) = \delta_{ij} \delta({\bf x}-{\bf y}), \nonumber\\ 
C^{-1}_{sB_j,B_i}({\bf x},{\bf y})&=& - C^{-1}_{B_i,sB_j}({\bf y},{\bf x}) = \delta_{ij} \delta({\bf x}-{\bf y}), \nonumber\\ 
C^{-1}_{sB^*_j,B_i^*}({\bf x},{\bf y})&=& - C^{-1}_{B_i^*,sB^*_j}({\bf y},{\bf x}) = \delta_{ij} \delta({\bf x}-{\bf y}), \nonumber\\
C^{-1}_{sA_0,A_0}({\bf x},{\bf y})&=& - C^{-1}_{A_0,sA_0}({\bf y},{\bf x}) = \frac{1}{m^2} \delta({\bf x}-{\bf y}), \nonumber\\ 
C^{-1}_{sA^*_0,A^*_0}({\bf x},{\bf y})&=& - C^{-1}_{A^*_0,sA^*_0}({\bf y},{\bf x}) = \frac{1}{m^2} \delta({\bf x}-{\bf y}), \nonumber\\ 
C^{-1}_{A_i,A^*_0}({\bf x},{\bf y})&=& - C^{-1}_{A^*_0,A_i}({\bf y},{\bf x}) = \frac{1}{m^2} D_{i} \delta({\bf x}-{\bf y}) , \nonumber\\ 
C^{-1}_{A_0,A^*_i}({\bf x},{\bf y})&=& - C^{-1}_{A^*_i,A_0}({\bf y},{\bf x}) = \frac{1}{m^2} D_{i} \delta({\bf x}-{\bf y}), \nonumber\\ 
C^{-1}_{E_i,B^*_j}({\bf x},{\bf y})&=& - C^{-1}_{B^*_j,E_i}({\bf y},{\bf x}) = \varepsilon_{ijk} D_{k}  \delta({\bf x}-{\bf y}), \nonumber\\ 
C^{-1}_{B_i,E^*_j}({\bf x},{\bf y})&=& - C^{-1}_{E^*_j,B_i}({\bf y},{\bf x}) = \varepsilon_{ijk} D_{k}\delta({\bf x}-{\bf y}).
\label{h30.0127}
\end{eqnarray}
The Dirac bracket for the fields $A_i,A^*_i,E_i,E^*_i$ reads 
\begin{eqnarray}
[ A_i({\bf x}), E^*_j({\bf y})]_D &=& -\delta_{ij} \delta({\bf x}-{\bf y}), \nonumber\\ 
{[ A^*_i({\bf x}), E_j({\bf y})]_D } &=& -\delta_{ij} \delta({\bf x}-{\bf y})
\label{h30.012701}
\end{eqnarray}
and all the other Dirac brackets involving the fields $A_i,A^*_i,E_i,E^*_i$ are zero. The Dirac bracket for the other fields can be derived from these by using the constraints ({\ref{h30.0124}}) and ({\ref{h30.0125}}).

Because we already have identified the true degrees of freedom, there is no need to perform the Maskawa and Nakajima canonical transformation explicitly. The theory is quantized by imposing the following commutation relations for the true degrees of freedom
\begin{eqnarray}
[{\widehat A}_i({\bf x}),{\widehat E}^*_j({\bf y})] &=& i [ A_i({\bf x}), E^*_j({\bf y})]_D = -i\delta_{ij} \delta({\bf x}-{\bf y}), \nonumber\\ 
{ [{\widehat A}^*_i({\bf x}),{\widehat E}_j({\bf y})]}&=& i[A^*_i({\bf x}), E_j({\bf y})]_D  = -i\delta_{ij} \delta({\bf x}-{\bf y}).
\label{h30.0128}
\end{eqnarray}
All the other fundamental commutation relations involving the operators ${\widehat A}_i,{\widehat E}^*_i,{\widehat A}^*_i,{\widehat E}_i$ and $A^*_i({\bf x})$ are zero. The commutation relations are realized by the representation 
\begin{eqnarray}
{\widehat A}_i({\bf x}) &=& A_i({\bf x}) ,  \qquad {\widehat E}^*_j({\bf x})= i \frac{\delta}{\delta A_j({\bf x})} , \nonumber\\
{\widehat A}^*_i({\bf x}) &=& A^*_i({\bf x}) ,  \qquad {\widehat E}_j({\bf x})= i \frac{\delta}{\delta A^*_j({\bf x})}.
\label{h30.0129}
\end{eqnarray}
In this representation the Hamiltonian operator reads
\begin{eqnarray}
{\widehat H} &=& -\int d^3 y d^3 z   h_{ij}({\bf y},{\bf z}) \frac{\delta}{\delta A_i({\bf y}) } \frac{\delta}{\delta A^*_j({\bf z}) }\nonumber\\
&&+  \int d^3 x \bigg[ eV_0({\bf x})\left(A^*_i({\bf x}) \frac{\delta}{\delta A^*_i({\bf x})}  - A_i({\bf x}) \frac{\delta}{\delta A_i({\bf x})}  \right) \nonumber\\
&&+ \frac{1}{2} G^*_{ij}({\bf x}) G_{ij}({\bf x}) +m^2 A^*_i({\bf x})A_i({\bf x})\bigg] .
\label{h30.0130000002}
\end{eqnarray}
The operators act on wave\-func\-tion\-als $\Psi(A_i,A^*_i,t) = \langle A_i,A^*_i | \Psi (t)\rangle$. The functional Schr\"o\-ding\-er equation for the wave\-func\-tion\-al is given by
\begin{equation}
i\frac{\partial \Psi(A_i,A^*_i,t)}{\partial t} = {\widehat H}\Psi(A_i,A^*_i,t).
\label{h30.014}
\end{equation}
This is the same quantum field theory that can be derived from quantizing the Proca theory. This shows the equivalence of the quantized DKP theory in the spin-1 representation and the quantized Proca theory. The same pilot-wave interpretation could hence be obtained by considering the quantized Proca theory from the outset.

\subsection{Pilot-wave interpretation}
In order to obtain the pilot-wave interpretation we consider the corresponding conservation equation 
\begin{equation}
\frac{\partial |\Psi|^2}{\partial t}  + \int{ d^3x \bigg( \frac{\delta J_{A_i}({\bf x})}{\delta A_i({\bf x})}+ \frac{\delta J_{A^*_i}({\bf x})}{\delta A^*_i({\bf x})}  \bigg)} =0 ,
\label{h30.01401}
\end{equation}
with
\begin{eqnarray}
J_{A_i}({\bf x}) &=& \frac{1}{2i}\int { d^3y h_{ij} ({\bf x},{\bf y}) \left(\Psi^*  \frac{\delta}{\delta A^*_j({\bf y})}\Psi  - \Psi \frac{\delta}{\delta A^*_j({\bf y})} \Psi^* \right)}\nonumber\\
&&- ieV_0({\bf x}) |\Psi|^2  A_i({\bf x}), \nonumber\\
J_{A^*_i}({\bf x}) &=& \frac{1}{2i}\int{ d^3y   h^*_{ij} ({\bf x},{\bf y}) \left(\Psi^*  \frac{\delta}{\delta A_j({\bf y})}\Psi - \Psi \frac{\delta}{\delta A_j({\bf y})} \Psi^* \right)}\nonumber\\
&&+ ieV_0({\bf x}) |\Psi|^2  A^*_i({\bf x}).
\label{h30.015}
\end{eqnarray}
In the pilot-wave interpretation, the conserved density of the field beables $A_i,A^*_i$ is given by $|\Psi|^2$ and the guidance equations for the fields are
\begin{equation}
{\dot A_i} =  J_{A_i}/ |\Psi|^2, \qquad  {\dot A}^*_i =  J_{A^*_i}/ |\Psi|^2.
\label{h30.016}
\end{equation}

\section{The electromagnetic field}\label{maxwell}
In this section we consider the pilot-wave interpretation of the quantized the electromagnetic field. We start with reconsidering the quantization of the electromagnetic field. The reason to do is because the two existing pilot-wave approaches, the one originated by Bohm \cite{bohm2} and the one by Valentini \cite{valentini92,valentini96}, find a natural home in two different ways of quantizing the electromagnetic field. Although these different ways of quantizing the electromagnetic field yield equivalent quantum theories, the corresponding pilot-wave interpretations are not equivalent. We will indicate some problems in the approach by Valentini, which, in our opinion, makes the original approach by Bohm favourable. 

This is the organization of the section. First, in Section \ref{hamiltonianformulationemfield}, we recall the Hamiltonian formulation of Max\-well's theory for the electromagnetic field. In order not to obscure the issue, we will not start from the Harish-Chandra theory (cf.\ Section \ref{masslessspin-0andspin-1harishchandra}), which is equivalent to Maxwell's theory at the level of classical field equations, but from the Maxwell form straight away. Most probably it presents no problem to show the equivalence for the quantized theories. The Hamiltonian formulation of Maxwell's theory can also be found in \cite{dirac64,sundermeyer82,henneaux91,weinberg95}. In Section \ref{constrasoperators} we then consider the quantization of the electromagnetic field by imposing the constraints as operator identities. We will work with the Coulomb gauge and then later in Section \ref{otuoftbi} we will discuss some other gauges. This approach of quantizing the electromagnetic field will lead to Bohm's original pilot-wave interpretation. In Section \ref{constrconsta} we then consider the quantization of the electromagnetic field by imposing constraints as conditions on states. This approach will lead to Valentini's pilot-wave approach.

\subsection{Hamiltonian formulation of the electromagnetic field}\label{hamiltonianformulationemfield}
The free Lagrangian density for the electromagnetic vector potential $V^{\mu}=(V_0,{\bf V})$ is given by 
\begin{eqnarray}
L_M &=& \int d^3 x \mathcal{L}_M = -\frac{1}{4}\int d^3 x  F^{\mu \nu}F_{\mu \nu} \nonumber\\
&=&\int d^3 x \left( \frac{1}{2} (\partial_0 V_i +\partial_i V_0 )(\partial_0 V_i +\partial_i V_0 )  -\frac{1}{4} F_{ij}F_{ij} \right)
\label{h30.017}
\end{eqnarray}
with $F^{\mu \nu} = \partial^\mu V^{\nu} - \partial^\nu V^{\mu}$ and $F_{ij} = \partial_i V_j - \partial_j V_i$. The Lagrangian is invariant under gauge transformations
\begin{equation}
V^{\mu} \to V^{\mu} - \partial^\mu  \theta.
\label{h30.01701}
\end{equation}
The equations of motion are
\begin{equation}
\partial_\mu F^{\mu \nu} = 0.
\label{h30.01702}
\end{equation}
The canonically conjugate momenta of the fields are 
\begin{eqnarray}
\Pi_{V_0} &=& \frac{\delta L }{\delta {\dot V_0} } = 0,\nonumber\\
\Pi_{V_i} &=& \frac{\delta L }{\delta {\dot V_i} } = (\partial_0 V_i +\partial_i V_0 ).
\label{h30.018}
\end{eqnarray}
Because $\Pi_{V_0}=0$, we have a primary constraint
\begin{equation}
\chi_1 = \Pi_{V_0}. 
\label{h30.019}
\end{equation}
The canonical Hamiltonian reads
\begin{equation}
H_C =  \int d^3 x \left( \frac{1}{2}\Pi_{V_i}\Pi_{V_i} + \frac{1}{4} F_{ij}F_{ij} - \Pi_{V_i}\partial_i V_0 \right).
\label{h30.0181}
\end{equation}
The total Hamiltonian reads
\begin{equation}
H_T = H_C + \int d^3 x u_1 \chi_{1}.
\label{h30.0181001}
\end{equation}
The consistency requirement that the primary constraint $\chi_1$ is conserved in time leads to the secondary constraint
\begin{equation}
\chi_2 = \partial_i \Pi_{V_i}, 
\label{h30.020}
\end{equation}
which is called the Gauss constraint. There are no further constraints and the field $u_1$ remains undetermined. Hence both constraints are first class constraints. The constraint  $\chi_1$ should be included in the Hamiltonian to yield the extended Hamiltonian
\begin{equation}
H_E = H_C + \int d^3 x u_1 \chi_{1} + \int d^3 x u_2 \chi_{2}.
\label{h30.0181001}
\end{equation}

In the next section, we will apply both approaches of dealing with first class constraints, as described in Section \ref{firstclassconstraints}. Let us first identify the true degrees of freedom. In order to do so we make use of the following property. If $F_i$ is a vector field (which, together its its spatial derivatives, vanishes sufficiently fast at spatial infinity), one can uniquely decompose it into a transversal part and a longitudinal part: 
\begin{equation}
F_i = F^T_i + F^L_i,
\label{h30.02510001}
\end{equation}
with 
\begin{eqnarray}
F^T_i &=&  \left(\delta_{ij} -  \frac{\partial_i \partial_j }{\nabla^2}\right) F_j, \nonumber\\
 F^L_i &=&  \frac{\partial_i \partial_j }{\nabla^2} F_j
\label{h30.02510002}
\end{eqnarray}
and where the operator $\nabla^{-2}$ acts as
\begin{equation}
\frac{1}{\nabla^2} f({\bf x}) = -\int d^3 y \frac{f({\bf y})}{4\pi|{\bf x}-{\bf y}|}.
\label{h30.0251}
\end{equation}
By using this decomposition for the fields $V_i$ and  $ \Pi_{V_i}$, one sees that transversal components $V^T_i$ and $\Pi^T_{V_i}$ live on the constraint space $\Pi_{V_0} =\partial_i \Pi_{V_i}=0$. One can also check that $V^T_i$ and $\Pi^T_{V_i}$ have Poisson brackets zero with the constraints $\chi_1$ and $\chi_2$. Because $\chi_1$ and $\chi_2$ were the generators of gauge transformations this means that the transversal fields $V^T_i$ and $\Pi^T_{V_i}$ are gauge invariant and hence they represent the true degrees of freedom. We can do a counting to see that the transversal fields represent {\em all} the true degrees of freedom. We have 8 canonical variables at each point in space and 4 constraints. That leaves us with 4 true degrees of freedom at each point in space and this is exactly the number of degrees of freedom of the transversal fields. The Hamiltonian $H$ on the constraint space can also be expressed in terms of $V^T_i$ and $\Pi^T_{V_i}$:
\begin{equation}
H =\frac{1}{2}  \int d^3 x \left(  \Pi^T_{V_i} \Pi^T_{V_i}  -  V^T_i \nabla^2 V^T_i  \right).
\label{h30.02502}
\end{equation}

\subsection{Constraints as operator identities: The Coulomb gauge} \label{constrasoperators}
\subsubsection{Commutation relations in the Coulomb gauge }
The two first class constraints $\chi_1$ and $\chi_2$ are generators of infinitesimal gauge transformations. In particular, the infinitesimal gauge transformations $V^{\mu} \to V^{\mu} + \delta V^{\mu}=  V^{\mu} - \partial^\mu  \varepsilon$, which are symmetry transformations of the Lagrangian equations of motion ({\ref{h30.01702}}), are generated by the linear combination $\int d^3 y \left( \chi_1(y)\varepsilon_1(y)+ \chi_2(y)\varepsilon_2(y) \right)$ with $\varepsilon_1= -\partial_0 \varepsilon$ and $\varepsilon_2= \varepsilon$, i.e.
\begin{equation}
\delta V^{\mu} (x) = \left[V^{\mu}(x),  \int d^3 y \left( \chi_1(y)\varepsilon_1(y)+ \chi_2(y)\varepsilon_2(y) \right) \right]_P = -\partial^\mu  \varepsilon(x),
\label{h30.02001}
\end{equation}
where the fields are taken at equal time, i.e.\  $x_0=y_0$. In fact there still exist gauge transformations generated by the constraints, which are not gauge symmetries of the Lagrangian equations of motion, see \cite[p.\ 134]{sundermeyer82}. This indicates that the set of infinitesimal gauge transformations of the Hamiltonian equations of motion (which are generated by the first class constraints) does not necessarily correspond to the set of infinitesimal gauge transformation of the Lagrangian equations of motion.

As indicated in Section \ref{firstclassconstraints} we can impose additional constraints, i.e.\ gauge constraints, so that the full set of constraints becomes second class. A suitable set of constraints is given by the Coulomb gauge 
\begin{eqnarray}
\chi_3 &=& \partial_i V_i, \label{h30.021}\\
\chi_4 &=& V_0. \label{h30.022}
\end{eqnarray}
The Coulomb gauge satisfies the requirements of an admissible gauge (cf.\ Section \ref{firstclassconstraints}). First, the Coulomb gauge can be attained with the gauge transformation ({\ref{h30.01701}}) with 
\begin{equation}
\theta = -\frac{1}{\nabla^2} \partial_i V_i 
\label{h30.02201}
\end{equation}
Second, if we restrict ourself to gauge transformations for which the function $\theta$ vanishes at spatial infinity, then the Coulomb gauge fixes the gauge uniquely. 

The quantization of the electromagnetic field in the Coulomb gauge, by imposing constraints as operator equations, is a textbook example of dealing with constraints. The treatment can for example be found in \cite[pp.\ 123-140]{sundermeyer82} and \cite[pp.\ 339-350]{weinberg95}. Here we repeat only the basic elements. In Section \ref{otuoftbi}, we will consider quantization in other gauges. 

The Dirac bracket can be calculated by using the inverse of the matrix $C_{NM}({\bf x},{\bf y})=[\chi_N({\bf x}),\chi_M({\bf y})]_P$, with $M,N=1,\dots,4$. The matrix $C$ has the following nonzero components
\begin{eqnarray}
C_{23}({\bf x},{\bf y})&=&-C_{32}({\bf x},{\bf y})   = \nabla^2 \delta({\bf x} - {\bf y}), \nonumber\\
C_{14}({\bf x},{\bf y})&=&-C_{41}({\bf x},{\bf y}) =  - \delta({\bf x} - {\bf y}).
\label{h30.023}
\end{eqnarray}
One can construct an inverse matrix $C^{-1}_{NM}({\bf x},{\bf y})$, which has the following nonzero components
\begin{eqnarray}
C^{-1}_{14}({\bf x},{\bf y})&=&-C^{-1}_{41}({\bf x},{\bf y}) = \delta({\bf x} - {\bf y}), \nonumber\\
C^{-1}_{23}({\bf x},{\bf y})&=&-C^{-1}_{32}({\bf x},{\bf y}) = -\frac{1}{\nabla^2}\delta({\bf x} - {\bf y}).
\label{h30.024}
\end{eqnarray}
Note that this inverse is not unique. There is an ambiguity in the matrix elements $C^{-1}_{23}({\bf x},{\bf y})$ and $C^{-1}_{32}({\bf x},{\bf y})$. They are both determined up to a function $g({\bf x},{\bf y})$ which satisfies $\nabla^2_x g({\bf x},{\bf y}) = \nabla^2_y g({\bf x},{\bf y}) = 0$. This ambiguity in the matrix $C^{-1}$ may lead to an ambiguity in the Dirac bracket and hence in the field commutators. However, the ambiguity for the inverse matrix $C^{-1}$ can be removed by considering the boundary conditions for the fields (they vanish sufficiently fast at spatial infinity). We shall not do this analysis here, but we refer the reader to \cite[pp.\ 65-72]{sundermeyer82} and \cite{steinhardt80} where the same issue is treated in the context of light-cone quantization. 

Note that there was no such an ambiguity in the Duffin-Kemmer-Petiau theory. In the Duffin-Kemmer-Petiau theory, the inverse of the matrix $C$ was always uniquely determined.

Using this inverse matrix $C^{-1}$ we obtain the following Dirac bracket for the fields 
\begin{eqnarray}
[V_i({\bf x}),\Pi_{V_j}({\bf y})]_D &=&  \left(\delta_{ij} -  \frac{\partial_i \partial_j }{\nabla^2}\right)\delta({\bf x} - {\bf y}),\label{h30.02401} \\ 
{[V_{i}({\bf x}),V_{j}({\bf y})]}_D &=&[\Pi_{V_i}({\bf x}),\Pi_{V_j}({\bf y})]_D =0 .
\label{h30.02401}
\end{eqnarray}
The Dirac brackets involving the fields $V_0$ and $\Pi_{V_0}$ are zero. 

Quantization proceeds by imposing the following commutation relations for the operators 
\begin{eqnarray}
[{\widehat V}_i({\bf x}),{\widehat \Pi}_{V_j}({\bf y})] &=& i \left(\delta_{ij} -  \frac{\partial_i \partial_j }{\nabla^2}\right)\delta({\bf x} - {\bf y}),\label{h30.02401} \\ 
{[{\widehat V}_{i}({\bf x}),{\widehat V}_{j}({\bf y})]} &=&[{\widehat \Pi}_{V_i}({\bf x}),{\widehat \Pi}_{V_j}({\bf y})] =0. 
\label{h30.025}
\end{eqnarray} 
The commutation relations involving the operators ${\widehat V_0}$ and ${\widehat \Pi}_{V_0}$ are zero. 

The rest of this section is divided in three parts: 
\begin{itemize}
\item
First, we consider in detail the construction of a class of Maskawa and Nakajima canonical transformations which enable us to separate the true degrees of freedom from the constraints.
\item
Second, we consider the functional Schr\"odinger equation in terms of the true degrees of freedom, together with the pilot-wave interpretation.
\item
Finally, we consider some explicit examples of Maskawa and Nakajima canonical transformations.
\end{itemize}

\subsubsection{The true degrees of freedom}\label{repcanvar}
We have to find a representation for the operators ${\widehat V}_{0},{\widehat V}_{i},{\widehat \Pi}_{V_{0}}$ and ${\widehat \Pi}_{V_{i}}$ so that the constraints and the commutation relations are satisfied. The operators ${\widehat V}_{0}$ and ${\widehat \Pi}_{V_{0}}$ are zero as constraints, so we only have to consider a representation for the operators ${\widehat V}_{i}$ and ${\widehat \Pi}_{V_{i}}$.

Inspired by the decomposition ({\ref{h30.02510001}}) into longitudinal and transversal part of fields, it would be tempting to use the representation
\begin{eqnarray}
{{\widehat V}_i}({\bf x}) &=& \left( \delta_{ij} -  \frac{\partial_i \partial_j}{\nabla^2} \right) V_j({\bf x}), \nonumber\\
{\widehat \Pi}_{V_i}({\bf x})&=& - i \left(\delta_{ij} -  \frac{\partial_i \partial_j }{\nabla^2}\right) \frac{\delta}{\delta V_j({\bf x}) } ,
\label{h30.026}
\end{eqnarray}
where $V_j({\bf x})$ is a real-valued three component field. This representation satisfies the constraints and the commutation relations ({\ref{h30.025}}). However, with this representation we would introduce superfluous degrees of freedom. This is because this representation keeps 6 degrees of freedom at each point in space, namely $V_i({\bf x})$ and $\delta /\delta V_i({\bf x})$. This means that the representation has two degrees of freedom in excess. The reason for the superfluous degrees of freedom is that the representation ({\ref{h30.026}}) is invariant under the transformations $V_i({\bf x}) \to V_i({\bf x}) + \partial_i \theta({\bf x})$ and is therefore not one-to-one.

In order to construct a correct representation, we will first explicitly perform the Maskawa-Nakajima (MN) canonical transformation. This canonical transformation will separate the true degrees of freedom from the constraints. Then we can use the standard representation for the operators corresponding to the true degrees of freedom. In order to find the MN canonical transformation we will work constructively. We will also try to keep some generality. In this way we will in fact end up with a class of possible MN canonical transformations.

Because the constraints $\chi_1=\Pi_{V_0}$ and $\chi_4=V_0$ already form a canonical pair they should not be involved in the MN canonical transformation of the other canonical variables $V_i$ and $ \Pi_{V_i}$. Because the Coulomb gauge is linear in the fields $V_i$, it is most simple to try to find a canonical transformation which is also linear in the fields. With ${\widetilde V}_i$ the new field variables, with corresponding momenta $\Pi_{{\widetilde V}_i}$, we then look for a transformation of the form
\begin{eqnarray}
V_i({\bf x}) &=& \int d^3 y K_{ij} ({\bf x},{\bf y})  {\widetilde V}_j({\bf y}), \nonumber\\
{\widetilde V}_i({\bf x}) &=& \int d^3 y K^{-1}_{ij} ({\bf x},{\bf y})  V_j({\bf y}),
\label{h30.02601}
\end{eqnarray}
with $K$ a matrix with inverse $K^{-1}$
\begin{eqnarray}
\int d^3 x K^{-1}_{ki} ({\bf z},{\bf x})  K_{ij} ({\bf x},{\bf y}) = \delta_{kj} \delta({\bf z} - {\bf y}),  \label{h30.0260101}\\
\int d^3 x K_{ki} ({\bf z},{\bf x})  K^{-1}_{ij} ({\bf x},{\bf y}) = \delta_{kj} \delta({\bf z} - {\bf y}) \label{h30.0260102}.
\end{eqnarray}
We further need a transformation of the momenta $\Pi_{V_i}$. For simplicity we construct a canonical transformation with corresponding generating function $F=0$. With $F=0$, it follows from ({\ref{h30.000910101}}) that
\begin{equation}
 \Pi_{{\widetilde V}_i} ({\bf x}) = \int d^3 y \Pi_{V_j} ({\bf y})  \frac{\delta V_j ({\bf y}) }{\delta {\widetilde V}_i({\bf x})}  =  \int d^3  y \Pi_{V_j} ({\bf y})   K_{ji} ({\bf y},{\bf x}) .
\label{h30.0260301}
\end{equation}
The inverse transformation reads
\begin{equation}
\Pi_{V_i} ({\bf x}) =  \int d^3 y \Pi_{{\widetilde V}_j} ({\bf y})  K^{-1}_{ji} ({\bf y},{\bf x}).
\label{h30.0260302}
\end{equation}
In this way, the transformation determined by ({\ref{h30.02601}}), ({\ref{h30.0260301}}) and ({\ref{h30.0260302}}) is a canonical transformation (and even a {\em point transformation}).

Now we further need to assure the the canonical transformation is a MN canonical transformation. We will construct a canonical transformation such that the remaining constraints $\partial_i V_i= \partial_i \Pi_{V_i}=0$ read ${\widetilde V}_3=\Pi_{{\widetilde V}_3}=0$ in terms of the new variables. The true degrees of freedom then will be ${\widetilde V}_1,{\widetilde V}_2,\Pi_{{\widetilde V}_1}$ and $\Pi_{{\widetilde V}_2}$. Therefore we further require that the transformations $K$ and $K^{-1}$ satisfy
\begin{eqnarray}
\partial_{x_i} K_{ij} ({\bf x},{\bf y}) &=& 0, \qquad {\textrm{for }} j=1,2 , \label{h30.026021} \\
\varepsilon_{ilk} \partial_{x_l} K_{kj} ({\bf x},{\bf y}) &=& 0, \qquad {\textrm{for }} j=3, \label{h30.026022} \\
\partial_{x_i} K^{-1}_{ji} ({\bf y},{\bf x}) &=& 0, \qquad {\textrm{for }} j=1,2 , \label{h30.026023} \\
\varepsilon_{ilk} \partial_{x_l} K^{-1}_{jk} ({\bf y},{\bf x}) &=& 0, \qquad {\textrm{for }} j=3. \label{h30.026024} 
\end{eqnarray}
This implies that the transversal parts of the canonical variables $V_i$ and $\Pi_{V_i}$ can be written respectively in terms of ${\widetilde V}_1$ and ${\widetilde V}_2$, and $\Pi_{{\widetilde V}_1}$ and $\Pi_{{\widetilde V}_2}$. The longitudinal parts of the canonical variables $V_i$ and $\Pi_{V_i}$ can be written respectively in terms of ${\widetilde V}_3$ and $\Pi_{{\widetilde V}_3}$. I.e.\ we have 
\begin{eqnarray}
V^T_i({\bf x}) &=&\sum^2_{j=1} \int d^3 y K_{ij} ({\bf x},{\bf y}) {\widetilde V}_j({\bf y}), \nonumber\\
V^L_i({\bf x}) &=& \int d^3 y K_{i3} ({\bf x},{\bf y}) {\widetilde V}_3({\bf y}), \nonumber\\
\Pi^T_{V_i} ({\bf x}) &=&  \sum^2_{j=1} \int d^3 y \Pi_{{\widetilde V}_j} ({\bf y})  K^{-1}_{ji} ({\bf y},{\bf x}), \nonumber\\
\Pi^L_{V_i} ({\bf x}) &=&  \int d^3 y \Pi_{{\widetilde V}_3} ({\bf y})  K^{-1}_{3i} ({\bf y},{\bf x}).
\label{h30.02603}
\end{eqnarray}
Because $K_{i3}$ and $K^{-1}_{3i}$ are irrotational, cf.\ ({\ref{h30.026022}}) and ({\ref{h30.026024}}), we can write
\begin{eqnarray}
K_{i3} ({\bf x},{\bf y}) &=& \partial_{x_i} U({\bf x},{\bf y}), \label{h30.0260303} \\
K^{-1}_{3i} ({\bf y},{\bf x}) &=& \partial_{x_i} {\bar U}({\bf y},{\bf x}). \label{h30.026030001}
\end{eqnarray}
From the expression ({\ref{h30.0260101}}) for $j=k=3$ it then follows that
\begin{equation}
\int d^3 x K^{-1}_{3i} ({\bf z},{\bf x}) K_{i3} ({\bf x},{\bf y})  = \int d^3 x   \left( \partial_{x_i}{\bar U} ({\bf z},{\bf x}) \right)   \left( \partial_{x_i}  U({\bf x},{\bf y})\right)  =\delta({\bf z} - {\bf y}). 
\label{h30.026030002}
\end{equation}
Because $U$ and ${\bar U}$ will make no appearance in the theory when the constraints are imposed as operator identities (because $U$ and ${\bar U}$ are only present in the longitudinal components of the fields), we can make some more assumptions on these matrices. We will assume that 
\begin{equation}
{\bar U} ({\bf z},{\bf x}) = -  \frac{1}{\nabla^2_x} U^{-1}({\bf z},{\bf x}),
\label{h30.026030003}
\end{equation}
with $U^{-1}$ the inverse of $U$, and that the boundary terms in ({\ref{h30.026030002}}) vanish after partial integration. Under these circumstances, the relation ({\ref{h30.026030002}}) is then satisfied.

From 
\begin{equation}
\partial_{i}  V_i({\bf x})=\partial_{i}  V^L_i({\bf x})=\int d^3 y \nabla^2_x U ({\bf x},{\bf y})  {\widetilde V}_3({\bf y})
\label{h30.026030004}
\end{equation}
it follows that ${\widetilde V}_3=0 \Rightarrow \partial_{i}  V_i =0$. On the other hand, from 
\begin{equation}
{\widetilde V}_3({\bf x})= - \int d^3 y \frac{\partial_{y_i}}{\nabla^2_y} U^{-1}({\bf x},{\bf y}) V_i({\bf y}) =\int d^3 y \frac{1}{\nabla^2_y} U^{-1}({\bf x},{\bf y}) \partial_{y_i}V_i({\bf y})
\label{h30.026030005}
\end{equation}
(for the last equality, we used the fact that the fields $V_i$ vanish sufficiently fast at spatial infinity), it follows that $\partial_{i}  V_i =0 \Rightarrow {\widetilde V}_3=0$. Similarly, we have that $\partial_i \Pi_{V_i}= 0 \Leftrightarrow { \Pi}_{{\widetilde V}_3}=0$. Hence, in terms of the new variables the constraints read ${\widetilde V}_3={ \Pi}_{{\widetilde V}_3}=0$. The true degrees of freedom are ${\widetilde V}_1,{\widetilde V}_2,\Pi_{{\widetilde V}_1}$ and $\Pi_{{\widetilde V}_2}$. As mentioned before the true degrees of freedom are gauge independent degrees of freedom. In this case this follows from the fact that ${\widetilde V}_1,{\widetilde V}_2,\Pi_{{\widetilde V}_1}$ and $\Pi_{{\widetilde V}_2}$ have zero Poisson brackets with ${\Pi}_{{\widetilde V}_3}$ and ${\Pi}_{{\widetilde V}_3}$ is the generator of gauge transformations in the new coordinates (recall that the first class constraint $\partial_i \Pi_{V_i}= 0$ reads ${ \Pi}_{{\widetilde V}_3}=0$ in the new coordinates).

We can conclude that the transformation determined by ({\ref{h30.02601}}), ({\ref{h30.0260301}}) and ({\ref{h30.0260302}}) satisfies the MN theorem. 

One can also explicitly check that the Dirac brackets ({\ref{h30.025}}) of the canonical variables $V_i({\bf x})$ and $\Pi_{V_j}({\bf y})$ equal the Poisson brackets restricted to the unconstrained variables ${\widetilde V}_i$ and $\Pi_{{\widetilde V}_i}$ with $i=1,2$. This is done by using the relation
\begin{eqnarray}
\sum^2_{i=1} \int d^3 x K_{ki} ({\bf z},{\bf x})  K^{-1}_{ij} ({\bf x},{\bf y}) &=& \delta_{kj} \delta({\bf z} - {\bf y}) - \int d^3 x K_{k3} ({\bf z},{\bf x})  K^{-1}_{3j} ({\bf x},{\bf y}) \nonumber\\
&=& \left(\delta_{kj} -  \frac{\partial_{y_k} \partial_{y_j} }{\nabla^2}\right)\delta({\bf z} - {\bf y}),
\label{h30.026062}
\end{eqnarray}
which is found from ({\ref{h30.0260101}}), ({\ref{h30.0260303}}), ({\ref{h30.026030001}}) and ({\ref{h30.026030003}}). 

\subsubsection{Functional Schr\"odinger equation and pilot-wave interpretation}
By applying the canonical transformation given by ({\ref{h30.02601}}) and ({\ref{h30.0260302}}) to the Hamiltonian ({\ref{h30.0181001}}), and by using the constraints ${\widetilde V}_3=0$ and ${ \Pi}_{{\widetilde V}_3}=0$, or by directly applying the canonical transformation to the Hamiltonian ({\ref{h30.02502}}), and by using the fact that the generating function is zero, we obtain the following Hamiltonian for the unconstrained variables ${\widetilde V}_i$ and $\Pi_{{\widetilde V}_i}$ $(i=1,2)$:
\begin{equation}
H  = \sum^{2}_{k,l=1} \! \int\!  d^3 y d^3 z \left(  \! h_{kl}({\bf y},{\bf z}) \Pi_{{\widetilde V}_k}({\bf y}) \Pi_{{\widetilde V}_l}({\bf z})\! + \!{\bar h}_{kl}({\bf y},{\bf z}) {\widetilde V}_k({\bf y})  {\widetilde V}_l({\bf z}) \! \right)\!,  \label{h30.0260501}
\end{equation}
with
\begin{eqnarray}
h_{kl}({\bf y},{\bf z}) &=& \frac{1}{2} \sum^{3}_{i=1}\int d^3 x K^{-1}_{ki} ({\bf y},{\bf x})   K^{-1}_{li} ({\bf z},{\bf x}),\label{h30.0260502}\\
{\bar h}_{kl}({\bf y},{\bf z}) &=&  \frac{1}{2} \sum^{3}_{i=1}\int d^3 x \varepsilon_{imn} \partial_{x_m} K_{nk}({\bf x},{\bf y}) \varepsilon_{irs} \partial_{x_r} K_{sl}({\bf x},{\bf z}).
\label{h30.026051}
\end{eqnarray}

Because the fields ${\widetilde V}_k$ and $\Pi_{{\widetilde V}_k}$ $(k=1,2)$ are unconstrained, the theory is quantized by using the standard commutation relation for the corresponding operators and hence we can use the standard representation
\begin{equation}
{\widehat {\widetilde V}}_k({\bf x}) = {\widetilde V}_k({\bf x}),\quad {\widehat \Pi}_{{\widetilde V}_k}({\bf x}) = -i\frac{\delta}{\delta {\widetilde V}_k({\bf x})}, \qquad {\textrm{for }} k=1,2.
\label{h30.026052}
\end{equation}
In this representation, the functional Schr\"o\-ding\-er equation for the wavefunctional $\Psi({\widetilde V}_1,{\widetilde V}_2,t) = \langle {\widetilde V}_1,{\widetilde V}_2 |\Psi (t)\rangle$ reads
\begin{eqnarray}
i\frac{\partial \Psi({\widetilde V}_1,{\widetilde V}_2,t)}{\partial t} &=& \sum^{2}_{k,l=1} \int d^3 y d^3 z \Big(  -h_{kl}({\bf y},{\bf z})\frac{\delta}{\delta {\widetilde V}_k({\bf y})}\frac{\delta}{\delta {\widetilde V}_l({\bf z})} + \nonumber\\
&&  {\bar h}_{kl}({\bf y},{\bf z}) {\widetilde V}_k({\bf y}) {\widetilde V}_l({\bf z}) \Big)\Psi({\widetilde V}_1,{\widetilde V}_2,t).
\label{h30.026053}
\end{eqnarray}
The corresponding continuity equation for the density $|\Psi({\widetilde V}_1,{\widetilde V}_2,t)|^2$ reads
\begin{equation}
\frac{\partial |\Psi|^2}{\partial t}  + \sum^{2}_{k=1} \int{ d^3x  \frac{\delta J_{{\widetilde V}_k}({\bf x})}{\delta {\widetilde V}_k({\bf x}) } }=0,
\label{h30.026054}
\end{equation}
with
\begin{equation}
J_{{\widetilde V}_k}({\bf x}) = \frac{1}{2i}\sum^{2}_{l=1}\int{ d^3 y h_{kl}({\bf x},{\bf y})\left(\Psi^*  \frac{\delta}{\delta {\widetilde V}_l({\bf y}) }\Psi  - \Psi \frac{\delta}{\delta {\widetilde V}_l({\bf y})} \Psi^* \right)}.
\label{h30.02605401}
\end{equation}
The pilot-wave interpretation is straightforward. The conserved density of the field beables ${\widetilde V}_1$ and ${\widetilde V}_2$ is given by $|\Psi|^2$ and the guidance equations for the fields are
\begin{equation}
{\dot {\widetilde V}}_k({\bf x}) =  J_{{\widetilde V}_k}({\bf x}) / |\Psi|^2,\qquad  \textrm{for } k=1,2.
\label{h30.026055}
\end{equation}

It is important to note that the fields ${\widetilde V}_k({\bf x})$ ($k=1,2$) do not necessary live in physical 3-space,{\footnote{This could perhaps be indicated more explicitly in the notation, e.g.\ by replacing the variable ${\bf x}$ by some other variable.}} but in some abstract space defined by the transformation ({\ref{h30.02601}}). The transversal part of the electromagnetic field potential which lives in physical 3-space can always be obtained by considering ({\ref{h30.02603}}). For the quantized Duffin-Kemmer-Petiau field, it was no problem to find true degrees of freedom which live in physical 3-space. This will be difficult to do for the electromagnetic field. In the following paragraph we consider some examples of possible representations. In the first example the fields ${\widetilde V}_k({\bf x})$ ($k=1,2$) will live in physical 3-space, in the second example not.

\subsubsection{Explicit examples of Maskawa and Nakajima canonical transformations}
{\bf Example one:} Assume the transformation matrix \label{exampleone}
\begin{equation}
\!K_{ij}({\bf x},{\bf y})  = 
\left\{\! \begin{array}{ll}
\delta_{ij} \delta({\bf x} \!-\! {\bf y})\! -\! \frac{1}{2}\delta_{i3} \partial_{x_j}\delta(x_1\!-\!y_1)\delta(x_2\!-\!y_2)\sgn(x_3\!-\!y_3) & \textrm{if } j\!=1,2\\
 \partial_{x_i} U({\bf x},{\bf y}) & \textrm{if } j\!=3
\end{array} \right.
\label{h30.026065}
\end{equation}
where $U$ is an arbitrary non-singular matrix which has the properties discussed in the previous section (cf.\ the paragraph containing equation ({\ref{h30.026030002}})) and `$\sgn$' the sign function. Note that $K$ satisfies the requirements ({\ref{h30.026021}}) and ({\ref{h30.026022}}). The inverse of $K$ is given by
\begin{equation}
K^{-1}_{ij}({\bf x},{\bf y}) = 
\left\{ \begin{array}{ll}
\left(\delta_{ij} -  \frac{\partial_{x_i} \partial_{x_j} }{\nabla^2}\right)\delta({\bf x} - {\bf y}) & \textrm{if } i=1,2\\
-\frac{\partial_{y_j}}{\nabla^2_{y}} U^{-1}({\bf x},{\bf y}) & \textrm{if } i=3
\end{array} \right. .
\label{h30.026066}
\end{equation}
With this transformation the transversal part of $V_i$ reads 
\begin{eqnarray}
V^T_1({\bf x}) &=& {\widetilde V}_1({\bf x}), \label{h30.02606601}\\
V^T_2({\bf x}) &=& {\widetilde V}_2({\bf x}), \label{h30.02606602}\\
V^T_3({\bf x}) &=& \frac{1}{2}\left(  \int^{+\infty}_{x_3} - \int^{x_3}_{-\infty}  \right) \left(  d s  \sum^2_{i=1} \partial_{x_i} {{\widetilde V}_i}(x_1,x_2,s)  \right)  .
\label{h30.026067}
\end{eqnarray}
The transversal momentum field reads
\begin{equation}
\Pi^T_{V_i}({\bf x})=   \sum^2_{k=1}\left(\delta_{ik} -  \frac{\partial_{x_i} \partial_{x_k} }{\nabla^2}\right) \Pi_{ {\widetilde V}_k}({\bf x}).
\label{h30.026068}
\end{equation}
The expression of the fields $V_i$ in terms of unconstrained variables given by ({\ref{h30.02606601}})-({\ref{h30.026067}}) was suggested by Weinberg \cite{weinberg95}. As shown in the preceding section we can write the Schr\"o\-ding\-er equation solely in terms of the unconstrained variables ${\widetilde V}_1$ and ${\widetilde V}_2$ with the representation ({\ref{h30.026052}}) and we can construct the corresponding pilot-wave interpretation. A pleasant feature is that the unconstrained variables ${\widetilde V}_1$ and ${\widetilde V}_2$ live in physical 3-space. But the Hamiltonian will display a highly nonlocal dependence on the unconstrained variables because the field $V^T_3({\bf x})$ depends nonlocally on ${\widetilde V}_1$ and ${\widetilde V}_2$.

However, there is a problem with this transformation which is left unmentioned by Weinberg and which prevents us from using it. Because 
\begin{equation}
\lim_{x_3 \to \pm \infty}  V^T_3({\bf x}) = \mp  \frac{1}{2} \int^{+\infty}_{-\infty} d s\left(  \sum^2_{k=1} \partial_{x_k} {{\widetilde V}_k}(x_1,x_2,s)  \right)
\label{h30.02606901}
\end{equation}
is not zero (unless further constraints are brought into play), our assumption that the fields vanish at infinity are not met. This in itself is not a problem, we could do with different boundary conditions. However, because $ V^T_3({\bf x})$ does not vanish at spatial infinity there appears an explicit infinity in the Hamiltonian ({\ref{h30.0260501}}) and hence in the equations of motion, which is intolerable. The infinity appears explicitly because for $k,l=1,2$, ${\bar h}_{kl}({\bf y},{\bf z})$ contains the term $ \frac{1}{2} \sum^2_{i=1}\int d^3 x \partial_{x_i} K_{3k} ({\bf x},{\bf y}) \partial_{x_i} K_{3l} ({\bf x},{\bf z})$ for which 
\begin{eqnarray}
\frac{1}{2} \sum^2_{i=1} \int d^3 x \partial_{x_i} K_{3k} ({\bf x},{\bf y}) \partial_{x_i}  K_{3l} ({\bf x},{\bf z}) &\sim& \int d x_3  \sgn(x_3-y_3)   \sgn(x_3-z_3) \nonumber\\
&=& \left( \int^{+\infty}_{-\infty}   -2   \int^{\textrm{max}(y_3,z_3)}_{\textrm{min}(y_3,z_3)}    \right)d x_3 \nonumber\\
&=& +\infty.
\label{h30.02606902}
\end{eqnarray}
This problem can not be solved by merely adding a function independent of $x_3$ to the right hand side of ({\ref{h30.026067}}). It can only be solved by adding further constraints or by introducing a cut-off in the Hamiltonian.\\

\noindent
{\bf Example two:} Let us first introduce two 3-vectors ${\boldsymbol \varepsilon}^j({\bf k})$, $j=1,2$, for each 3-vector ${\bf k}$, such that the vectors ${\boldsymbol \varepsilon}^1({\bf k}),{\boldsymbol \varepsilon}^2({\bf k}),{\bf k}/k $ forms a orthonormal triad. In other words the following conditions should be satisfied: orthogonality 
\begin{equation}
\sum^3_{m=1} {\varepsilon}^j_m({\bf k}) {\varepsilon}^i_m({\bf k})= \delta_{ji},\quad i,j=1,2,
\label{h30.028}
\end{equation}
the transversality condition
\begin{equation}
{\bf k} \cdot {\boldsymbol \varepsilon}^j({\bf k}) = 0
\label{h30.030}
\end{equation}
and the completeness relation
\begin{equation}
\sum^2_{j=1} {\varepsilon}^j_m({\bf k}) {\varepsilon}^j_n({\bf k})= \delta_{mn} - \frac{k_m k_n}{k^2}.
\label{h30.029}
\end{equation}
We also demand that 
\begin{equation}
{\boldsymbol \varepsilon}^j({\bf k}) = {\boldsymbol \varepsilon}^j(-{\bf k}).
\label{h30.02701}
\end{equation}

For example, we could take the following choice for the vectors ${\boldsymbol \varepsilon}^1({\bf k})$ and ${\boldsymbol \varepsilon}^2({\bf k})$. For each $k_3 \ge 0$ we take
\begin{equation}
{\boldsymbol \varepsilon}^j({\bf k}) = R({\bf k}/k,{\bf e}_1)\left( \begin{array}{c}
0 \\
\frac{1}{\sqrt{2}} \\ 
(-1)^{j} \frac{1}{\sqrt{2}} 
\end{array} \right)
\quad \textrm{for } k_3 \ge 0 ,
\label{h30.027}
\end{equation}
where $R({\bf k}/k,{\bf e}_1)$ is the rotation matrix that carries the unit vector ${\bf e}_1 = (1,0,0)^T$ to the unit vector ${\bf k}/k$. For $k_3 < 0$ the vectors are defined by imposing the condition ${\boldsymbol \varepsilon}^j({\bf k}) = {\boldsymbol \varepsilon}^j(-{\bf k})$.

We are now ready to define the transformation matrix as 
\begin{equation}
K_{ij}({\bf x},{\bf y}) = 
\left\{ \begin{array}{ll}
\frac{1}{(2\pi)^3} \int d^3 k e^{i{\bf k} \cdot ({\bf x} - {\bf y})} {\varepsilon}^j_i({\bf k}) & \textrm{if } j=1,2\\
\partial_{x_i} U({\bf x},{\bf y}) & \textrm{if } j=3
\end{array} \right.
\label{h30.031}
\end{equation}
where $U$ is an arbitrary non-singular matrix which has the properties discussed in the previous section (cf.\ the paragraph containing equation ({\ref{h30.026030002}})). Note that $K$ is a real matrix because of ({\ref{h30.02701}}) and that $K$ satisfies the requirements ({\ref{h30.026021}}) and ({\ref{h30.026022}}). The inverse of $K$ is given by
\begin{equation}
K^{-1}_{ij}({\bf x},{\bf y}) = 
\left\{ \begin{array}{ll}
\frac{1}{(2\pi)^3} \int d^3 k e^{i{\bf k} \cdot ({\bf x} - {\bf y})} {\varepsilon}^i_j({\bf k})  & \textrm{if } i=1,2\\
-\frac{\partial_{y_j}}{\nabla^2_{y}} U^{-1}({\bf x},{\bf y}) & \textrm{if } i=3
\end{array} \right. .
\label{h30.032}
\end{equation}
In this representation we have the pleasant feature that 
\begin{eqnarray}
h_{kl}({\bf y},{\bf z}) &=& \frac{1}{2} \delta ({\bf y}-{\bf z}),  \nonumber\\
{\bar h}_{kl}({\bf y},{\bf z}) &=& \frac{1}{2} \nabla^2 ({\bf y}-{\bf z}) . 
\label{h30.033}
\end{eqnarray}
Hence the Schr\"o\-ding\-er equation ({\ref{h30.026053}}) reads
\begin{equation}
i\frac{\partial \Psi({\widetilde V}_1,{\widetilde V}_2,t)}{\partial t} = \frac{1}{2} \sum^2_{l=1} \int d^3 x \left( -\frac{\delta^2}{\delta {\widetilde V}_l({\bf x})^2} -  {\widetilde V}_l({\bf x})\nabla^2 {\widetilde V}_l({\bf x}) \right)\Psi({\widetilde V}_1,{\widetilde V}_2,t).
\label{h30.034}
\end{equation}
This is the same Schr\"o\-ding\-er equation that would be obtained for two noninteracting massless uncharged spin-0 particles.{\footnote{Although we did not treat the massless spin-0 case explicitly, it can be obtained form the massive spin-0 theory (which was considered in Section \ref{Massive spin-0}) simply by putting the mass $m$ equal to zero. Note that the massless spin-1 case cannot be obtained by doing this in the Proca theory.}} The only important diffe\-rence is that here, the fields ${\widetilde V}_k({\bf x})$ ($k=1,2$) do not live in physical 3-space. The corresponding guidance equations ({\ref{h30.026055}}) are
\begin{equation}
{\dot {\widetilde V}}_l = \frac{1}{2i|\Psi|^2} \left(\Psi^*  \frac{\delta}{\delta {\widetilde V}_l }\Psi  - \Psi \frac{\delta}{\delta {\widetilde V}_l} \Psi^* \right),\qquad  \textrm{with } l=1,2.
\label{h30.035}
\end{equation}

We can rewrite the Schr\"o\-ding\-er equation and the corresponding pilot-wave interpretation by considering the Fourier expansion  
\begin{equation}
{\widetilde V}_l({\bf x}) = \frac{1}{\sqrt{(2\pi)^3}} \int d^3 k q_l({\bf k}) e^{i{\bf k} \cdot {\bf x}},\qquad  \textrm{with } l=1,2
\label{h30.03501}
\end{equation}
and with $q_l({\bf k})=q^*_l(-{\bf k})$, because the fields ${\widetilde V}_l$ are real. In terms of the Fourier modes $q_l({\bf k})$, the Schr\"o\-ding\-er equation ({\ref{h30.034}}) becomes the Schr\"o\-ding\-er equation 
\begin{equation}
i\frac{\partial \Phi(q_1,q_2,t)}{\partial t} = \frac{1}{2} \sum^2_{l=1} \int d^3 k \left( -\frac{\delta^2}{\delta q_l({\bf k}) \delta q^*_l({\bf k})} + k^2 q_l({\bf k}) q^*_l({\bf k}) \right)\Phi(q_1,q_2,t)
\label{h30.03502}
\end{equation}
for the wave\-func\-tion\-al $\Phi(q_l,t)$. If $\Psi({\widetilde V}_1,{\widetilde V}_2,t)$ is a solution of the Schr\"o\-ding\-er equation ({\ref{h30.034}}) then we can construct the following solution 
\begin{equation}
\Phi(q_l,t) \sim \Psi \left( (2\pi)^{-3/2} \int d^3 k q_l({\bf k}) e^{i{\bf k} \cdot {\bf x}},t\right)
\label{h30.0350201}
\end{equation}
for the Schr\"o\-ding\-er equation ({\ref{h30.03502}}).

We can also write $V^T_i({\bf x})$, the transversal component of the vector-potential, directly in terms of the modes $q_l({\bf k})$:
\begin{equation}
V^T_i({\bf x}) =\frac{1}{\sqrt{(2\pi)^3}} \sum^2_{l=1} \int d^3 k q_l({\bf k}){\varepsilon}^l_i({\bf k})  e^{i{\bf k} \cdot {\bf x}}.
\label{h30.03504}
\end{equation}
This relation could of course be used to construct a Maskawa and Nakajima canonical transformation which directly yields us the true degrees of freedom $q_l({\bf k})$, without the need to introduce the variables ${\widetilde V}_l({\bf x})$. This is in fact what most textbooks on quantum field theory do. 

The representation in terms of the $q_l({\bf k})$ is certainly the most transparent picture one in the free case. But when the electromagnetic field is coupled to a charged matter field then the Hamiltonian density becomes nonlocal. I.e.\ the value of the Hamiltonian density at the momentum ${\bf k}$ will depend on the value of fields at other momenta ${\bf l} \neq {\bf k}$. Therefore in the coupled case, the representation loses some of its attractiveness. 

The corresponding guidance equations in terms of the Fourier modes read
\begin{equation}
{\dot q}_l({\bf k})   = \frac{1}{2i|\Phi|^2} \left(\Phi^*  \frac{\delta}{\delta {\dot q}_l({\bf k})  }\Phi  - \Phi \frac{\delta}{\delta {\dot q}_l({\bf k}) } \Phi^* \right),\qquad  \textrm{with } l=1,2.
\label{h30.03503}
\end{equation}
Here, we recognize the pilot-wave interpretation for the electromagnetic field that was originally presented by Bohm \cite{bohm2} and which was further developed by Kaloyerou \cite{kaloyerou85,kaloyerou94,kaloyerou96}.{\footnote{In fact there is a slight difference between the model presented in \cite{bohm2,kaloyerou85} and the model presented in \cite{kaloyerou94}. In the model in \cite{kaloyerou94} the fields satisfy $q_1({\bf k}) = - q^*_1(-{\bf k})$ and $ q_2({\bf k}) = q^*_2(-{\bf k})$, whereas in the model presented in \cite{bohm2,kaloyerou85} the fields satisfy $q_l({\bf k}) = q^*_l(-{\bf k})$. The model presented here hence corresponds with the model of Bohm. Of course the transition from one model to the other is a triviality.}} 

Bohm and Kaloyerou directly used the expansion in terms of the transversal Fourier modes of the electromagnetic potential and a similar expansion for the transversal part of the momentum field, in order for the fields to satisfy respectively the Coulomb gauge constraint and the Gauss constraint. Only afterwards, the commutation relations for the transversal Fourier modes are then imposed. This is in accordance with the treatment that can be found in many textbooks. The Dirac approach of dealing with constraints justifies these arguments.

In \cite{kaloyerou94,kaloyerou96} Kaloyerou also took a different approach to a pilot-wave interpretation. Instead of passing to Fourier modes, Kaloyerou devised a pilot-wave interpretation in which he introduced field beables corresponding to the vector potential ${\bf V}({\bf x})$. However, we think this approach is not justified (or at least incomplete). The reason is the following. In order to find a representation for the field operators ${\widehat V}_{i}$ and ${\widehat \Pi}_{V_i}$ he replaced the transversal $\delta$-function, i.e.\ $(\delta_{ij} -  \partial_i \partial_j /\nabla^2 )\delta({\bf x} - {\bf y})$, on the right hand side of equation ({\ref{h30.02401}}) by $\delta_{ij} \delta({\bf x}-{\bf y})$. Kaloyerou argued that this replacement is justified because the same equations are obtained when the vector potential is expressed in terms of normal modes. By replacing the transverse $\delta$-function, the commutation relations for the operators become indeed simple and a representation is easily found. However, Kaloyerou's argument is not correct. It is a direct consequence of the Gauss constraint and the Coulomb gauge constraint that we need the transverse $\delta$-function in ({\ref{h30.02401}}). By replacing the transverse $\delta$-function by $\delta({\bf x}-{\bf y})$, it is in fact implicitly assumed that the Gauss constraint and the Coulomb gauge do not apply. But without the Gauss constraint and the Coulomb gauge constraint we cannot dismiss the longitudinal modes in the Fourier expansions of the field operators ${\widehat V}_i$ and ${\widehat \Pi}_{V_i}$, whereas we can do this if we have the Gauss constraint and the Coulomb gauge constraint. Nevertheless, when Kaloyerou makes the Fourier expansion of the fields $V_i$ and $\Pi_{V_i}$, as presented above, he correctly dismisses the longitudinal components.

Kaloyerou also discusses the gauge invariance of the pilot-wave model. We will turn to this issue in the next section.

In \cite{kaloyerou85,kaloyerou94}, Kaloyerou further discusses in detail the application of this pilot-wave interpretation to various typical quantum phenomena involving the electromagnetic field. In particular a detailed account was given for the photo-electric and Compton effect, along the lines of Bohm \cite{bohm2}. Recently Kaloyerou also applied the pilot-wave interpretation to describe the Mach-Zehnder Wheeler delayed-choice experiment \cite{kaloyerou03}.

\subsection{Quantization in other gauges}\label{otuoftbi}
\subsubsection{Admissible gauges}
The Coulomb gauge is a natural gauge for the quantization of the electromagnetic field. It allows us to identify the transversal components of the electromagnetic vector potential as the true degrees of freedom and longitudinal component of the vector potential as the gauge degree of freedom. There also exist other gauges which allow for the quantization of the electromagnetic field by means of Dirac's prescription. These gauges were called the {\em admissible} gauges in Section \ref{firstclassconstraints}. A gauge is admissible if it is attainable by a sequence of infinitesimal gauge transformations and if it uniquely fixes the gauge. Sometimes remaining gauge invariance can be removed restricting the possible gauge transformations ({\ref{h30.01701}}) by requiring boundary condition for the function $\theta(x)$ (for example in the Coulomb gauge we require that $\theta(x)$ vanishes at spatial infinity). 

If a gauge is admissible then we can perform the Dirac method of quantization by adding the gauge to the set of first class constraints. For a finite number of degrees of freedom, the true degrees of freedom that may be found by performing a Maskawa and Nakajima canonical transformation are unique up to a canonical transformation. As a result, for a finite number of degrees of freedom, the Hamiltonian formulation that is obtained by imposing gauge constraints does not depend on the particular choice of gauge and hence there might only be one ambiguity in the canonical quantization procedure, which is the operator ordering ambiguity. Although it is to be expected that the true degrees of freedom are also unique for a system described by an infinite number of degrees of freedom, we have not seen an explicit statement or proof of this. But even if we take it for granted that the Hamiltonian formulation for a system described by an infinite number of degrees of freedom does not depend on the particular choice of gauge, there still may appear other ambiguities when applying the canonical quantization procedure. Apart from the operator ordering problem there is also the problem that one can use representations which yield unitarily inequivalent quantum theories \cite[pp.\ 53-55]{haag96}. This is a peculiarity of quantum field theory which is not present when quantizing a system with a finite number of degrees of freedom. 

Different quantum field theories will of course lead to different pilot-wave interpretations. However, even equivalent quantum field theories may lead to different pilot-wave interpretations. We have already seen that there is an ambiguity in identifying the guidance equations. But also, different representations, which may yield equivalent quantum theories, may lead to pilot-wave interpretations with inequivalent ontologies. For example, in non-relativistic quantum theory one can use the configuration representation or the momentum representation, which are equivalent at the quantum level. But, as shown by Brown and Hiley, the corresponding pilot-wave interpretations are not equivalent \cite{brown00}. 

For certain classes of representations, may we explicitly show that the corresponding quantum theories and the corresponding pilot-wave interpretations are equivalent. An example is given in Appendix \ref{appendixaaa}, where we consider the quantum theories which arise by using different transformation matrices $K$, which were used in the Maskawa and Nakajima canonical transformation ({\ref{h30.02601}}), ({\ref{h30.0260301}}) and ({\ref{h30.0260302}}), and we show that they are equivalent on the level of quantum theory and on the level of the pilot-wave interpretation.

In the rest of this section, we will leave this issue of uniqueness aside. In the next paragraph we will consider some frequently used gauges and look whether they may naturally lead to a pilot-wave interpretation. We will see that none of the discussed gauges is straightforwardly amenable for developing a pilot-wave interpretation. In the second to next paragraph, we make a note on the gauge invariance of the pilot-wave model. 

\subsubsection{Some examples of frequently used gauges}
A first example is the {\em Lorentz gauge}: $\partial_\mu V^{\mu} = 0$. This gauge is often used because it is explicitly Lorentz covariant. However, the Lorentz gauge contains $\partial_0 V_0$ which is not expressible in terms of the conjugate momenta and hence this gauge is not suitable for the Dirac procedure of quantization. Nevertheless, the Dirac procedure could be maintained in this particular case by introducing fermionic fields (the ghost fields) \cite[p.\ 119]{sundermeyer82}. However, because it is difficult at present to construct a pilot-wave interpretation for fermionic fields (see following chapter), we will not pursue this approach. 

A second example is the {\em axial gauge}: $V_3 = \Pi_{V_3} + \partial_3 V_0 = 0$. This is an example of an admissible gauge. The electromagnetic potential is uniquely fixed by this gauge if we restrict ourself to gauge transformations ({\ref{h30.01701}}) for which the function $\theta$ vanishes at spatial infinity. The Dirac brackets are very simple in this case; for the variables $V_1,V_2,\Pi_{V_1}$ and $\Pi_{V_2}$, the Dirac bracket equals the Poisson bracket \cite[p.\ 142]{sundermeyer82}. However, using this gauge leads to explicit infinities in the Hamiltonian \cite{schwinger63,bars78},{\footnote{Although these papers concern non-Abelian gauge theories, some of the content can be applied to the Abelian case as well.}} in a similar way as we encountered in example one in previous section (on p.\ \pageref{exampleone}). Various suggestions of how these problems may be overcome can be found in \cite{yao64,girotti82}. In \cite{chodos78,simoes86}, the problem is treated in the context of non-Abelian gauge theories ($SU(N)$, $N>1$, Yang-Mills theories).

Another example of an admissible gauge is the {\em superaxial gauge}. This gauge was presented by Girotti and Rothe \cite{girotti82} as a solution for the infinities appearing in the Hamiltonian in the axial gauge. The superaxial gauge reads
\begin{eqnarray}
V_1(x_0,x_1,x_2,x_3^{(0)}) &=& V_2(x_0,x_1^{(0)},x_2,x_3^{(0)}) = V_3(x_0,x_1,x_2,x_3) = 0, \nonumber\\
V_0(x_0,x_1,x_2,x_3)&=& \int^{x_1}_{x^{(0)}_1} d x'_1   \Pi_{V_1}(x_0,x'_1,x_2,x^{(0)}_3)\nonumber\\
&& +\! \int^{x_2}_{x^{(0)}_1} d x'_2  \Pi_{V_2}(x_0,x^{(0)}_1,x'_2,x^{(0)}_3) \nonumber\\
&&+ \int^{x_3}_{x^{(0)}_3} d x'_3\Pi_{V_3}(x_0,x_1,x_2,x'_3),
\label{h30.04101}
\end{eqnarray}
where ${\bf x}^{(0)}$ is some fixed point. This gauge picks a unique representative out of the equivalence class of gauge equivalent fields and can be attained with the gauge transformation ({\ref{h30.01701}}) with 
\begin{eqnarray}
\theta &=&\! \int^{x_1}_{x^{(0)}_1} d x'_1 V_1(x_0,x'_1,x_2,x^{(0)}_3) +\! \int^{x_2}_{x^{(0)}_1} d x'_2  V_2(x_0,x^{(0)}_1,x'_2,x^{(0)}_3) \nonumber\\
&+&\! \int^{x_3}_{x^{(0)}_3} d x'_3 V_3(x_0,x_1,x_2,x'_3) -\! \int^{x_0}_{x^{(0)}_0} d x'_0 V_0(x'_0,x_1^{(0)},x_2^{(0)},x_3^{(0)}) .
\label{h30.04102}
\end{eqnarray}
The resulting equal-time commutation relations for the field operators are
\begin{equation}
[{\widehat V}_i ({\bf x}), {\widehat \Pi}_{V_j}({\bf y}) ] = i \delta_{ij}\delta({\bf x} - {\bf y}) - i \partial_{x_i} r_j ({\bf x}, {\bf y})
\label{h30.04103}
\end{equation}
with 
\begin{eqnarray}
r_j ({\bf x}, {\bf y}) &=& \delta_{1j} \Delta(x_1,x^{(0)}_1;y_1) \delta(x_2 - y_2) \delta(x^{(0)}_3 - y_3) \nonumber\\
&& + \delta_{2j}  \delta(x^{(0)}_1 - y_1) \sgn(x_2 - y_2)  \delta(x^{(0)}_3 - y_3)/2 \nonumber\\
&& + \delta_{3j}  \delta(x_1 - y_1) \delta(x_2 - y_2) \Delta(x_3,x^{(0)}_3;y_3), \nonumber\\
\Delta(x,x^{(0)};y) &=& \int^{x}_{x^{(0)}} d x' \delta(x' - y).
\label{h30.04104}
\end{eqnarray}
The commutation relations involving $ {\widehat \Pi}_{V_0}$ are zero and the commutation relations involving $ {\widehat V}_0$ can be obtained from the commutation relations above by using the operator equivalents of the gauge constraints (\ref{h30.04101}). For the superaxial gauge it is suggestive to take $V^{}_1$ and $V^{}_2$, with $V_1(x_1,x_2,x_3^{(0)}) = V_2(x_1^{(0)},x_2,x_3^{(0)})=0$, as the true degrees of freedom. This leads to the following natural representation. Take
\begin{eqnarray}
{\widehat V}^{}_k ({\bf x}) &=& V^{}_k ({\bf x})\qquad k=1,2,\nonumber\\
{\widehat V}^{}_3 ({\bf x}) &=& 0,
\label{h30.04105}
\end{eqnarray}
where $V_1(x_1,x_2,x_3^{(0)}) = V_2(x_1^{(0)},x_2,x_3^{(0)})=0$. Because the representation should be compatible with the commutation relations ({\ref{h30.04103}}), we find 
\begin{eqnarray}
{\widehat \Pi}^{}_1({\bf x}) &=& -i \frac{\delta   }{\delta V_1({\bf x}) } + i\delta(x^{(0)}_3 - x_3)  \int d x'_3 \frac{\delta }{\delta V_1(x_1,x_2,x'_3) } \nonumber\\
&& -  i \partial_{x_2} \delta(x^{(0)}_3 - x_3) \int d  x'_1 d x'_3 \Delta(x'_1,x^{(0)}_1;x_1) \frac{\delta }{\delta V_2(x'_1,x_2,x'_3) } \nonumber\\
{\widehat \Pi}^{}_2({\bf x}) &=& -i \frac{\delta   }{\delta V_2({\bf x}) } +  i \delta(x^{(0)}_1 - x_1) \delta(x^{(0)}_3 - x_3) \int d  x'_1 d x'_3  \frac{\delta }{\delta V_2(x'_1,x_2,x'_3) }, \nonumber\\
{\widehat \Pi}^{}_3({\bf x}) &=& -i \sum^2_{k=1} \int  d x'_3 \partial_{x_k} \Delta(x'_3,x^{(0)}_3;x_3) \frac{\delta }{\delta V_1(x_1,x_2,x'_3) },
\label{h30.04106}
\end{eqnarray}
where it is understood that functional derivatives with respect to the fields $V_1(x_1,x_2,x_3^{(0)})$ and $V_2(x_1^{(0)},x_2,x_3^{(0)})$ are put zero.

Although the commutation relations ({\ref{h30.04103}}) and the constraints are satisfied in this representation, it is not possible to use it, because this representation leads to explicit infinities in the Hamiltonian (this is a direct consequence of the $\delta$-functions appearing in the conjugate momenta). 

Recall that the superaxial gauge was introduced to deal with the infinities in the Hamiltonian for the axial gauge. Although the superaxial gauge does not lead to explicit infinities in the Hamiltonian operator, the infinities in the Hamiltonian reappear if we take the most obvious representation given by ({\ref{h30.04105}}) and ({\ref{h30.04106}}).

We can conclude that admissible gauges are not always suitable to construct a quantum theory or they may have associated representation that are not suitable. Without a suitable quantum theory it is then of course not possible to devise a pilot-wave interpretation.

\subsubsection{Note on gauge invariance of the pilot-wave interpretation}
In \cite{kaloyerou94,kaloyerou96}, Kaloyerou made some comments on the gauge invariance of the pilot-wave interpretation. Kaloyerou posed two questions:{\footnote{These questions are cited from \cite{kaloyerou94}.}
\begin{itemize}
\item
According to our ontology, are physical results, i.e.\ expectation values of field observables gauge invariant?
\item
Does the gauge freedom conflict with attributing ontological significance to the potentials?
\end{itemize}
Kaloyerou answers the first question affirmative by recalling the equivalence, at the standard quantum mechanical level, of the Gupta-Bleuler formalism and the formalism that arises in the Coulomb gauge. Because the Gupta-Bleuler formalism leads to gauge invariant expectation values, then also the formalism in the Coulomb gauge leads to gauge invariant expectation values. Because the pilot-wave model, devised for the quantum formalism in the Coulomb gauge, produces the same statistics as the standard quantum interpretation, the pilot-wave model yields gauge invariant expectation values of field observables. However, there is in fact no need to compare the quantum theory in the Coulomb gauge to the Gupta-Bleuler formalism. From our analysis it follows that the quantum theory, although it is devised by using the Coulomb gauge, was formulated solely in terms of gauge invariant degrees of freedom, i.e.\ degrees of freedom which commute with the generators of gauge transformations, and hence all the predictions of this quantum theory are gauge invariant. Because the pilot-wave model is equivalent to the standard interpretation at the empirical level, its empirical predictions are also gauge invariant. 

The second question was inspired by the situation for classical electromagnetism. In classical electromagnetism different potentials may correspond to the same physical situation. Therefore one can either adopt the position to attach an ontological status to all the potentials, but in physical measurements the potential may only be revealed up to a gauge transformation. Or one could adopt the position to attach an ontological status only to the potentials that satisfy a particular gauge. Now the second question does in fact not apply to the pilot-wave model presented above, because in this model beables were only introduced for gauge invariant variables. The question is meaningful for pilot-wave models where beables are introduced for gauge variables as well, as in Valentini's model, which is to be discussed in the next section. However, we do not favor such models because, as we will see in the following section, these models tend to lead to non-normalizable densities for the field beables.

\subsection{Quantization with constraints as conditions on the state}\label{constrconsta}
\subsubsection{Quantization and functional Schr\"odinger equation}
Instead of adding further gauge constraints to the set of first class constraints $\{\chi_1,\chi_2\}$ in order to quantize the electromagnetic field, we can also proceed another way. As explained in Section \ref{firstclassconstraints}, we can use the standard commutation relations for the fields, i.e.
\begin{eqnarray}
[{\widehat V}_{0}({\bf x}),{\widehat \Pi}_{V_{0}}({\bf y})] &=& i[V_{0} ({\bf x}), \Pi_{V_{0}}({\bf y})]_P= i \delta({\bf x} - {\bf y}) \nonumber\\
{[{\widehat V}_{i}({\bf x}),{\widehat \Pi}_{V_{j}}({\bf y})]} &=& i[V_{i} ({\bf x}), \Pi_{V_{j}}({\bf y})]_P= i \delta_{ij}\delta({\bf x} - {\bf y})
\label{h30.042}
\end{eqnarray}
and impose the constraints as conditions on states,
\begin{equation}
{\widehat \chi}_1 |\Psi \rangle = {\widehat \Pi}_{V_{0}}|\Psi \rangle = 0, \quad {\widehat \chi}_2 |\Psi \rangle =\partial_i {\widehat \Pi}_{V_{i}}|\Psi \rangle = 0.  
\label{h30.043}
\end{equation}
The states $ |\Psi \rangle $ which satisfy these constraint equations are then the physical states. Because the constraints $\chi_1$ and $\chi_2$ were generators of gauge transformations, the conditions ({\ref{h30.043}}}) mean that physical states are gauge invariant.

Because we have the commutation relations ({\ref{h30.042}}), we can use the standard representation for the field operators in the Schr\"o\-ding\-er picture, i.e.\
\begin{eqnarray}
{\widehat V}_{0}({\bf x}) &=& V_{0} ({\bf x}),\quad {\widehat \Pi}_{V_{0} }({\bf x}) = -i\frac{\delta}{\delta V_{0} ({\bf x})}, \nonumber\\
{\widehat V}_{i}({\bf x}) &=& V_{i} ({\bf x}),\quad {\widehat \Pi}_{V_{i} }({\bf x}) = -i\frac{\delta}{\delta V_{i} ({\bf x})}.
\label{h30.044}
\end{eqnarray}
By using this representation for the extended Hamiltonian, given in ({\ref{h30.0181001}}), we obtain the following functional Schr\"o\-ding\-er equation for the wave\-func\-tion\-al $\Psi(V_{0},V_i,t) = \langle V_{0}, V_i |\Psi(t) \rangle$
\begin{equation}
i\frac{\partial \Psi}{\partial t}= \int d^3 x \left( -\frac{1}{2}\frac{\delta^2}{\delta  V_{i}\delta  V_{i}} + \frac{1}{4}F_{ij}F_{ij} -i V_0 \partial_j \frac{\delta}{\delta  V_{j}} -iu_1 \frac{\delta}{\delta  V_{0}} -iu_2 \partial_j \frac{\delta}{\delta  V_{j}}   \right)\Psi.
\label{h30.045}
\end{equation}
The physical states further have to satisfy ({\ref{h30.043}}), i.e.\ 
\begin{eqnarray}
\frac{\delta}{\delta  V_{0}} \Psi&=&0, \label{h30.046}\\
\partial_i \frac{\delta}{\delta  V_{i}} \Psi&=&0.
\label{h30.047}
\end{eqnarray}
The first constraint implies that $\Psi$ does not depend on $V_0$. This means that $\Psi$ is invariant under transformations of the form
\begin{equation}
V_0({\bf x}) \to V^{\theta_0}_0({\bf x}) = V_0({\bf x}) +  \theta_0({\bf x})
\label{h30.0471}
\end{equation}
with $\theta_0$ an arbitrary space-dependent function. The second constraint implies that $\Psi$ is invariant under time independent gauge transformations, i.e. that $\Psi$ is invariant under transformations
\begin{equation}
V_i({\bf x}) \to V^\theta_i({\bf x}) = V_i({\bf x}) + \partial_i \theta({\bf x}), 
\label{h30.048}
\end{equation}
with $\theta$ an arbitrary space dependent function. 

Hence, for a physical state $\Psi$, the functional Schr\"o\-ding\-er equation ({\ref{h30.045}}) can be written as 
\begin{equation}
i\frac{\partial \Psi}{\partial t}=  \int d^3 x \left( -\frac{1}{2}\frac{\delta^2}{\delta  V_{i}\delta  V_{i}} + \frac{1}{4}F_{ij}F_{ij}  \right)\Psi.
\label{h30.049}
\end{equation}

\subsubsection{Pilot-wave interpretation}
There is a corresponding conservation equation
\begin{equation}
\frac{\partial |\Psi|^2}{\partial t}  +   \int d^3x  \left(\frac{\delta J_{V_{0}}({\bf x})}{\delta  V_{0} ({\bf x}) } + \frac{\delta J_{V_{i}}({\bf x})}{\delta  V_{i} ({\bf x}) }  \right) =0, 
\label{h30.050}
\end{equation}
with
\begin{eqnarray}
J_{ V_0}({\bf x}) &=& 0, \label{h30.052} \\
J_{ V_j}({\bf x}) &=& \frac{1}{2i} \left(\Psi^*  \frac{\delta}{\delta V_j({\bf x}) }\Psi  - \Psi \frac{\delta}{\delta V_j({\bf x}) } \Psi^* \right). 
\label{h30.051} 
\end{eqnarray}
It would now be tempting to construct a pilot-wave interpretation, inspired by this conservation equation, by taking the guidance equations for the field beables $V_0$ and $V_{i}$ as
\begin{equation}
{\dot V}_{0}({\bf x}) =  J_{V_{0} }({\bf x}) / |\Psi|^2,\quad {\dot V}_{i}({\bf x}) =  J_{V_{i} }({\bf x}) / |\Psi|^2 .
\label{h30.053}
\end{equation}
If the field beable $V_{0}$ is discarded, it is in fact simply constant in time by ({\ref{h30.052}}), this is exactly the approach to a pilot-wave interpretation by Valentini \cite{valentini92,valentini96,valentini04}. Though, Valentini derived his pilot-wave interpretation by considering different starting principles (this is the `3+1' view). To be more precise, from the constrained dynamics point of view, the quantization scheme implicitly used by Valentini is a mixture of the two schemes described in Section \ref{firstclassconstraints}. First, the {\em temporal gauge} $V_0=0$ is chosen. In accordance with the first scheme explained in Section \ref{firstclassconstraints}, the constraints $V_0=\Pi_{V_0}=0$ are then treated as operator identities, so that the fields  $V_0$ and $\Pi_{V_0}$ will make no further appearance in the theory. Second, the unconstrained canonical commutation relations are imposed on the fields $V_i$ and $\Pi_{V_i}$ and the remaining constraint $\chi_2=\partial_i V_i=0$ is then imposed as a condition on states, in accordance with the second scheme explained in Section \ref{firstclassconstraints}. 

\subsubsection{Problem with non-normalizable field beable densities}
There is a problem with this pilot-wave approach. The density $|\Psi(V_0,V_{i},t)|^2$ of field beables is not normalizable with respect to the variables $V_0$ and $V_{i}$. This is because $\Psi$ does not depend on $V_0$ and is further invariant under time independent gauge transformations ({\ref{h30.048}}). Hence the integral is proportional to the volume of the gauge group (by the gauge group we mean the group of transformations determined by ({\ref{h30.0471}}) and ({\ref{h30.048}})), which is infinite. I.e.\ we have
\begin{equation}
\int {\mathcal D} V_0 \left( \Pi^3_{j=1} {\mathcal D} V_j \right) |\Psi(V_0,V_i,t)|^2 \sim \int {\mathcal D} V_0 \int {\mathcal D} \theta \sim \int {\mathcal D}  \theta_0 \int {\mathcal D} \theta \sim \infty.
\label{h30.054}
\end{equation}

Nevertheless, there is an easy way out to this problem. Let us make a change in field variables from $V_{i}$ to ${\widetilde V}_i$ by the transformation
\begin{equation}
V_i({\bf x}) = \int d^3 y K_{ij} ({\bf x},{\bf y})  {\widetilde V}_j({\bf y}),
\label{h30.054001}
\end{equation}
with $K$ a matrix which satisfies the properties discussed in Section \ref{repcanvar}. We can use this transformation to rewrite the functional Schr\"odinger equation ({\ref{h30.049}}) and the constraints ({\ref{h30.0471}}) and ({\ref{h30.048}}) as equations for a wavefunctional $\Psi'(V_0,{\widetilde V}_i)$. 

Let us first look at the constraints. As before, the constraint ({\ref{h30.0471}}) implies that $\Psi'$ is independent of $V_0$. For the second constraint, we use the relation
\begin{equation}
\frac{\delta }{\delta V_i} ({\bf x}) =  \int d^3 y   K^{-1}_{ji} ({\bf y},{\bf x}) \frac{\delta }{\delta {\widetilde V}_j({\bf y})}
\label{h30.054002}
\end{equation}
to obtain
\begin{eqnarray}
\partial_{x_i} \frac{\delta}{\delta  V_{i}({\bf x}) } \Psi=0 & \Leftrightarrow & \int d^3 y \partial_{x_i} K^{-1}_{ji} ({\bf y},{\bf x}) \frac{\delta}{\delta {\widetilde V}_j({\bf y}) } \Psi' =0          \nonumber\\
& \Leftrightarrow &  \int d^3 y U^{-1} ({\bf y},{\bf x}) \frac{\delta}{\delta {\widetilde V}_3({\bf y}) } \Psi' =0         \nonumber\\
& \Leftrightarrow & \frac{\delta}{\delta {\widetilde V}_3({\bf y}) } \Psi' =0 .
\label{h30.08}
\end{eqnarray}
Hence we find the constraints imply that $\Psi'$ is independent of $V_0$ and ${\widetilde V}_3$, i.e.\ $\Psi'= \Psi'({\widetilde V}_1,{\widetilde V}_2,t)$. 

If we rewrite the functional Schr\"odinger equation ({\ref{h30.049}}) as an equation for $\Psi'({\widetilde V}_1,{\widetilde V}_2,t)$ we obtain
\begin{eqnarray}
i\frac{\partial \Psi'({\widetilde V}_1,{\widetilde V}_2,t)}{\partial t} &=& \sum^{2}_{k,l=1} \int d^3 y d^3 z \Big(  -h_{kl}({\bf y},{\bf z})\frac{\delta}{\delta {\widetilde V}_k({\bf y})}\frac{\delta}{\delta {\widetilde V}_l({\bf z})} + \nonumber\\
&&  {\bar h}_{kl}({\bf y},{\bf z}) {\widetilde V}_k({\bf y}) {\widetilde V}_l({\bf z}) \Big)\Psi'({\widetilde V}_1,{\widetilde V}_2,t),
\label{h30.054003}
\end{eqnarray}
with $h_{kl}({\bf y},{\bf z})$ and ${\bar h}_{kl}({\bf y},{\bf z})$ as defined in ({\ref{h30.0260502}}) and ({\ref{h30.026051}}). In other words we obtain the functional Schr\"odinger equation ({\ref{h30.026053}}), which was the functional Schr\"o\-ding\-er equation in terms of unconstrained variables when we quantized the electromagnetic field in the Coulomb gauge. 

Because the fields $V_0$ and ${\widetilde V}_3$ make no appearance in the theory anymore we can in fact safely forget about them. We should only introduce beables corresponding to the fields ${\widetilde V}_1$ and ${\widetilde V}_2$ and not to the fields $V_0$ and ${\widetilde V}_3$. It would not even be very meaningful to introduce beables for the fields $V_0$ and ${\widetilde V}_3$. They would just remain constant with time. In addition, by dismissing the fields $V_0$ and ${\widetilde V}_3$ we do not encounter a problem anymore with infinities when normalizing the density of field beables $|\Psi'|^2$. The infinity would appear if we would integrate the density over the fields $V_0$ and ${\widetilde V}_3$. 

Of course the pilot-wave theory that we obtain by dismissing the fields $V_0$ and ${\widetilde V}_3$ is exactly the pilot-wave theory that was obtained in the previous section. We think this is the most natural approach to a pilot-wave interpretation for the electromagnetic field. By getting rid of the gauge degrees of freedom we do not have a problem with normalizing the densities of field beables. In addition, we saw that introducing beables for gauge degrees of freedom is rather meaningless because the constraints imply that these beables remain constant in time.

\subsubsection{Note on the definition of the inner product}
We saw that keeping gauge degrees of freedom when developing a pilot-wave interpretation leads to non-normalizable densities of field beables. A related problem arises in standard quantum field theory. When the inner product of two gauge invariant states $|\Psi_1\rangle$ and  $|\Psi_2\rangle$ were to be defined as 
\begin{equation}
\langle \Psi_1 | \Psi_2 \rangle = \int {\mathcal D} V_0 \left( \Pi^3_{j=1} {\mathcal D} V_j \right)  \Psi^*_1 \Psi_2,
\label{h30.05401}
\end{equation}
then this product would be infinite for the same reasons. In addition, if expectation values of operators would be defined in a similar way, one would encounter ambiguities \cite{kakudo83,partovi84,hatfield84,rossi84,bialynicki-birula84}. This would for example be the case for the following expectation value 
\begin{equation}
\langle \Psi | [{\widehat V}_i({\bf x}),\partial_j {\widehat \Pi}_{V_j}({\bf y})] |\Psi \rangle.
\label{h30.05402}
\end{equation}
By using ({\ref{h30.043}}) one finds that the expectation value is zero. If on the other hand the expectation value is calculated with ({\ref{h30.042}}) then one finds that this quantity is different from zero.

In standard quantum field theory, these problems are solved by introducing a measure $\mu(V)$ on the fields \cite{partovi84,hatfield84,rossi84,bialynicki-birula84}. The measure is found by applying the Faddeev-Popov trick. With the Faddeev-Popov trick, the gauge volume can explicitly be factored out from the integral in ({\ref{h30.05401}}). The remaining part then represent the integration of $ \Psi^*_1 \Psi_2$ over gauge independent variables, which yields a finite number.

It is instructive to apply the Faddeev-Popov formalism explicitly. Suppose a gauge ${\bar \chi}_3(V_i)=0$ which picks a unique representative from each equivalence class of fields that are connected by time independent gauge transformations. Then
\begin{equation}
1=\Delta({\bar \chi}_3(V_j)) \int {\mathcal D} \theta \delta \big({\bar \chi}_3 \big(V_i^\theta \big) \big)
\label{h30.055}
\end{equation}
with 
\begin{equation}
\Delta({\bar \chi}_3(V_j))= \bigg| \det \bigg( \frac{\delta {\bar \chi}_3(V_j^\theta)({\bf x})}{\delta \theta({\bf y})} \bigg) \bigg| \Bigg|_{{\bar \chi}_3(V_i^\theta)=0} 
\end{equation}
the Faddeev-Popov determinant, which is gauge invariant. Suppose similarly a gauge ${\bar \chi}_4(V_0)=0$ which picks a unique representative from each equivalence class of fields that are connected by the transformations ({\ref{h30.0471}}), then 
\begin{equation}
1=\Delta({\bar \chi}_4(V_0)) \int {\mathcal D} \theta_0 \delta \big({\bar \chi}_4 \big(V^{\theta_0}_0 \big) \big),
\label{h30.058}
\end{equation}
with
\begin{equation}
\Delta({\bar \chi}_4(V_0)) = \bigg| \det \bigg( \frac{\delta {\bar \chi}_4(V^{\theta_0}_0)({\bf x})}{\delta \theta_0({\bf y})} \bigg) \bigg| \Bigg|_{{\bar \chi}_4(V^{\theta_0}_0)=0} .
\label{h30.05801}
\end{equation}
By substituting ({\ref{h30.055}}) and ({\ref{h30.058}}) in the inner product ({\ref{h30.1101}}), we can write
\begin{eqnarray}
&&\int {\mathcal D} V_0 \left( \Pi^3_{j=1} {\mathcal D} V_j \right) \Psi^*_1 \Psi_2  \nonumber\\
&&=\int {\mathcal D} \theta  {\mathcal D} \theta_0  {\mathcal D} V_0 \left( \Pi^3_{j=1} {\mathcal D} V_j \right)  \Psi^*_1 \Psi_2 \Delta({\bar \chi}_3(V_i)) \Delta({\bar \chi}_4(V_0))\delta \big({\bar \chi}_3 \big(V_i \big) \big) \delta \big({\bar \chi}_4 \big(V_0 \big) \big)\nonumber\\
&&= \left( \int {\mathcal D} \theta  {\mathcal D} \theta_0 \right) \int {\mathcal D} V_0 \left( \Pi^3_{j=1} {\mathcal D} V_j \right) \Psi^*_1 \Psi_2 \Delta({\bar \chi}_3(V_i))\Delta({\bar \chi}_4(V_0)) \nonumber\\
&& \quad\quad\quad\quad \quad\quad\quad\quad \times \delta \big({\bar \chi}_3 \big(V_i \big) \big) \delta \big({\bar \chi}_4 \big(V_0 \big) \big).
\label{h30.059}
\end{eqnarray}
The last equality arises because $\Psi_1$, $\Psi_2$, the measure ${\mathcal D} V_0 \left( \Pi^3_{j=1} {\mathcal D} V_j \right)$ and the Faddeev-Popov determinants are invariant under the gauge transformations ({\ref{h30.0471}}) and ({\ref{h30.048}}). In this way, we have been able to separate the infinite gauge part from the integral. We can now define the measure 
\begin{equation}
\mu(V)= \frac{\delta \big({\bar \chi}_3 \big(V_i \big) \big) \delta \big({\bar \chi}_4 \big(V_0 \big) \big)}{\int {\mathcal D} \theta  {\mathcal D} \theta_0}
\label{h30.060}
\end{equation}
and a new inner product 
\begin{eqnarray}
\langle \Psi_1 | \Psi_2 \rangle&=&\int {\mathcal D} V_0 \left( \Pi^3_{j=1} {\mathcal D} V_j \right) \mu(V)  \Psi^*_1 \Psi_2 \nonumber\\
&=& \int {\mathcal D} V_0 \left( \Pi^3_{j=1} {\mathcal D} V_j \right)  \Psi^*_1 \Psi_2 \Delta({\bar \chi}_3(V_i)) \Delta({\bar \chi}_4(V_0)) \delta\big({\bar \chi}_3 \big(V_i \big) \big) \delta \big({\bar \chi}_4 \big(V_0 \big) \big) \nonumber\\
&=&    \int \left( \Pi^3_{j=1} {\mathcal D} V_j \right)   \Psi^*_1 \Psi_2 \Delta({\bar \chi}_3(V_i))    \delta\big({\bar \chi}_3 \big(V_i \big) \big)
\label{h30.061}
\end{eqnarray}
which is finite. With the introduction of the measure $ \mu(V)$ also the ambiguities with expectation values such as ({\ref{h30.05402}}) are removed \cite{partovi84,hatfield84,rossi84,bialynicki-birula84}.

We can for example use the Coulomb gauge for ${\bar \chi}_3$, i.e.\ ${\bar \chi}_3=\partial_i V_i$. In order to perform the integral in the inner product in ({\ref{h30.05401}}) explicitly, we can make a transition to the new variables ${\widetilde V}_i$, $i=1,2,3$, by performing the transformation ({\ref{h30.054001}}). First we define
\begin{equation}
N_k \Psi'_k(V_0,{\widetilde V}_i,t) = \Psi_k \left(V_0, \sum^3_{j=1} \int d^3 y K_{ij} ({\bf x},{\bf y})  {\widetilde V}_j({\bf y}),t \right)
\label{h30.07}
\end{equation}
for $k=1,2$. The $N_k$ are normalization constants which will be determined later. With $\Psi_1$ and $\Psi_2$ being gauge invariant states we know that $\Psi'_1$ and $\Psi'_2$ will not depend on $V_0$ and ${\widetilde V}_3$.    

If we now make use of the identities
\begin{equation}
 \Delta({\bar \chi}_3(V_i)) =\big| \det \big( \nabla^2  \delta ({\bf x} - {\bf y})  \big)\big|, 
\label{h30.09}
\end{equation}
and
\begin{equation}
\delta(\partial_i V_i) = \frac{\delta({\widetilde V}_3 ) }{ \big| \det \big( \nabla^2_xU({\bf x},{\bf y})  \big)\big|},
\label{h30.10}
\end{equation}
we can write the inner product in ({\ref{h30.05402}}) as
\begin{equation}
\langle \Psi_1 | \Psi_2 \rangle = \frac{N_1 N_2 \big|\det K\big|  }{\big|\det  U\big|} \int {\mathcal D} {\widetilde V}_1 {\mathcal D} {\widetilde V}_2 \Psi'_1( {\widetilde V}_1, {\widetilde V}_2,t)^*\Psi'_2( {\widetilde V}_1, {\widetilde V}_2,t).
\label{h30.11}
\end{equation}

In particular the norm of gauge invariant wavefunctionals $\Psi$ reads
\begin{equation}
\langle \Psi | \Psi \rangle = \frac{N^2 \big|\det K\big|  }{\big|\det U\big|} \int {\mathcal D} {\widetilde V}_1 {\mathcal D} {\widetilde V}_2 |\Psi'( {\widetilde V}_1, {\widetilde V}_2,t)|^2.
\label{h30.1101}
\end{equation}
with $\Psi'$ defined similarly as in ({\ref{h30.05402}}). If we take $N^2_k = \big|\det K \big|  /\big|\det  U\big|$, then the wavefunctionals $\Psi'({\widetilde V}_1,{\widetilde V}_2,t)$ are normalized with respect to the variables ${\widetilde V}_1$ and ${\widetilde V}_2$. From this expression for the norm of a state, we see that the natural definition for the density of field beables ${\widetilde V}_1$ and ${\widetilde V}_2$ is $|\Psi'( {\widetilde V}_1, {\widetilde V}_2,t)|^2$. The pilot-wave theory that follows by considering the continuity equation for this density is of course the one we presented before.

\section{Scalar quantum electrodynamics}\label{scalarqed}
In this section we present a pilot-wave field interpretation for a quantized bosonic field interacting with a quantized electromagnetic field. In Sections \ref{Massive spin-0} and \ref{Massive spin-1}, we have shown that the quantized DKP theory coupled to a non-quantized electromagnetic field is the same as the quantized Klein-Gordon or quantized Proca theory coupled to a non-quantized electromagnetic field. This equivalence is also true if the electromagnetic field is quantized. Therefore we do not not bother to start from the coupled DKP theory here. Instead we start with the coupled Klein-Gordon theory (called scalar quantum electrodynamics). The Proca theory can be treated completely analogously and will therefore not be discussed here.

For the electromagnetic field the quantization is only slightly different compared to the free case. This is because the constraints for the electromagnetic field differ. After the discussion on the quantization of the theory we present the corresponding pilot-wave interpretation.

\subsection{Quantization in the Coulomb gauge}
The Lagrangian is given by the sum of the minimally coupled Klein-Gordon Lagrangian and the free Maxwell Lagrangian 
\begin{equation}
L = \int d^3 x {\mathcal L} = \int d^3 x \left(D^*_{\mu} \phi^* D^{\mu} \phi   -m\phi^*\phi -\frac{1}{4} F^{\mu \nu}F_{\mu \nu} \right).
\label{h30.12}
\end{equation}
The equations of motion are 
\begin{eqnarray}
&\partial_{\mu} F^{\mu \nu}= s^{\nu}_{KG},& \\
&D_{\mu}D^{\mu}\phi + m^2 \phi = 0,\quad D^*_{\mu}D^{\mu*} \phi^* + m^2 \phi^* = 0&
\label{h30.1201}
\end{eqnarray}
with $s^{\mu}_{KG}=ie\left(\phi^*D^{\mu}\phi - \phi D^{\mu*} \phi^* \right)$ the Klein-Gordon charge current. In the following we will not write the subscript `$KG$' anymore. The Lagrangian and hence the equations of motion are invariant under the gauge transformations
\begin{eqnarray}
&\phi \to e^{ie \theta} \phi, \quad \phi^* \to e^{-ie \theta} \phi^*,&\nonumber\\
&V^{\mu} \to V^\mu -  \partial^\mu  \theta.& 
\label{h30.1203}
\end{eqnarray}

The canonically conjugate momenta read
\begin{eqnarray}
&\Pi_{\phi} = \frac{\delta L }{\delta {\dot \phi}  } = D^*_0 \phi^{*} , \quad \Pi_{\phi^*} = \frac{\delta L }{\delta {\dot \phi^*}  } = D_0 \phi &\nonumber\\
&\Pi_{V_0} = \frac{\delta L }{\delta {\dot V_0} } = 0, \quad \Pi_{V_i} = \frac{\delta L }{\delta {\dot V_i} } = (\partial_0 V_i +\partial_i V_0 ).&
\label{h30.1204}
\end{eqnarray}
We can read of that we have one primary constraint, i.e.\
\begin{equation}
\chi_1 = \Pi_{V_0}.
\label{h30.13003}
\end{equation}
The canonical Hamiltonian is given by
\begin{eqnarray}
H_C &=&   \int d^3 x \left( \Pi_{\phi}\Pi_{\phi^*} + \left(D^*_{i} \phi^{*}\right)  D_{i} \phi + m^2 \phi^*\phi +ie V_0 \left( \phi^{*}\Pi_{\phi^*}- \phi \Pi_{\phi} \right)\right)\nonumber\\
&&+\int d^3 x \left( \frac{1}{2}\Pi_{V_i}\Pi_{V_i} + \frac{1}{4} F_{ij}F_{ij} - \Pi_{V_i}\partial_i V_0 \right).
\label{h30.13}
\end{eqnarray}
The total Hamiltonian reads
\begin{equation}
H_T = H_C + \int d^3 x u_1 \chi_{1}.
\label{h30.13004}
\end{equation}
The requirement that $\chi_1$ is conserved leads to the secondary constraint
\begin{equation}
\chi_2 = \partial_i \Pi_{V_i} + s_0.
\label{h30.13005}
\end{equation}
In the constraint the charge density is to be considered in terms of momentum phase-space variables, i.e.\ $s_0 =ie \left(  \phi^{*} \Pi_{\phi^*} - \phi \Pi_{\phi}\right) $. There are no further constraints. The requirement that $\chi_2$ is conserved in time is identically fulfilled and hence does not determine the field $u_1({\bf x})$. Hence we are left with two first class constraints and the constraint $\chi_2$ can be added to the total Hamilonian to yield the extended Hamiltonian
\begin{equation}
H_E = H_C + \int d^3 x u_1 \chi_{1} + \int d^3 x u_2 \chi_{2}.
\label{h30.13004}
\end{equation}

We can use the Coulomb gauge to quantize the system. In this case the Coulomb gauge reads \cite[p.\ 151]{sundermeyer82}
\begin{eqnarray}
\chi_3 &=& \partial_i V_i ,\label{h30.13005001} \\
\chi_4 &=&  V_0 +  \frac{1}{\nabla^2}s_0.\label{h30.13006}
\end{eqnarray}
As in the case of the free electromagnetic field this is an admissible gauge. It can be obtained by the transformation ({\ref{h30.01701}}) with 
\begin{equation}
\theta = -\frac{1}{\nabla^2} \partial_i V_i .
\label{h30.130050010001}
\end{equation}

The Dirac bracket can be calculated by using the inverse of the matrix $C_{NM}({\bf x},{\bf y})=[\chi_N({\bf x}),\chi_M({\bf y})]_P$, with $M,N=1,\dots,4$. The matrix $C$ has the following nonzero components
\begin{eqnarray}
C_{14}({\bf x},{\bf y})&=&-C_{41}({\bf x},{\bf y}) = -\delta({\bf x} - {\bf y}), \nonumber\\
C_{23}({\bf x},{\bf y})&=&-C_{32}({\bf x},{\bf y}) = {\nabla^2}\delta({\bf x} - {\bf y}).
\label{h30.1501}
\end{eqnarray}
The inverse reads{\footnote{As in the case of the free electromagnetic field, the inverse is unique by considering the boundary conditions for the fields.}}
\begin{eqnarray}
C^{-1}_{14}({\bf x},{\bf y})&=&-C^{-1}_{41}({\bf x},{\bf y}) = \delta({\bf x} - {\bf y}), \nonumber\\
C^{-1}_{23}({\bf x},{\bf y})&=&-C^{-1}_{32}({\bf x},{\bf y}) = -\frac{1}{\nabla^2}\delta({\bf x} - {\bf y}).
\label{h30.1502}
\end{eqnarray}
We do not give the Dirac brackets explicitly, instead we will directly present the commutation relations for the operators below.

The Hamiltonian which will generate the dynamics of the fields is derived from the canonical Hamiltonian ({\ref{h30.13}}) by imposing the constraints:
\begin{eqnarray}
H &=&   \int d^3 x \Big( \Pi_{\phi^*} \Pi_{\phi} + \left(D^{T*}_{i} \phi^{*}\right)  D^T_{i} \phi + m^2 \phi^*\phi  +\frac{1}{2}\Pi^T_{V_i} \Pi^T_{V_i} - \frac{1}{2}V^T_i \nabla^2 V^T_i \Big) \nonumber\\ 
&& + \frac{1}{2}\int d^3 x d^3 y \frac{ s_0({\bf x}) s_0({\bf y})}{4\pi |{\bf x} -{\bf y}|}.
\label{h30.15001}
\end{eqnarray}
where
\begin{equation}
D^T_{i} = \partial_{i} - ieV^T_i .
\label{h30.2101}
\end{equation}
We have hereby used the decomposition of the fields in longitudinal and transversal components as defined in ({\ref{h30.02510001}}). The longitudinal part of the field $V_i$ is zero because of the Coulomb constraint $\partial_i V_i=0$ and the longitudinal part of the field $\Pi_{V_i}$ can be expressed in terms of the charge density as 
\begin{equation}
\Pi^L_{V_i}   =  \frac{\partial_i \partial_j }{\nabla^2} \Pi_{V_j} =  -\frac{\partial_i  }{\nabla^2} s_0.
\label{h30.15001001}
\end{equation}

Let us now quantize the theory by associating operators to the field variables. The equal time commutation relations for the field operators are found using the Dirac bracket. The commutation relations for the operators corresponding to the matter field read
\begin{equation}
[{\widehat \phi}({\bf x}),{\widehat \Pi}_{\phi}({\bf y})] = [{\widehat \phi}^*({\bf x}),{\widehat \Pi}_{\phi^*}({\bf y})] = i\delta({\bf x} - {\bf y})
\label{h30.15001001001}
\end{equation}
The other commutation relations between the field operators ${\widehat \phi},{\widehat \Pi}_{\phi},{\widehat \phi}^*,{\widehat \Pi}_{\phi^*}$ are zero. The commutation relations for the operators corresponding to the electromagnetic field read
\begin{eqnarray}
&[{\widehat V}_i({\bf x}),{\widehat \Pi}_{V_j}({\bf y})] = i \left(\delta_{ij} -  \frac{\partial_i \partial_j }{\nabla^2}\right)\delta({\bf x} - {\bf y}),&\nonumber\\  
&{[{\widehat V}_{i}({\bf x}),{\widehat V}_{j}({\bf y})]} =[{\widehat \Pi}_{V_i}({\bf x}),{\widehat \Pi}_{V_j}({\bf y})] =0.&
\label{h30.18}
\end{eqnarray}
The commutation relations involving the operator ${\widehat \Pi}_{V_0}$ are zero. The commutation relations of the operator ${\widehat V}_0$ and other operators corresponding to the electromagnetic field are also zero. If $R$ is a functional of the canonical variables of the matter field, then we have the following commutation relations
\begin{eqnarray}
&[{\widehat R},{\widehat V}_i({\bf x})] = 0, \qquad [{\widehat R}, {{\widehat V}_0}({\bf x})]=  -[{\widehat R},\frac{1 }{\nabla^{2}} {\widehat s_0}({\bf x})],&\nonumber\\  
&{[{\widehat R},{\widehat \Pi}_{V_i}({\bf x})]} =[{\widehat R},\partial_i {{\widehat V}_0}({\bf x})]=-\left[{\widehat R},\frac{\partial_i }{\nabla^{2}} {\widehat s_0}({\bf x})\right].&
\label{h30.19}
\end{eqnarray}

If we make use of the transformation matrix $K$ defined in Section \ref{constrasoperators}, a general representation for the field operators, consistent with the commutation relations and the constraints, is given by 
\begin{eqnarray}
{\widehat  \phi}  ({\bf x}) &=&  \phi ({\bf x}), \qquad \;\: {\widehat \Pi }_{\phi} ({\bf x}) = -i\frac{\delta }{\delta \phi({\bf x})}, \nonumber\\
{\widehat  \phi}^{*} ({\bf x}) &=& \phi^{*} ({\bf x}), \qquad  {\widehat \Pi }_{\phi^*} ({\bf x}) = -i\frac{\delta }{\delta \phi^*({\bf x})}, \nonumber\\
{\widehat V}_i({\bf x}) &=& \sum^2_{k=1} \int d^3 y K_{ik} ({\bf x},{\bf y}) {\widetilde V}_k({\bf y}),\nonumber\\
{\widehat \Pi}_{V_i}({\bf x}) &=& {\widehat \Pi}^T_{V_i}({\bf x}) + {\widehat \Pi}^L_{V_i}({\bf x}),\nonumber\\
{\widehat \Pi}^T_{V_i}({\bf x})  &=& -i \sum^2_{k=1} \int d^3 y  K^{-1}_{ki} ({\bf y},{\bf x}) \frac{\delta }{\delta  {\widetilde V}_k({\bf y})},\nonumber\\
{\widehat \Pi}^L_{V_i}({\bf x}) &=& \partial_i {\widehat V_0} ({\bf x}) =-\frac{\partial_i }{\nabla^2} {\widehat s_0}({\bf x}),\nonumber\\
{\widehat s_0} ({\bf x})&=&e \left(  \phi^* ({\bf x}) \frac{\delta}{\delta \phi^* ({\bf x})} - \phi ({\bf x}) \frac{\delta}{\delta \phi ({\bf x})}  \right).
\label{h30.20}
\end{eqnarray}
With this representation, the operator Hamiltonian becomes
\begin{eqnarray}
{\widehat H} &= & \int d^3 x \left(- \frac{\delta}{\delta \phi^* } \frac{\delta}{\delta \phi} + \left(D^{T*}_{i} \phi^{*}\right)  D^T_{i} \phi + m^2|\phi|^2 \right) \nonumber\\
&& +\sum^{2}_{k,l=1} \int d^3 y d^3 z \left( - h_{kl}({\bf y},{\bf z}) \frac{\delta}{\delta {\widetilde V}_k({\bf y})}\frac{\delta}{\delta {\widetilde V}_l({\bf z})} + {\bar h}_{kl}({\bf y},{\bf z}) {\widetilde V}_k({\bf y})  {\widetilde V}_l({\bf z}) \right)\nonumber\\
&&+\frac{1}{2}\int d^3 x d^3 y \frac{{\widehat s}_0({\bf x}){\widehat s}_0({\bf y})}{4\pi |{\bf x} -{\bf y}|}
\label{h30.21}
\end{eqnarray}
with
\begin{equation}
D^T_{x_i} = \partial_{x_i} - ie \sum^2_{k=1} \int d^3 y K_{ik} ({\bf x},{\bf y}) {\widetilde V}_k({\bf y}).
\label{h30.2101}
\end{equation}
The Schr\"o\-ding\-er equation for the wave\-func\-tion\-al $\Psi(\phi,\phi^*,{\widetilde V}_1,{\widetilde V}_2,t)$ reads
\begin{equation}
i\frac{\partial \Psi}{\partial t} = {\widehat H}\Psi.
\label{h30.25}
\end{equation}

Note that the Hamiltonian density depends nonlocally on the true degrees of freedom, even when the theory is written in the momentum representation. This is not only due to the Coulomb interaction (this is the last term in the Hamiltonian), but also due to the presence of the term $\left(D^{T*}_{i} \phi^{*}\right)  D^T_{i} \phi$ in the Hamiltonian.

\subsection{Pilot-wave interpretation}
Now we are ready to present the pilot-wave interpretation. The conservation equation corresponding to the functional Schr\"odinger equation ({\ref{h30.25}}) reads
\begin{equation}
\frac{\partial |\Psi|^2}{\partial t}  +  \int{ d^3x \left( \frac{\delta J_{\phi}({\bf x})}{\delta \phi({\bf x})}+ \frac{\delta J_{\phi^*}({\bf x})}{\delta \phi^*({\bf x})} + \sum^{2}_{k=1}  \frac{\delta J_{{\widetilde V}_k}({\bf x})}{\delta {\widetilde V}_k({\bf x}) }  \right)} =0,
\label{h30.26}
\end{equation}
with
\begin{eqnarray}
J_{\phi}({\bf x})  &=& \frac{1}{2i} \Big(\Psi^* \frac{\delta}{\delta \phi^*({\bf x})}\Psi - \Psi \frac{\delta}{\delta \phi^*({\bf x})} \Psi^* \Big) , \nonumber\\
&&+ \frac{e^2}{2i} \int d^3 y \frac{1}{4\pi |{\bf x} -{\bf y}|} 
\bigg( 
\frac{\delta \Psi^*}{\delta \phi({\bf y})} \phi({\bf x})\phi({\bf y}) \Psi
+ \Psi^* \phi({\bf x})\phi^*({\bf y}) \frac{\delta \Psi}{\delta \phi^*({\bf y})} \nonumber\\
&& -\Psi^*  \phi({\bf x})\phi({\bf y}) \frac{\delta \Psi}{\delta \phi({\bf y})}
- \frac{\delta \Psi^*}{\delta \phi^*({\bf y})} \phi({\bf x})\phi^*({\bf y}) \Psi
\bigg), \nonumber\\
J_{\phi^*}({\bf x}) &=& J_{\phi}^*({\bf x}), \nonumber\\
J_{{\widetilde V}_k}({\bf x}) &=& \frac{1}{2i}\sum^{2}_{l=1}\int{ d^3 y h_{kl}({\bf x},{\bf y})\left(\Psi^*  \frac{\delta}{\delta {\widetilde V}_l({\bf y}) }\Psi  - \Psi \frac{\delta}{\delta {\widetilde V}_l({\bf y})} \Psi^* \right)}, 
\label{h30.2601}
\end{eqnarray}
with $k=1,2$. 

In the pilot-wave interpretation the conserved density of the field beables $\phi,\phi^*,{\widetilde V}_1,{\widetilde V}_2$ is given by $|\Psi|^2$ and the guidance equations for the fields are
\begin{equation}
{\dot \phi} =  J_{\phi}/ |\Psi|^2, \qquad  {\dot \phi}^* =  J_{\phi^*}/ |\Psi|^2, 
\label{h30.27}
\end{equation}
\begin{equation}
{\dot {\widetilde V}}_k =  J_{{\widetilde V}_k} / |\Psi|^2, \qquad {\textrm{for }} k=1,2.
\label{h30.271}
\end{equation}

\subsection{Conclusion}
We see that we can give a pilot-wave interpretation for a quantized scalar field coupled to a quantized electromagnetic field. In the pilot-wave interpretation we only have introduced beables corresponding to true degrees of freedom. This approach has to be contrasted with Valentini's approach \cite{valentini92,valentini96}, where also beables are introduced corresponding to gauge degrees of freedom. However, just as in the free case this leads to densities of the field beables which are not normalizable.

We have used the Coulomb gauge to quantize the Maxwell field, but of course, other admissible gauges could equally well be used. An example of an admissible gauge is the superaxial gauge \cite{girotti82}. However, as we have seen already in the free case, the most obvious representation for the superaxial gauge leads to infinities in the Hamiltonian. Another interesting gauge is the unitary gauge, which can be used for the treatment of the Abelian Higgs model. We could in fact give a pilot-wave account for spontaneous symmetry breaking by adding the Higgs potential to the Lagrangian of the scalar field and by quantizing the system in unitary gauge. But we shall not do it here.

\section{A note on the quantization of non-Abelian gauge theories}\label{anoteonthequantizationnon-Abeliangaugetheories}
We have seen that in order to construct a pilot-wave interpretation for a constrained system it seems essential to isolate the true degrees of freedom. This presented no problem for the electromagnetic field. It is no problem to find an admissible gauge for the electromagnetic field. These admissible gauges then in turn suggest the use of some particular set of true degrees of freedom. The situation is different for non-Abelian gauge theories ($SU(N)$, $N > 1$, Yang-Mills theories). 

It seems difficult to find a admissible gauge for non-Abelian gauge theories. The Coulomb gauge for example does not uniquely fix the gauge. As shown by Gribov \cite{gribov78}, there remain gauge equivalent fields which satisfy the Coulomb gauge (and which are non-perturbative of nature). As was shown later this is not a problem particular for the Coulomb gauge. It was namely shown that there exists no continuous gauge for (non-Abelian) Yang-Mills theories on a compactified space or space-time which uniquely fixes the gauge \cite{singer78,chodos80}. Although we are not dealing with a compact space or space-time this theorem is powerful because space could be treated as compactified if the fields have suitable boundary conditions, for example if all the fields vanish at infinity. Hence if we want a continuous gauge, which uniquely fixes the gauge, then the fields (i.e.\ the vector potentials and the corresponding momenta) behave non-trivially at spatial infinity.  

Because some gauges, such as the Coulomb gauge, can be used to fix the gauge locally, a possible solution could be to restrict the configuration space of the fields to a certain subset called the {\em fundamental modular region} such that the gauge picks a unique representative in this subset \cite{vanbaal95,vanbaal97}.

The axial gauge is suitable for the Dirac procedure. However, this gauge leads to explicit infinities in the Hamiltonian. The superaxial gauge \cite{simoes86}, which does not yield infinities in the Hamiltonian, does not bring us any further either, because the corresponding natural representation leads, make the infinities reappear in the Hamiltonian, just as in the case of Maxwell's theory (cf.\ Section \ref{otuoftbi}).

The difficulty in finding an admissible gauge now leads to problems if we want to quantize the theory either by imposing constraints as operator identities or by imposing constraints as conditions on states. In the first method this is because the matrix which components are the Poisson brackets of the first class constraints $[\chi_i,\chi_j]_P$ is then not invertible. This blocks the construction of the Dirac bracket and hence it is unclear what commutation relations should be used for the operators. In the second method one can impose commutation relations for the operators, by taking the constraints as conditions on the states. The functional Schr\"odinger equation is then easily found. However, the difficulties arise with the construction of an inner product. In order to render the inner product finite, the gauge volume should be separated out of the functional integral and in order to accomplish this by performing the Faddeev-Popov trick, a admissible gauge is needed.
 
The issue of finding an admissible gauge and hence of finding the true degrees of freedom, is already a problem for the standard interpretation. Of course these problems persist when we want to construct a pilot-wave approach. Valentini, for example, considered an approach where not only beables are introduced for the true degrees of freedom, but also for gauge variables \cite{valentini04}. However, just as in the Abelian case, cf.\ Section \ref{constrconsta}, the density of field beables is not normalizable.

\section{The measurement process in terms of field beables}\label{localizedparticlesinthepwfieldinterpretation}
In Section \ref{themeasurementprocess} we gave the pilot-wave description of a measurement process in terms of particle beables. The description in terms of field beables proceeds along similar lines. In the field interpretation, the wavefunctionals develop non-overlapping branches in the configuration space of fields. The field beables then enter one of the branches, and under suitable conditions (the branches should not overlap again at a later time) the empty branch may be dismissed for the future description of the system. In the standard interpretation, this would then be referred to as the collapse of the wavefunctional.

Let us consider this in some more detail. Suppose we have a system described by the wavefunctional $\Psi^{(s)}(\phi)$. In a measurement situation the system couples to a measurement apparatus. We write the wavefunctional of the apparatus as $\chi^{(a)}({\tilde \phi})$, where the argument ${\tilde \phi}$ represents all the field degrees of freedom of the apparatus. During the measurement, the total wavefunctional, of the system and apparatus, then evolves as 
\begin{equation}
\Psi^{(s)}(\phi) \chi^{(a)}({\tilde \phi}) \to \sum_i  \Psi^{(s)}_i(\phi) \chi^{(a)}_i({\tilde \phi})  .
\label{h30.33}
\end{equation}
If the different terms $\Psi^{(s)}_i(\phi) \chi^{(a)}_i({\tilde \phi})$ are non-overlapping in the configuration space of fields $(\phi,{\tilde \phi})$ and if they remain non-overlapping in the future, the field beables are effectively guided by one of the wavefunctionals, say $\Psi^{(s)}_k(\phi) \chi^{(a)}_k({\tilde \phi})$. The empty wavefunctionals can then be dismissed from the future description of the field beables and we have an effective collapse. This is completely analogous to the situation in non-relativistic quantum theory. There is, however, one issue that needs to be addressed. In non-relativistic quantum theory it was guaranteed, for a general measurement situation, that different terms in a macroscopic superposition were non-overlapping in the configuration space, because the different macroscopic states generally correspond to systems localized at different regions in physical space (you can for example think of states which correspond to a macroscopic needle pointing in different directions). This was straightforward to show. Now with a field ontology this is not so straightforward anymore. Are states corresponding corresponding to macroscopic systems non-overlapping in the configuration space of {\em fields}?

A natural approach would be to consider quantum states which correspond to systems which are localized at distinct regions in physical 3-space and to look whether these states are non-overlapping in the configuration space of fields. Valentini addressed this question for non-relativistic one-particle states \cite{valentini92,valentini04}. It is interesting to consider his analysis here.

Valentini considered the real Klein-Gordon field. By letting the Klein-Gordon field operator ${\hat \phi} ({\bf x})$ act on the ground state $|0  \rangle $, one-particle states $|{\bf x} \rangle \sim {\hat \phi} ({\bf x}) |0  \rangle $ are constructed. These states obey the Klein-Gordon equation (because the field operator ${\hat \phi} ({\bf x})$ obeys the Klein-Gordon equation). But only in the non-relativistic limit these states represent strictly localized particles. This is because only in the non-relativistic limit the different states $|{\bf x} \rangle$ become orthogonal. In the field basis $|\phi  \rangle $, we have  $\langle\phi  |{\bf x} \rangle = \phi({\bf x})\langle\phi  | 0  \rangle$ with $\langle\phi  | 0  \rangle$  the wavefunctional of the vacuum. 

If we now consider a low energy state $|\Psi \rangle$, which contains only one particle, then we can expand the corresponding wavefunctional as
\begin{equation}
\Psi(\phi)= \langle\phi  |\Psi \rangle = \int d^3 x \langle \phi  |{\bf x} \rangle  \langle {\bf x}|\Psi \rangle =  \langle\phi  | 0  \rangle \int d^3 x \phi({\bf x}) \Psi ({\bf x}) ,
\label{h30.32}
\end{equation}
with $\Psi ({\bf x})= \langle {\bf x}|\Psi\rangle$ an amplitude which obeys the non-relativistic Schr\"odinger equation. So, a non-relativistic particle which would be described by the wavefunction $\Psi ({\bf x})$ in non-relativistic quantum theory, is described by the wavefunctional $\Psi(\phi)$ given in ({\ref{h30.32}}) in quantum field theory. 

Valentini showed that the probability density $|\Psi(\phi)|^2$ reaches its maximum for a field $\phi({\bf x})$ which mimics the non-relativistic wavefunction $\Psi ({\bf x})$. For example if $\Psi ({\bf x})$ happens to be real then $\phi({\bf x})$ is proportional to $\Psi ({\bf x})$. Suppose now that the probability density $|\Psi(\phi)|^2$ is sharply peaked around the field configuration $\phi({\bf x})= \Psi ({\bf x})$. Let us further consider two wavefunctionals $\Psi_1(\phi)$ and $\Psi_2(\phi)$ which describe a non-relativistic particle localized at different regions in physical 3-space, i.e.\ 
\begin{equation}
\Psi_i(\phi) =  \langle\phi  | 0  \rangle \int d^3 x \phi({\bf x}) \Psi_i ({\bf x}), \quad i=1,2  ,
\label{h30.3201}
\end{equation}
where the non-relativistic wavefunctions  $\Psi_1 ({\bf x})$ and $\Psi_2 ({\bf x})$ have a support at distinct regions in physical 3-space. Now, with the assumption that the probability densities $|\Psi_i(\phi)|^2$ are sharply peaked around the field configurations $\phi({\bf x})= \Psi_i ({\bf x})$ it is clear that the wavefunctionals $\Psi_i(\phi)$ are non-overlapping in the configuration space of fields. 

So, if it could be shown that the probability densities $|\Psi_i(\phi)|^2$ are sharply peaked around the field with maximum probability, wavefunctionals which correspond to particles which are localized at different regions in physical 3-space would be non-overlapping in the configuration space of fields. And under these circumstances it is to be expected that wavefunctionals corresponding to different macroscopic systems, such as macroscopic pointer needles, are also non-overlapping. However, this need not to be the case at all. I.e.\ the different wavefunctionals $\chi^{(a)}_i$ of the apparatus need not to correspond with to different configurations in physical 3-space in order to be non-overlapping. It is in for example sufficient to have non-overlap if the wavefunctionals $\chi_i$ are peaked at different values for the electromagnetic field. We think it might be interesting to consider coherent states for the electromagnetic field in this respect.

\chapter{Field beables for fermionic quantum field theory}\label{chapter5}
\section{Introduction}
Little work has yet appeared on the construction of a pilot-wave interpretation for fermionic field theory with fields as beables. Bohm, Hiley and Kaloyerou argued that the anti-commutation relations of fermionic fields do not permit a pilot-wave approach in which the beables are continuous fields \cite{bohm872,bohm5}. According to their view, the field beable interpretation should only be adopted for bosons. For fermions Bohm {\em et al.}\  prefer the particle beable approach. However, in the previous chapter, we gave arguments why we do not favour this approach. One of the main reasons is that the model of Bohm {\em et al.}\ requires the notion of a Dirac sea, which makes the model only suitable for quantum electrodynamics. 

In 1988, shortly after Bohm {\em et al.}\ argued against a field beable approach to fermionic quantum fields, Holland presented such a model \cite{holland881,holland}. Holland presented his model for the quantized non-relativistic Schr\"o\-ding\-er field, but the model can be straightforwardly extended to any fermionic field theory. The beables in Holland's model are the Euler angles at each point in momentum space. 

Later, in 1992, Valentini presented a pilot-wave interpretation for the quantized Van der Waerden theory (which describes relativistic spin-$1/2$ fields) \cite{valentini92,valentini96}. In Valentini's model the beables are anti-commuting fields, also called Grassmann fields. It is on this last approach by Valentini that we will focus in this chapter. We will argue that this approach is untenable. We will see that it is no problem to write a fermionic field theory in the functional Schr\"o\-ding\-er picture. However, the quantity which would be identified as the probability of the Grassmann fields, is itself an element of the Grassmann algebra and hence cannot be interpreted as a probability. In addition, it is hard to introduce meaningful guidance equations for Grassmann fields. 

Because Valentini's pilot-wave approach to fermionic fields is often quoted as a valid alternative (for more than ten years by now) we devote a chapter to it, explaining in detail where the problems are situated. Instead of using the quantized Dirac theory or the equivalent Van der Waerden theory to treat relativistic spin-$1/2$ fields, we will start with the quantized non-relativistic Schr\"o\-ding\-er theory. Apart from notational simplicity, it has the additional advantage that it can be quantized using both Fermi-Dirac statistics and Bose-Einstein statistics. We can then clearly indicate the analogies and differences between bosonic and fermionic quantization. Our approach to introduce a pilot-wave model with Grassmann beables slightly differs from Valentini's original approach (nevertheless we face the same problems). In Section \ref{relationtovalentiniswork}, we compare our approach to Valentini's one. Although the model of Holland hence seems to be the only alternative for a field beable approach for the moment, we do not elaborate on this model. 

In the previous chapter we also expressed the opinion that a field beable approach to quantum field theory seems the most natural approach, but this by no means excludes the possibility of a particle beable approach. In fact such an approach was even presented by Bell who treated fermionic field theory on a lattice \cite{bell86}. Later D\"urr {\em et al.}\ constructed a continuum version and applied it to quantum electrodynamics (where particle beables are only introduced for the fermions), see \cite{durr04} and references therein. These models differ form `ordinary' pilot-wave models in the sense that they also include an element of stochasticity. These models were therefore termed {\em Bell-type} models by D\"urr {\em et al.} We do not consider these {\em Bell-type} models further either in this thesis. 

There is also a particle model by Colin \cite{colin031,colin032,colin033}, who also tried to find a continuum version of the Bell model. Colin ended up with a deterministic pilot-wave type model, instead of a stochastic one. This model is in fact the same as the one originally presented by Bohm for the Dirac equation, in which beables are introduced for all the particles in the Dirac sea. Contrary to the model of D\"urr {\em et al.}, Colin's model relies hence on the existence of a Dirac sea and therefore this model might perhaps not be extendible to other type of interactions, such as weak interaction.

\section{The quantized non-relativistic Schr\"o\-ding\-er theory}
The Lagrangian for the non-relativistic Schr\"o\-ding\-er theory in one spatial dimension reads
\begin{equation}
L = \int {d} x \left( \frac{i\hbar}{2} \left( \psi^* \partial_t  \psi  - \psi \partial_t  \psi^*   \right) - \frac{\hbar^2}{2m} \partial_x  \psi^* \partial_x  \psi \right).
\label{h30.001}
\end{equation}
Dirac's method of quantization is simple in this case \cite{sundermeyer82} (one only encounters second class constraints) and the resulting quantum field theory can be found in many textbooks. Therefore there is no need to repeat the analysis here. We directly present the well-known expression for the Hamiltonian operator
\begin{equation}
\widehat{H} = - \frac{\hbar^2}{2m} \int {d} x \widehat{\psi}^* \partial^2_x \widehat{\psi}
\label{h30.1}
\end{equation}
and the bosonic and fermionic commutation relations for the field operators $\widehat{\psi}$ and $\widehat{\psi}^*$: 
\begin{equation}
[\widehat{\psi}(x),\widehat{\psi}^*(y)]_{\pm} = \delta(x -y), \quad  [\widehat{\psi}(x),\widehat{\psi}(y)]_{\pm}=[\widehat{\psi}^*(x),\widehat{\psi}^*(y)]_{\pm}=0.
\label{h30.2}
\end{equation}
Here $[.,.]_{-}$ denotes the commutator (bosonic quantization) and $[.,.]_{+}$ the anti-commutator (fermionic quantization). For notational convenience, we write the theory in terms of Fourier components
\begin{eqnarray}
\widehat{\psi}(x) &=& \frac{1}{\sqrt{2\pi}} \int {d} k \widehat{a}(k) e^{ikx}, \nonumber\\
\widehat{\psi}^*(x) &=& \frac{1}{\sqrt{2\pi}} \int {d} k \widehat{a}^{\dagger}(k) e^{-ikx}.
\label{h30.3}
\end{eqnarray}
The Hamiltonian operator (\ref{h30.1}) then becomes
\begin{equation}
\widehat{H} = \int {d} k E(k) \widehat{a}^{\dagger}(k) \widehat{a}(k)
\label{h30.4}
\end{equation}
with $E(k) = \hbar^2 k^2/2m$ and the commutation relations (\ref{h30.2}) then imply
\begin{equation}
[\widehat{a}(k),\widehat{a}^{\dagger}(k')]_{\pm} = \delta(k-k'), \quad  [\widehat{a}(k),\widehat{a}(k')]_{\pm}=[\widehat{a}^{\dagger}(k),\widehat{a}^{\dagger}(k')]_{\pm}=0.
\label{h30.5}
\end{equation}
We now deal with the bosonic and fermionic case separately.

\section{Bosonic quantization of the non-relativistic \\ Schr\"o\-ding\-er equation}
As shown in the previous chapter, the construction of a pilot-wave interpretation presents no difficulty in the case of bosonic quantization. The quantized non-relativistic Schr\"o\-ding\-er field with Bose-Einstein statistics was already treated by Holland \cite{holland881,holland}. It is this model that we recall here, the only difference is that we use take a continuous momentum space instead of a discretized one.

The bosonic commutation relations $[\widehat{a}(k),\widehat{a}^{\dagger}(k')]_{-}=\delta(k-k')$ can be realized with the representation
\begin{eqnarray}
\widehat{a}(k) &=& \frac{1}{{\sqrt 2}} \left(  q(k)  + \frac{\delta}{\delta  q(k) } \right),\nonumber\\
\widehat{a}^{\dagger}(k) &=& \frac{1}{{\sqrt 2}}  \left(  q(k)  - \frac{\delta}{\delta  q(k) }\right).
\label{h31} 
\end{eqnarray}
By substituting these relations for $\widehat{a}(k)$ and $\widehat{a}^{\dagger}(k)$ in the Hamiltonian (\ref{h30.4}) we obtain{\footnote{The irrelevant infinite $c$-number term could be omitted in the Hamiltonian.}}
\begin{equation}
H = \frac{1}{2}\int {d} k E(k) \left(- \frac{\delta^2}{\delta q^2(k)} +  q^2(k)  - \delta(0) \right).
\label{h31.01}
\end{equation}
The Schr\"o\-ding\-er equation for the wavefunctional $\Psi(q(k),t)$ (which is complex valued) then reads
\begin{equation}
i\hbar \frac{\partial \Psi}{\partial t} = \frac{1}{2}\int {d} k E(k) \left(- \frac{\delta^2}{\delta q^2(k)} +  q^2(k) - \delta(0)   \right) \Psi.
\label{h31.1}
\end{equation}
The corresponding conservation equation is 
\begin{eqnarray}
\frac{\partial |\Psi|^2}{\partial t } \! \! &+&\! \!  \int {d} k \frac{\delta J_{q} (k)}{\delta q(k)}  = 0,\nonumber\\
J_{q} (k)&=& \frac{E(k)}{2i\hbar} \left(\Psi^* \frac{\delta }{\delta  q(k) } \Psi - \Psi \frac{\delta }{\delta  q(k) } \Psi^* \right).
\label{h31.101}
\end{eqnarray}
In the pilot-wave interpretation the density of the fields $q(k)$ is given by $|\Psi|^2$ and the guidance equation reads
\begin{equation}
\dot{q}(k) = J_{q}(k) / |\Psi|^2.
\label{h32}
\end{equation}

\section{Fermionic quantization of the non-relativistic Schr\"o\-ding\-er equation}
\subsection{Functional Schr\"odinger representation}
Let us now try to apply the same scheme in the fermionic case. The first thing to do, is to find a representation for $\widehat{a}(k)$ and $\widehat{a}^{\dagger}(k)$ in terms of certain variables and differential operators with respect to these variables, such that the anti-commutation relations $[\widehat{a}(k),\widehat{a}^{\dagger}(k')]_{+}=\delta(k-k')$ are satisfied. With such a representation, we can then obtain a functional Schr\"o\-ding\-er equation. A possible representation is the one in terms of Euler angles which was used by Holland \cite{holland881,holland}. Other possible representations which are more commonly used nowadays are written in terms of Grassmann numbers \cite{floreanini88,hallin95}. In this thesis we only focus on this representation. 

For each wavenumber $k$, we introduce a Grassmann number{\footnote{For the definitions and properties concerning Grassmann numbers we refer to Appendix \ref{appb}.}} $\eta(k)$ and its conjugate $\eta^{\dagger}(k)$, which satisfy 
\begin{equation}
[\eta(k),\eta(l)]_{+}=[\eta(k),\eta^{\dagger}(l)]_{+}=[\eta^{\dagger}(k),\eta^{\dagger}(l)]_{+}=0.
\label{h32.1}
\end{equation}
The anti-commutation relations for the creation and annihilation operators are then realized in the representation \cite{floreanini88}{\footnote{Other representations can be used as well. An alternative representation is for example $\widehat{a}(k)=\eta(k) $, $a^{\dagger}(k)= \overrightarrow{\delta} / \delta \eta(k)$ \cite{duncan87}. Note that this representation is one-to-one, whereas the representation ({\ref{h33}}) is not. However, because we will encounter more severe obstructions when trying to formulate a pilot-wave interpretation using a Grassmann representation, we will not further consider this fact.}}
\begin{eqnarray}
\widehat{a}(k) &=& \frac{1}{\sqrt{2}} \left( \eta(k) + \frac{ \overrightarrow{\delta} }{ \delta \eta^{\dagger}(k)} \right),\nonumber\\
\widehat{a}^{\dagger}(k) &=& \frac{1}{\sqrt{2}}  \left( \eta^{\dagger}(k) + \frac{ \overrightarrow{\delta} }{\delta \eta(k)}\right).
\label{h33} 
\end{eqnarray}
By substituting these relations in the second quantized Hamiltonian (\ref{h30.4}), we obtain the Hamiltonian
\begin{equation}
H \!= \frac{1}{2} \int \!\!{d} k E(k)\! \left(\!\frac{ \overrightarrow{\delta} }{\delta \eta(k)} \frac{ \overrightarrow{\delta} }{\delta \eta^{\dagger}(k)}  \!+\! \eta^{\dagger}(k) \frac{ \overrightarrow{\delta} }{\delta \eta^{\dagger}(k)}\!-\! \eta(k) \frac{ \overrightarrow{\delta} }{\delta \eta(k)} \! +\! \eta^{\dagger}(k) \eta(k) \!+\! \delta(0) \! \right)\!\!
\label{h33.1}
\end{equation}
which leads to the following Schr\"o\-ding\-er equation for the wavefunctional $\Psi( \eta,\eta^{\dagger},t)$ 
\begin{eqnarray}
i \hbar \frac{\partial \Psi}{\partial t} &=& \frac{1}{2} \int {d} k E(k) \Big(\frac{ \overrightarrow{\delta} }{\delta \eta(k)} \frac{ \overrightarrow{\delta} }{\delta \eta^{\dagger}(k)}  + \nonumber\\
&&\eta^{\dagger}(k) \frac{ \overrightarrow{\delta} }{\delta \eta^{\dagger}(k)}- \eta(k) \frac{ \overrightarrow{\delta} }{\delta \eta(k)}  + \eta^{\dagger}(k) \eta(k) + \delta(0)  \Big)\Psi.
\label{h34.1}
\end{eqnarray}
We want to stress the fact that the wavefunctional is an element of the Grassmann algebra with generators $\eta(k)$ and $\eta^{\dagger}(k)$, and that hence the wavefunctional is not a complex valued functional. This fact is a direct consequence of the representation in terms of Grassmann numbers for the creation and annihilation operators.

The conventional inner product of two Grassmann valued functionals $\Psi_1$ and $\Psi_2$ is defined by \cite{floreanini88}{\footnote{In the literature other definitions can be encountered, e.g.\ Hallin and Liljenberg \cite{hallin95} use a Grassmann valued inner product instead of a complex valued inner product. However, all these definitions boil down to the same expressions for probability amplitudes at the end.}}
\begin{equation}
\langle \Psi_1 | \Psi_2 \rangle = \int \mathcal{D}\eta^{\dagger} \mathcal{D} \eta \Psi^*_1 \Psi_2 = \langle \Psi_2 | \Psi_1 \rangle^*
\label{h35}
\end{equation}
with $\Psi^*$ the dual of $\Psi$ (and not the Hermitian conjugate of $\Psi$ which is denoted by a dagger) given by
\begin{equation}
\Psi^*(\eta,\eta^{\dagger},t) = \int \mathcal{D}{\bar \eta}^{\dagger} \mathcal{D} {\bar \eta} \exp({\bar \eta}\eta^{\dagger}+ {\bar \eta}^{\dagger} \eta) \Psi^{\dagger}({\bar \eta},{\bar \eta}^{\dagger},t),
\label{h36}
\end{equation}
with $\mathcal{D} {\bar \eta} = \prod_k {d} {\bar \eta}(k)$ and $\Psi^{\dagger}$ the Hermitian conjugate of $\Psi$. We used the notation ${\bar \eta} \eta^{\dagger} = \int {d} k{\bar \eta}(k)  \eta^{\dagger}(k)$. 

At this stage, it is important not to confuse $\psi(\eta,\eta^{\dagger},t)$ with $\la \eta \eta^{\dagger} | \psi(t) \ra$. The wavefunctional $\psi(\eta,\eta^{\dagger},t)$ is an element of the Grassmann algebra (in order to satisfy the Schr\"o\-ding\-er equation (\ref{h34.1})) and because $\la \eta \eta^{\dagger} | \psi(t) \ra$ is an inner product it is a complex number. This difference was also stressed in \cite{floreanini88,kiefer94}.

\subsection{Problem to identify a suitable density of field beables}
We can now turn to the question whether this functional Schr\"odinger picture admits for a pilot-wave interpretation. In order to identify a candidate for the probability of the beables we can consider the norm of a wavefunctional. Such an approach is likely to guarantee equivalence between standard fermionic quantum field theory and a possible pilot-wave interpretation. The norm of a wavefunctional reads
\begin{eqnarray}
\langle \Psi | \Psi \rangle &=&  \int \mathcal{D}\eta^{\dagger} \mathcal{D}\eta \Psi^* \Psi \nonumber\\
&=& \int \mathcal{D}\eta^{\dagger} \mathcal{D}\eta \frac{1}{2} \left(\Psi^* \Psi +\Psi^{\dagger} (\Psi^*)^{\dagger}\right) \nonumber\\
&=& \int \mathcal{D}\eta^{\dagger} \mathcal{D}\eta P(\eta,\eta^{\dagger},t)
\label{h37}
\end{eqnarray}
where 
\begin{equation}
 P(\eta,\eta^{\dagger},t) =  \frac{1}{2} (\Psi^* \Psi +\Psi^{\dagger} (\Psi^*)^{\dagger})= P(\eta,\eta^{\dagger},t)^{\dagger}.
\label{h38}
\end{equation}
Hence, $P(\eta,\eta^{\dagger},t)$ would be the most natural candidate for the probability density. However, $P$ is an element of the Grassmann algebra and hence not a real positive number. This means that $P(\eta,\eta^{\dagger},t)$ can not be interpreted as a probability density of field beables. This is a first problem we encounter when we try to construct a pilot-wave interpretation. There is no clear candidate for the probability density of field beables. 

The core of the problem is that the configuration space of Grassmann fields is trivial, it consist of just one configuration $(\eta(k),\eta^{\dagger}(k))$. The wavefunctional $\Psi(\eta,\eta^{\dagger},t)$ is a mapping this one configuration $(\eta(k),\eta^{\dagger}(k))$ to the Grassmann algebra. This situation has to be contrasted with the situation in the bosonic case. In the bosonic case the configuration space of fields is the space of smooth functions which consists hence of more than one configuration. The wavefunctional $\Psi(q(k),t)$ for a bosonic system is a mapping from this configuration space of smooth functions $(q(k))$ to the complex numbers. From this point of view it is clear that is futile to introduce a probability distribution on the the configuration space of Grassmann fields. 

For the same reason it is unclear how to introduce a meaningful guidance equation for the Grassmann fields. Because the configuration space of Grassmann fields consists only of one configuration, trajectories in this configuration space are trivial. 

\subsection{Problem to construct a well defined guidance equation}
Although we already anticipated the problems in constructing a guidance equation for the Grassmann fields in the previous section, it is still instructive to make an explicit attempt. For this purpose we will treat the quantity $P$ {\em formally} as a density of Grassmann field beables. As usual we try to identify a guidance equation by considering the continuity equation for $P$. 

After a rather tedious calculation, one can obtain the following conservation equation for $\Psi^*(\eta,\eta^{\dagger},t) \Psi(\eta,\eta^{\dagger},t)$:
\begin{eqnarray}
 \frac{\partial \Psi^* \Psi}{\partial t} &+& \int {d} k \Bigg(\frac{I_{\eta}(k) \overleftarrow{\delta}}{\delta \eta(k)} + \frac{ \overrightarrow{\delta} I_{\eta^{\dagger}}(k)}{\delta \eta^{\dagger}(k)} \Bigg) = 0, \nonumber\\
I_{\eta}(k) &=& \frac{iE(k)}{2\hbar}\Bigg( \Psi^* \Psi \eta(k) -  \frac{1}{2}\frac{ \overrightarrow{\delta} {\bar \Psi}^*  }{\delta \eta^{\dagger}(k)}{\bar \Psi} + \frac{1}{2}{\Psi}^* \frac{ \overrightarrow{\delta} {\bar \Psi}  }{\delta \eta^{\dagger}(k)} \Bigg),  \nonumber\\
I_{\eta^{\dagger}}(k) &=& \frac{iE(k)}{2\hbar}\Bigg(-\eta^{\dagger}(k) \Psi^* \Psi -  \frac{1}{2} \frac{ {\bar \Psi}^*  \overleftarrow{\delta}  }{\delta \eta(k)} \Psi + \frac{1}{2}{\bar \Psi}^* \frac{ {\bar \Psi} \overleftarrow{\delta}  }{\delta \eta(k)} \Bigg), 
\label{h39}
\end{eqnarray}
where ${\bar \Psi} = \Psi_e - \Psi_o$ with $\Psi_e$ and $\Psi_o$ respectively the even part and the odd part of $\Psi=\Psi_e + \Psi_o$. Hence the conservation equation for $ P(\eta,\eta^{\dagger},t)$ is
\begin{equation}
 \frac{\partial P(\eta,\eta^{\dagger},t)}{\partial t} + \int {d} k \Bigg(\frac{ \overrightarrow{\delta} }{\delta \eta^{\dagger}(k)}\left(I^{\dagger}_{\eta(k)} + I_{\eta^{\dagger}(k)} \right) +\left(I_{\eta(k)} + I^{\dagger}_{\eta^{\dagger}(k)} \right) \frac{ \overleftarrow{\delta}}{\delta \eta(k)}  \Bigg)=0.
\label{h310}
\end{equation}

With $P$ formally interpreted as a `probability density', the guidance equations for the fields $\eta(k),\eta^{\dagger}(k)$ should look something like
\begin{eqnarray}
\dot{\eta}(k) &=& P(\eta,\eta^{\dagger},t)^{-1} \left(I_{\eta(k)} + I^{\dagger}_{\eta^{\dagger}(k)} \right),  \nonumber\\
\dot{\eta}^{\dagger}(k) &=& \left(I^{\dagger}_{\eta(k)} + I_{\eta^{\dagger}(k)} \right)  P(\eta,\eta^{\dagger},t)^{-1}.
\label{h311} 
\end{eqnarray}
In order for these guidance equations to be well defined, we have to address some problems:
\begin{enumerate}
\item
Not every element of the Grassmann algebra has an inverse for the multiplication (see Appendix \ref{appb}) and hence we must ensure that $P^{-1}$ is always well defined. This is the case if and only if $\Psi^{-1}$ is well defined. For example the ground state of the Schr\"o\-ding\-er wave equation $\Psi_v$ is given by \cite{kiefer94}
\begin{equation}
\Psi_v = N \exp \left( \int {d} k \eta(k) \eta^{\dagger}(k) \right).
\label{h34.51}
\end{equation}
Because $\int \mathcal{D}\eta^{\dagger} \mathcal{D} \eta \prod_k \left( \eta(k) \eta^{\dagger}(k) \right)\Psi_v =N \neq 0$ the ground state has an inverse (cf.\ Appendix \ref{appb}). This inverse is given by 
\begin{equation}
\Psi^{-1}_v = N^{-1} \exp \left( -\int {d} k \eta(k) \eta^{\dagger}(k) \right).
\label{h312.1}
\end{equation}
The first excited state state, describing one particle with energy $E_l$, has the form
\begin{equation}
\Psi_l = a^{\dagger}_l \Psi_v e^{-iE_lt/\hbar}= \sqrt{2} \eta^{\dagger}_l \Psi_v e^{-iE_lt/\hbar} .
\label{h34.52}
\end{equation}
But because this state is proportional to $\eta^{\dagger}_l$ it has no inverse. Further excited states will exhibit the same problem. 

A possible way to regularize these excited states is by taking a suitable superposition with the ground state. Consider for example the first excited state $\Psi_l$, which we superpose with the ground state as follows
\begin{equation}
\Psi^{\epsilon}_l = \sqrt{1 - \epsilon}\Psi_l +  \sqrt{\epsilon} \Psi_v. 
\label{h314}
\end{equation}
The inverse of the state $\Psi^{\epsilon}_l$ is now well defined because
\begin{equation}
\int \mathcal{D}\eta^{\dagger} \mathcal{D} \eta \prod_k \eta(k) \eta^{\dagger}(k) \Psi^{\epsilon}_l = \sqrt{\epsilon}\int \mathcal{D}\eta^{\dagger} \mathcal{D} \eta \prod_k \eta(k) \eta^{\dagger}(k)  \Psi_v =\sqrt{\epsilon} \neq 0.
\label{h315}
\end{equation}
In the limit $\epsilon \to 0$ the state $\Psi^{\epsilon}_l$ will approach $\Psi_l$. Because $\Psi_l$ and $\Psi_v$ are orthogonal with respect to the inner product (\ref{h35}), the probability of finding the state $\Psi_l$ is given by $1-\epsilon$ and the probability of finding the ground state is $\epsilon$.
\item
Because $\dot{\eta}(k)$ and $\dot{\eta}^{\dagger}(k)$ should be anti-commuting variables, the right hand sides of (\ref{h311}) should be odd elements of the Grassmann algebra. This can only be accomplished if the wavefunctional $\Psi$ is an even element of the Grassmann algebra. This implies that $\Psi$ is a superposition of wavefunctionals which describe an even number of particles. This is because the ground state of the Schr\"o\-ding\-er wave equation $\Psi_v$ is even. The state $\Psi_l$ which is the first excited state with energy $E_l$ describes one particle and is an odd element of the Grassmann algebra. Generally, applying $n$ different creation operators on the ground state, multiplies the groundstate with $n$ Grassmann numbers. Hence, only states describing an even number of particles are even and only these states will lead to well defined guidance equations.  
\item
The third and most important problem arises when we try to make sense of the `guidance equations' (\ref{h311}) as differential equations. This is basically due to the problem that the configuration space of Grassmann fields only contains one configuration $(\eta(k),\eta^{\dagger}(k))$.

A possible way to make sense of the guidance equations as differential equations could be by introducing a Grassmann algebra which is generated by $u(k),u^{\dagger}(k)$, where there are as many $u(k)$'s as there are $\eta(k)$'s, and by expressing the ${\eta(k)}$'s as time dependent superpositions of odd elements of the basis:  
\begin{eqnarray}
{\eta(k)}\!\! &=&\!\! \int {d} l_1 f^{10}(k;l_1;t) u(l_1) +  \int {d} l_1 f^{01}(k;l_1;t) u^{\dagger}(l_1) \nonumber\\
&+& \int {d} l_1 {d} l_2 {d} l_3 f^{21}(k;l_1,l_2,l_3;t)u(l_1)u(l_2) u^{\dagger}(l_3) + \dots
\label{h34.50}
\end{eqnarray}
where every $f^{mn}(k;l_1,\dots,l_{m+n})$ is, for every $k$, a time dependent distributions. The guidance equations (\ref{h311}) are then well defined as differential equations for the coefficients $f^{mn}$. By this construction we have in fact introduced a nontrivial configuration space of fields. The fields are now the coefficients $f^{mn}$ which are distributions. But it is unclear how to proceed from this point. It is unclear what the probability distribution on the configuration space of fields $f^{mn}$ should look like.
\end{enumerate}

In conclusion, we see that we encounter problems with the construction of a pilot-wave interpretation for the functional Schr\"odinger equation in the Grassmann representation. The main problems have to do with the fact that the configuration space of Grassmann fields consists only of a single configuration. Hence it makes no sense to introduce a probability density on this configuration space nor does it make sense to introduce dynamics on this configuration space.

\subsection{Relation to Valentini's work}\label{relationtovalentiniswork}
It was originally believed by Valentini that a pilot-wave interpretation could only be given for field theories for which the field equations contain second-order time derivatives, such as the Van der Waerden spin-$1/2$ equation \cite{valentini92,valentini96}. However, as shown in the preceding chapter, we could give a pilot-wave interpretation for the Duffin-Kemmer-Petiau theory which is first order in time. Also the non-relativistic Schr\"o\-ding\-er equation is first-order in time and as shown above, the pilot-wave interpretation could be adopted equally well for this first-order theory when using the Bose-Einstein statistics. In fact, one can even show that the quantized Van der Waerden theory and the quantized Dirac theory can be transformed into one another by a canonical transformation and are hence equivalent (as should be expected because they are equivalent already on the first quantized level).

Valentini did not start with the inner product (\ref{h35}) either to arrive at the `probability density'. He considered the conservation equation for $\Psi^{\dagger} \Psi$, which has the following form in the case of the non-relativistic Schr\"o\-ding\-er theory  
\begin{eqnarray}
\frac{\partial \Psi^{\dagger} \Psi}{\partial t} &+& \int {d} k  \left(\frac{C_{\eta}(k) \overleftarrow{\delta}}{\delta \eta(k)} + \frac{ \overrightarrow{\delta} C_{\eta}^{\dagger}(k) }{\delta \eta^{\dagger}(k)} \right) = 0, \nonumber\\
C_{\eta}(k) &=& \frac{iE(k)}{2\hbar}\left( \Psi^{\dagger} \Psi \eta(k) -  \frac{ \overrightarrow{\delta} {\bar \Psi}^*  }{\delta \eta^{\dagger}(k)}{\bar \Psi} + \Psi^* \frac{ \overrightarrow{\delta} {\bar \Psi}  }{\delta \eta^{\dagger}(k)} \right). 
\label{h34.2}
\end{eqnarray}
In the pilot-wave interpretation the conserved quantity $\Psi^{\dagger} \Psi$ would then be the probability density of the field beables and the guidance equations for these field beables would then read
\begin{eqnarray}
\dot{\eta}(k) &=& (\Psi^{\dagger} \Psi)^{-1} C_{\eta}(k), \nonumber\\
\dot{ \eta}^{\dagger}(k) &=&  C^{\dagger}_{\eta} (k)(\Psi^{\dagger} \Psi)^{-1}.
\label{h34.3}
\end{eqnarray}
However, apart from the fact that $\Psi^{\dagger} \Psi$ is also an element of the Grassmann algebra, it is in general not normalizable. Integrating $\Psi^{\dagger} \Psi$ over the configuration space of Grassmann fields can yield zero for a nonzero wavefunctional. Even the quantity  $\Psi^{\dagger} \Psi$ can be zero for a nonzero wavefunctional. These problems are not present for the quantity $P$ defined in (\ref{h38}). This is also the reason why the inner product was defined as (\ref{h35}).

 \chapter{On Peres' statement ``opposite momenta lead to opposite directions'', decaying systems and optical imaging}\label{chapter6}
\section{Introduction}
In this chapter we will consider two things.{\footnote{The results of this chapter are published in \cite{struyve041}.}} First, we consider the question to which extent {\em opposite momenta} lead to {\em opposite directions} for the fragments of a decaying quantum system. We hereby improve an analysis by Peres \cite{peres201}. According to Peres, there are only two sources for deviation from perfect angular alignment for a two-particle system with total momentum zero. We will argue that there is also another contribution to the deviation from angular alignment, which is due to the uncertainty of the location of the source. It will appear that Peres' estimation for the angular deviation only applies in, what we will call, the large time or large distance regime. This is the regime where the two particles have traveled a large distance from the source. In the small time or small distance regime, the other contribution to the deviation becomes dominant. Peres applied his analysis to two different experiments \cite{peres201,peres99}, the thought experiment of Popper \cite{popper34,popper82,tarozzi85pop} and the {\em optical imaging} experiment reported by Pittman {\em et al.}\ \cite{pittman95}. We will argue that Popper's thought experiment occurs in the small time regime and hence Peres' analysis should not be applied there. On the other hand, Peres' analysis can be correctly applied to the experiment of Pittman {\em et al.}, which can be seen as occurring in the large time regime. 

Second, we will reconsider the experiment by Pittman {\em et al.}, from another point of view, namely pilot-wave theory. It is clear that one of the main merits of the pilot-wave interpretation is that it provides an observer independent description of quantum phenomena. Another merit of the pilot-wave interpretation is that it can be used to visualize quantum processes (just consider the many examples in Holland's book \cite{holland}). Now, the paper by Pittman {\em et al.} contains drawings of {\em conceptual} photon trajectories. These trajectories, which are derived with `usual' geometrical optics, are not the real paths of the photons (because, according to quantum theory, photons do not exist as localized entities between two measurements), but merely serve as a tool to visualize the experiment. However, when we calculate the trajectories predicted by pilot-wave theory they coincide with these conceptual trajectories. In this way the trajectories predicted by pilot-wave theory can serve as a theoretical basis for otherwise rather ad hoc drawings. Note however that we derive the trajectories for the massive particle equivalent of the experiment of Pittman {\em et al.}; it would take us too far to derive them for example from the Harish-Chandra theory (which is equivalent to Maxwell's theory).{\footnote{Although the trajectory model that can be derived for the Harish-Chandra theory cannot serve as a valid interpretation for photons, it serves well for illustrative purposes as indicated in Chapter \ref{chapter3}.}}

In the following section we start with recalling Popper's thought experiment and the optical imaging experiment of Pittman {\em et al.}. In Section \ref{oppositemomentaoppositedirections} we consider the question to which extent `opposite momenta' lead to `opposite directions'. In section \ref{pwidescriptionofadecayingsystem} we first consider a simplified pilot-wave description of a decaying system and then in Section \ref{pwidescriptionpittmanexperiment} we consider the pilot-wave description of the massive particle equivalent to the experiment of Pittman {\em et al.}

\section{On Popper's experiment}
In 1934, Popper proposed his experiment which aimed to test the general validity of quantum mechanics \cite{popper34}. Popper assumes a source $S$ from which pairs of particles are emitted in opposite directions. Two observers, say Alice and Bob, are located at opposite sides of the source, both equipped with an array of detectors. If Alice puts a screen with a slit in her way of the particles, she will observe a diffraction pattern behind the screen. According to Popper, quantum mechanics will also predict a diffraction pattern on the other side of the source, where Bob is located, when coincidence counts are considered. This is because every measurement by Alice is in fact a virtual position measurement of the correlated particle on Bob's side, leading to an increased momentum uncertainty for Bob's particle as well. This is the same diffraction pattern that would be observed when a physical slit was placed on Bob's side. Popper, who declared himself a metaphysical realist, found this idea of `virtual scattering' absurd and predicted no increased momentum uncertainty for Bob's measurement due to Alice's position measurement. He therefore saw his proposed experiment as a possible test against quantum mechanics and in favor of his realist vision in which particles have at each time well defined positions and momenta; particles for which the Heisenberg uncertainty, for example, is only a lower, statistical limit of scatter.

Unfortunately, to describe the setup of the experiment, Popper occasionally invoked classical language, which veiled some severe problems which could obstruct a practical realization of his experiment (for an extensive discussion see Peres \cite{peres99}). For example, Popper writes: ``We have a source $S$ (positronium, say) from which pairs of particles that have interacted are emitted in opposite directions. We consider pairs of particles that move in opposite directions \ldots''. It is the validity of this statement, which appears to be a very delicate issue, that is one of the topics we shall deal with in this chapter.

Of course, when we consider a decaying system at rest in classical mechanics, the fragments will have opposite momenta  
\begin{equation}
{\bf p}_1 + {\bf p}_2 = {\bf 0}
\label{h4.1}
\end{equation}  
and if we take the place of decay of the system as the centre of our coordinate system the positions of the two fragments will satisfy $m_1 {\bf x}_{1}+ m_2 {\bf x}_{2} = {\bf 0}$, so the fragments will be found in opposite, isotropically distributed directions. If the fragments have equal masses, then they will be found at opposite places, relative to the centre of the coordinate system. 

But these properties do not hold in quantum mechanics. Suppose that a system has a two-particle wavefunction $\psi$ which has a sharp distribution at ${\bf p}_1 + {\bf p}_2 = {\bf 0}$, i.e.\ both $|\la \hat{ p}_{1j}+\hat{ p}_{2j} \ra|$ and $\De (\hat{ p}_{1j}+\hat{ p}_{2j}) $ are small for every component $j$ of the momentum vectors $\hat{{\bf p}}_{1}+\hat{{\bf p}}_{2}$. Then according to the uncertainty relations 
\begin{equation}
\De (\hat{ p}_{1j}+\hat{ p}_{2j}) \De (m_1 \hat{ x}_{1j}+ m_2 \hat{ x}_{2j}) \ge \frac{\hbar}{2} (m_1 + m_2), 
\label{h4.2}
\end{equation} 
the distribution of $m_1 x_{1j}+ m_2 x_{2j}$ will be broad for every component $j$. In the case of equal masses, the inequalities in (\ref{h4.2}) imply that opposite momenta are incompatible with opposite positions. In particular, at the moment of decay, the inequalities imply that opposite momenta of the particles are incompatible with the latter being both located at the origin of the coordinate system. It was even shown by Collett and Loudon \cite{collett872} that this initial uncertainty on the location of the source implies that Popper's experiment is inconclusive. 

Although the original proposal of Popper's experiment can hence not be performed practically, due to the fact that opposite momenta are incompatible with opposite positions, the intention of Popper's proposal can be maintained if we have a two-particle system which displays some form of entanglement in the position coordinates.{\footnote{The position entanglement should not be exact, otherwise, as shown by Short \cite{short00}, both of the observers would observe an infinite momentum spread.}} One is then in principle able to test experimentally whether one of the particles will experience an increased momentum spread due to a position measurement (within a slit width) of the correlated particle, i.e.\ we would be able to test a possible `virtual scattering'. Such a form of position entanglement was obtained with the phenomenon of {\em optical imaging} \cite{pittman95}. By making use of optical imaging, Kim and Shih \cite{kim99} were able to perform an experiment in the spirit of Popper's original proposal.

Let us briefly review this experiment by Pittman {\em et al.}. The experiment uses momentum correlated photons resulting from spontaneous parametric down conversion (see Fig.\ \ref{figure1}). 
\begin{figure}[t]
\begin{center}
\epsfig{file=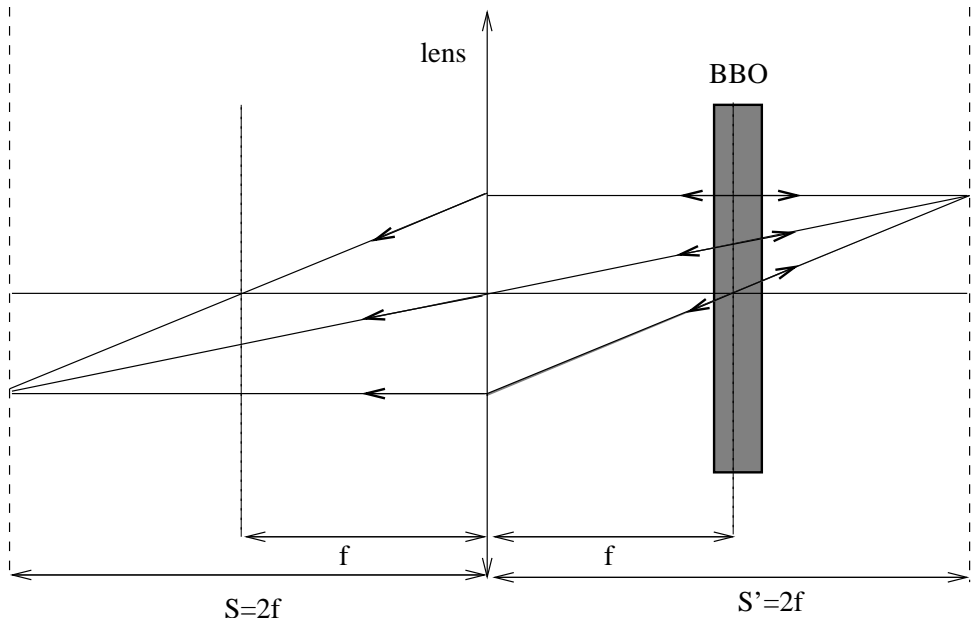}
\end{center}
\caption{Unfolded schematic of the experiment. Particle $1$ (the signal photon) is moving to the right and particle $2$ (the idler photon) is moving to the left. We have put $S=S'=2f$. Similar schematics are shown in the paper by Pittman {\em et al.} \cite{pittman95}. In the paper the trajectories represent conceptual photon trajectories. As we show in Section \ref{pwidescriptionpittmanexperiment}, this same picture follows from the pilot-wave description of the massive particle equivalent of the experiment by Pittman {\em et al.}}
\label{figure1}
\end{figure}   
\noindent
In this process a pump photon incident on a nonlinear beta barium borate (BBO) crystal leads to the creation of a signal and an idler photon. Due to momentum conservation, the sum of the momenta of these photons has to equal the momentum of the pump photon.  This results in the momentum entanglement of the two photons, because the momenta of the idler and signal photon can be combined in an infinite number of ways to equal the momentum of the pump photon. In order to avoid a momentum spread due to spatial confinement, the width of the pump beam is sufficiently large. Hence, the exact place of creation within the BBO crystal is unknown. In the same way, the energies of the created photons add up to the energy of the pump photon. In the experiment, the signal and idler photons are sent in two different directions where coincidence records may be performed by two photon counting detectors. A convex lens, with focal length $f$, is placed in the signal beam in order to turn the momentum correlation of the created photons into spatial correlation. In front of the detector for the signal beam an aperture is placed at a distance $S$ from the lens. By placing the detector for the idler beam at a distance $S'$ from the lens, prescribed by the Gaussian thin lens equation, i.e.\
\begin{equation}
\frac{1}{S} + \frac{1}{S'} = \frac{1}{f}
\label{h4.23}
\end{equation}
and scanning in the transverse plane of the idler beam, an image of this aperture is observed in the coincidence counts. This image obeys the classical lens equations in the following sense. If a classical point-like light source would be placed in the plane of the aperture, where the signal photon was detected, it would have an image where the idler photon was detected. This is the spatial correlation of the photons.

Finally, although we will not further deal with the experiment of Kim and Shih, it is interesting to note that unfortunately they failed in their original intention to perform Popper's gedankenexperiment. As was shown by Short \cite{short00}, the diameter of the incoming beam of pump photons was still to small to guarantee perfect momentum entanglement of the parametric down converted photons and this blurred the predicted results.   

\section{Opposite momenta and opposite directions}\label{oppositemomentaoppositedirections} 
Peres gave an analysis of the extent to which opposite momenta lead to opposite directions \cite{peres201}. He argues that the inequalities in (\ref{h4.2}) do not exclude a priori the possibility that opposite momenta of particles lead to opposite directions (instead of opposite positions) where the particles will be found when a measurement is performed; on the contrary, the operator equivalent of (\ref{h4.1}) would even lead to an observable alignment of the detection points of the two particles. We will indicate that Peres' analysis is correct in the large time regime, but in the small time regime there is an additional source for the deviation from perfect alignment which is not mentioned by Peres.

Let us recall Peres' arguments. To discuss the possible angular alignment of momentum entangled particles, Peres considers a non-relativistic wavefunction describing massive particles. According to Peres, the reason for angular correlation of momentum entangled photons (as in the experiment of Pittman {\em et al.}) is the same as in the considered massive case.

The momentum correlated particles can be assumed to result from a decaying system at rest. The decaying system can then be described by the wavefunction
\begin{equation}
\psi({\bf x}_1,{\bf x}_2,t) = \int {d^3 p_1  d^3 p_2 F({\bf p}_1 , {\bf p}_2) e^{i({\bf p}_1 \cdot {\bf x}_1 +  {\bf p}_2  \cdot {\bf x}_2 - Et)/\hbar}   }
\label{h4.3}
\end{equation}
where the momentum distribution $F$ is peaked around ${\bf p}_1 + {\bf p}_2 = {\bf 0}$ and around the rest energy $E_0$ of the decaying system. According to Peres the opposite momenta of the particles lead to opposite directions where the particles will be found when a measurement is performed. He shows this by applying the stationary phase method. The main contribution to the integral in ({\ref{h4.3}}) comes from values ${\bf p}_1$ and ${\bf p}_2$ for which ${\bf p}_1 + {\bf p}_2 \simeq {\bf 0}$. Because of the rapid oscillations of the phase in the integrand in ({\ref{h4.3}}), the integral will be appreciably different from zero only if the phase is stationary with respect to the six integration variables ${\bf p}_1$ and ${\bf p}_2$ in the vicinity of ${\bf p}_1 + {\bf p}_2 = {\bf 0}$, i.e.\ 
\begin{equation}
\fr{\pa S}{\pa {\bf p}_i} + {\bf x}_i - \fr{\pa E}{\pa  {\bf p}_i} t = {\bf 0}, \qquad  i=1,2
\label{h4.4}
\end{equation}
where $S$ is the phase of $F({\bf p}_1 , {\bf p}_2)$ measured in units of $\hbar$, i.e.\
\begin{equation} 
F({\bf p}_1 , {\bf p}_2) = |F({\bf p}_1 , {\bf p}_2)| e^{iS({\bf p}_1 , {\bf p}_2)/\hbar}
\label{h4.5}
\end{equation}
and the equations ({\ref{h4.4}}) have to be evaluated for ${\bf p}_1 + {\bf p}_2 = {\bf 0}$. The equations ({\ref{h4.4}}) then determine the conditions on $ {\bf x}_i$ in order to have a non-zero $\psi$ (and $|\psi|^2$).

Peres then introduces spherical coordinates to describe ${\bf p}_i$ and ${\bf x}_i$, and varies the phase $S$ with respect to the six spherical variables of ${\bf p}_i$: ($p_i=|{\bf p}_i|,\phi_i,\th_i)$. Peres further assumes that the phase $S$ obeys
\begin{equation}
\pa S/ \pa {\bf p}_i = {\bf 0}
\label{h4.8}
\end{equation}
in the vicinity of ${\bf p}_1 + {\bf p}_2 = {\bf 0}$. This would restrict the place of decay near the origin of the coordinate system, because of (\ref{h4.4}). By varying with respect to the momentum angles, Peres obtains that the phase is stationary if ${\bf p}_i$ and ${\bf x}_i$ have the same direction. Because $F$ is peaked at ${\bf p}_1 + {\bf p}_2 = {\bf 0}$, this results in 
\begin{eqnarray}
\th'_1 + \th'_2 &=& \pi  \nonumber\\
|\phi'_1 -  \phi'_2| &=& \pi
\label{h4.7}
\end{eqnarray}
where $(x_i=|{\bf x}_i|,\phi'_i,\th'_i)$ are the spherical coordinates of ${\bf x}_i$. These equations show that the two particles can only be detected at opposite directions relative to the centre of our coordinate system. This is because the integral in ({\ref{h4.3}}) (and hence $|\psi|^2$) would only be appreciably different from zero if the vectors ${\bf x}_i$ obey ({\ref{h4.7}}). 

Peres mentions two causes for deviation from perfect angular alignment. The first is a transversal deviation of the order $\sqrt{ht/m}$ due to the spreading of the wavefunction, which was recognized as the {\em standard quantum limit} \cite{caves85}. The second is an angular spread of the order $\De (\hat{ p}_{1j}+\hat{ p}_{2j})/p_i$. Below, we show that there is another cause for deviation which arises from the uncertainty on the source and which is particularly important in the `small distance' or `small time' regime. 

First, we want to note that there is, apart from the conditions on $\th'_i$ and $\phi'_i$, also a condition on the variables $x_i$, which is not mentioned by Peres. This condition is obtained by varying the phase of the integrand with respect to $p_i$, having in mind the previous result that ${\bf p}_i$ and ${\bf x}_i$ have the same direction. If we define  $v_i  = d E / d p_i$, then the additional condition reads
\begin{equation}
x_i = v_i t,
\label{h4.12}
\end{equation}
where the $v_i$ have to be evaluated for ${\bf p}_1 + {\bf p}_2 = {\bf 0}$. Because $\psi$ obeys the non-relativistic Schr\"odinger equation, $E=\frac{p^2_1}{2m_1} + \frac{p^2_2}{2m_2}$, with $m_i$ the masses of the particles. In this way ({\ref{h4.12}})  becomes 
\begin{equation}
x_1 = \frac{p_1}{m_1} t, \quad x_2 = \frac{p_2}{m_2} t.
\label{h4.9}
\end{equation}
Using ${\bf p}_1 + {\bf p}_2 = {\bf 0}$ one obtains
\begin{equation}
x_1 m_1 = x_2 m_2.
\label{h4.13}
\end{equation}
Note that we have not yet used the fact that $F$ is peaked around a certain energy $E_0$, as is required in the case of a decaying system at rest. As Peres notes in his paper, a restriction of the energy to $E_0$ further restricts the momenta of the particles to satisfy $p^2_1 = p^2_2 = 2 E_0 m_1 m_2/(m_1 + m_2)$. By combining (\ref{h4.7}) and (\ref{h4.13}), we obtain that the joint detection probability has a maximum for the classically expected relation $m_1 {\bf x}_1 + m_2 {\bf x}_2 = {\bf 0}$. Hence, Peres' statement `opposite momenta lead to opposite directions' may be replaced by a stronger statement, namely that the opposite momenta lead to a maximum detection probability for $m_1 {\bf x}_1 + m_2 {\bf x}_2 = {\bf 0}$. 

Let us now consider the possible sources for deviation from this classical relation. Classically one can, in theory, make both quantities $\De (\hat{ p}_{1j}+\hat{ p}_{2j})$ and $\De (m_1 \hat{ x}_{1j} + m_2 \hat{ x}_{2j})$ as small as wanted. Quantum mechanically one can at best prepare the system, such that initially the equality in
\begin{equation}
\De (\hat{ p}_{1j}+\hat{ p}_{2j}) \De (m_1 \hat{ x}_{1j} + m_2 \hat{ x}_{2j}) \ge \frac{\hbar}{2} (m_1 + m_2 )
\label{h4.14.11}
\end{equation} 
is reached. Note that this equation implies that in case of opposite momenta, the particles can not depart from a confined, fixed source. 

Because the operator $\hat{ p}_{1j} + \hat{ p}_{2j}$ commutes with the free Hamiltonian, the variance of the momentum operator $\hat{ p}_{1j}+\hat{ p}_{2j}$ is stationary. The variance of $\hat{ x}_{1j}+\hat{ x}_{2j}$ however, will in general increase with time due to the spreading of the wavefunction. This can be seen if we write down the expression for the free evolution of the operator $m_1 \hat{ {\bf x}}_{1} + m_2 \hat{ {\bf x}}_{2}$ in the Heisenberg picture
\begin{equation}
m_1 \hat{ {\bf x}}_{1}(t) + m_2 \hat{{\bf x}}_{2}(t) = m_1 \hat{{\bf x}}_{1}(0) + \hat{{\bf p}}_{1}(0)t + m_2 \hat{ {\bf x}}_{2}(0) +\hat{{\bf p}}_{2}(0)t
\label{h4.20}
\end{equation}
The variance of this operator for an arbitrary component $j$ is
\begin{eqnarray}
&&\De \big( m_1 \hat{ x}_{1j}(t) + m_2 \hat{ x}_{2j}(t) \big)^2 \nonumber\\
&&= \De \big( m_1 \hat{ x}_{1j}(0) + m_2 \hat{ x}_{2j}(0) \big)^2 + \De \big( \hat{ p}_{1j}(0)+\hat{ p}_{2j}(0) \big)^2 t^2\nonumber\\
&&+ \Big< \left[ \big( m_1 \hat{ x}_{1j}(0) + m_2 \hat{ x}_{2j}(0) \big), \big( \hat{ p}_{1j}(0)+\hat{ p}_{2j}(0) \big) \right]_+ \Big> t \nonumber\\
&&-2 \big< m_1 \hat{ x}_{1j}(0) + m_2 \hat{ x}_{2j}(0)  \big> \big<  \hat{ p}_{1j}(0)+\hat{ p}_{2j}(0) \big>t
\label{h4.21}
\end{eqnarray}
where the brackets $[,]_+$ denote the anti-commutator. If we assume a distribution $F$ which is real and symmetric, i.e.\ $F({\bf p}_1 , {\bf p}_2) = F(-{\bf p}_1 , -{\bf p}_2)$ then the last two terms in ({\ref{h4.21}}) are both zero. So the variance of $ m_1 \hat{ x}_{1j}(t) + m_2 \hat{ x}_{2j}(t) $ increases with time
\begin{equation}
\De \big( m_1 \hat{ x}_{1j}(t) + m_2 \hat{ x}_{2j}(t) \big)^2  =  \De \big( m_1 \hat{ x}_{1j}(0) + m_2 \hat{ x}_{2j}(0) \big)^2 + \De \big( \hat{ p}_{1j}(0)+\hat{ p}_{2j}(0) \big)^2 t^2 .
\label{h4.22.1}
\end{equation}
This leads to an increasing deviation from the relation $m_1 {\bf x}_1 + m_2 {\bf x}_2 = {\bf 0}$. 

Thus there is always an interplay between opposite momenta and opposite directions which is expressed in (\ref{h4.14.11}) and (\ref{h4.22.1}). We argue that one should study (\ref{h4.22.1}), where the variances at $t=0$ in the right hand side of the expression are limited by the Heisenberg uncertainty in (\ref{h4.14.11}), to determine to which extent we can speak of possible angular alignment. 

Let us now see how Peres' estimates for deviation from perfect alignment come about and in which regime they are important. Assume for convenience that $m_1 = m_2=m$. The transversal deviation $L(t)$ can be taken of the order $\De \big( \hat{ x}_{1j}(t) +  \hat{ x}_{2j}(t) \big)$. There are two contributions to this transversal deviation. The first is $\De \big( \hat{ x}_{1j}(0) +  \hat{ x}_{2j}(0) \big) = L(0)$ and is important for small times. The second contribution is $\De \big( \hat{ p}_{1j}+\hat{ p}_{2j} \big) t/m$ which becomes important for larger times. The angular deviation $\theta$ may be derived from $\tan(\theta) = L(t)/R(t)$, where $R(t)= pt/m$ is the distance that both particles have traveled. For small times one has $\tan(\theta) \simeq \De \big( \hat{ x}_{1j}(0) +  \hat{ x}_{2j}(0) \big)m/pt$ and for large times one has $\tan(\theta) \simeq \De \big( \hat{ p}_{1j} +  \hat{ p}_{2j} \big)/p$. Hence, for large times we obtain the estimate of deviation mentioned by Peres. From the relations (\ref{h4.14.11}) and (\ref{h4.22.1}) one can also easily derive the standard quantum limit
\begin{eqnarray}
\De \big( \hat{ x}_{1j}(t) +  \hat{ x}_{2j}(t) \big)^2 &\ge& 2 \De \big(  \hat{ x}_{1j}(0)+ \hat{ x}_{2j}(0) \big)   \De \big( \hat{ p}_{1j}(0)+\hat{ p}_{2j}(0) \big) t/m  \nonumber\\
 &\ge& 2\hbar t/m.
\label{h4.22.2}
\end{eqnarray}
However, this uncertainty is misleading because for small times it neglects the contribution arising from the uncertainty on the source $\De \big( m_1 \hat{ x}_{1j}(0)+ m_2 \hat{ x}_{2j}(0) \big)$. Especially in the considered case of nearly opposite momenta, this contribution will be large because $\De \big( \hat{ p}_{1j}(0)+\hat{ p}_{2j}(0) \big)$ is small. 

In summary, we see that Peres gave causes for deviation which only apply in the large time regime. These causes are in perfect agreement with the `scattering into cones' theorem which states that for every cone $C$ in $\mathbb{R}^m$ with apex in the origin 
\begin{equation}
\lim_{t \to \infty} \int_C d^m x |\psi(x,t)|^2 = \int_C   d^m p |\phi(p)|^2
\label{h4.22.201}
\end{equation}
with $\phi(p)$ the momentum wave function \cite{newton}. This means that the probability that in the infinite future the particles will be found in the cone $C$ is equal to the probability that their momenta lie in the same cone. In the small time regime the uncertainty on the source gives the major contribution to deviation. 

Let us give a simple example. We can use a Gaussian distribution to represent the momentum correlation
\begin{equation}
F({\bf p}_1 , {\bf p}_2) \sim e^{-\frac{({\bf p}_1 + {\bf p}_2)^2}{\si}}.
\label{h4.10}
\end{equation}
The smaller the value of $\si$, the better the momentum correlation between the two fragments. In the limit $\si \to 0$ this distribution approaches the Dirac $\delta$-distribution. Note that this distribution is not peaked around a certain energy $E_0$ as should be required for a decaying system at rest. In Appendix \ref{appc} it is explained why we can leave this restriction on the energy aside without changing the main result. It will follow that a reasonable energy width, peaked around $E_0$, will imply only a minor broadening of the wavefunction. The wavefunction at $t=0$ is
\begin{equation}
\psi({\bf x}_1 , {\bf x}_2,0) \sim \de({\bf x}_1 - {\bf x}_2) e^{-{\bf x}^2_1 \si /4 \hbar^2}.
\label{h4.11}
\end{equation}   
Thus clearly $\psi$ represents a decaying system because initially ${\bf x}_1 = {\bf x}_2$. But for small values of $\si$ (when $F$ is peaked around ${\bf p}_1 + {\bf p}_2 = {\bf 0}$) the probability of finding the particles at $t=0$ at some configuration is totally smeared out, though the probability has a maximum at the origin of the coordinate system. Note that although the relation $\pa S/ \pa {\bf p}_i = {\bf 0}$ is satisfied, this condition does not restrict the place of decay near the origin of the coordinate system, as was assumed by Peres. So, in the small time regime, it may be hard to speak of possible opposite movements of particles relative to the origin because we cannot exactly say (at least without measurement) where the decay of the system took place. As time increases, the deviation from the detection probability peak at $m_1 {\bf x}_1 + m_2 {\bf x}_2 = {\bf 0}$ will even increase with time as was shown above. However, because the distance from the particles to the source increases, the uncertainty of the source will become less important for the angular deviation; in the large time regime the angular deviation will then be dominated by the momentum uncertainty.

In the following section, we will show that we can retain the classical picture of a decaying system in the previous example, in both the small time and the large time regime, when it is described by pilot-wave theory. For example, in the case of the wavefunction considered above, the particle beables will depart near each other and will move along opposite directions. However, the place of departure will vary from pair to pair over an extended region. The more this initial region is confined, the less perfect the momentum entanglement will be, and the less perfect the opposite movements of the pilot-wave particles will be. 

When do we have a transition between the small time and the large time regime? We could say that the transition between the small time regime and the large time regime occurs at time $T$ when both contributions to $L(t)$ are equally large, i.e.\ $L(0) = \De \big( \hat{ p}_{1j}(0)+\hat{ p}_{2j}(0) \big) T/m$ (for fragments with equal masses). If we have $\De \big( \hat{ p}_{1j}(0)+\hat{ p}_{2j}(0) \big)\De \big( \hat{ x}_{1j}(0)+\hat{ p}_{xj}(0) \big) \simeq \hbar$ then the transition time is given by $T \simeq  L(0)^2 k_c / c$, with $k_c$ the wavenumber corresponding to the Compton wavelength of the particles. 

If we now consider the experiment performed by Pittman {\em et al.}, then Peres' analysis can be applied. This is because the experiment can be seen to occur in the large distance regime because of presence of the lens. In some sense the lens can be seen as projecting the angular correlation at infinity, to finite distances (at distances $2f$ from the lens). The better the momentum correlation, the better the angular correlation is at infinity or the better the optical imaging is. 

It is interesting to use the data of this experiment to give an example where the transition can be situated between the large distance and the small distance regime for a system which displays strong momentum correlation. The wavelength of the pump photon is $351.1nm$ and the width of the pump beam is $L(0)= 2mm$. If we now assume that the equality is reached for the Heisenberg uncertainty then, in the absence of a lens, the transition time is of the order $T \simeq L(0)^2 k_c / c$. The distance traveled by the photons at that time would be $R \simeq  L(0)^2 k_c \simeq 70m $. This means that if we would create momentum entangled photons via spontaneous parametric down conversion, then the angular deviation would be dominated by the error arising from the uncertainty of the source within a distance $R\simeq 70m$.

The deviation from perfect alignment resulting from the uncertainty on the source is also important in Popper's experimental proposal. It can easily be seen that Popper's experiment cannot occur in the large time regime. If the particles would travel large distances (and hence obtain good angular correlation), the virtual slit (which is of the order of the transversal deviation) would be too large to have virtual diffraction of the particles on Bob's side. Hence, Popper's original experimental proposal should be considered in the small time regime. However, in this regime the uncertainty on the source becomes the most important contribution to the angular deviation and as follows from the analysis of Collett and Loudon, this uncertainty makes a detectable virtual diffraction impossible. This also implies that in discussing Popper's experiment one should be careful with statements such as ``\ldots the allowed deviation from perfect alignment is of the order of $\De |p_1+ p_2|/|p_1 - p_2|$, which is much too small to be of any consequence in the present discussion.' and ``\ldots nearly perfect alignment can be taken for granted, \ldots" \cite{peres99}.

\section{Pilot-wave description of a decaying system}\label{pwidescriptionofadecayingsystem}
In this section, we give a simplified description of a system consisting of two particles with opposite momenta, resulting from a decaying system at rest, in terms of the pilot-wave interpretation.{\footnote{Our simplified pilot-wave approach to a decaying system is to be distinguished from the one studied by Y.\ Nogami {\em et al.} \cite{nogami00}, where a decaying system is represented by a particle that leaks out from a region surrounded by a repulsive potential barrier.}} We use the wavefunction in ({\ref{h4.3}}) with momentum distribution 
\begin{equation}
F({\bf p}_1 , {\bf p}_2) = N \de ({\bf p}_1 + {\bf p}_2)  e^{-\alpha p^2_1 / \hbar}
\label{h4.24}
\end{equation}
where $N$ is a normalization factor. The parameter $\al$ sets the scale of the initial separation of the two particles, as will be seen soon. A small $\al$ will correspond to the considered physical situation of a decaying system. The parameter is introduced in order to avoid singularities arising from the $\delta$-distribution when calculating the pilot-wave trajectories later on. Although a system that decays from rest has a certain fixed total energy, we omit this energy restriction, just as in Section \ref{oppositemomentaoppositedirections}, for reasons explained in Appendix \ref{appc}. Note that the exponential factor in (\ref{h4.24}) does not restrict the value of the total energy for a small $\al$. 

The wavefunction corresponding to the distribution $F$ is
\begin{equation}
\psi({\bf x}_1 , {\bf x}_2,t) = N \bigg(\frac{\pi \hbar}{\al + it/2\mu} \bigg)^{3/2} e^{-({\bf x}_1 - {\bf x}_2)^2 / 4 \hbar (\al + \frac{it}{2\mu} )}
\label{h4.25}
\end{equation}
with $\mu$ the reduced mass of the fragments: $\frac{1}{\mu} = \frac{1}{m_1} +\frac{1}{m_2}$. At $t=0$ the probability distribution is
\begin{equation}
|\psi({\bf x}_1 , {\bf x}_2,0)|^2 = N^2 \bigg(\frac{ \pi \hbar}{\al}\bigg)^3  e^{-({\bf x}_1 - {\bf x}_2)^2 / 2 \hbar \al}.
\label{h4.26}
\end{equation}
It follows that a small value for $\al$ corresponds to the considered physical situation of a decaying system. However, the place of decay is unknown. This is a consequence of the Heisenberg uncertainty, as explained in the Section \ref{oppositemomentaoppositedirections}. 

The trajectories ${\bf x}_j(t)$ of the particle beables are found by solving the guidance equations  
\begin{equation}
\frac{d {\bf x}_j}{d t} = \frac{1}{m_j} \frac{\textrm{Re} \left( \psi^*({\bf x}_1 , {\bf x}_2,t) {\hat {\bf p}}_j \psi({\bf x}_1 , {\bf x}_2,t) \right)}{|\psi({\bf x}_1 , {\bf x_2},t)|^2 } .
\label{h4.27}
\end{equation}
Because $(\hat {\bf p}_1 + \hat {\bf p}_2) \psi = 0$ the trajectories of the particles satisfy
\begin{equation}
\frac{d}{d t} (m_1 {\bf x}_1 + m_2 {\bf x}_2) = {\bf 0}.
\label{h4.28}
\end{equation}
This shows that the particles have opposite speeds and thus move in opposite directions. Integration of the differential equations (\ref{h4.27}) leads to
\begin{eqnarray}
{\bf x}_1(t) &=& {\bf c}_1 + {\bf c}_2 \sqrt{t^2/4\mu^2 + \al^2} \nonumber\\
{\bf x}_2(t) &=& {\bf c}_1 - {\bf c}_2 \sqrt{t^2/4\mu^2 + \al^2}
\label{h4.29}
\end{eqnarray}
where ${\bf c}_1$ and ${\bf c}_2$ are arbitrary constant vectors. It follows that the particles also move along straight lines. At $t=0$ the probability distribution $|\psi|^2$ is sharply peaked at ${\bf x}_1 = {\bf x}_2$ (for small values of $\al$) and hence the particle beables will depart near each other. As follows from (\ref{h4.29}), their further propagation proceeds along straight lines, in the direction of their connecting line. Thus opposite momenta lead to opposite directions of movement for pilot-wave particles. But their place of departure is located within an extended area, in order to preserve momentum correlation. 

By using pilot-wave theory, we are thus able to retain part of the classical picture of a decaying system at rest. Note the similarity in language with the one used by Popper to describe his experiment. The difference is that Popper assumed the particles to depart from a confined region (which is however incompatible with opposite momenta in quantum mechanics). 

\section{Pilot-wave description of the experiment of Pittman {\em et al.}}\label{pwidescriptionpittmanexperiment}
Although the experiment of Pittman {\em et al.}\ can be correctly explained with quantum optics, we will provide a pilot-wave account of the experiment, when it is `translated' into its massive particle equivalent. One of the reasons to use pilot-wave theory is that it justifies the conceptual photon trajectories drawn by Pittman {\em et al.}\  \cite{pittman95}. I.e.\ the photon trajectories coincide with the trajectories of pilot-wave particles in the massive particle equivalent of the experiment. This pilot-wave approach is to be contrasted with the explanation in terms of `usual' geometrical optics used by Pittman {\em et al.}. In quantum optics, these paths are usually regarded as a visualization of the different contributions to the detection probabilities.  

Because there is at present no satisfactory pilot-wave interpretation in terms of particle beables for photons, see Chapter \ref{chapter3}, we will follow Peres' point of departure and we will consider the non-relativistic massive particle wavefunction in (\ref{h4.3}) which can then be seen as describing the massive particle equivalent of the experiment by Pittman {\em et al.} The spontaneous parametric down conversion source then corresponds to a decaying system at rest, resulting in two energy and momentum correlated fragments. We will assume the total momentum of the fragments to be zero, instead of some fixed value corresponding to the initial momentum of the total system (which would represent the momentum of the pump photon). This assumption corresponds to the `unfolded' schematic introduced by Pittman {\em et al.}\ \cite{pittman95} (which is displayed in Fig.\ \ref{figure1}). In this way we can use the momentum distribution $F$ defined in the previous section
\begin{equation}
F({\bf p}_1 , {\bf p}_2) = N \de ({\bf p}_1 + {\bf p}_2)  e^{-\alpha p^2_1 / \hbar}.
\label{h4.24.0000000003}
\end{equation}

In the preceding section, we described the free evolution after the decay of the system. The unknown place of decay in the massive particle case corresponds to the unknown place of creation within the BBO crystal in the photon case, due to the width of the pump beam. To complete the pilot-wave description of the massive particle equivalent of optical imaging, we just have to describe the system's interaction with the lens. In classical optics we can use ray optics to describe the action of the lens on an impinging light beam \cite{bornwolf80}. The rays are such that the Gaussian thin lens equations are satisfied. Two generic examples, which we will need later on, are the following. The effect of the lens on a plane wave is to turn it into a converging wave, with focus in the focal plane, such that the corresponding rays obey the lens equations. The characteristics of the converging wave are then determined by the momentum of the incoming plane wave and the focal length. A second example is a spherical wave, representing a point source. If we assume that the light source is located in a plane at a distance $S$ from the lens, then the spherical wave will turn into a converging wave with focus in the plane at a distance $S'$ from the lens so that $1/S + 1/S' = 1/f$ and the source, the image and the centre of the lens will be aligned. In massive particle quantum physics the equivalent of optical lenses are electrostatic or magnetic lenses. These electromagnetic lenses are generally used to collimate or focus beams of charged particles. This field of research is usually called {\em optics of charged-particle beams} or the {\em theory of charged-particle beams through electromagnetic systems}. Most of the literature deals with the classical description of the particles and only recently the quantum mechanical approach has been studied, see for example Hawkes and Kasper \cite{hawkes96}, and Khan and Jagannathan \cite{khan95} and references therein. Here, we will not consider the detailed analysis of particles passing through such electromagnetic lenses, and use directly, in the spirit of de Broglie, the analogy with classical optics. For example, we can describe the action of an electromagnetic lens as turning a quantum mechanical plane wave into a Gaussian wave (we can take this as the analogue of the converging wave in classical optics, because a Gaussian wave is contracting before expanding), determined by the momentum of the incoming wave and the focal length. This analogy is very appealing because the rays in classical optics can be `identified' with the pilot-wave trajectories. This is because in the one-particle case, the curves determined by the normals of the wavefronts of the quantum mechanical wavefunction are just the possible trajectories of the particle beables.{\footnote{For a non-relativistic particle the guidance equation can be written as $d {\bf x} /d t = {\boldsymbol \nabla}S/m$, where $S$ is the phase of the wavefunction measured in units $\hbar$, i.e.\ $\psi=|\psi|\exp({iS/\hbar})$ \cite{debroglie60,bohm1}.}} If we apply this to our decaying system, then every plane wave of the particle impinging on the lens, say particle two, in the integral in (\ref{h4.3}) is turned into a particular Gaussian wave. The resulting wave is then
\begin{equation}
\psi'({\bf x}_1,{\bf x}_2,t) = \int {d^3 p_1  d^3 p_2 F({\bf p}_1 , {\bf p}_2) e^{i({\bf p}_1 \cdot {\bf x}_1  - p^2_1t/2m_1)/\hbar}  } G({\bf x}_2,{\bf p}_2,f) 
\label{h4.30}
\end{equation}
where $G$ represents the Gaussian wave. This wave is guiding the particle beables after particle beable two passed the lens. To avoid unnecessary mathematical complications when calculating the trajectories implied by the wave (\ref{h4.30}), we assume that the place of decay of the system is somewhere in the middle between the lens and the detector on the right (where the idler photon arrives in the experiment of Pittman {\em et al.}). This corresponds to a BBO crystal placed in the middle instead of it placed near the lens, as in the experiment. When particle beable two arrives in the vicinity of the lens, particle beable one will arrive in the vicinity of the transversal plane on the right, where the detectors are placed. Suppose that the particle beable is detected in the transversal plane at the position ${\bf a}$. Due to the correlation of the detector and the particle on the right, particle beable two will be effectively guided by the wavefunction{\footnote{In the standard interpretation of quantum mechanics, this process is the collapse of the wavefunction. The pilot-wave description of this process was given in Section \ref{themeasurementprocess}.}}
\begin{eqnarray}
\psi_2({\bf x}_2,t) &\sim&  \int {d^3 p e^{\frac{i}{\hbar}\big({\bf p} \cdot ({\bf a} -  {\bf x}_2) -   {p}^2 t/2m_2  \big) -\al p^2/\hbar}   }\nonumber\\ 
&\sim& \bigg(\frac{\pi \hbar}{\al + it/2m_2} \bigg)^{3/2} e^{-({\bf a} - {\bf x}_2)^2 / 4 \hbar (\al + \frac{it}{2m_2} )}.
\label{h4.32}
\end{eqnarray}
The phase of this wave is 
\begin{equation}
S({\bf x}_2,t) = \frac{t({\bf a} - {\bf x}_2)^2}{8m_2 \al^2 + t^2/m_2} - \frac{3 \hbar}{2} \tan^{-1} (t/2m_2 \alpha).
\label{h4.33}
\end{equation}
So, the wavefronts of the guiding wave of particle two are spheres with centre in ${\bf a}$. Because the detectors are placed in transversal planes at distances $S$ and $S'$ from the lens, with $S$ and $S'$ obeying the Gaussian lens equation (\ref{h4.23}), this wave will result, after propagation through the lens, in a converging wave with focus in the plane at a distance $S$ from the lens and where the focus is determined by the Gaussian thin lens equations. Hereby we used again the analogy with classical Gaussian optics. If for example $S=S'$ and if the centre of the lens is taken as the origin of our coordinate system, then the focus will be at $-{\bf a}$ (see Fig.\ 1). As a result, particle beable two will be detected in the focus of the wave. Because we used a Gaussian to describe the converging wave, the trajectories will not be straight lines, but will be curved (for images see Holland \cite{holland} p162). The curvature will depend on the width in the focus of the Gaussian. In the limit of a zero width, however, the trajectories will approach straight lines, directed from the lens towards the focus of the wave. When the coincidence detections are considered, it will appear as if particle beable two departs from the place of detection of particle beable one.

This completes the analysis in terms of the pilot-wave interpretation of the phenomenon of optical imaging. Before the fragments reach the lens, they move along straight lines from the place of decay. Note that this place of decay is not fixed, in order to guarantee the momentum correlation ${\bf p}_1 + {\bf p}_2 = {\bf 0}$. When one of the particles reaches the lens, its direction of movement will change in accordance with the classical thin lens approximations. We assumed hereby that the place of decay is centred between the right detector and the lens. It can be expected, although it is not proven, that a random place of decay (for example near the lens) will lead to the same results in the pilot-wave description of the experiment.  

\section{Conclusion}  
In conclusion, we showed that Peres' analysis concerning the question to what extent opposite momenta lead to opposite directions, is only valid in the large distance regime. In the small time regime there is an additional source of angular deviation. On the other hand the statement `opposite momenta lead to opposite directions' is true in the  language of pilot-wave theory. I.e.\ the particle beables travel in opposite directions when the wavefunction has eigenvalue zero for the total momentum operator (however from an unknown place of departure). We also showed that pilot-wave trajectories coincide with the pictures of conceptual trajectories present in the paper Pittman {\em et al.} Hence, pilot-wave theory could be used to mathematically underpin these conceptual trajectories.

Note that the experiment of Kim and Shih is very illustrative for the need for perfect momentum correlation of the photons, or equivalently that there must be very little restriction on the place of creation of the photons to create a perfect image. This is because Kim and Shih failed in their original intention to perform Popper's gedankenexperiment, due to the restricted diameter of the pump beam used in the experiment. The imperfect momentum correlation then led to an imperfect optical image \cite{short00}. This is immediately obvious when we consider our pilot-wave description of optical imaging, because if the momentum correlation is imperfect, the pilot-wave particles will not move in opposite directions before the system reaches the lens.

\chapter{Conclusion and outlook}
In this thesis we studied the de Broglie-Bohm pilot-wave interpretation of quantum theory. We mainly focussed on the question to which extent it is possible to provide a pilot-wave interpretation for quantum theory. 

In the context of non-relativistic quantum theory it is no problem to construct a pilot-wave interpretation. It is straightforward to extend the pilot-wave interpretation of de Broglie and Bohm for a spinless particle to arbitrary spin. In fact there even exists more than one approach. In the model we presented here, the beables are point-particles without any spin properties.

Problems arise when we try to transcript this pilot-wave model to relativistic wave equations. In fact these problems do not only arise when we try to develop a pilot-wave interpretation. Problems already arise when we try to transcript the quantum mechanical interpretation of non-relativistic quantum theory to relativistic wave equations. In the first place, there is the problem of identifying a future-causal current which can be interpreted as a particle probability current. Closely related to this problem is the problem of defining a positive definite inner product. Then there is also the problem of interpreting the negative energy states. Only in restricted cases these problems can be solved (e.g.\ as in the spin-1/2 Dirac theory coupled to the electromagnetic field). As is well known these problems led to the conception of quantum field theory. The problems which blocked a quantum mechanical interpretation of relativistic wave equations are not present in quantum field theory. One has a positive definite inner product and there are no problems with the interpretation of negative energy states. Particles and anti-particles can be freely created and annihilated without the need of considering negative energy states. Hence in the pilot-wave approach we should not stick to relativistic wave equations either, but focus on quantum field theory instead.

In quantum field theory, field operators take over the role of the particle operators in non-relativistic quantum theory. This suggests that the notion of fields is more fundamental than the notion of particles for high energy physics. Therefore, fields seem to be the most natural beables in the pilot-wave approach. This seems to be confirmed at least in the case of bosonic quantum field theory. For bosonic quantum field theory it was straightforward to construct a pilot-wave interpretation in terms of field beables. We could construct a pilot-wave theory for massive spin-0 fields and spin-1 fields. Even for gauge theories we could find a pilot-wave interpretation. We took an approach in which only beables are introduced for gauge independent variables, as in the model by Bohm and Kaloyerou. Valentini presented an other approach in which beables are also introduced for gauge variables. But we argued that this last approach leads to non-normalizable densities of field beables. In addition the guidance equations for the gauge variables are rather meaningless because they just express that these variables are stationary. We also indicated that for non-Abelian gauge theories a pilot-wave interpretation is in principle possible along the same lines as in the Abelian case (i.e.\ the electromagnetic field case), the only problem is the identification of the gauge invariant degrees of freedom. 

In the case of fermionic quantum field theory the construction of a pilot-wave interpretation is not so straightforward. We elaborated on an idea of Valentini to use a representation for the field operators in terms of Grassmann fields. In this representation a functional Schr\"odinger picture can be obtained. However, we found that it was not possible to devise a pilot-model where the beables are Grassmann fields, at least not in the way Valentini originally suggested. 

Is a pilot-wave interpretation in terms of field beables therefore impossible? It seems not, because there is still a model by Holland, where the beables correspond to rotators in each point of the configuration space (which can be physical 3-space or momentum space). However, this model brings with it new questions which have yet to be answered. Is the ontology for fermionic fields in terms of rotator beables compatible with the ontology for bosonic fields in terms of `ordinary' field beables? Or can one construct a rotator ontology for bosonic fields as well? Another question is whether wavefunctionals which correspond to distinct macroscopic states non-overlapping in the configuration space of rotator fields? Admittedly, this last question is in fact not yet sufficiently addressed either in the context of `ordinary' field beables. 

Another possibility is that we should go along with a particle ontology, instead of with a field ontology, for fermionic field theory. This is an approach taken by D\"urr, Goldstein, Tumulka and Zangh\`\i. The Bell-type model they introduced for quantum electrodynamics introduces point particles as beables, but only for the fermions and not for the electromagnetic field. It is the question whether this type of model can be extended so that particle beables are introduced for the bosonic fields as well. It could also be that beables need not be introduced for bosons at all. Perhaps introducing beables corresponding to fermions is sufficient to solve the measurement problem. It would then be interesting to see the extension of this model to account for weak and strong interactions.

We think that the construction of a coherent pilot-wave interpretation for fermionic field theory is the main issue to be addressed in pilot-wave theory. This could be done by further elaborating on the model of Holland. Or if one allows stochasticity in the model, one could start from the model by D\"urr {\em et al.} In any case, we think that a coherent pilot-wave interpretation for fermionic fields must be possible. The reason is that the standard interpretation for fermionic fields does not encounters problems, i.e.\ a Hilbert space can be set up together with an associated operator formalism. This is contrary to the case of relativistic wave equations where the problems in providing a pilot-wave interpretation were directly related to problems already rooted in the standard interpretation. With a coherent pilot-wave interpretation for fermionic fields and together with the pilot-wave interpretation for massive spin-0 and massive spin-1 presented in this thesis and with a pilot-wave interpretation for non-Abelian gauge theories (which presents no problem in principle), we could in principle be given a pilot-wave interpretaton for the {\em standard model}, which represents to this day the credo of high energy quantum physics.

\newpage

\noindent \\ 

\appendix
\chapter{Representation of the Kemmer-Duffin-Petiau matrices}
\label{appa}
We adopt the matrix representation used in \cite{ghose96}, where the
$\beta^i$ correspond to $-i\beta^i$ in \cite{kemmer39}. For spin-0 the $\beta^{\mu}$ matrices read
\begin{eqnarray}
\beta^0 = \left( 
\begin{array}{c}
0 \\
0 \\
0 \\
0 \\
i \\
\end{array} 
\begin{array}{c}
0 \\
0 \\
0 \\
0 \\
0 \\
\end{array} 
\begin{array}{c}
0 \\
0 \\
0 \\
0 \\
0 \\
\end{array} 
\begin{array}{c}
0 \\
0 \\
0 \\
0 \\
0 \\
\end{array} 
\begin{array}{c}
-i\\
0 \\
0 \\
0 \\
0 \\
\end{array} 
\right),\quad
\beta^1 = \left( 
\begin{array}{c}
0 \\
0 \\
0 \\
0 \\
0 \\
\end{array} 
\begin{array}{c}
0 \\
0 \\
0 \\
0 \\
-i \\
\end{array} 
\begin{array}{c}
0 \\
0 \\
0 \\
0 \\
0 \\
\end{array} 
\begin{array}{c}
0 \\
0 \\
0 \\
0 \\
0 \\
\end{array} 
\begin{array}{c}
0 \\
-i \\
0 \\
0 \\
0 \\
\end{array} 
\right),
\nonumber\\
\beta^2 = \left( 
\begin{array}{c}
0 \\
0 \\
0 \\
0 \\
0 \\
\end{array} 
\begin{array}{c}
0 \\
0 \\
0 \\
0 \\
0 \\
\end{array} 
\begin{array}{c}
0 \\
0 \\
0 \\
0 \\
-i \\
\end{array} 
\begin{array}{c}
0 \\
0 \\
0 \\
0 \\
0 \\
\end{array} 
\begin{array}{c}
0 \\
0 \\
-i \\
0 \\
0 \\
\end{array} 
\right),\quad
\beta^3 = \left( 
\begin{array}{c}
0 \\
0 \\
0 \\
0 \\
0 \\
\end{array} 
\begin{array}{c}
0 \\
0 \\
0 \\
0 \\
0 \\
\end{array} 
\begin{array}{c}
0 \\
0 \\
0 \\
0 \\
0 \\
\end{array} 
\begin{array}{c}
0 \\
0 \\
0 \\
0 \\
-i \\
\end{array} 
\begin{array}{c}
0 \\
0 \\
0 \\
-i \\
0 \\
\end{array} 
\right).
\label{a1}
\end{eqnarray}
The $\gamma$ matrix used in the massless spin-0 theory reads
\begin{eqnarray}
\gamma = \left( 
\begin{array}{c}
1 \\
0 \\
0 \\
0 \\
0 \\
\end{array} 
\begin{array}{c}
0 \\
1 \\
0 \\
0 \\
0 \\
\end{array} 
\begin{array}{c}
0 \\
0 \\
1 \\
0 \\
0 \\
\end{array} 
\begin{array}{c}
0 \\
0 \\
0 \\
1 \\
0 \\
\end{array} 
\begin{array}{c}
0 \\
0 \\
0 \\
0 \\
0 \\
\end{array} 
\right).
\label{a1.1}
\end{eqnarray}

For spin-1 the $\beta^{\mu}$ matrices read
\begin{eqnarray}
&&\!\!\!\!\!\!\!\!\! \beta^0 = \left( 
\begin{array}{c}
0 \\
0 \\
0 \\
0 \\
0 \\
0 \\
i \\
0 \\
0 \\
0 \\
\end{array} 
\begin{array}{c}
0 \\
0 \\
0 \\
0 \\
0 \\
0 \\
0 \\
i \\
0 \\
0 \\
\end{array} 
\begin{array}{c}
0 \\
0 \\
0 \\
0 \\
0 \\
0 \\
0 \\
0 \\
i \\
0 \\
\end{array} 
\begin{array}{c}
0 \\
0 \\
0 \\
0 \\
0 \\
0 \\
0 \\
0 \\
0 \\
0 \\
\end{array} 
\begin{array}{c}
0 \\
0 \\
0 \\
0 \\
0 \\
0 \\
0 \\
0 \\
0 \\
0 \\
\end{array} 
\begin{array}{c}
0 \\
0 \\
0 \\
0 \\
0 \\
0 \\
0 \\
0 \\
0 \\
0 \\
\end{array} 
\begin{array}{c}
-i \\
0 \\
0 \\
0 \\
0 \\
0 \\
0 \\
0 \\
0 \\
0 \\
\end{array} 
\begin{array}{c}
0 \\
-i \\
0 \\
0 \\
0 \\
0 \\
0 \\
0 \\
0 \\
0 \\
\end{array} 
\begin{array}{c}
0 \\
0 \\
-i \\
0 \\
0 \\
0 \\
0 \\
0 \\
0 \\
0 \\
\end{array} 
\begin{array}{c}
0 \\
0 \\
0 \\
0 \\
0 \\
0 \\
0 \\
0 \\
0 \\
0 \\
\end{array} 
\right),\;
\beta^1 = \left( 
\begin{array}{c}
0 \\
0 \\
0 \\
0 \\
0 \\
0 \\
0 \\
0 \\
0 \\
i \\
\end{array} 
\begin{array}{c}
0 \\
0 \\
0 \\
0 \\
0 \\
0 \\
0 \\
0 \\
0 \\
0 \\
\end{array} 
\begin{array}{c}
0 \\
0 \\
0 \\
0 \\
0 \\
0 \\
0 \\
0 \\
0 \\
0 \\
\end{array} 
\begin{array}{c}
0 \\
0 \\
0 \\
0 \\
0 \\
0 \\
0 \\
0 \\
0 \\
0 \\
\end{array} 
\begin{array}{c}
0 \\
0 \\
0 \\
0 \\
0 \\
0 \\
0 \\
0 \\
i \\
0 \\
\end{array} 
\begin{array}{c}
0 \\
0 \\
0 \\
0 \\
0 \\
0 \\
0 \\
-i \\
0 \\
0 \\
\end{array} 
\begin{array}{c}
0 \\
0 \\
0 \\
0 \\
0 \\
0 \\
0 \\
0 \\
0 \\
0 \\
\end{array} 
\begin{array}{c}
0 \\
0 \\
0 \\
0 \\
0 \\
-i \\
0 \\
0 \\
0 \\
0 \\
\end{array} 
\begin{array}{c}
0 \\
0 \\
0 \\
0 \\
i \\
0 \\
0 \\
0 \\
0 \\
0 \\
\end{array} 
\begin{array}{c}
i \\
0 \\
0 \\
0 \\
0 \\
0 \\
0 \\
0 \\
0 \\
0 \\
\end{array} 
\right),
\nonumber\\
&&\!\!\!\!\!\!\!\!\! \beta^2 = \left( 
\begin{array}{c}
0 \\
0 \\
0 \\
0 \\
0 \\
0 \\
0 \\
0 \\
0 \\
0 \\
\end{array} 
\begin{array}{c}
0 \\
0 \\
0 \\
0 \\
0 \\
0 \\
0 \\
0 \\
0 \\
i \\
\end{array} 
\begin{array}{c}
0 \\
0 \\
0 \\
0 \\
0 \\
0 \\
0 \\
0 \\
0 \\
0 \\
\end{array} 
\begin{array}{c}
0 \\
0 \\
0 \\
0 \\
0 \\
0 \\
0 \\
0 \\
-i \\
0 \\
\end{array} 
\begin{array}{c}
0 \\
0 \\
0 \\
0 \\
0 \\
0 \\
0 \\
0 \\
0 \\
0 \\
\end{array} 
\begin{array}{c}
0 \\
0 \\
0 \\
0 \\
0 \\
0 \\
i \\
0 \\
0 \\
0 \\
\end{array} 
\begin{array}{c}
0 \\
0 \\
0 \\
0 \\
0 \\
i \\
0 \\
0 \\
0 \\
0 \\
\end{array} 
\begin{array}{c}
0 \\
0 \\
0 \\
0 \\
0 \\
0 \\
0 \\
0 \\
0 \\
0 \\
\end{array} 
\begin{array}{c}
0 \\
0 \\
0 \\
-i \\
0 \\
0 \\
0 \\
0 \\
0 \\
0 \\
\end{array} 
\begin{array}{c}
0 \\
i \\
0 \\
0 \\
0 \\
0 \\
0 \\
0 \\
0 \\
0 \\
\end{array} 
\right),\;
\beta^3 = \left( 
\begin{array}{c}
0 \\
0 \\
0 \\
0 \\
0 \\
0 \\
0 \\
0 \\
0 \\
0 \\
\end{array} 
\begin{array}{c}
0 \\
0 \\
0 \\
0 \\
0 \\
0 \\
0 \\
0 \\
0 \\
0 \\
\end{array} 
\begin{array}{c}
0 \\
0 \\
0 \\
0 \\
0 \\
0 \\
0 \\
0 \\
0 \\
i \\
\end{array} 
\begin{array}{c}
0 \\
0 \\
0 \\
0 \\
0 \\
0 \\
0 \\
i \\
0 \\
0 \\
\end{array} 
\begin{array}{c}
0 \\
0 \\
0 \\
0 \\
0 \\
0 \\
-i \\
0 \\
0 \\
0 \\
\end{array} 
\begin{array}{c}
0 \\
0 \\
0 \\
0 \\
0 \\
0 \\
0 \\
0 \\
0 \\
0 \\
\end{array} 
\begin{array}{c}
0 \\
0 \\
0 \\
0 \\
-i \\
0 \\
0 \\
0 \\
0 \\
0 \\
\end{array} 
\begin{array}{c}
0 \\
0 \\
0 \\
i \\
0 \\
0 \\
0 \\
0 \\
0 \\
0 \\
\end{array} 
\begin{array}{c}
0 \\
0 \\
0 \\
0 \\
0 \\
0 \\
0 \\
0 \\
0 \\
0 \\
\end{array} 
\begin{array}{c}
0 \\
0 \\
i \\
0 \\
0 \\
0 \\
0 \\
0 \\
0 \\
0 \\
\end{array} 
\right).
\label{a2}
\end{eqnarray}
The $\gamma$ matrix used in the massless spin-0 theory reads
\begin{eqnarray}
\gamma = \left( 
\begin{array}{c}
1 \\
0 \\
0 \\
0 \\
0 \\
0 \\
0 \\
0 \\
0 \\
0 \\
\end{array} 
\begin{array}{c}
0 \\
1 \\
0 \\
0 \\
0 \\
0 \\
0 \\
0 \\
0 \\
0 \\
\end{array} 
\begin{array}{c}
0 \\
0 \\
1 \\
0 \\
0 \\
0 \\
0 \\
0 \\
0 \\
0 \\
\end{array} 
\begin{array}{c}
0 \\
0 \\
0 \\
1 \\
0 \\
0 \\
0 \\
0 \\
0 \\
0 \\
\end{array} 
\begin{array}{c}
0 \\
0 \\
0 \\
0 \\
1 \\
0 \\
0 \\
0 \\
0 \\
0 \\
\end{array} 
\begin{array}{c}
0 \\
0 \\
0 \\
0 \\
0 \\
1 \\
0 \\
0 \\
0 \\
0 \\
\end{array} 
\begin{array}{c}
0 \\
0 \\
0 \\
0 \\
0 \\
0 \\
0 \\
0 \\
0 \\
0 \\
\end{array} 
\begin{array}{c}
0 \\
0 \\
0 \\
0 \\
0 \\
0 \\
0 \\
0 \\
0 \\
0 \\
\end{array} 
\begin{array}{c}
0 \\
0 \\
0 \\
0 \\
0 \\
0 \\
0 \\
0 \\
0 \\
0 \\
\end{array} 
\begin{array}{c}
0 \\
0 \\
0 \\
0 \\
0 \\
0 \\
0 \\
0 \\
0 \\
0 \\
\end{array} 
\right).
\label{a3}
\end{eqnarray}

\chapter{Class of representations which leads to equivalent pilot-wave interpretations for the electromagnetic field}   \label{appendixaaa}
Suppose we have two different canonical transformations represented by $K$ and $K'$ (cf.\ Section \ref{repcanvar}) which allow for a separation of the true degrees of freedom from the constraints in the Coulomb gauge:  
\begin{eqnarray}
 V_i({\bf x}) &=& \int d^3 y K_{ij} ({\bf x},{\bf y})  {\widetilde V}_j({\bf y})= \int d^3 y K'_{ij} ({\bf x},{\bf y})  {\widetilde V}'_j({\bf y}), \nonumber\\
\Pi_{V_i}({\bf x}) &= &\int d^3 y \Pi_{{\widetilde V}_j}({\bf y}) K^{-1}_{ji} ({\bf y},{\bf x})  =\int d^3 y \Pi_{{\widetilde V}'_j}({\bf y}) K'^{-1}_{ji} ({\bf y},{\bf x})  .
\label{apph30.036}
\end{eqnarray}
We assume that the constraints are ${\widetilde V}_3=\Pi_{{\widetilde V}_3} =0$ for the set of unprimed variables and ${\widetilde V}'_3=\Pi_{{\widetilde V}'_3} =0$ for the set of primed variables. Quantization then proceeds by imposing the standard commutation relations for the true degrees of freedom ${\widetilde V}_k,\Pi_{{\widetilde V}_k}$, $k=1,2$ for the set of unprimed variables or ${\widetilde V}'_k,\Pi_{{\widetilde V}'_k}$, $k=1,2$ for the set of primed variables. These commutation relations can then be realized by the standard representation
\begin{equation}
{\widehat {\widetilde V}}_k({\bf x}) = {\widetilde V}_k({\bf x}),\quad {\widehat \Pi}_{{\widetilde V}_k}({\bf x}) = -i\frac{\delta}{\delta {\widetilde V}_k({\bf x})}, \qquad {\textrm{for }} k=1,2,
\label{h30.037}
\end{equation}
and similarly for the primed variables. 

In terms of respectively the unprimed and the primed variables, the matrices $h_{kl}$ and ${\bar h}_{kl}$ in the Hamiltonian ({\ref{h30.026051}}) will respectively depend on the unprimed and primed transformations $K$ and $K^{-1}$. The different Hamiltonians can then be transformed into each other. This can be done by using the transformation ({\ref{apph30.036}}) but then applied to the fields which are used in the representation ({\ref{apph30.037}}).{\footnote{Note that we have used the same notation for the classical fields (i.e.\ the unquantized fields) and the fields in the representation ({\ref{apph30.037}}). But despite the same notation, these fields have a different meaning.}} 

The transformation between the primed and unprimed fields which are used in the above representation read
\begin{eqnarray}
{\widetilde V}'_k({\bf x}) &=& \sum^2_{l=1} \int d^3 y  T_{kl} ({\bf x},{\bf y}) {\widetilde V}_l({\bf y}), \nonumber\\
{\widetilde V}_k({\bf x}) &=& \sum^2_{l=1} \int d^3 y  T^{-1}_{kl} ({\bf x},{\bf y}) {\widetilde V}'_l({\bf y}),\nonumber\\
\frac{\delta }{\delta {\widetilde V}_k({\bf x})} &=& \sum^2_{j=1} \int d^3 y  T_{kl}({\bf y},{\bf x}) \frac{\delta }{\delta {\widetilde V}'_l({\bf y})},\nonumber\\
\frac{\delta }{\delta {\widetilde V'}_k({\bf x})} &=& \sum^2_{j=1} \int d^3 y  T^{-1}_{kl}({\bf y},{\bf x}) \frac{\delta }{\delta {\widetilde V}_l({\bf y})}
\label{apph30.039}
\end{eqnarray}
with $k=1,2$ and
\begin{eqnarray}
T_{kl} ({\bf x},{\bf y}) &=& \sum^3_{i=1} \int d^3 z  K'^{-1}_{ki} ({\bf x},{\bf z}) K_{il} ({\bf z},{\bf y}), \nonumber\\
T^{-1}_{kl} ({\bf x},{\bf y}) &=& \sum^3_{i=1} \int d^3 z  K^{-1}_{ki} ({\bf x},{\bf z}) K'_{il} ({\bf z},{\bf y}).
\label{apph30.037}
\end{eqnarray}
We have hereby used the properties ({\ref{h30.026021}})-({\ref{h30.026024}}) of $K$ and $K'$ to show identities like
\begin{eqnarray}
\sum^3_{i=1} \int d^3 z  K'^{-1}_{ki} ({\bf x},{\bf z}) K_{i3} ({\bf z},{\bf y}) &=& \sum^3_{i=1} \int d^3 z  K'^{-1}_{ki} ({\bf x},{\bf z}) \partial_{z_i} U ({\bf z},{\bf y}) \nonumber\\
 &=&- \sum^3_{i=1} \int d^3 z  \left(\partial_{z_i}  K'^{-1}_{ki} ({\bf x},{\bf z}) \right) U ({\bf z},{\bf y}) \nonumber\\
&=& 0 , \quad {\textrm{for }} k=1,2.
\label{apph30.037}
\end{eqnarray}
We have hereby assumed that the boundary terms vanishes after the partial integration. Identities like these are the reason why the transformations are independent of the constraint variables ${\widetilde V}_3,\Pi_{{\widetilde V}_3}$ and ${\widetilde V}'_3,\Pi_{{\widetilde V}'_3} $, as should be. 

If the normalized wavefunction $\Psi({\widetilde V}_i)$ is a solution to the functional Schr\"odinger equation ({\ref{h30.026053}}) where the functions $h_{kl}$ and ${\bar h}_{kl}$ depend on the unprimed transformations $K$ and $K^{-1}$, then the wavefunctional 
\begin{equation}
\Psi'({\widetilde V}'_j) = N \Psi \left( \sum^2_{j=1} \int d^3 y T^{-1}_{ij} ({\bf x},{\bf y}) {\widetilde V}'_j({\bf y})\right)= N \Psi({\widetilde V}_i)
\label{h30.041}
\end{equation}
satisfies the functional Schr\"odinger equation ({\ref{h30.026053}}) where the Hamiltonian depends on the primed transformations. The normalization constant is determined by the Jacobian of the transformation, i.e.\ $|N|^2=\det T ^{-2}$. 

Hence the quantum theories in terms of primed and unprimed fields are equivalent. The pilot-wave interpretations that may be constructed in terms of the two sets of unconstrained variables are also equivalent; the field beables ${\widetilde V}_l({\bf x},t)$ and ${\widetilde V}'_l({\bf x},t)$ can be transformed to each other by application of the same transformation ({\ref{apph30.039}}) as the classical fields.

\newpage

\noindent \\

\chapter{The Grassmann algebra: definitions and properties} \label{appb}
\section{The Grassmann algebra}
An algebra over the complex numbers whose generators $\eta(k)$, labeled by $k$, satisfy 
\begin{equation}
\left[\eta(k),\eta(l)\right]_{+}= 0 ,
\label{appb.001}
\end{equation}
is called a {\em Grassmann algebra} \cite{berezin66}. The generators are called {\em Grassmann numbers}. An element of the Grassmann algebra which respectively commutes or anti-commutes with all the other elements of the Grassmann algebra is respectively called an {\em even} or {\em odd} element of the Grassmann algebra. 

The left derivatives $\frac{ \overrightarrow{\delta}   }{\delta \eta(k)}$ and right derivatives $\frac{ \overleftarrow{\delta}   }{\delta \eta(k)}$ are defined by
\begin{eqnarray}
&& \left[\eta(k),\frac{ \overrightarrow{\delta} }{\delta \eta(l)}\right]_{+}= \left[\eta(k),\frac{ \overleftarrow{\delta} }{\delta \eta(l)}\right]_{+} =\delta(k-l), \nonumber\\
&& \left[\frac{ \overrightarrow{\delta} }{\delta \eta(k)},\frac{ \overleftarrow{\delta} }{\delta \eta(l)}\right]_{+}= \left[\frac{ \overrightarrow{\delta} }{\delta \eta(k)},\frac{ \overrightarrow{\delta} }{\delta \eta(l)}\right]_{+}=\left[\frac{ \overleftarrow{\delta} }{\delta \eta(k)},\frac{ \overleftarrow{\delta} }{\delta \eta(l)}\right]_{+}=0. 
\label{appb.002}
\end{eqnarray}
With this definition we have 
\begin{eqnarray}
\frac{ \overrightarrow{\delta} }{\delta \eta(k)} \left( AB \right) &=&  \frac{ \overrightarrow{\delta}A  }{\delta \eta(k)}   B + (-1)^{P_A} A\frac{ \overrightarrow{\delta}B }{\delta \eta(k)}, \nonumber\\
 \left( AB \right) \frac{ \overleftarrow{\delta} }{\delta \eta(k)} &=&  (-1)^{P_B}  \frac{ A \overleftarrow{\delta} }{\delta \eta(k)}  B +  A \frac{ B \overleftarrow{\delta} }{\delta \eta(k)},
\label{appb.003}
\end{eqnarray}
where $P_A$ equals $0$ or $1$, depending on whether the element $A$ is respectively even or odd.

With the symbols ${d}  \eta(k) $ which satisfy 
\begin{equation}
\left[\eta(k),{d}  \eta(l) \right]_{+}=\left[{d} \eta(k),{d}  \eta(l) \right]_{+} = 0
\label{appb.0031}
\end{equation}
we may define the integrals 
\begin{equation}
\int {d}  \eta(k) = 0 ,\quad \int \eta(k) {d}  \eta(k) = 1.
\label{appb.0032}
\end{equation}
The integral over an arbitrary element of the Grassmann algebra is defined by linearly extending the definitions ({\ref{appb.0032}}).

For each element of the Grassmann algebra $f$ we may also introduce a conjugate $f^{\dagger}$ which has the properties
\begin{equation}
(f_1 f_2)^{\dagger}= f_2^{\dagger}f_1^{\dagger},\quad \left( f^{\dagger}\right)^{\dagger}  = f,\quad  \left[\eta(k),\eta^{\dagger}(l)\right]_{+}=\left[\eta^{\dagger}(k),\eta^{\dagger}(l)\right]_{+}=0.
\label{appb.004}
\end{equation}
and which is such that when the conjugate is restricted to the subspace of complex numbers, it is simply the complex conjugate. As a corollary we have
\begin{equation}
\left( \frac{ \overrightarrow{\delta} }{\delta \eta(k)} f \right)^{\dagger} = f^{\dagger}\frac{ \overleftarrow{\delta} }{\delta \eta^{\dagger}(k)}, \quad  \left( f \frac{ \overleftarrow{\delta} }{\delta \eta(k)}  \right)^{\dagger} = \frac{ \overrightarrow{\delta} }{\delta \eta^{\dagger}(k)}f^{\dagger}
\label{appb.00401}
\end{equation}
and similarly for derivatives with respect to the fields $\eta^{\dagger}(k)$.

\section{The inverse of an element of the Grassmann algebra}
Suppose an infinite dimensional complex Grassmann algebra generated by the Grassmann numbers $\{\eta(k),\eta^{\dagger}(k)\}$ with $\eta^{\dagger}(k)$ the conjugate of $\eta(k)$, which satisfy the anti-commutation relations 
\begin{equation}
\left[\eta(k),\eta(l)\right]_{+}=\left[\eta(k),\eta^{\dagger}(l)\right]_{+}=\left[\eta^{\dagger}(k),\eta^{\dagger}(l)\right]_{+}=0.
\label{appb.1}
\end{equation}
The inverse element for the multiplication can not be defined for any element of the Grassmann algebra, e.g.\ if the Grassmann number $\eta(k)$ would have an inverse then this would be in contradiction with the property $\eta(k)^2 = 0$. However, there is a class of elements for which we can define an inverse. Consider therefore the following decomposition of an arbitrary element $f$ of the Grassmann algebra as $f = c + g$, where $c$ is a complex number with 
\begin{equation}
\int \mathcal{D}\eta^{\dagger} \mathcal{D} \eta \prod_k \eta(k) \eta^{\dagger}(k) f =c.
\label{appb.2}
\end{equation}
This last equation implies that $g$ can be written as $g=\sum_i g_i$ with $g^2_i=0$. For every element $f$ of the Grassmann algebra for which $c \neq 0$, one can define the unique inverse element $f^{-1}$ as
\begin{equation}
f^{-1} = \frac{1}{c} \sum^{+\infty}_{l = 0} \Big(-\frac{g}{c}\Big)^l, 
\label{appb.3} 
\end{equation}
for which $f^{-1}f=ff^{-1} =1$. The inverse of a product satisfies
\begin{equation}
(fg)^{-1} = g^{-1} f^{-1}. 
\label{appb.31} 
\end{equation}
For an element $f$ for which $c = 0$, no such elements $f^{-1}$ exist.

\newpage

\noindent \\

\chapter{Note on the energy restriction for a decaying system}\label{appc}
Conservation of energy requires that the energy of the total system equals the energy of the decaying system $E_0$. If this decaying system is initially at rest, $E_0$ will be the rest energy of the system. Suppose now that we take a $\delta$-distribution for this energy restriction i.e.\ 
\begin{equation}
F({\bf p}_1 , {\bf p}_2) =f({\bf p}_1,{\bf p}_2) \delta(E - E_0)
\label{h4.50}
\end{equation}
where $f$ determines the momentum correlation (this is for example the distribution in (\ref{h4.10}) or (\ref{h4.24})). If we take a distribution $f$ which satisfies $f({\bf p}_1 , {\bf p}_2) = f^*(-{\bf p}_1 , -{\bf p}_2)$, then the probability currents of the two particles are both zero for all times, i.e.\
\begin{equation}
{\bf j}_i= \frac{\hbar}{m_i} \textrm{Im} \left(\psi^* {\bf  \nabla}_{{\bf x}_i} \psi \right) = {\bf 0}, \qquad i=1,2.
\label{h4.50.1}
\end{equation}
This implies that the particles show no evolution. In pilot-wave theory this corresponds with particle beables that stand still, because the speeds are defined as $d {\bf x}_i / d t = {\bf j}_i/|\psi|^2 $.  As a result, the considered distribution in (\ref{h4.50}) does not actually represent a decaying system. We can resolve this problem by allowing a finite energy width centred around $E_0$. Nevertheless, it will follow that, if the wavefunction displays strong momentum correlation in the sense that ${\bf p}_1 + {\bf p}_2 = {\bf 0}$, then the restriction to a small energy width only involves a minor broadening of the wavefunction, which implies that we can leave the restriction on the total energy aside for our qualitative analysis. 

For the momentum distribution $f$ we will take the distribution in (\ref{h4.24}), i.e.\ $f({\bf p}_1 , {\bf p}_2) \sim \de ({\bf p}_1 + {\bf p}_2)  e^{-\alpha p^2_1 / \hbar}$. The restriction on the total energy is accomplished by integrating over values for $({\bf p}_1 , {\bf p}_2)$ for which $E_- \le E \le E_+$, for a certain minimum energy value $E_-$ and a certain maximum energy value $E_+$. We therefore define the following function
\begin{displaymath}
 \textrm{disc}(E_{\pm})({\bf p}_1 , {\bf p}_2) = \left\{  \begin{array}{ll} 1 & \textrm{if} \quad  \frac{p^2_1}{2m_1} + \frac{p^2_2}{2m_2} \le E_{\pm} \\ 
0   & \textrm{otherwise} \end{array} \right. .
\label{h4.51}
\end{displaymath}
The momentum distribution then becomes
\begin{equation}
F({\bf p}_1 , {\bf p}_2) =f({\bf p}_1,{\bf p}_2)\big[ \textrm{disc}(E_{+}) - \textrm{disc}(E_{-}) \big].
\label{h4.52}
\end{equation}
The resulting wavefunction of the system is then
\begin{eqnarray}
\psi({\bf x}_1 , {\bf x}_2,t) &=& \int f({\bf p}_1,{\bf p}_2)\big[ \textrm{disc}(E_{+}) - \textrm{disc}(E_{-}) \big]  e^{i({\bf p}_1 \cdot {\bf x}_1 +  {\bf p}_2  \cdot {\bf x}_2 - Et)/\hbar}  d {\bf p}_1  d {\bf p}_2   \nonumber\\
   &\sim& \int e^{i{\bf p} \cdot ({\bf x}_1 - {\bf x}_2)/\hbar - (it/2\mu + \al)p^2/\hbar}  \big[ \textrm{disc}'(E_{+}) - \textrm{disc}'(E_{-}) \big] d {\bf p}
\label{h4.53}
\end{eqnarray}
where
\begin{displaymath}
 \textrm{disc}'(E_{\pm})({\bf p}) = \left\{   \begin{array}{ll} 1 & \textrm{if} \quad p^2/2\mu \le E_{\pm} \\ 
0   & \textrm{otherwise} \end{array} \right. .
\label{h4.54}
\end{displaymath}
If we write (\ref{h4.53}) as a Fourier transform then we can apply the convolution theorem
\begin{eqnarray}
&&\psi({\bf x}_1 , {\bf x}_2,t) \nonumber\\
&&\sim \mathcal{F}^+_{ \{( {\bf x}_1 -  {\bf x}_2)/\hbar \}} \big( e^{- (it/2\mu + \al)p^2/\hbar}\big) \otimes \mathcal{F}^+_{\{ ( {\bf x}_1 -  {\bf x}_2)/\hbar \}} \big(\textrm{disc}'(E_{+}) - \textrm{disc}'(E_{-})\big) \nonumber\\
&&\sim h(x,t)  \otimes {\bar g}(x),
\label{h4.55}
\end{eqnarray}
where 
\begin{eqnarray}
h(x,t) &=& \mathcal{F}^+_{ \{( {\bf x}_1 -  {\bf x}_2)/\hbar \}} \big( e^{- (it/2\mu + \al)p^2/\hbar}\big), \nonumber\\
{\bar g}(x) &=& \mathcal{F}^+_{ \{( {\bf x}_1 -  {\bf x}_2)/\hbar \}}\big(\textrm{disc}'(E_{+}) - \textrm{disc}'(E_{-})\big), \nonumber\\
x &=&  |{\bf x}_1 -  {\bf x}_2|.
\label{h4.55.10}
\end{eqnarray}
The first function in the convolution in (\ref{h4.55}) is just the wavefunction in (\ref{h4.25}),
\begin{equation}
h(x,t) \sim \bigg(\frac{\pi \hbar}{\al + it/2\mu} \bigg)^{3/2} e^{-({\bf x}_1 - {\bf x}_2)^2 / 4 \hbar (\al + \frac{it}{2\mu} )} .
\label{h4.55.1}
\end{equation}
If we define $a_{\pm} =2 \pi \sqrt{E_{\pm} 2\mu}/ \hbar $, then the second function in the convolution in (\ref{h4.55}) becomes
\begin{equation}
{\bar g}(x)  \sim g(x) =\big[a_+ J_1 ( a_+ x  )- a_- J_1(a_- x )\big] /x  
\label{h4.57}
\end{equation}
where $J_1$ is the first order spherical Bessel function.

We present now two ways to show that the restriction on the energy can be relinquished, without changing the qualitative analysis. The first way proceeds as follows. In order to evaluate $g(x)$, we will substitute some reasonable values for $E_+$ and $E_-$ in (\ref{h4.57}). In addition we will assume that the fragments have equal masses so that we can put $2 \mu = m$, with $m$ the mass of one fragment. For $E_+$ we will take one percent of the rest mass of the total system in order to avoid the relativistic regime, $E_+ = 0.02mc^2$. We will take an energy gap of $0.001E_+$, so that $E_- = 0.999E_+$. In this way $a_+ \approx 5.58309/\lambda_c$ and $a_- \approx 5.58023/\lambda_c$, where $\lambda_c$ is the Compton wavelength of the fragments. In Fig.\ \ref{figure2} the function $g(x) = \big[a_+ J_1 ( a_+ x  )- a_- J_1(a_- x )\big] /x$ is plotted for $x$ in units of the Compton wavelength $\lambda_c$.
\begin{figure}[t]
\begin{center}
\epsfig{file=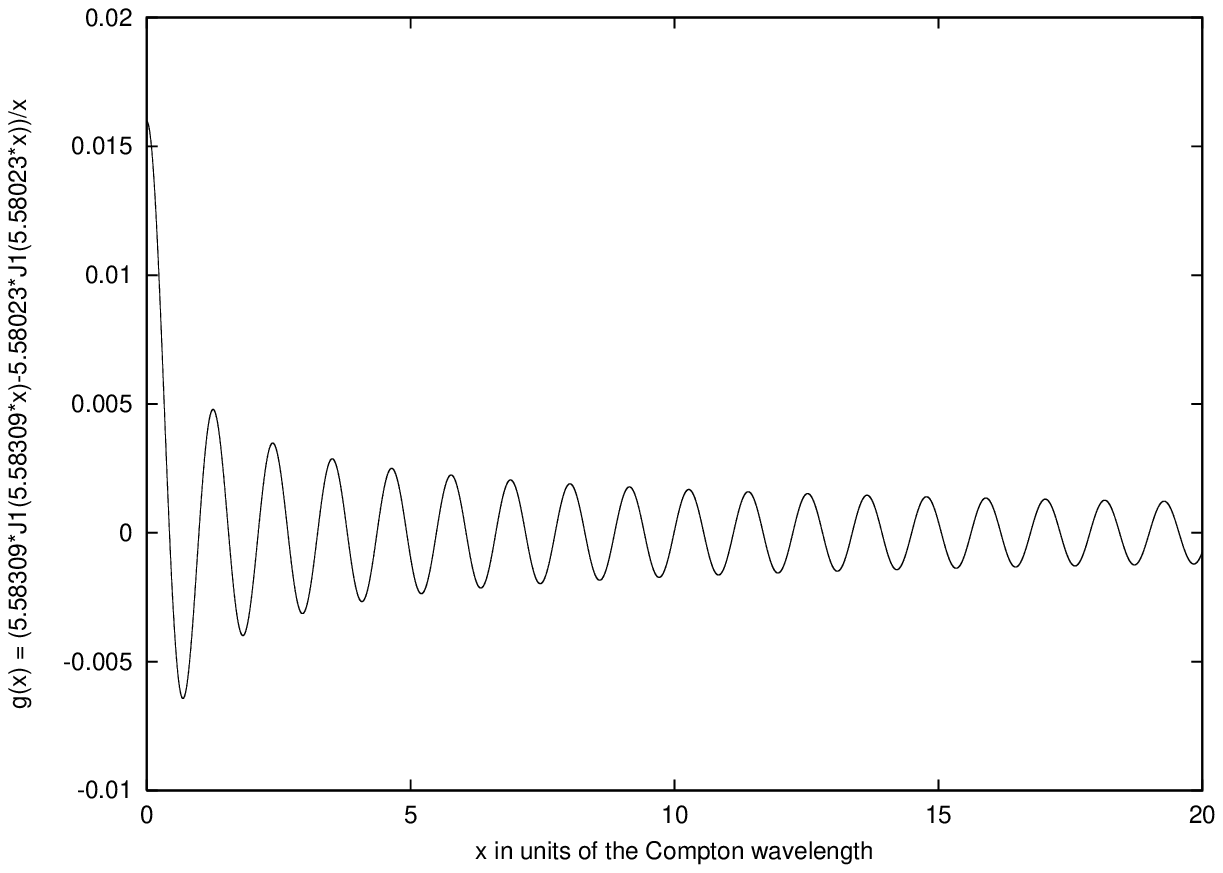}
\end{center}
\caption{The function $g(x) = \big[5.58309  J_1 ( 5.58309 x  )- 5.58023 J_1(5.58023  x )\big] /x$ is plotted for $x$ in units of the Compton wavelength $\lambda_c$.}
\label{figure2}
\end{figure}   

The figure shows that $g(x)$ is a rapidly oscillating function with a peak at $x = 0$. We now give the distribution of $h(x,t)$ at $t=0$ a width of the order of twenty times the Compton wavelength, which can be done by adjusting $\al$. Recall that the function $h(x,t)$ was in fact the wavefunction of the system if we did not restrict the energy. So, the width of $h(x,t)$ is in fact the measure of the initial nearness of the fragments, which is then of the order of twenty times the Compton wavelength. Then due to the rapid oscillation, the main contribution in the convolution will arise only from the peak in $g(x)$ at $x = 0$. This peak will result in only a small broadening of $h(x,0)$ so that $h(x,0) \otimes g(x)  \approx h(x,0)$. 

Because the width of $h(x,t)$ only increases with time, this approximation will become more precise with time. In conclusion, we can put the unnormalized wavefunction equal to
\begin{equation}
\psi({\bf x}_1 , {\bf x}_2,t) \approx \bigg(\frac{\pi \hbar}{\al + it/2\mu} \bigg)^{3/2} e^{-({\bf x}_1 - {\bf x}_2)^2 / 4 \hbar (\al + \frac{it}{2\mu} )}.
\label{h4.58}
\end{equation}
The larger the energy gap or the larger the rest energy, the more rapid the oscillation of $g(x)$ will be and the more peaked $g(x)$ will be at $x=0$, and as a result the more the approximation is valid.
 
A second way to achieve this result is to assume that $h(x,t)$ is very narrowly peaked at $x=0$ for $t=0$, so that $h(x,0) \approx \delta(x)$. This is the case if $\al$ approaches zero. As a result $h(x,0) \otimes g(x) \approx g(x)$. Thus in this case the non-normalized initial probability distribution is $g(x)^2$. This distribution is plotted in Fig.\ \ref{figure3} with again $x$ in units of the Compton wavelength. This figure shows that most of the probability is concentrated within a few times the Compton wavelength.
\begin{figure}[t]
\begin{center}
\epsfig{file=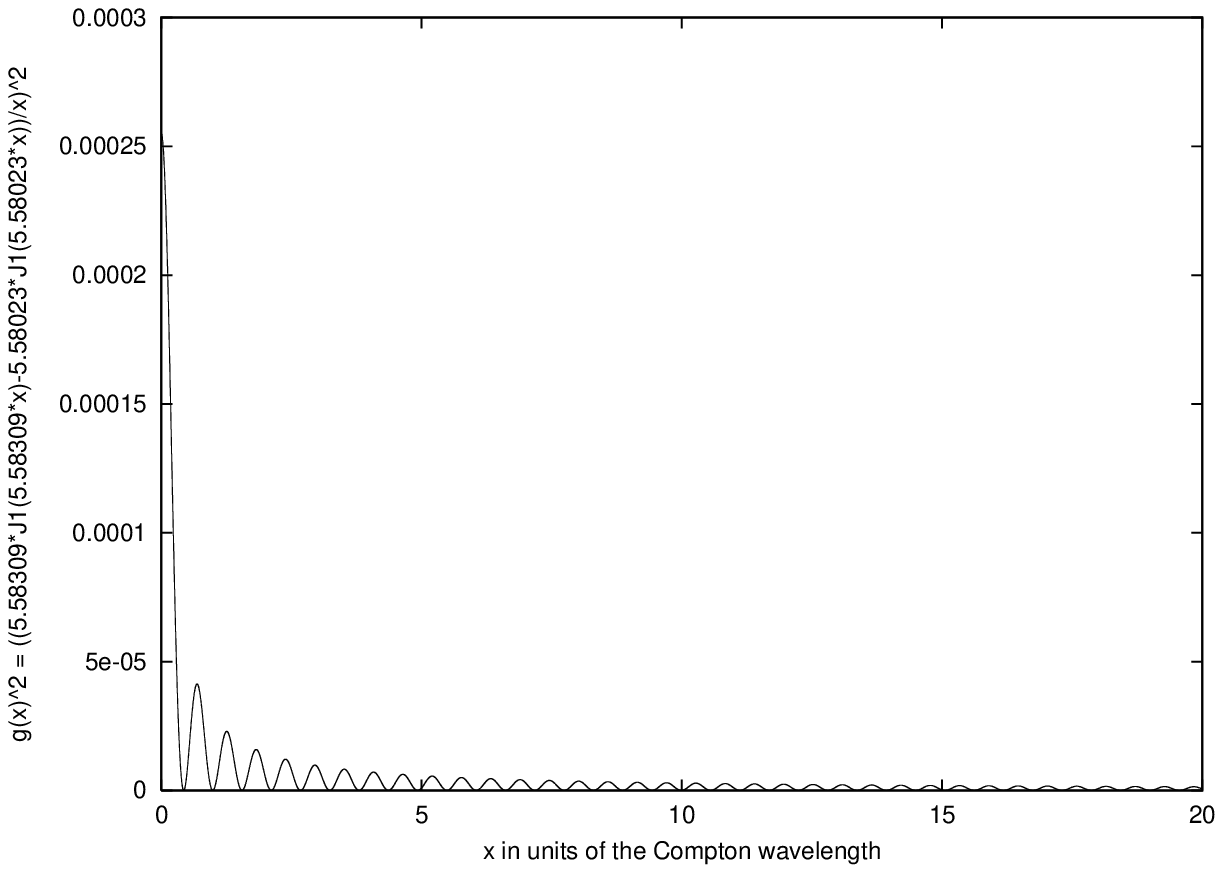}
\end{center}
\caption{The function $g(x)^2 = \frac{\big[5.58309 J_1 ( 5.58309 x)- 5.58023 J_1(5.58023 x )\big]^2}{x^2}$ is plotted for $x$ in units of the Compton wavelength $\lambda_c$.}
\label{figure3}
\end{figure}   
This implies that the particle beables depart from a very narrow region, only a few times the Compton wavelength in diameter, from each other. Because the wavefunction $\psi({\bf x}_1 , {\bf x}_2,t)$ in (\ref{h4.55}) only depends on the difference $|{\bf x}_1 - {\bf x}_2|$, the velocities (as defined in (\ref{h4.27})) will be opposite and the particle beables will travel along straight lines in opposite directions.

In conclusion, we have shown that the energy restriction does not put a restriction on the pilot-wave description of the system. The only effect of the energy restriction is a minor broadening of the probability density.

\newpage
\noindent \\


\end{document}